\pdfoutput=1
\documentclass[fullpage]{uiucthesis}

\usepackage{mathptmx}
\usepackage{fullpage}
\usepackage{spacetime}
\usepackage{amsmath, amssymb, amsfonts, amstext, amsthm}
\usepackage[pdftex]{graphicx}
\usepackage{natbib}
\usepackage{url}
\usepackage[breaklinks,hypertexnames=false]{hyperref}
\usepackage{currvita}

\newif\iffig
\figtrue

\iffig
\def\fig#1{fig/#1}
\def\mov#1{mov/#1}
\else
\def\fig#1{fig/dummy-fig}
\def\mov#1{fig/dummy-mov}
\fi

\def\parasc*#1{\paragraph*{\textsc{#1}}}

\def\thesisspacing{\onehalfspacing}

\newif\ifproofs
\proofstrue

\newif\ifonlyabstract
\onlyabstractfalse

\begin{document}


\title{Spacetime Meshing for Discontinuous Galerkin Methods}
\author{Shripad Vidyadhar Thite}
\department{Computer Science}
\schools{%
B.E., University of Poona, 1997\\
M.S., University of Illinois at Urbana-Champaign, 2001%
}

\phdthesis
\degreeyear{2005}
\advisor{Prof.\@ Jeff Erickson}
\maketitle


\frontmatter

\begin{abstract}
Important applications in science and engineering, such as modeling
traffic flow, seismic waves, electromagnetics, and the simulation of
mechanical stresses in materials, require the high-fidelity numerical
solution of hyperbolic partial differential equations (PDEs) in space
and time variables.  Many interesting physical problems involve
nonlinear and anisotropic behavior, and the PDEs modeling them exhibit
discontinuities in their solutions.  Spacetime discontinuous Galerkin
(SDG) finite element methods are used to solve such PDEs arising from
wave propagation phenomena.

To support an accurate and efficient solution procedure using SDG
methods and to exploit the flexibility of these methods, we give a
meshing algorithm to construct an unstructured simplicial spacetime
mesh over an arbitrary simplicial space domain.  Our algorithm is the
first spacetime meshing algorithm suitable for efficient solution of
nonlinear phenomena in anisotropic media using novel discontinuous
Galerkin finite element methods for implicit solutions directly in
spacetime.  Given a triangulated $d$-dimensional Euclidean space
domain $\sp$ (a simplicial complex) and initial conditions of the
underlying hyperbolic spacetime PDE, we construct an unstructured
simplicial mesh of the $(d+1)$-dimensional spacetime domain $\sp
\times [0,\infty)$.  Our algorithm uses a near-optimal number of
spacetime elements, each with bounded temporal aspect ratio for any
finite prefix $\sp \times [0,T]$ of spacetime.  Unlike Delaunay
meshes, the facets of our mesh satisfy gradient constraints that allow
interleaving the construction of the mesh by adding new spacetime
elements, and computing the solution within the new elements
immediately and in parallel.  Our algorithm is an advancing front
procedure that constructs the spacetime mesh incrementally, in a
fashion similar to the Tent Pitcher algorithm of \Ungor{} and
Sheffer~\cite{ungor00tentpitcher}, later extended by Erickson
\etal{}~\cite{erickson02building}.  We extend previous work which was limited by conservative assumptions; in doing so, we solve the algorithmic
problem of predicting discontinuous changes in the geometric
constraints that must be satisfied by the mesh.

In 2D$\times$Time, our algorithm simultaneously adapts the size and
shape of spacetime tetrahedra to a spacetime error indicator.  We
build on and extend the mesh adaptivity
technique~\cite{abedi04spacetime} of Abedi and others including this
author.  We are able to incorporate more general front modification
operations, such as edge flips and limited mesh smoothing.  Our new
algorithm is the first to simultaneously adapt to nonlinear and
anisotropic response of the underlying medium, thus unifying mesh
adaptivity with earlier work by the author.  The key insight of our
unified algorithm is to exploit the geometric structure of the
so-called \emph{cone constraints} that impose gradient constraints on
certain facets of the mesh.  We present our algorithm as a
two-parameter optimization problem where the objective is to maximize
the \emph{height} of spacetime tetrahedra.  The parameters that guide
the algorithm can be chosen adaptively by the algorithm at each step.
Our unified algorithm can choose to refine the mesh to increase the
\emph{temporal aspect ratio} of spacetime elements, independently of
the numerical error indicator.  Our new adaptive algorithm can better
optimize the shape of spacetime elements and is therefore promising to
produce better quality meshes.

The front is coarsened by merging adjacent facets, whether or not they
were created by a previous refinement.  We show how to make tent
pitching less greedy about maximizing the progress at each step so
that a subset of the front can be made coplanar for coarsening in the
next step.  This algorithm can be used in arbitrary dimensions to make
the entire front conform to a target time $T \ge 0$.  The main
contribution of this extension is to retain the progress guarantee of
earlier algorithms up to a constant factor.

In higher spatial dimensions $d \ge 3$, we are able to guarantee
limited mesh smoothing to improve in practice the quality of the mesh.
Though we are unable to extend our adaptive algorithm to spatial
dimensions $d \ge 3$, we state the major open problem that remains.

Empirical evidence confirms that improving the quality of the spatial
projection of each front leads to better quality spacetime elements.
We give an algorithm in arbitrary dimensions to smooth the front as by
pitching inclined tentpoles, and by performing edge flips and edge
contractions in 2D$\times$Time, so that the quality of future
spacetime elements can be improved in practice.  We strengthen the
progress constraints to anticipate future changes in the front
geometry. This algorithm represents recent progress towards a meshing
algorithm in 2D$\times$Time to track moving domain boundaries and
other singular surfaces such as shock fronts.

Additionally, we are able to prove theoretical bounds on the
worst-case \emph{temporal aspect ratio} of spacetime elements, while
at the same time providing parameters to the algorithm that improve
either the number or the aspect ratio of elements in practice.  The
number of elements in our mesh is provably close to that in a size
optimal mesh that also satisfies certain constraints required of any
mesh suitable for solution by SDG methods.

\end{abstract}

\begin{dedication}
To all my friends and co-travellers,
who have made this journey almost more fun than the destination.%
\end{dedication}

\chapter*{Acknowledgments}
Thanks to the other members of the CPSD spacetime group with whom I
have collaborated directly---Reza Abedi, Jonathan Booth, Shuo-Heng
Chung, Jeff Erickson, Yong Fan, Michael Garland, Damrong Guoy, Robert
Haber, Morgan Hawker, Mark Hills, Sanjay Kale, Jayandran Palaniappan,
John Sullivan, and Yuan Zhou---especially Jeff Erickson and Robert
Haber.  I thank Prof.\@ Haber for driving the whole group to solve
more and more interesting problems, and for giving me the freedom to
pursue my own interests within the larger project.  Alper \Ungor{} has
been a great source of encouragement and has renewed my enthusiasm
every time we met. Thanks, Alper!

I have been enormously fortunate to have Professors Michael Loui and
Jeff Erickson as my M.S.\@ and Ph.D.\@ advisors respectively.  I thank
them for their time and effort, and for holding me up to such high
standards of research and professional service.  I cannot thank
Prof.\@ Loui enough for many substantial discussions on ethics in
research and teaching.  I hope to carry those ideals with me
throughout my professional career.  I thank Jeff for his continuous
financial support and the several conference trips (especially where I
did \emph{not} have a paper!) for which he paid.  One day, I hope to
write a paper written as clearly as, with figures drawn as well as,
his ``worst'' paper.

Sariel Har-Peled and Edgar Ramos, together with Jeff who started it
all, made computational geometry a happening thing at UIUC.  Thank you
all very much for your wonderful talks in the theory and computational
geometry seminars and in classes, and for attracting so many excellent
visitors to the middle of the cornfields.

Thanks to Mitch Harris and Ari Trachtenberg for being such great
friends, colleagues, and role models.  I couldn't have asked for
better people to look up to.  David Bunde and Erin Wolf Chambers were
a great help---you guys were always there to give me feedback when I
asked.  Thank you for the time and effort you spared for me.  Thanks
to Afra Zomorodian for being around (if only briefly) and for being so
cool.  Lenny Pitt and Cinda Heeren let me participate in MathManiacs
occasionally---I had never done anything like it before and it was an
awesome experience.  Professors West and Kostochka were the reason I
liked the Math department so much.  The combinatorics group is the
friendliest group of researchers I have known (and that includes you,
Radhika and Andr\'e!).  Thanks!

Thanks also to the anonymous referees for the 20th ACM Symposium on
Computational Geometry (SoCG) and the 13th International Meshing
Roundtable (IMR~2004), where papers related to this thesis appeared,
for their valuable feedback.  I also gave a talk about the results in
this thesis at the 21st European Workshop on Computational Geometry
(EuroCG) at TU-Eindhoven, the Netherlands, in March 2005; I would like
to thank the students and faculty at Eindhoven for their hospitality.
I completed the final revision of this thesis at TU-Eindhoven during
the first few weeks of my postdoc position.  I would like to
especially thank Mark de~Berg for giving me the time and freedom to do
so.

I gratefully acknowledge funding from the Center for Process
Simulation and Design (CPSD) at the University of Illinois at
Urbana-Champaign.  CPSD is funded primarily by the National Science
Foundation under the ITR initiative, with joint sponsorship by the
Division of Materials Research, and the Directorate for Computer and
Information Science and Engineering.  My research was supported in
part by NSF ITR grant DMR 01-21695.

\tableofcontents                     
\listoffigures


\mainmatter

\chapter{Introduction}
\label{sec:intro}

In this thesis, we study spacetime meshing to support fast and accurate
simulation of certain physical phenomena by Spacetime Discontinuous
Galerkin (SDG) finite element methods.  We give algorithms to
construct spacetime meshes.  We prove guarantees on the performance of
the algorithm.  We prove worst-case bounds on the size and quality of
the individual elements in our meshes.  We want our meshes to be
efficient and lead to accurate solutions in practice.  Our meshes
support an efficient solution strategy by SDG methods because they
satisfy certain geometric constraints.  Our meshing algorithms
exploit the flexibility of novel SDG methods.

Mesh generation is an important problem in computational geometry.
Delaunay meshes~\cite{deberg00computationalgeometry} are widely used
in engineering applications.  Solving engineering problems using a
mesh imposes geometric requirements on the shape and size of mesh
elements~\cite{shewchuk02whatis}.  The efficiency of the solution
technique depends on the number and distribution of elements in the
mesh.  Delaunay meshes are known to nicely satisfy many such
requirements in practice and have very good theoretical properties
such as size optimality.  Our spacetime meshes are not Delaunay meshes
but our spacetime meshing problem is motivated by similar geometric
requirements and by a similar desire for efficiency.  We confront and
solve many challenges unique to meshing a non-Euclidean spacetime
domain.  We do so by building on previous
research~\cite{ungor00tentpitcher,ungor02phd,erickson02building,erickson05building}
and insight due to several researchers.  The research in this thesis
is part of a larger project at the Center for Process Simulation and
Design (CPSD) at the University of Illinois at Urbana-Champaign.

The algorithms in this thesis are being implemented by members of
CPSD, including the author, and tested with actual DG solvers on
practical simulation problems to illustrate the usefulness and
applicability of the research.  The reader is referred to other
publications~\cite{abedi04spacetime,abedi04adaptiveDG,abedi05adaptive} for
empirical results.


\section{Background}
\label{sec:background}

In this section, we will review basic concepts about hyperbolic
partial differential equations (PDEs) and physical phenomena modeled
by hyperbolic PDEs, to motivate the application of this research to
solving simulation problems in science and engineering.  The reader is
referred to classical
textbooks~\cite{strang86appliedmath,heath02scientificcomputing,folland95introduction,whitham74linear}
for a review of the theory of PDEs, hyperbolic problems, and the
finite element method.  There are also several online resources.  Jim
Herod of Georgia Tech has notes and Maple worksheets
online~\cite{mapletools}.  See also MathWorld~\cite{mathworld99pde}
and links therein.  Introductory texts on PDEs include those by Tveito
and Winther~\cite{tveito98introduction},
Folland~\cite{folland95introduction},
Cooper~\cite{cooper98introduction}, and
Thomas~\cite{thomas99numerical}.

Simulation problems in mechanics consider the behavior of an object or
region of space over time.  Scientists and engineers use conservation
laws and hyperbolic partial differential equations (PDEs) to model
transient, wave-like phenomena propagating over time through the domain
of interest.  PDEs are mathematical models of physical phenomena,
often derived by applying a fundamental principle such as conservation
of mass, momentum, or energy.  Example applications are numerous,
including, for instance, the equations of elastodynamics in seismic
analysis and the Euler equations for compressible gas dynamics.
Computational mechanics is concerned with the accurate computer
simulation of physical phenomena.

An example of a hyperbolic PDE is the one-dimensional wave equation
\[
 \frac{\partial^2 u}{\partial t^2}
=
 \omega^2
 \frac{\partial^2 u}{\partial x^2}
\qquad
 \text{for~$0 \le x \le L$ and~$t \ge 0$}
\]
with boundary conditions
\[
 u(0,t) = u_0
\qquad
 u(L,t) = u_L
\]
and initial conditions
\[
 u(x,0) = f(x)
\qquad
 \rest{\frac{\partial u}{\partial t}}{t=0} = g(x)
\]
This PDE describes, for instance, the vibration of a taut string when
it is plucked.  Here, $u(x,t)$ is the displacement of point~$x$ at
time~$t$ and~$\omega$, a constant, is the \emph{wavespeed}, i.e., the
speed at which the wave travels along the string.

A PDE is \emph{homogeneous} if the trivial function $u \equiv 0$ is a solution.

A PDE is \emph{linear} if the PDE and the initial or boundary conditions
do not include any product of the dependent variable~$u$ or its partial
derivatives.  A PDE that is not linear is \emph{nonlinear}. For example,
\begin{itemize}
\item First-order linear PDE:~$u_t + c u_x = 0$
\item Second-order linear PDE:~$u_{xx} + u_{yy} = \phi(x,y)$ for an arbitrary
function~$\phi$ of the independent variables only.
\end{itemize}
A linear PDE has the property that if~$u_1$ and~$u_2$ both satisfy the
PDE, then so does every linear combination $\alpha u_1 + \beta u_2$ This
property is called the \emph{principle of superposition}.

A nonlinear equation is \emph{semi-linear} if the coefficients of the
highest-order derivatives are functions of the independent variables
$x_1$, $x_2$, $\ldots$, $x_n$ only.  Examples of semi-linear PDEs are
the following:
\begin{itemize}
\item $(x+3) u_x + xy^2 u_y = u^3$
\item $x u_{xx} + (xy + y^2) u_{yy} + u u_x + u^2 u_y = u^4$
\end{itemize}

A nonlinear PDE of order~$m$ that is not semi-linear is
\emph{quasi-linear} if the coefficients of the derivatives of order~$m$
depend only on the independent variables and on derivatives of order
less than~$m$, e.g.,
\[
 (1+u_y^2) u_{xx} - 2 u_x u_y u_{xy} + (1+u_x^2) u_{yy} = 0
\]

A nonlinear PDE of order~$m$ is \emph{fully nonlinear} if the
highest-order derivatives of~$u$ appear nonlinearly in the equation.

The following examples illustrate the distinction between the
different types of PDEs:
\begin{itemize}
\item $u_t + u_x = 0$ is homogeneous linear
\item $u_t + x u_x = 0$ is homogeneous linear
\item $u_{xx} + u_{yy} = 0$ is homogeneous linear
\item $u_t + x^2 u_x = 0$ is homogeneous linear
\item $u_t + u_{xxx} + u u_x = 0$ is homogeneous semi-linear
\item $u_t + u_x + u^2 = 0$ is homogeneous semi-linear
\item $u_{xx} + u_{yy} = x^2 + y^2$ is inhomogeneous linear
\item $u_t + u u_x = 0$ is not semi-linear, but it is quasi-linear
\item $u_x^2 + u_y^2 = 1$ is inhomogeneous fully nonlinear
\end{itemize}

Many interesting physical problems are modeled by second-order PDEs.
The general form of a linear second-order PDE is
\[
 a u_{xx} + b u_{xt} + c u_{tt} + d u_x + e u_t + f u + g(x,t) = 0
\]
This PDE is classified as (i)~parabolic if $b^2 - 4ac = 0$,
(ii)~hyperbolic if $b^2 - 4ac > 0$, and (iii)~elliptic if $b^2 - 4ac <
0$.

\subsection{Numerical methods}

It is very difficult and often impossible to obtain analytical
solutions of PDEs arising in practical problems.  Therefore,
approximate solutions are obtained by numerical techniques.  The
choice of numerical method depends on the physical behavior of the
system, i.e., on the type of PDE that models the system.

Several numerical techniques can be used for numerical solutions of
partial differential equations.  They are classified on the basis of
the discretization method by which the continuum mathematical model is
approximated.  These include the finite difference method, finite
volume method (FVM), meshfree methods, and finite element method
(FEM).

The finite element (FEM) method is a vastly popular numerical
technique for solving PDEs.  The FEM method requires a mesh of the
domain. There is a vast body of literature on mesh generation.  In
Section~\ref{sec:meshgeneration}, we will briefly mention the results
most relevant to our problem to place our meshing algorithms in
context.

Discrepancies can arise between the actual physics and the approximate
numerical solution of the PDE modeling the physics.  The sources of
such discrepancies include errors in modeling physical phenomena as
well as errors due to computation such as discretization error and
numerical roundoff error.

Some errors are caused because of the discretization of the domain.
Errors are also caused by truncation and rounding due to finite
precision of the computations.  The \emph{order of accuracy} of a
numerical method is the exponent of the largest order term in the
Taylor series expansion of the difference between the analytical
solution and the approximate solution.  Numerical instability occurs
when the error term grows so fast that it dominates the actual
solution.  Both accuracy and stability affect the computational time
required to converge (if at all) to the approximate solution.
Therefore, it is important to minimize error and maximize stability.
Since there is always a tradeoff between accuracy and efficiency of
any numerical method, the most desirable methods are those for which
the running time increases slowly as the acceptable error
decreases.

\subsubsection{Galerkin method}

The Galerkin method~\cite{mathworld99galerkin} provides a unifying
basis from which several numerical methods, like finite element and
finite difference methods, can be derived.  The Galerkin approximation
to the analytical solution of a PDE is an approximate numerical
solution that is the unique best approximation in a certain sense to
the exact solution.  We refer the reader to other resources, for
instance the online lecture notes by the Chalmers Finite Element
Center~\cite{chalmers03fem}, for a detailed discussion of the Galerkin
method.

\subsubsection{Discontinuous Galerkin method}

The discontinuous Galerkin (DG) method, originally developed by Reed
and Hill in 1973 for neutron transport problems and first analyzed by
Le Saint and Raviart in 1975, is used to solve ordinary
differential equations and hyperbolic, parabolic, and elliptic partial
differential equations.

The DG method is a technique that uses discontinuous basis functions
to formulate a Galerkin approximation.  Given a mesh of the analysis
domain, the DG method approximates the solution within each element by
a function from a low-dimensional vector space of functions, e.g., as
a linear combination of basis functions like polynomials.  For a pair
of adjacent mesh elements, the approximate solution computed in the
interior of the elements does not have to agree on their common
boundary.

The DG method has many desirable properties that have made it popular.
For example (i)~it can sharply capture solution discontinuities
relative to a computational mesh; (ii)~it simplifies adaptation since
inter-element continuity is neither required for mesh refinement and
coarsening, nor for $p$-adaptivity; (iii)~it conserves the appropriate
physical quantities (e.g., mass, momentum, and energy) on an
element-by-element basis; (iv)~it can handle problems in complex
geometries to high order; (v)~regardless of order, it has a simple
communication pattern to elements sharing a common face that
simplifies parallel computation.  If the solution field has a
discontinuity in the form of a \emph{shock}, SDG methods can exactly
capture this discontinuity if the mesh facets are perfectly aligned
with the shock surface.  On the other hand, with a discontinuous
basis, the DG method produces more unknowns for a given order of
accuracy than traditional finite element or finite volume methods,
which may lead to some inefficiency.  The DG method is harder when
applied to unstructured meshes; in particular, it is harder to
formulate limiting strategies to reduce spurious oscillations when
high-order methods are used.

For a thorough discussion of DG methods and a review of the state of
the art, we refer the reader to the book by Cockburn, Karniadakis, and
Shu~\cite{cockburn00discontinuous}.

\subsubsection{Spacetime discontinuous Galerkin methods}

The first discontinuous Galerkin formulation for spacetime problems,
introduced by Reed and Le Saint \etal{}, was discontinuous in time
only.  The spacetime domain was divided into slabs by constant-time
planes.  The solution was assumed to be continuous within a slab and
jump conditions at the inter-slab boundaries were used to enforce the
appropriate level of continuity between adjacent slabs.  The spacetime
basis functions of this so-called \emph{time-discontinuous Galerkin
method} were continuous in spacetime except across the constant-time
planes that separated two consecutive slabs.  The spacetime mesh was,
therefore, required to conform to the inter-slab boundaries.

Richter~\cite{Richter94} introduced the first spacetime-discontinuous
Galerkin finite element method, which permitted basis functions to be
discontinuous in spacetime across all element boundaries, for the wave
equation.  Yin \etal{}~\cite{YinASHT99,YinASHT00,yin02thesis} proposed
the first spacetime-discontinuous Galerkin method for linear
elastodynamics.  Abedi
\etal{}~\cite{abedi04adaptiveDG,abedi05spacetime,abedi05adaptive} give
an improved formulation.  Since the SDG method allows every pair of
adjacent elements to have basis functions that are discontinuous across
their common boundary, the SDG method can solve fully unstructured and
even nonconforming meshes.  The flexibility of SDG methods, in terms of
the types of mesh that are solvable, is exploited by the meshing
algorithms in this thesis.

\subsection{Mesh generation}
\label{sec:meshgeneration}

Given a domain of interest, the mesh generation problem is to
decompose the domain into simple cells, called \emph{elements}, each
of constant complexity.  Such a decomposition is called a
\emph{mesh}.  Elements consist of vertices, edges, ridges, and facets
of appropriate dimensions.  Linear, bilinear, or trilinear elements
are often used; convex cells (e.g., simplices) are sometimes
preferred.  A \emph{triangulation} is a decomposition into
simplices~\cite{ziegler95lectures}.  In general, a mesh may contain
hexahedra, prisms, pyramids, and other simple shapes, possibly in the
same (hybrid) mesh.

If the domain is bounded, the mesh must conform to the boundary,
i.e., the domain boundary must be contained in the union of a subset
of the mesh faces.  A curved domain boundary cannot be captured
exactly by a mesh consisting of linear elements; in this case, the
mesh must conform to the boundary approximately in a piecewise linear
fashion.

An important question is whether a domain can be meshed at all and, if
so, with how many elements.  It is well-known that every simple
polygon in the plane can be triangulated~\cite{garey78triangulating},
in fact in linear
time~\cite{chazelle91triangulating,amato00lineartime}.  In dimensions
three and higher, additional vertices, called \emph{Steiner points},
may be required; in general, it is NP-complete to determine if one or
more Steiner points are required and to compute the minimum number of
Steiner points~\cite{ruppert92difficulty}.

The accuracy and stability of numerical methods are affected by the
geometric shape of mesh elements.  Therefore, it is important to
generate a mesh with only good quality elements.  The notion of
element quality depends on the numerical
method~\cite{shewchuk02whatis}.

When the domain is Euclidean, several popular metrics are used to define
the quality of the mesh elements and of the mesh as a whole.  Popular
objectives are to optimize at least one of the following quality
measures for every element in the mesh: minimize ratio of
circumradius to length of shortest edge, maximize smallest angle,
minimize largest angle, etc.  Different metrics appear to work
better than others depending on the application.  See
Shewchuk~\cite{shewchuk02whatis} for a comprehensive study of quality
metrics.  In 2D, all these metrics are equivalent to each other up to
constant factors; this is not true in higher dimensions.  In 2D,
Delaunay triangulations~\cite{deberg00computationalgeometry} have
guaranteed quality~\cite{cheng00sliver,cheng03quality}.

For anisotropic problems, quality metrics should also be anisotropic.
Typically, the local anisotropic nature of the problem is used to
define a local quality measure at each point in the domain.  The
problem now is to generate a mesh such that for every point in the
domain, the quality of the element that contains the point is at least
as good as the desired quality metric at that point.  Labelle and
Shewchuk~\cite{labelle03anisotropic} propose a definition of an
anisotropic Delaunay triangulation and an algorithm to construct such
a triangulation of guaranteed quality.

Bern, Eppstein, Gilbert~\cite{bern94provablygood} use quadtrees to
construct triangulations with extra Steiner vertices added to the
original input.  They present the first algorithm to triangulate a
planar point set and polygons, with all angles bounded away from zero,
using a number of triangles within a constant of optimal.  In higher
dimensions, they are able to triangulate point sets such that the
resulting triangulation has no small solid angles and the number of
its simplices is within a constant factor of the optimal number.  The
result of Bern \etal{} was generalized to higher dimensions by
Mitchell and Vavasis~\cite{mitchell00quality}.

A $d$-dimensional \emph{structured mesh}, also known as a (structured)
\emph{grid}, is a cubical complex~\cite{ziegler95lectures} with the
property that every $k$-dimensional cell belongs to exactly $2^{d-k}$
$d$-dimensional cells; only cells on the boundary of the complex are
excepted from this rule.  Thus, all interior nodes are incident on the
same number of elements.  For example, a structured quadrilateral grid
in 2D has each interior node incident on four quads; a structured
hexahedral grid in 3D has each interior node incident on eight
hexahedra.  Because the combinatorial structure of such a grid is so
regular, simple data structures suffice to store and manipulate the
grid.  Unstructured meshes relax the uniformity requirement on the
node degree.

Mesh smoothing and mesh refinement are commonly used to improve the
quality of a given mesh, for instance, a Delaunay triangulation of a
given domain.

Mesh smoothing, also called $r$-refinement, adjusts the locations of
the vertices of the mesh while keeping the combinatorics of the mesh
unchanged.  Smoothing serves to improve the quality of the elements at
least locally~\cite{amenta99optimal} but global convergence to some
optimum quality is not guaranteed.  Mesh smoothing can be applied to
both structured and unstructured meshes.

Mesh refinement, also called $h$-refinement, is the operation of
introducing additional Steiner points to alter the mesh and improve
its quality.  Delaunay
refinement~\cite{chew89guaranteed,ruppert95delaunay}, is used when the
objective it to obtain a Delaunay
triangulation~\cite{deberg00computationalgeometry} of the domain where
each element has some guaranteed quality.  Delaunay refinement
proceeds by destroying bad quality elements by inserting their
circumcenters and updating the Delaunay triangulation.  The refinement
process eventually terminates but in 3D does not eliminate so-called
\emph{slivers}.  Weighted Delaunay refinement~\cite{cheng03quality}
attempts to remove slivers in the next step.  If the mesh is required
to conform to a piecewise linear boundary and if the boundary contains
an acute dihedral angle, it is not known whether weighted Delaunay
refinement can produce a good-quality triangulation without slivers.

Delaunay refinement is an incremental algorithm to improve the quality
of a simplicial mesh, for instance by inserting circumcenters of
bad-quality simplices, and updating the mesh to maintain the Delaunay
property.  See the seminal papers by Chew~\cite{chew89guaranteed} and
by Ruppert~\cite{ruppert95delaunay} on the quality and optimality of
the triangulations resulting from Delaunay refinement by circumcenter
insertion.  For Delaunay refinement using alternatives to circumcenter
insertion, see Rivara's longest edge
bisection~\cite{rivara97longestedge}, sink insertion by Edelsbrunner
and Guoy~\cite{edelsbrunner01sink}, and \Ungor{}'s Off-Center
insertion algorithm~\cite{ungor04offcenter}.

Another popular technique to mesh a domain is to begin with a surface
mesh of the boundary of the domain and extend it incrementally into
the interior of the domain to get a volume mesh.  Such an approach is
called an \emph{advancing front} method.  Advancing front techniques
are popular due to their simplicity, but suffer from a lack of
provable guarantees; for instance, an advancing front algorithm may
create degenerate or inverted elements~\cite{seveno97towards}, it may
create non-triangulable volumes and fail~\cite{bajaj99tetrahedral}, or
it may require resorting to ad-hoc techniques and heuristics where
different parts of the front collide.

Meshes of complicated domains frequently consist of millions or
billions of elements.  Representing and storing large meshes in memory
is complicated because memory in real-world computers is organized in
a hierarchy ranging from fast online storage to slow offline devices,
and the amount of fastest memory is limited.  Exploiting the spatial
and temporal locality exhibited by most algorithm to compactly store a
large mesh in a way that supports efficient access is an active and
exciting area of
research~\cite{cignoni03externalmemory,yoon05cacheoblivious}.
Minimizing memory usage is of concern to us too.  Our algorithms
minimize memory requirements by storing in main memory only a small
subset of the mesh, i.e., the elements adjacent to the advancing
front, instead of the entire spacetime mesh.

Multiprocessor computers and computer networks are available to solve
large-scale problems.  Parallel algorithms for mesh
generation~\cite{spielman04timecomplexity} and parallel numerical
methods take advantage of such computer systems by sophisticated
techniques.  Our meshing algorithms are highly parallelizable and the
discontinuous nature of the SDG method makes it possible to compute
the solution in unrelated subdomains simultaneously in parallel.

Automatic generation of good quality finite element meshes---with at
most minimal user interaction---remains a topic of active research
interest~\cite{owen99phd}.  In the spacetime domain, we have only
begun addressing the many challenges with many exciting avenues to
explore in the future.

\subsection{Causality}

The phenomenon that characterizes hyperbolic problems is causality,
the fairly intuitive notion of cause and effect, stimulus and
response, commonly observed in daily life.  When a pebble is dropped
into a pond, the disturbance radiates outwards in expanding circles
from the point of origin.  The result is waves traveling on the
surface of the water, traveling with a bounded wavespeed.  An example
in one-dimensional space is the vibration of a taut string fixed at
both ends when it is plucked.  Let~$L$ be the length of the string.
The displacement $u(x,t)$ of the point of the string at coordinate~$x$
about its mean position at time~$t$ is described by a second-order
hyperbolic PDE, the wave equation $u_{tt} - \omega^2 u_{xx} = 0$ with
the initial conditions $u(x,0) = 0$ for every $x \in [0,L]$ and
$u(0,t) = u(L,t) = 0$ for every $t \ge 0$.  The wavespeed~$\omega$ is
the rate at which the initial stimulus travels to other points of the
string.  The propagation of influence with a bounded wavespeed within
a spatial domain, a subset of~$\mathbb{E}^d$, can be visualized as a
relation between points in the spacetime domain one dimension higher,
i.e., $\mathbb{E}^d \times \Real$.


\subsubsection{Influence and dependence}

Points in spacetime are partially ordered by \emph{causality}---a
point~$\fp$ \emph{influences} another point~$\fq$, written as $\fp
\prec \fq$ if and only if changing the physical parameters at~$\fp$
could possibly change the result at~$\fq$.  A point~$\fq$
\emph{depends} on~$\fp$, written as $\fq \succ \fp$, if and only if
$\fp$ influences~$\fq$, i.e., $\fp \prec \fq \Leftrightarrow \fq \succ
\fp$.  The \emph{domain of influence} of~$\fp$ is the set of points~$\fq$ such
that $\fp \prec \fq$.  Symmetrically, the \emph{domain of dependence}
of a point~$\fp$ is the set of points~$\fr$ such that~$\fp \succ \fr$.
We say that two distinct points~$\fp$ and~$\fq$ are \emph{independent}
if and only if neither $\fp \prec \fq$ nor $\fq \prec \fp$.

A \emph{characteristic} of a PDE through a given point~$\vec{x}$ is
the iso-surface along which~$u$ satisfies an ordinary differential
equation (ODE).  Thus, characteristics are curves or surfaces or
hypersurfaces, of dimension~${n-1}$ or less, in the space spanned by the
$n$ independent variables.  Each characteristic corresponds to a
\emph{characteristic equation} which is an ordinary differential
equation (ODE) or a system of ODEs.  The \emph{slope} of a
characteristic at~$\fp$ is the slope of the tangent to the
characteristic at~$\fp$.  The method of characteristics for solving a
given PDE changes coordinates from $(x,t)$ to a new coordinate system
$(x(s),s)$ in which the PDE becomes an ordinary differential equation
(ODE) in the independent variable~$s$.  The variable~$s$ is a
parameter that varies along the characteristic hypersurface.

For a hyperbolic spacetime PDE, a characteristic hypersurface through
a point~$\fp$ represents the flow of information through the space
domain with time.  For hyperbolic PDEs such as the wave equation
$u_{tt} - \omega^2 u_{xx} = 0$ the solution at $p=(x_p,t_p)$
influences the solution at every future point in the wedge bounded by
the characteristic lines through~$p$ with slope~$\pm \omega$.  The
information at~$p$ propagates into the future with a uniform bounded
speed~$\omega$ in every direction.

A curve $\Gamma = (x,y(x))$ is \emph{characteristic} for the
second-order PDE
\[
  a \, u_{xx} + b \, u_{xy} + c \, u_{yy} 
    + d \, u_x + e \, u_y + f \, u + g(x,y) = 0
\]
if
\begin{equation}
 \frac{dy}{dx} = \frac{b \pm \root{b^2 - 4ac}}{2a}
\label{eqn:pde:roots}
\end{equation}
along~$\Gamma$.  The three different classes of second-order PDEs have
three different types of characteristics.  Hyperbolic PDEs have $b^2 -
4ac > 0$, so Equation~\ref{eqn:pde:roots} has two real solutions and
there are two characteristic curves through every point $(x,y)$.  For
example, the wave equation $u_{xx} - u_{yy} = 0$ is hyperbolic and has
two characteristic lines given by $y = \pm x$.

The solution to a PDE can have discontinuities.  If two
characteristics intersect at a point~$\vec{x}$, the two values for the
solution~$u$ at~$\vec{x}$ obtained by back-tracing the characteristics
can be different.  Such discontinuities are called \emph{shocks}.
Shocks can occur even when the initial conditions are smooth.

For spacetime hyperbolic PDEs, the wavespeed everywhere is finite,
i.e., information propagates through the space domain with a bounded
wavespeed.


\subsubsection{Cones of influence and dependence}

The \emph{slope} $\S(\fp)$ at a point~$\fp$ is a quantity computed by
the numerical solver to bound from above the speed at which any
change in the parameters at~$\fp$ propagates through the spacetime domain.
For instance,~$\S(\fp)$ can be computed as the maximum slope of
every characteristic through~$\fp$.  The \emph{wavespeed} at~$\fp$,
denoted by $\omega(\fp)$, is the reciprocal of the slope, i.e.,
\[
  \omega(\fp) = \frac{1}{\S(\fp)}
\]
Thus, the slope (or wavespeed) at every point in the domain defines a
scalar field over the domain.

\begin{figure}[t]\centering\small
\includegraphics[height=2.2in]{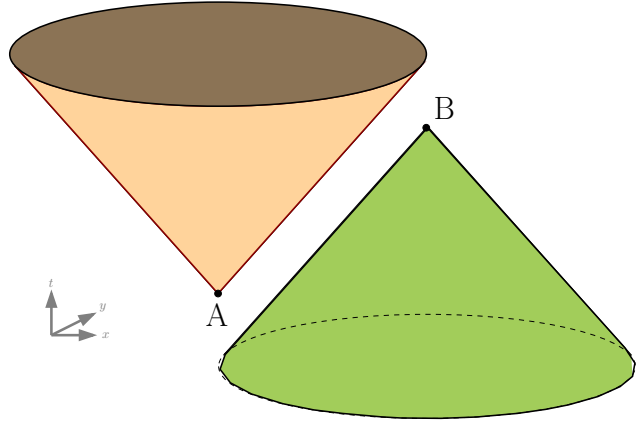}
\caption{Points~$\fa$ and~$\fb$ are independent
because~$\fb$ lies outside the cone of influence of~$\fa$ and~$\fa$
lies outside the cone of dependence of~$\fb$}
\label{fig:independent}
\end{figure}

The \emph{cone of influence} of~$\fp$, denoted by $\cone^+(\fp)$, is a
circular cone with apex at~$\fp$, axis in the positive time direction,
and slope equal to $\S(\fp)$.  The cone $\cone^+(\fp)$ is a closed
full-dimensional subset of the spacetime domain.  Symmetrically, the
\emph{cone of dependence} of~$\fp$, denoted by $\cone^-(\fp)$, is a
circular cone with apex at~$\fp$, axis in the negative time direction,
and slope equal to~$\S(\fp)$.  See Figure~\ref{fig:independent}.

When the underlying medium is anisotropic, due to nonlinear response,
waves propagate faster in some spatial directions than in others.  In
such cases, cones of influence need not have circular cross-sections.
We will postpone a discussion of this anisotropic situation until
later in this section.


\subsubsection{No-focusing assumption}

We make the following assumption about how well the cone of influence
$\cone^+(\fp)$ at a point~$\fp$ approximates the domain of influence
of~$\fp$.  We assume that the slope of the cone of influence at a
point~$\fp$ in the future can be estimated from the slopes of the
cones of influence of all points~$\fq$ such that $\fp \in
\cone^+{\fq}$.  In other words, we assume that~$\fp$ is influenced
only by points~$\fq$ such that $\fp \in \cone^+{\fq}$.  This property
is formally stated as Axiom~\ref{axiom:nofocusing}.

\begin{figure*}[t]\centering\small
\includegraphics[height=3in]{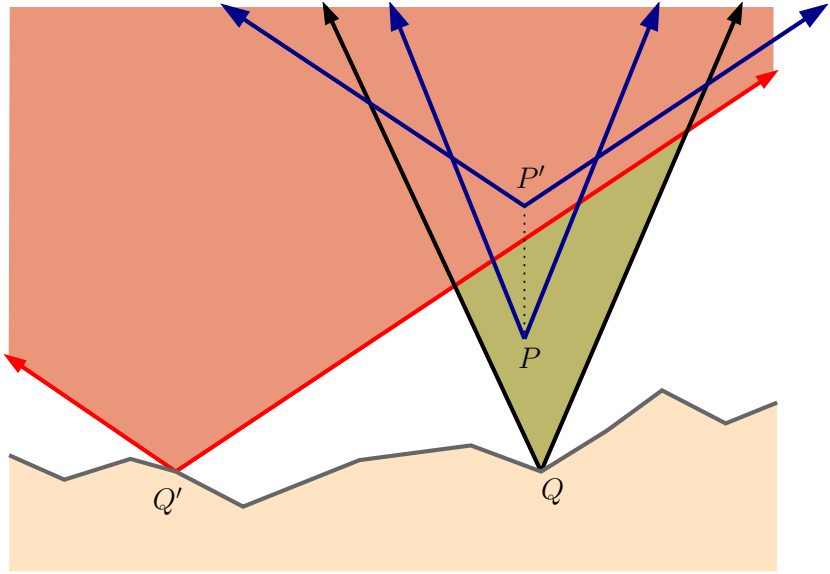}
\caption{No-focusing means that the wavespeed at a point in the future
can be estimated from the wavespeeds of points on the current front.}
\label{fig:nofocusing}
\end{figure*}

\begin{axiom}[No focusing]
  For every point~$\fp$ in the spacetime domain, the slope $\S(\fp)$ is
  bounded by the minimum and maximum slope of every cone of influence
  containing~$\fp$. Thus,
  \[
    \min_{\text{$\fp \in \cone^+(\fq)$}} \{ \S(\fq) \}
  \quad \le \quad
    \S(\fp)
  \quad \le \quad
    \max_{\text{$\fp \in \cone^+(\fq)$}} \{ \S(\fq) \}
  \]
\label{axiom:nofocusing}
\end{axiom}

In Figure~\ref{fig:nofocusing}, point~$\fp$ is influenced by point
$\fq$, hence $\S(\fp) = \S(\fq)$ is possible.  Suppose~$\fp$ is advanced
in time to~$\fp'$.  The point~$\fp'$ is influence by both~$\fq$ and~$\fq'$.  Hence, $\S(\fp) = \min\{\S(\fq),\allowbreak{}
\S(\fq')\}$ is possible.

If a point $\fp=(p,t)$ is in a cone of influence~$C$, then every point
$\fp'=(p,t')$ where $t' > t$ is also in~$C$.  See
Figure~\ref{fig:nofocusing}.  Together with
Axiom~\ref{axiom:nofocusing}, we obtain the following monotonicity
lemma.

\begin{lemma}
  For an arbitrary fixed point~$p \in \mathbb{E}^d$, let $\fp=(p,t)$
  and $\fp'=(p,t')$ be two points in spacetime with the same spatial
  projection~$p$ and time coordinates~$t$ and~$t'$ respectively.  If
  $t' \ge t$, then $\S(\fp') \le \S(\fp)$.
\label{lemma:monotonicity}
\end{lemma}

\parasc*{Explanation of the no-focusing assumption:
Axiom~\ref{axiom:nofocusing}}

The causal slope~$\S(\fp)$ is a first-order approximation to the slope
of the characteristic curves/surfaces through~$\fp$; the slope of the
cone of influence at~$\fp$ in every direction is less than or equal to
the slope of all the characteristics through~$\fp$ in that direction.
Axiom~\ref{axiom:nofocusing}, which holds in the absence of focusing,
allows us to conservatively estimate the causal slope at every point
in spacetime where the solution has not been computed by our algorithm
yet.  Cones of influence are locally conservative approximations to
the domains of influence.  A sufficiently conservative cone can be
chosen that is a valid approximation of the actual domain of influence
for the finite step size chosen by our algorithm to advance the
solution.  Thus, in Figure~\ref{fig:independent}, points~$\fa$ and
$\fb$ are independent because neither influences the other.

We assume only that the no-focusing assumption of
Axiom~\ref{axiom:nofocusing} holds everywhere in the spacetime
domain~$\spt$.  Our algorithms guarantee correctness of the solution
only in the absence of focusing.

\parasc*{Asymmetric cones}

Due to nonlinear response and anisotropy of the underlying medium,
waves can propagate with different speeds in different directions.
Therefore, in general, the wavespeed at a point~$\fp$ is a function of
the spatial direction~$\vec{n}$, and hence, cones of influence and
dependence have non-circular cross-sections.  Our assumption of
no-focusing in this anisotropic context can be stated as the following
axiom.

\begin{axiom}[Anisotropic no-focusing]
  For every point~$\fp$ in the spacetime domain, the slope at~$\fp$ in
  an arbitrary spatial direction~$\vec{n}$, denoted by $\S_{\vec{n}}(\fp)$, is
  bounded by the minimum and maximum slope in the same spatial
  direction~$\vec{n}$ of every cone of influence containing~$\fp$.
  Thus, for every spatial direction~$\vec{n}$, we have
  \[
    \min_{\text{$\fp \in \cone^+(\fq)$}}
      \left\{ \S_{\vec{n}}(\fq) \right\}
    \quad \le \qquad
      \S_{\vec{n}}(\fp)
    \quad \le \quad
    \max_{\text{$\fp \in \cone^+(\fq)$}}
      \left\{ \S_{\vec{n}}(\fq) \right\}
  \]
\label{axiom:anisotropic:nofocusing}
\end{axiom}


\section{Spacetime meshing}
\label{sec:intro2spacetime}

We saw in Section~\ref{sec:background} that wave propagation is
modeled by hyperbolic partial differential equations (PDEs) in both
space and time variables, for instance, the wave equation $u_{tt} -
\omega^2 u_{xx} = 0$ in 1D space $\times$ time.  The
\emph{wavespeed}~$\omega$, the speed at which changes in physical
parameters at a point $(x,t)$ propagate to other points in the domain,
may be a function of~$x$ and~$t$ as well as of~$u$ and its
derivatives.  The spacetime discontinuous Galerkin (SDG) method
approximates the solution within each spacetime element of a
mesh of the spacetime domain as a linear combination of simple
basis functions such as polynomials.  The SDG method allows basis
functions to be discontinuous across element boundaries---adjacent
elements need not agree on the solution along their common boundary.

The spacetime DG method motivates our meshing problem.  In this
chapter, we will give an overview of the goal of this thesis---to give
provably correct algorithms for generating efficient spacetime meshes
suitable for DG solvers.  We will describe the challenges that are
encountered.  We will enumerate the assumptions under which our
algorithms and the meshes they produce satisfy the stated goal.  We
conclude this chapter with a summary of our results and of previous
research on spacetime meshing.

This section defines the basic notions and vocabulary used throughout
the rest of the thesis.  We will remind the reader of our assumptions
and refer back to this section whenever necessary.


\subsubsection{Patches}

The notions of influence and dependence extend naturally to arbitrary
subsets of spacetime.  For arbitrary elements~$A$ and~$B$, not
necessarily distinct, of a spacetime mesh, we say that~$A$
\emph{influences}~$B$, written as $A \preceq B$ if and only if some
point $a \in A$ influences some point $b \in B$.  If $A \preceq B$,
then~$A$ must be solved no later than~$B$, otherwise the solution in
$B$ is not valid.  Elements~$A$ and~$B$ are \emph{coupled} if and only
if both $A \preceq B$ and $B \preceq A$.  A pair of coupled elements
must be solved together, i.e., as a simultaneous system of equations.
If neither $A \preceq B$ nor $B \preceq A$, then we say~$A$ and~$B$
are \emph{independent}.

A \emph{patch}~$\Pi$ in a spacetime mesh~$\spt$ is a set of elements
that must be solved simultaneously.  More formally, a patch~$\Pi$ is a
set of one or more elements with the following properties:
\begin{enumerate}
\item for every $A \in \Pi$ there exists $B \in \Pi$ such that~$A$ and
$B$ are coupled (dependency condition);
\item for every $A \in \Pi$, if there exists $B \in \spt$ such that~$A$ and~$B$ are coupled then $B \in \Pi$ (maximality condition).
\end{enumerate}
The \emph{size} of a patch is the number of elements in it.

We have seen that causality defines a partial order over points in
spacetime.  We say that a subset~$\spt'$ of spacetime is
\emph{causal} if and only if~$\spt'$ is an anti-chain in the partial
order, i.e., if and only if no two points of~$\spt'$ influence each
other.  The boundary of a patch~$\Pi$ is necessarily causal because
otherwise elements outside the patch would be coupled with elements in
the patch, violating the maximality of~$\Pi$.  Patches are therefore
partially ordered by causal dependence.

A spacetime mesh~$\spt$ can be solved patch-by-patch in an order that
respects the partial order of patches.  The total computation time of
such a patch-wise solution strategy is the sum over every patch~$\Pi$
in the mesh~$\spt$ of the time to solve~$\Pi$.  We say that a
patch-wise solution strategy is \emph{efficient} if the total
computation time is proportional to the number of elements in the
mesh.  Our meshing algorithms support an efficient solution strategy
because both (i)~the maximum size of a patch and (ii)~the number of
patches in a given spacetime volume are bounded.  In a mesh
constructed by other algorithms, either the maximum size of a patch or
the number of patches or both are unbounded; furthermore, the solver
must explicitly compute the partial order of patches or solve all
patches together as a large coupled system.  Therefore, a generic
purpose meshing algorithm is inefficient in general; specialized
algorithms are necessary for efficiently meshing in spacetime.

\subsection{Terminology and notation}
\label{sec:notation}

Before we proceed to the technical details of our spacetime meshing
problem and a summary of the results in this thesis, we define some
basic concepts and introduce the reader to the notation used in the
rest of this thesis.

First, we review some basic concepts.  See the
references~\cite{ziegler95lectures} for further details.

A \emph{simplex} of dimension~$k$, also called a $k$-simplex, is the
convex hull of ${k+1}$ affinely independent points.  A simplex is a
closed set of points.  We identify a simplex with its vertices.  Let
$p_0p_1p_2{\ldots}p_k$ be a $k$-simplex.  Every subset of
$\{p_0,p_1,p_2,\ldots,p_k\}$ defines a \emph{face} of the simplex.
Simplices of dimension $0$, $1$, $2$, and~$3$ are called vertices,
segments, triangles, and tetrahedra respectively.  For a $k$-simplex
$S$, let $\aff{S}$ denote the affine hull of~$S$, i.e., the set of all
linear combinations of vertices of~$S$;  the affine hull $\aff{S}$ is
a \emph{$(k-1)$-flat}~\cite{deberg00computationalgeometry}.  For
example, the affine hull $\aff{pq}$ of two points~$p$ and~$q$ is the line
$pq$.

A \emph{simplicial complex} is a collection~$\mathcal{C}$ of
simplices with the following properties: (i)~if a simplex $S \in
\mathcal{C}$ then every face of~$S$ is in~$\mathcal{C}$, and (ii)~two
simplices in~$\mathcal{C}$ either do not intersect, or their
intersection is a simplex of smaller dimension which is their common
face of maximal dimension.  The empty simplex, whose dimension is
$-1$, is a face of every simplex.  The dimension of the simplicial
complex~$\mathcal{C}$ is the highest dimension of any simplex in the
collection~$\mathcal{C}$.  For a $d$-dimensional simplicial complex,
the ${(d-1)}$-dimensional faces are called the \emph{facets}.  A
simplicial complex is a special type of
\emph{cell complex}~\cite{ziegler95lectures}.

Two faces of a complex are \emph{incident} if one is included in the
other.  Given a simplicial complex~$\mathcal{C}$, the \emph{star} of a
face~$F$, denoted by $\st(F)$, is the sub-complex consisting of all
simplices incident on~$F$, and all their faces.  The star of~$F$ is a
closed set.  The \emph{link} of a face~$F$, denoted by $\link(F)$, is
the sub-complex consisting of all faces~$G$ of simplices in $\st(F)$
such that $G \cap F = \emptyset$.

A particular type of simplicial complex is a triangulation.  A
\emph{$d$-dimensional triangulation} is a $d$-dimensional simplicial
complex such that all maximal faces are $d$-simplices.  Note that a
triangulation need not form a manifold.  In a $d$-triangulation, every
${(d-1)}$-face~$F$ belongs to one or more $d$-simplices; in the former
case, $F$ is \emph{a boundary facet}, and in the latter case, $F$ is
\emph{an interior facet}.  All faces of a boundary facet are
\emph{boundary faces}; all other faces are
\emph{interior faces}.  For example, if~$v$ is a vertex of a
$2$-dimensional triangulation, then $\st(v)$ consists of the vertex
$v$, all edges incident on~$v$, and all triangles incident on~$v$
together with their edges and vertices; $\link(v)$ is a set of edges
which form a closed loop if $\st(v)$ is homeomorphic to a disk.

\medskip

Next, we define the terms and notation used to describe our spacetime
meshing problem.

A $d$-dimensional \emph{space mesh} is a $d$-dimensional
triangulation.  We visualize the ${(d+1)}$-dimensional spacetime domain
with the time axis drawn vertically so that time increases upwards.
We use uppercase letters like $\fp$, $\fq$, $\fr$ to denote points in
spacetime and corresponding lowercase letters like $p$, $q$, $r$ to
denote their spatial projections.  

A \emph{front} is a maximal subset of spacetime that forms an
anti-chain in the partial order of dependence; i.e., a front~$\tau$ is
a maximal set of points such that no two points of~$\tau$ influence
each other.  An example of a front is the set of all points with a
given time coordinate~$T$, whose graph is the horizontal constant-time
plane $t=T$.

The \emph{front}~$\tau$ defines a piecewise linear function $\tau :
\mathbb{E}^d \to \Real$.  Equivalently, the front~$\tau$ is a
$d$-dimensional piecewise linear
terrain~\cite{deberg00computationalgeometry}, a subset of
$\mathbb{E}^d \times \Real$.  We will not distinguish between these
two equivalent descriptions of the front---one as a function and the
other as a set of points.  Each point~$\fp$ on the front~$\tau$ can be
written as $\fp=(p,\tau(p))$ where~$p$ is the spatial projection of
$\fp$.  For a real~$T$ and front~$\tau$, we use $\tau \ge T$ to mean
that every point $\fp=(p,\tau(p))$ of~$\tau$ satisfies $\tau(p) \ge
T$, i.e., the entire front~$\tau$ has achieved or exceeded time~$T$.
Arbitrary fronts~$\tau$ and~$\tau'$ satisfy $\tau' \ge \tau$ if and
only if $\tau'(p) \ge \tau(p)$ for every point~$p$ such that
$\fp=(p,\tau(p)) \in \tau$ and $\fp'=(p,\tau'(p)) \in \tau'$.

We will frequently assume that the front~$\tau$ consists of a single
$k$-simplex.  In this case, we will not distinguish between the
simplex and the corresponding time function $\tau: \mathbb{E}^k \to
\Real$ whose graph is the $k$-flat in spacetime spanned by the
simplex.  In such a scenario, since~$\tau$ is a linear function, its
gradient, denoted by $\grad{\tau}$, is the same
everywhere---$\grad{\tau}$ is a vector in $\Real^k$ in the direction
of steepest ascent and its $L_2$-norm is denoted by
$\norm{\grad{\tau}}$.

The front at the beginning of each iteration of our algorithm
represents the frontier of both the incremental mesh construction as
well as the solution.  Thus, if $\fp=(p,\tau(p))$ is a point of the
front~$\tau$ then~$p$ belongs to the space domain at time $\tau(p)$,
i.e., $p \in \sp_{\tau(p)}$; the solution at~$\fp$ has been computed;
and the set of points $\fp^+=(p,t^+)$ such that $t^+ >
\tau(p)$ remain to be meshed and the solution there is unknown.

A front is causal if and only if for every facet~$F$ of the front all
characteristics cross~$F$ from one side to the other.  For an element
with a facet on the front~$\tau$, this facet is an \emph{outflow}
facet of the element because influence travels from the interior of
the element to the exterior across this facet.  Equivalently, a front
$\tau$ is causal if and only if every point $\fp=(p,\tau(p))$ depends
only on points below the front, i.e., if $\fp \succ \fq$ then $\fq$
can be written as $\fq=(q,t_q)$ where $t_q < \tau(q)$.  Symmetrically,
every point of the front~$\tau$ influences only points above the
front.  By our earlier definition, every front is automatically
causal.

For a simplex (of any dimension)~$S$ of the mesh~$\sp$, let
$\rest{\tau}{S}$ denote the time function~$\tau$ restricted to~$S$ and
extended to the affine hull of~$S$; in other words, $\rest{\tau}{S}$
is a linear function that coincides with~$\tau$ for every point of
$S$.

Let $\tau_i : \sp \to \Real$ denote the front after the~$i$th step of
the algorithm; $\tau_0$ is the initial front.  At every step $\tau_i
\to \tau_{i+1}$, our algorithm will explicitly triangulate the new
front~$\tau_{i+1}$.  Thus, every front constructed by our algorithm
will be a triangulated terrain.  For every~$i$, the front~$\tau_i$ is
a terrain whose facets are $d$-simplices.

A local minimum of the front~$\tau$ is a vertex
$p$ such that $\tau(p) \le \tau(q)$ for every vertex~$q$ that is a
neighbor of~$p$.  Every front has a local minimum because it has at
least one global minimum which is also a local minimum.  When the
current front~$\tau$ is clear from the context, for every point $p \in
\sp$ we use~$\fp$ to denote the corresponding point on the front,
i.e., $\fp=(p,\tau(p))$.

For an arbitrary simplex $\fp_0\fp_1\fp_2{\ldots}\fp_d$
of the front~$\tau$, we say that a vertex, say~$\fp_0$, is a
\emph{lowest vertex} of the simplex if $\tau(p_0) \le \min_{0 \le i
\le d} \tau(p_i)$.  Note that a
simplex may have one or more lowest vertices.  We say that a facet,
say $\fp_1\fp_2{\ldots}\fp_d$ is a \emph{highest facet} if the
opposite vertex~$\fp_0$ is a lowest vertex.  Note that a simplex
may have more than one highest facets.

We say that a front~$\tau'$ is obtained by advancing a vertex
$\fp=(p,\tau(p))$ of~$\tau$ by $\dt \ge 0$ if $\tau'(p) = \tau(p) +
\dt$ and for every other vertex $q \ne p$ we have $\tau'(q) =
\tau(q)$.  For every front~$\tau$, vertex~$p$, and real~$\dt \ge 0$,
let $\tau'=\next(\tau,p,\dt)$ denote the front obtained from~$\tau$ by
advancing~$p$ by~$\dt$; the functions~$\tau$ and~$\tau'$ coincide
everywhere outside~$\st(p)$.

For a simplex~$\triangle$, we use $\S(\triangle)$ to denote the
minimum slope $\S(\fp)$ over all points~$\fp$ of~$\triangle$; we let
$\omega(\triangle)$ denote $1/\S(\triangle)$.  Let~$\minS$ denote
$\min_{\fp \in \spt} \{\S(\fp)\}$ and~$\maxS$ denote $\max_{\fp \in
\spt} \{\S(\fp)\}$.  We assume that $0 < \minS \le
\maxS < \infty$.  Thus, the slope $\S(\fp)$ at an
arbitrary point $\fp$ is bounded: $0 < \minS \le
\S(\fp) \le \maxS < \infty$.

For a vertex~$\fp$ of the front~$\tau$, define the quantity~$w_p$ in
the spatial projection as
\[
  w_p := \min_{\triangle \in link(p)} \dist(p,\aff{\triangle}).
\]

\begin{figure}\centering\small
\includegraphics[height=2in]{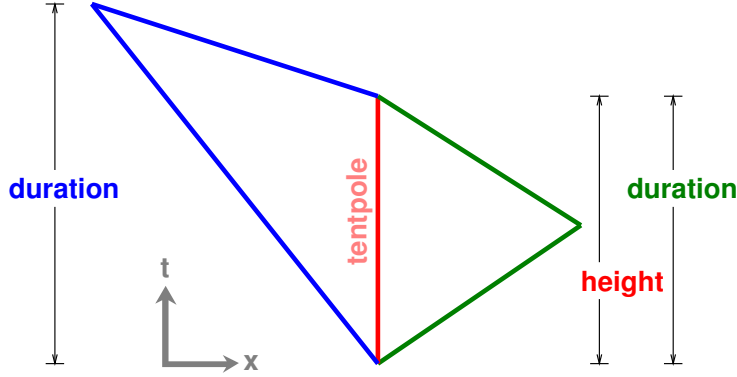}
\caption[Temporal aspect ratio of a spacetime element]{The temporal
aspect ratio of a spacetime element is the ratio of its height to its
duration.  The height of both spacetime triangles in the figure is
equal to the height of the tentpole they share.  The spacetime
triangle on the right has a better (larger) temporal aspect ratio than
the triangle on the left.}
\label{fig:temporalaspectratio}
\end{figure}

For an arbitrary subset~$K$ of $\mathbb{E}^d \times \Real$, the
\emph{height} of~$K$ is the length of the longest time interval
contained in the closure of~$K$, and the \emph{duration} of~$K$ is the
length of the shortest time interval containing the interior of~$K$.
The \emph{temporal aspect ratio} of a spacetime element is the ratio
of the height of the element to its duration.  See
Figure~\ref{fig:temporalaspectratio}.  Note that the temporal aspect
ratio is always in the range~$(0,1]$ with a larger value corresponding
to a ``better'' element.  The temporal aspect ratio of an arbitrary
subset of spacetime is unchanged if all time coordinates are
uniformly translated and scaled by a non-zero constant.

\subsection{Meshing objectives}

We want a mesh of the spacetime domain~$\spt$ consisting of spacetime
simplices or \emph{elements} such that every spacetime element has at
least one causal outflow facet, a necessary condition for the
numerical problem to be well-posed~\cite{jegdic04convergence}.  The
sizes of the spacetime elements are controlled by a spacetime error
indicator.  The spacetime elements must form a \emph{weak simplicial
complex}; this property allows a simple strategy to compute integrals
over the common intersection of any two spacetime elements---the
intersection is an entire face of at least one of the two elements and
has a natural parameterization that facilitates integration.
Spacetime DG solvers are particularly suited to solve such weakly
conforming meshes because of their discontinuous formulation.

\newpage

\begin{definition}[Weak simplicial complex]
  A \emph{weak simplicial complex} is a collection~$\mathcal{C}$ of
  simplices with the following properties:
  \begin{enumerate}
  \item if a simplex $S \in \mathcal{C}$, then every face of~$S$
  is in~$\mathcal{C}$; and
  \item the intersection of every two simplices of~$\mathcal{C}$ is a
  face of at least one of them.
  \end{enumerate}
\label{def:2d:weakcomplex}
\end{definition}

We have seen that an efficient solution strategy is to solve the mesh~$\spt$
patch-by-patch in an order that respects the partial order of patches.
Every spacetime mesh can be partitioned into patches---giving a
\emph{causal partition}---and can be solved by this strategy.  The
total computation cost is the sum of the cost of solving each
individual patch.  A generic mesh of spacetime cannot be efficiently
solved because either the size of a patch or the number of patches or
both are not bounded.  Our objective is to minimize the total
computation cost.  We give a meshing algorithm that incrementally
constructs the spacetime mesh patch-by-patch explicitly in the order
of causal dependence.  We prove that both the size of each patch and
the total number of patches in the mesh of a given spacetime volume
are bounded; hence, the total computation cost is bounded.

We give an advancing front algorithm such that for every $T \in
\Real^{\ge 0}$ there exists a finite integer $k \ge 0$ such that the
front~$\tau_k$ after the $k$th iteration of the algorithm satisfies
$\tau_k \ge T$.  The input to our problem is the initial front
$\tau_0$ and the initial conditions of the PDE.  We obtain the
spacetime mesh by triangulating the spacetime volume, called a
\emph{tent}, between each pair of successive fronts~$\tau_i$ and
$\tau_{i+1}$ such that each spacetime element has at least one facet
on the front~$\tau_{i+1}$, which by definition is causal.  The
elements in a tent are causally dependent and must be solved as a
coupled system.  We produce an efficient mesh such that the number of
elements in a coupled system is small, depending on the maximum degree
of the initial space mesh.

Our mesh is also efficient in the sense that elements are
non-degenerate in the sense that each element has temporal aspect
ratio bounded from below.  We expect that the non-degeneracy guarantee
means that in practice the numerical solution within each element
can be computed with acceptable accuracy.

Our goal is to support an accurate solution of the underlying PDE as
well as to reduce total computation time.  It is challenging enough to
devise meshing algorithms that provide a theoretical guarantee that
the meshing algorithm would terminate after creating a finite number
of non-degenerate elements.  In addition, our goal is to provide an
algorithm that is easy to implement and that performs significantly
better in practice than the theoretical guarantee, for instance by a
good choice of parameters to the algorithm and possibly using
different heuristics.  At the same time, we will never compromise on
the ability to prove correctness of all algorithms described in this
thesis.  Some aspects of the meshing algorithm go beyond this basic
goal and improve its performance in practice.

\subsubsection{Advancing front spacetime meshing}

\begin{figure}\centering
\includegraphics[height=2.8in]{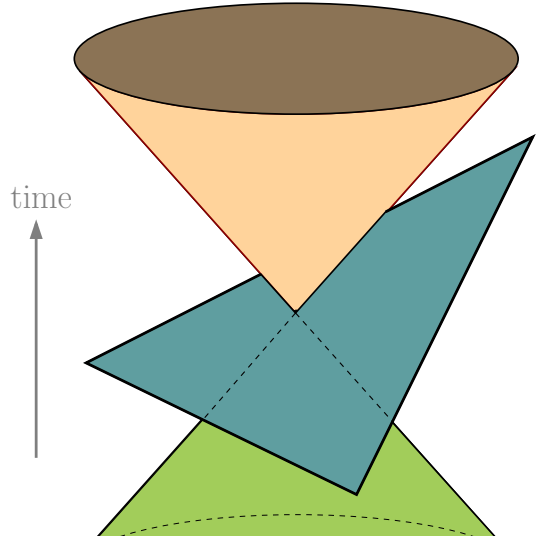}
\caption{A causal triangle separates the cones of influence
and dependence at every point on the face.}
\label{fig:causalface}
\label{fig:doublecone}
\end{figure}

\begin{figure}\centering\footnotesize
\includegraphics[height=3in]{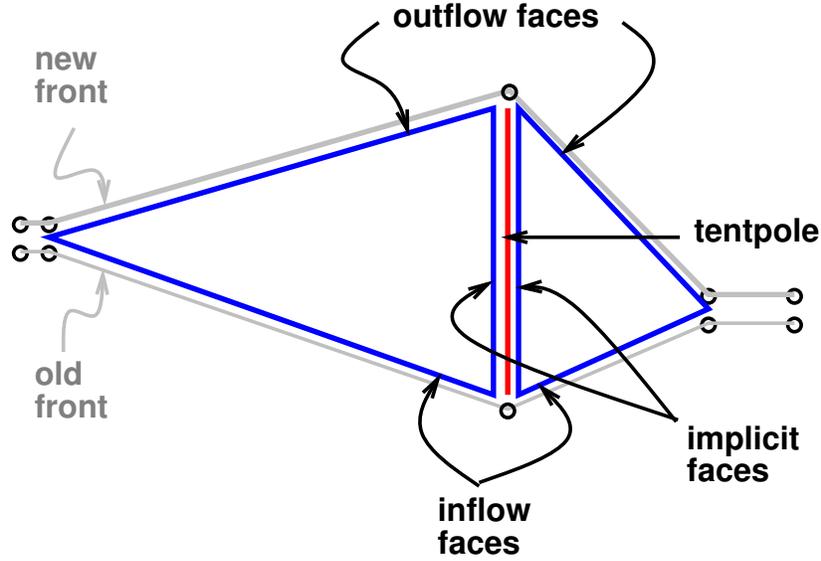}
\caption{Cross-section of a patch of tetrahedra; the inflow and
outflow faces are causal. Time increases up the page.}
\label{fig:patch}
\end{figure}

Given a simplicial mesh of some bounded domain $\sp \subset
\mathbb{E}^d$, we give a meshing algorithm to incrementally construct a
simplicial mesh of the spacetime domain using an advancing front
method.  The algorithm advances the front by moving a vertex forward
in time, thus also advancing the local neighborhood $\st(v)$, and
adding simplices in the volume between the old and the new fronts.
The inflow and outflow boundaries of each patch
(Figure~\ref{fig:patch}) are causal by construction, i.e., each
boundary facet~$F$ separates the cone of influence from the cone of
dependence for every point on~$F$ (Figure~\ref{fig:causalface}).
Equivalently, for every point~$\fp$ on~$F$ we have $\norm{\grad F} \le
\S(\fp)$.  If the outflow boundaries of a patch are
causal, every point in the patch depends only on other points in the
patch or points of inflow elements adjacent to the inflow boundaries
of the patch.  Therefore, the solution within the patch can be
computed as soon as the patch is created, given only the inflow data
from adjacent inflow elements and initial or boundary data where
appropriate.  The elements within a patch are causally dependent on
each other and must be solved as a coupled system.  Patches with no
causal relationship can be solved independently.  To minimize
undesirable numerical dissipation and the number of patches, we would
like the boundary facets of each patch to be as close as possible to
the causality constraint without violating it.

We require that for every target time value~$T$ the algorithm will
compute in a finite number of steps a mesh of the spacetime volume
$\sp \times [0,T]$ and the solution everywhere in this volume.  The
target time~$T$ may not be known \textsl{a priori} because it depends on
the evolving physics.

Our algorithm ensures that every spacetime element in a patch has one
facet on the new front, which is a causal outflow facet.

\parasc*{\textsc{Triangulating a tent}}

Our advancing front algorithm repeatedly advances a local neighborhood~$N$ of the current front~$\tau$ to a corresponding neighborhood~$N'$
of a new front~$\tau'$, where $\bd{N} = \bd{N'}$ and $\tau = \tau'$
everywhere in $\tau \setminus N$.  For instance, our algorithm pitches
a local minimum of the current front creating a new front which is the
input to the next iteration of the algorithm. \emph{Pitching} a vertex
$\fp$ of the front~$\tau$ means advancing $\fp=(p,\tau(p))$ to
$\fp'=(p,\tau'(p))$ where~$\tau$ and~$\tau'$ are the old and the new
fronts respectively. Pitching~$\fp$ means advancing the neighborhood
$N=\st(\fp)$ to $N'=\st(\fp')$ where $\bd{N}=\bd{N'}=\link(\fp)$.  The
\emph{progress} is defined to be the time difference $\tau'(p) -
\tau(p)$.

The volume between the new front and the old front is
called a \emph{tent} and the edge~$\fp\fp'$ is called the
\emph{tentpole}.  The tent is partitioned into spacetime elements
(simplices); the set of elements in a tent forms a patch.
Let~$F'$ denote an arbitrary facet on the new front~$\tau'$. The tent
is triangulated so that for each such facet~$F'$ the patch contains
the element corresponding to the convex hull of~$F'$ and the vertex
$\fp$, the bottom of the tentpole.

For each point~$\fu$ in the tent, the ray $\fp\fu$ intersects a
simplex~$F'$ on the new front~$\tau'$. Therefore,~$\fu$ is contained
in the convex hull of $F' \cup \{\fp\}$.  Hence, the set of elements
in a patch partition the tent.

By construction, each element~$E$ in the patch is the convex hull of
$F' \cup \{\fp\}$ for some facet~$F'$ on the new front~$\tau'$.  By
construction,~$F'$ is a causal facet of~$E$. For every point~$\fq$ on
$F'$ the cone of influence of~$\fq$ is entirely in the future and does
not intersect the element~$E$; therefore, $F'$ is an outflow facet of
$E$.  Hence, every element in a patch has a causal outflow face.

Also, the intersection of every two elements is either (i)~empty, (ii)~a
common vertex, (iii)~the tentpole of a patch containing both elements,
or (iii)~the entire facet of at least one of the two elements.  If the
two elements belong to the same patch with tentpole~$\fp\fp'$, then
they share the vertex~$\fp$; they also share either the tentpole
$\fp\fp'$ or a common facet, or they share a common implicit facet
containing the tentpole.  On the other hand, if one element is an
inflow into the other tetrahedron then their common intersection is
either the entire outflow face of the first element or the entire
inflow face of the second element.

When we generalize to other operations that advance the front we will
ensure that these properties are always satisfied.  For instance, an
edge flip in 2D$\times$Time creates a patch with a single tetrahedron,
which has two causal inflow facets and two causal outflow facets.

\subsection{Meshing challenges}

We say that a front~$\tau$ is \emph{valid} if there exists a positive
real~$\delta$ bounded away from zero such that for every $T \in
\Real^{\ge 0}$ there exists a sequence of fronts $\tau$, $\tau_1$, $\tau_2$,
$\ldots$, $\tau_k$ where $\tau_k \ge T$, each front in the sequence
obtained from the previous front by advancing some vertex by~$\delta$.
The minimum value of~$\delta$ for which we can guarantee that our
algorithms construct only valid fronts is the \emph{progress
guarantee} of our algorithm.  The progress guarantee is the same as
the worst-case tentpole height.  All our theoretical results, such as
finite termination, worst-case temporal aspect ratios of spacetime
elements, and near-size-optimality follow from the progress guarantee.
What makes the definition of a valid front nontrivial is the
requirement that all fronts be causal.  The main difficulty in
characterizing valid fronts arises when the wavespeed at a given point
in the space domain increases discontinuously and unpredictably over
time.  In spatial dimensions $d \ge 2$, an additional complication is
the changing front geometry due to adaptive refinement and coarsening,
and mesh smoothing.

\subsection{Summary of results}

The main contribution of our work is an efficient incremental
spacetime meshing algorithm that adapts quickly to changing geometric
constraints, that supports an efficient parallelizable solution
strategy, and that has provable worst-case guarantees on the temporal
aspect ratio of spacetime elements.

\begin{theorem}
  Suppose we are given a simplicial mesh $\sp \in \mathbb{E}^d$ as
  well as initial and boundary conditions of the underlying spacetime
  hyperbolic PDE.  Our algorithm builds a simplicial mesh~$\spt$ of
  the spacetime domain $\sp \times [0,\infty)$ that satisfies all the
  following criteria:

  \begin{enumerate}

  \item the elements of~$\spt$ form a \emph{weak simplicial complex}
  that conforms to the initial mesh~$\sp$;

  \item each spacetime element has at least one causal outflow facet;

  \item for every $T \ge 0$ the spacetime volume $\sp \times [0,T]$ is
  contained in the union of a finite number of simplices of~$\spt$;

  \item the minimum \emph{temporal aspect ratio} of every spacetime
  element is bounded from below.

  \end{enumerate}

  Additionally, in 2D$\times$Time, our algorithm adapts the size and
  duration of spacetime elements to a spacetime error indicator.  We
  guarantee that the diameter of each tetrahedron is no larger than
  that allowed by the spacetime error indicator.  Provided the error
  indicator does not reject any tetrahedra smaller than a bounded
  minimum size, our algorithm terminates with a finite mesh of $\sp
  \times [0,T]$ for every target time $T \ge 0$.

\label{thm:mainresult}
\end{theorem}

Given a triangulation~$\sp$ of the space domain and a target time~$T$,
we say that a simplicial spacetime mesh of $\sp
\times [0,T]$ is \emph{solvable} if (i)~each spacetime element has both a
causal inflow facet and a causal outflow facet; and (ii)~for every
point~$x$ in the spatial projection~$\Delta$ of each spacetime
element, the diameter of~$\Delta$ does not exceed the diameter of the
simplex of~$\sp$ containing~$x$.

\begin{theorem}
  Let~$\he$ be an arbitrary constant, a parameter to our algorithm, in
  the range $0 < \he \le \half$. The size of the mesh constructed by
  our algorithm is $\tilde{O}\left(\frac{1}{\he^2}\right)$ times the
  minimum size of any solvable mesh of the spacetime volume $\sp
  \times [0,T]$, where the hidden term in the $\tilde{O}$-notation
  depends on the dimension~$d$ and the worst-case spatial geometry of
  each front.
\end{theorem}

\begin{figure}
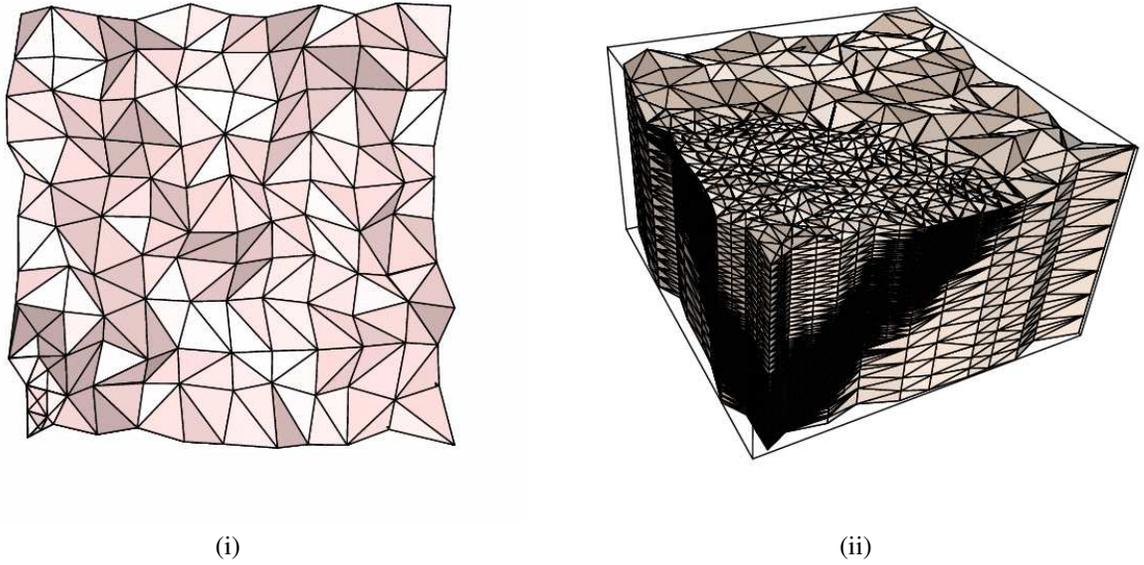
\centering
\begin{tabular}{cc}
\includegraphics[width=0.48\textwidth]{\fig{spacemesh2d}}
&
\includegraphics[width=0.48\textwidth]{\fig{spacetimemesh2d}}\\
(i) & (ii)
\end{tabular}
\caption[An unstructured tetrahedral spacetime mesh over a triangulated
  2D grid]{Given (i)~a triangulated 2D space mesh, our algorithm
  constructs (ii)~an unstructured tetrahedral spacetime mesh.  Time
  increases upwards in (ii).  In this example, the wavespeed at any
  point in spacetime is one of two distinct values: the maximum
  wavespeed occurs inside a circular cone where the tentpoles are
  shortest, the minimum wavespeed occurs everywhere else.  The size of
  spacetime elements adapts to both changing wavespeed to maintain
  causality as well as to a simulated error metric which depends on
  the temporal aspect ratio.}
\label{fig:2d:example}
\end{figure}

An example of a tetrahedral mesh of 2D$\times$Time constructed by our
algorithm is given in Figure~\ref{fig:2d:example}.  Additional
examples of meshes constructed using our algorithm appear in the
references~\cite{abedi04spacetime,abedi04adaptiveDG,abedi05adaptive,thite04nonlocal}.

\subsection{Previous work}

This thesis is part of an ongoing long-term project at the Center for
Process Simulation and Design (CPSD), a multi-disciplinary research
group at the University of Illinois (UIUC).  See
\url{http://www.cpsd.uiuc.edu/} for more about CPSD.

\begin{figure*}[t]
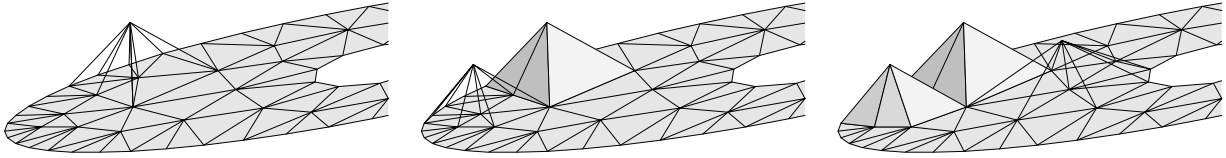
\centering\small
\begin{tabular}{ccc}
\includegraphics[width=.31\textwidth]{\fig{tent1}}&
\includegraphics[width=.31\textwidth]{\fig{tent2}}&
\includegraphics[width=.31\textwidth]{\fig{tent3}}
\end{tabular}
\caption{The first few tents pitched by Tent
Pitcher~\cite{erickson02building}.}
\label{fig:tents}
\end{figure*}

Building on ideas from earlier specialized algorithms, \Ungor{} and
Sheffer~\cite{us-ptstm-02} and Erickson
\etal~\cite{erickson02building} developed the first algorithm to build
graded spacetime meshes over arbitrary simplicially meshed spatial
domains, called Tent Pitcher.  Unlike most traditional approaches, the
Tent Pitcher algorithm does not impose a fixed global time step on the
mesh, or even a local time step on small regions of the mesh.  Rather,
it produces a fully unstructured simplicial spacetime mesh, where the
duration of each spacetime element depends on the local feature size
and quality of the underlying space mesh.  See Erickson
\etal{}~\cite{erickson02building} for sample meshes constructed by
Tent Pitcher.  Figure~\ref{fig:tents} from the paper by Erickson
\etal{}~\cite{erickson02building} illustrates the first few tent
pitching steps of the algorithm.

The original Tent Pitcher algorithm proposed by \Ungor{} and
Sheffer~\cite{ungor00tentpitcher} applied to one- and two-dimensional
space domains.  The algorithm could guarantee progress only if the
input triangulation contained only angles less than 90 degrees and if
the wavespeed did not increase or only increased smoothly.  \Ungor{}
and Sheffer referred to the causality constraint as an angle
constraint and assumed that this angle constraint was
Lipschitz-continuous.  They implemented Tent Pitcher and
investigated several different heuristics to improve its performance.

Erickson \etal~\cite{erickson02building} extended Tent Pitcher to
arbitrary spatial domains, even those with obtuse angles, in arbitrary
dimensions by imposing additional constraints, called \emph{progress
constraints}.  The progress constraint limits the amount of progress
in time when some vertex of the simplex is pitched, in addition to the
limit imposed by causality.  The progress constraint is a function of
the shape of the simplex and is necessary to guarantee progress only
when the initial space mesh contains a non-acute angle.

The Tent Pitcher algorithm due to \Ungor{} and
Sheffer~\cite{ungor00tentpitcher,ungor02phd} and extended by Erickson
\etal{}~\cite{erickson02building} applied to the case where the
wavespeed at a given point is either constant, decreasing, or
increasing smoothly as a Lipschitz function.  When the wavespeed
changes, the previous algorithms take the global upper bound on the
wavespeed and use that as a conservative upper bound on the wavespeed
at every point.  Limiting the progress at each step by a function of
the global maximum wavespeed is unnecessarily restrictive.  One would
like an algorithm that adapts to increasing wavespeeds so that fewer
spacetime elements, and therefore less computation time, are required
to mesh a given volume.  We develop such an algorithm, an extension to
Tent Pitcher, in Chapter~\ref{sec:nonlinear}.

The size of spacetime elements affects the numerical accuracy of the
approximate solution.  Where the exact solution is constant or
changing very little, large elements suffice to capture this small
variation.  Parts of the domain where the solution changes a lot need
to be meshed with smaller elements.  In Chapter~\ref{sec:adaptive} we
extend Tent Pitcher to make the size of spacetime elements adaptive to
\emph{a posteriori} error estimates.

Tent Pitcher is a good alternative to standard time-stepping methods.
Global time-marching schemes advance the solution from one time value
to another everywhere in the domain at once.  Computing the solution
for the new time value in this way has at least one of two
disadvantages---either the maximum possible time step is constrained
by the worst-quality element in the space mesh, or an arbitrarily
large coupled system has to be solved simultaneously.  The former
would correspond to the constraint that the height of each tentpole
erected by Tent Pitcher must be the same, and the latter would occur
if the causality constraint were not satisfied by Tent Pitcher.  Thus,
the advancing front approach avoids both limitations associated with
time-marching schemes. (Some recent advances have been made on
adaptive explicit time-stepping methods for
ODEs~\cite{eriksson03timestepping}.)

\subsection*{Chapter summary}

In this chapter, we gave a concise introduction to the properties of
hyperbolic partial differential equations (PDEs) as they concern this
thesis.  We briefly discussed the challenging problem of generating
good quality meshes to support fast and accurate numerical solutions
to PDEs.  We pointed out the advantages of spacetime discontinuous
Galerkin (SDG) methods over conventional finite element methods.  For
more about DG methods, the reader is referred to the
survey~\cite{cockburn00discontinuous} of discontinuous Galerkin
methods and their applications.

We outlined the specific problem of meshing in spacetime to support
SDG methods for solving hyperbolic PDEs, such as those modeling wave
propagation.  Meshing directly in spacetime presents unique challenges
not encountered in classical meshing problems.  It is hard to
formulate geometric criteria for a good quality spacetime element and
a good quality spacetime mesh, a problem that is already complicated
and application-dependent in Euclidean domains.  SDG methods, due to
their discontinuous formulation, permit nonconforming meshes and more
general mesh adaptivity operations.  We will exploit this increased
flexibility in developing meshing algorithms, extensions to the Tent
Pitcher algorithm, in the remaining chapters.

An advancing front algorithm for generating a spacetime mesh, as
exemplified by Tent Pitcher~\cite{ungor02phd,erickson05building}, has
several benefits.  Elements are added to the mesh in small subsets
called patches.  Each element in a patch has at least one outflow
facet.  Since the front at every step is causal, the front represents
all the information that is required for future iterations of the
algorithm.  The elements in a patch are coupled but two different
patches erected over the same front are independent.  Thus, the
solution procedure is highly parallelizable.  Solving each patch is
equivalent to performing a small finite element computation limited to
the patch.  Memory requirements are low because patches are created in
the order of causal dependence and hence can be discarded when they
are no longer adjacent to the current front.  The computation time
required for the simulation is reduced due to the limited dependency
only between elements in the same small patch and because the number
of patches is bounded.


\chapter{Basic advancing front meshing algorithm}
\label{sec:linear}

Tent Pitcher was the first
provably correct spacetime meshing algorithm to support an efficient
patch-wise solution strategy.  Tent Pitcher was proposed by \Ungor{}
and Sheffer~\cite{ungor00tentpitcher} and later extended by Erickson
\etal{}~\cite{erickson02building,erickson05building}.  Our advancing
front meshing algorithms are extensions to and augmentations of Tent
Pitcher.  In this chapter, we give an alternate derivation and proof
of correctness of the Tent Pitcher algorithm.  The style of the
derivation in this chapter anticipates extensions to nonlinear
problems and mesh adaptivity, problems that are considered in later
chapters.  Also, the alternate derivation in this chapter fixes an
error in the paper by Erickson \etal{}~\cite{erickson02building} for
the case of obtuse front triangles, an oversight that was corrected in
a subsequent journal paper~\cite{erickson05building}.

Our objective in this chapter is to set up a general framework for
advancing front spacetime meshing algorithms and highlight Tent Pitcher
as one specific member of this class of algorithms.  The more
sophisticated meshing algorithms described in later chapters fit in
this general framework.

The Tent Pitcher algorithm is the first instance of an advancing front
algorithm to mesh directly in spacetime.  The algorithm proceeds by
advancing a local neighborhood of the front in every step.  The
algorithm is simple because the geometric constraints that limit the
amount of progress at each step are linear or quadratic functions of
the coordinates.  For linear problems, the wavespeed~$\omega$ is a
fixed constant, determined by material properties, or can be
conservatively bounded by the global maximum wavespeed.  When the
wavespeed is constant, the amount of progress made by the front in
each local advancing step is limited only by local constraints.

\section{Problem statement}

The input is the initial front~$\tau_0$, which is a triangulation of
the space domain at time $t=0$, and the causal slope $\S$
(equivalently, the wavespeed $\omega = 1/\S$), which is determined by
the initial conditions of the PDE.  In this chapter, we assume that
the wavespeed~$\omega$ is a constant, either because the underlying
hyperbolic PDE is linear or because~$\omega$ is the global maximum
wavespeed.  We give an advancing front algorithm such that for every
$T \in \Real^{\ge 0}$ there exists a finite integer $k \ge 0$ such
that the front~$\tau_k$ after the $k$th iteration of the algorithm
satisfies $\tau_k \ge T$, which means that the entire front has
achieved or passed time~$T$.  The volume between each consecutive pair
of fronts~$\tau_i$ and~$\tau_{i+1}$ is triangulated to give a patch.

We will generate meshes with geometrically non-degenerate elements.
Specifically, we prove that each element has temporal aspect ratio
bounded from below.  Recall from Chapter~\ref{sec:intro} that the
temporal aspect ratio of an element is the ratio of its height to its
duration.

We say that a front~$\tau$ is \emph{valid} if for every $T \in
\Real^{\ge 0}$ there exists a finite sequence of fronts $\tau$, $\tau_1$,
$\tau_2$, $\ldots$, $\tau_k$ where $\tau_k \ge T$, each front in the
sequence obtained from the previous front by advancing some local
neighborhood.  Therefore, our problem can be defined as that of
constructing a succession of valid fronts such that the volume between
successive fronts is triangulated with a small number of spacetime
elements (simplices) each with bounded temporal aspect ratio.

\begin{definition}[Valid front]
  A front $\tau$ is \emph{valid} if there exists a positive real $\dt$
  bounded away from zero such that for every $T \in \Re^{\ge 0}$ there
  exists a finite sequence of fronts $\tau$, $\tau_1$, $\tau_2$,
  $\ldots$, $\tau_k$ where $\tau_k \ge T$, each front in the sequence
  obtained from the previous front by advancing some local
  neighborhood by $\dt$.
\label{def:validfront}
\end{definition}

The Tent Pitcher algorithm due to \Ungor{} and Sheffer, and Erickson
\etal{}, is obtained by making the following assumptions:
\begin{enumerate}
\item The front $\tau \equiv 0$, or more generally any constant time
  function $\tau \equiv T$, is valid.
\item The new front $\tau' \in \next(\tau)$ is obtained by advancing a single
  vertex of $\tau$ by a nonnegative amount, i.e., $N = \st(\fp)$ for
  some vertex $\fp$ of $\tau$.
\end{enumerate}

We say that a front~$\tau'$ is obtained by advancing a vertex~$\fp$
of~$\tau$ by~$\dt \ge 0$ to the vertex $\fp'$ of the front $\tau'$ if
$\tau'(p) = \tau(p) + \dt$, for every other vertex~$q \ne p$ we
have~$\tau'(q) = \tau(q)$, and every simplex incident on $\fp'$ is in
one-to-one correspondence with a simplex of $\tau$ incident on $\fp$.
For any front $\tau$, vertex $p$, and real $\dt \ge 0$, let
$\tau'=\next(\tau,p,\dt)$ denote the front obtained from $\tau$ by
advancing $p$ by $\dt$.

We say that a front~$\tau'$ is obtained by advancing a neighborhood
$N$ of~$\tau$ by~$\dt \ge 0$ to the neighborhood $N'$ of the front
$\tau'$ if $\dt = \max_{\fp=(p,\tau(p)) \in N}\,\{\tau'(p) -
\tau(p)\}$ and if $\tau \setminus N$ and $\tau' \setminus
N'$ coincide.

\parasc*{Our solution}

For one-dimensional space domains, we prove that every causal front is
valid.  In higher dimensions, we define \emph{progressive} fronts, and
we prove that if a front is progressive, then it is valid.  We give an
algorithm, a modification of Tent Pitcher, that given any progressive
front $\tau_i$ constructs a next front $\tau_{i+1}$ such that
$\tau_{i+1}$ is progressive.  The volume between $\tau_i$ and
$\tau_{i+1}$ is partitioned into simplices.  The new front
$\tau_{i+1}$ is obtained by advancing a local neighborhood of $\tau_i$
by a positive amount bounded away from zero.  The algorithm can be
parallelized in a straightforward manner to solve several patches
simultaneously by lifting any independent set of neighborhoods in
parallel.  Whenever the algorithm chooses to lift a local minimum
vertex, it is guaranteed to be able to lift it by at least $\minT > 0$
where $\minT$ is a function of the input and bounded away from zero.

In this chapter, we will restrict front advancing operations to tent
pitching, i.e., advancing $N = \st(p)$, for some vertex $p$, to $N'$
such that $\bd{N}=\bd{N'} = \link(p)$.  Since pitching a vertex does
not change the spatial projection, the triangulation of every front is
isomorphic to the initial space mesh.  In this sense, we say that the
meshing algorithm in this chapter is nonadaptive because the spatial
projection of every front is the initial triangulation of the space
domain.  In this chapter, we will assume that the underlying PDE is
linear and that $\S$ denotes the causal slope, the reciprocal of the
global wavespeed, a constant throughout spacetime.  Alternatively, if
the wavespeed is not constant, let $\S$ denote $\minS$, a lower bound
on the slope of any cone of influence in any spatial direction.

\section{Meshing in 1D$\times$Time}
\label{sec:linear:1d}

We begin by describing the linear nonadaptive Tent Pitcher algorithm
to construct spacetime meshes over one-dimensional space domains.  The
input space mesh, denoted by $\sp$, is a one-dimensional simplicial
complex, i.e., a set of vertices with pairs of vertices, say $p$ and
$q$, connected by a segment $pq$.  Let $\abs{pq}$ denote the length of
the segment $pq$, i.e., the Euclidean distance between $p$ and $q$.

At each step $i$, our algorithm chooses to advance an arbitrary local
minimum $p$ of the causal front $\tau_i$ to construct the new causal
front $\tau_{i+1}$ such that $\tau_{i+1}(p)$ is maximized.  Since
every front has at least one local minimum vertex, e.g., the global
minimum, the algorithm has a nonempty subset of candidate vertices to
pitch at each step.

Let $\fa\fb$ be an arbitrary segment of the front $\tau_{i+1}$. Then,
$\fa\fb$ is causal if and only if the gradient of the time function
$\tau_{i+1}$ restricted to $\aff{ab}$ is less than the slope $\S(\fa\fb)$,
i.e., if and only if
\begin{equation}
  \norm{\grad\rest{\tau_{i+1}}{ab}} 
:=
  \frac{\abs{\tau_{i+1}(b) - \tau_{i+1}(a)}}{\abs{ab}}
<
  \S(\fa\fb).
\label{eqn:1d:causalityconstraint}
\end{equation}

The new time coordinate $\tau'(p)$ of $p$ is constrained because every
segment $\fp'\fq$ of the new front is constrained as in
Equation~\ref{eqn:1d:causalityconstraint} such that $\tau'(p) <
\tau(q) + \abs{pq} \S$ for every segment $pq \in \st(p)$.

Note that the inequality in Equation~\ref{eqn:1d:causalityconstraint}
is a strict inequality, so the set of all $\tau_{i+1}(p)$, such that
the front $\tau_{i+1}$ is causal, is an open, convex set, whose
supremum value is not feasible.  Therefore, to maximize
$\tau_{i+1}(p)$, we compute the supremum time value $\supT$ that
satisfies the causality constraint of
Equation~\ref{eqn:1d:causalityconstraint} and then take any time value
less than $\supT$ but close enough to $\supT$ not to cause numerical
errors due to loss of precision.

\begin{figure*}[t]\centering
\begin{quote}
\textbf{Input:}  A one-dimensional space mesh $\sp \subset \mathbb{E}^1$\\[1em]

\textbf{Output:} A triangular mesh $\spt$ of $\sp \times [0,\infty)$\\[1em]

The initial front $\tau_0$ is $\sp \times \{0\}$, 
corresponding to time $t=0$ everywhere in space.\\[1em]

Repeat for $i = 0, 1, 2, \ldots$:

\begin{enumerate}

\item Advance in time an arbitrary local minimum vertex $\fp=(p,\tau_i(p))$
of the current front $\tau_i$ to $\fp'=(p,\tau_{i+1}(p))$ such that
$\tau_{i+1}$ is causal and $\tau_{i+1}(p)$ is maximized.

\item Partition the spacetime volume between $\tau$ and $\tau_{i+1}$
into a patch of triangles, each sharing the tentpole edge $\fp\fp'$.

\item Call the numerical solver to compute the solution everywhere in
the spacetime volume between $\tau_i$ and $\tau_{i+1}$ as well as the
\aposteriori{} error estimate.  The solution on $\tau_i$ is the inflow
information to the solver.

\end{enumerate}
\end{quote}
\caption{Advancing front algorithm in 1D$\times$Time.}
\label{fig:1d:alg}
\hrule
\end{figure*}

Figure~\ref{fig:1d:alg} describes our linear nonadaptive advancing
front meshing algorithm in 1D$\times$Time.  The solution everywhere in
spacetime is computed patch-by-patch in an order consistent with the
partial order of dependence until a particular target time $T \ge 0$
has been met.

In the parallel setting, we repeatedly choose an independent set of
local minima of the front $\tau_i$, equal to the number of processors,
to be advanced in time simultaneously.  The resulting patches can be
solved independently.  If a patch is accepted, the local neighborhood
of the front is advanced without any conflicts with other patches.

It is easy to see that each spacetime triangle has one causal inflow
facet (a segment) on the old front $\tau_i$ and one causal outflow
facet on the new front $\tau_{i+1}$.

\subsection{Progress guarantee}

It remains to show that the algorithm creates only a finite number of
triangles to mesh the volume $\sp \times [0,T]$ for any target time $T
\ge 0$ and that the minimum temporal aspect ratio of each triangle is
bounded.  Both requirements follow from the following lower bound on
the height of each tentpole.  Recall from Section~\ref{sec:notation}
that, for a vertex $\fp$ of the one-dimensional front $\tau$, the
distance $w_p$ is the minimum distance of $p$ from its nearest
neighbor in the spatial projection of $\tau$.

\begin{theorem}
  Let $\tau$ be a causal front and let $p$ be an arbitrary local
  minimum of $\tau$.  Then, for every $\dt$ such that $0 \le \dt < w_p
  \S$, the front $\tau' = \next(\tau,p,\dt)$ is causal.
\label{thm:1d:nextiscausal}
\end{theorem}

\ifproofs\begin{proof}[Proof of Theorem~\ref{thm:1d:nextiscausal}]
  Only the segments of the front incident on $\fp=(p,\tau(p))$ advance
  along with $p$.  Consider an arbitrary segment $pq$ incident on $p$.
  Since $p$ is a local minimum, we have $\tau(q) \ge \tau(p)$.  We
  have
\begin{align*}
\tau'(p)
&\le \tau(p) + \dt\\
&<   \tau(p) + w_p \S\\
&\le \tau(q) + \abs{pq} \S(\fp'\fq)
\end{align*}
  Therefore, the slope of the segment $\fp'\fq$ is less than
  $\S(\fp'\fq)$ and hence $\fp'\fq$ is causal.
\end{proof}\fi

Since each front $\tau$ is causal, the difference between the maximum
and the minimum time coordinate of points of $\tau$ must be smaller
than the diameter of the space domain $\sp$ times the slope $\S$ due
to causality.  Therefore, the algorithm must eventually advance the
global minimum vertex of $\tau$.  Thus, we have the following theorem.

\begin{theorem}
  For every $i \ge 0$, if the front $\tau_i$ is causal then $\tau_i$ is
  valid.
\label{thm:1d:causalisvalid}
\end{theorem}

\ifproofs\begin{proof}[Proof of Theorem~\ref{thm:1d:causalisvalid}]
Let $\minW$ denote the smallest length of the spatial projection
of a segment on any front.  Since causality limits the gradient of the
front, every vertex is pitched eventually when it becomes a local
minimum; specifically, a vertex becomes a global minimum after only a
finite number of iterations of the advancing front algorithm.  By
Theorem~\ref{thm:1d:nextiscausal}, each iteration advances the front
in time by a bounded finite amount.  Therefore, for any target time $T
\ge 0$, after a finite number of iterations, the entire front has
advanced past $T$.  Specifically, we show next that after at most
\[
  \ceil{\frac{n (T + \diam(\sp) \maxS)}{\minW \S}}
\]
iterations, the entire front has advanced up to or beyond the target
time $T$.  Here, $n$ denotes the maximum number of vertices and
$\minW$ denotes the minimum spatial length of any segment on any front
constructed by our algorithm.  Let $\minT = \minW \S$.

Consider step $i+1$ of the algorithm.  By
Theorem~\ref{thm:1d:nextiscausal} the front $\tau_{i+1}$ such that
$\tau_i(p) \le \tau_{i+1}(p) < \tau_i(p) + \minT$ is causal.
Therefore, we have shown that if $\tau_i$ is causal then there is a
front $\tau_{i+1} = \next(\tau_i,p,\dt)$ such that $\tau_{i+1}$ is
causal for every $\dt \in [0,\minT)$.  Note that $\sum_{p \in V(\sp)}
\tau_{i+1}(p) = \minT +
\sum_{p \in V(\sp)} \tau_i(p)$.  By induction on $i$, and because
$\S$ is finite and $\sp$ is bounded, there exists a finite $k \ge
i$ such that the front $\tau_k$ satisfies
\[
  \sum_{p \in V(\sp)} \tau_k(p) \ge n (T + \diam(\sp) \S)
\]
  for any real $T$.  Hence,
\[
  \max_{p \in V(\sp)} \tau_k(p)
\ge
  \frac{1}{n} \left( \sum_{p \in V(\sp)} \tau_k(p) \right)
\ge
  T + \diam(\sp) \S
\]
  Since $\tau_k$ is causal
\[
  \left( \max_{p \in V(\sp)} \tau_k(p) \right)
- \left( \min_{p \in V(\sp)} \tau_k(p) \right)
\le
  \diam(\sp) \S.
\]
Since $\max_{p \in V(\sp)} \tau_k(p) \ge T + \diam(\sp) \S$, it
follows that $\min_{p \in V(\sp)} \tau_k(p) \ge T$ and so $\tau_i$ is
valid.
\end{proof}\fi

In general, the Tent Pitcher algorithm is free to pitch any vertex,
not just local minima; however, the progress guarantee of
Theorem~\ref{thm:1d:nextiscausal} applies only to local minima.  Any
causal front can be input as the initial front $\tau_0$; thus, the
solution can be saved after each step and the algorithm can be
restarted from the last front.


\section{Meshing in 2D$\times$Time}
\label{sec:linear:2d}

In this section, we consider our advancing front algorithm in
2D$\times$Time.

\subsection{Need for progress constraints}

It has been observed by \Ungor{} and Sheffer and also by Erickson
\etal{} that in spatial dimension $d \ge 2$ the causality constraint
is not enough to guarantee that spacetime elements created by Tent
Pitcher are non-degenerate.  With causality constraints alone, it was
shown by \Ungor{} and Sheffer~\cite{ungor00tentpitcher}, and by
Erickson \etal{}~\cite{erickson02building} that if the space mesh
contains an obtuse or a right triangle then Tent Pitcher will
eventually construct a front such that no further progress is possible
while maintaining causality.  This is the case even for linear
PDEs where the wavespeed everywhere in spacetime is a fixed
constant.

Our algorithms advance a neighborhood $N$ of the front $\tau$ to the
neighborhood $N'$ of a new front $\tau'$ where $\bd{N} = \bd{N'}$.  It
is necessary that the lower-dimensional simplices of $\bd{N}$ satisfy
gradient constraints stricter that the causal cone constraint.

\begin{figure}
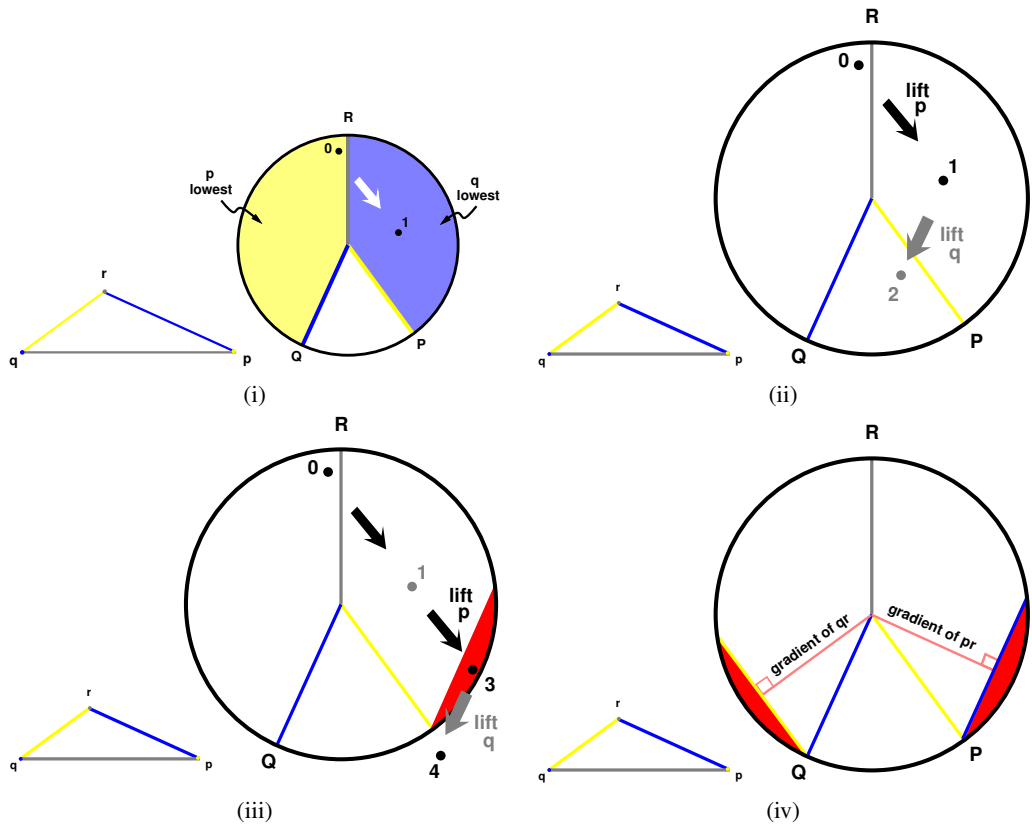
\centering\small
\begin{tabular}{cccc}
\includegraphics[width=.4\textwidth]{\fig{circle-1}}
&
\includegraphics[width=.4\textwidth]{\fig{circle-2}}
\\
(i) & (ii)\\
\includegraphics[width=.4\textwidth]{\fig{circle-3}}
&
\includegraphics[width=.4\textwidth]{\fig{circle-4}}\\
(iii) & (iv)
\end{tabular}
\caption[Vector diagram]{Vector diagram: Triangle $\fp\fq\fr$ is causal if
its gradient vector is inside the circular disc of radius $\S$.  To
ensure progress in the \emph{next} step, the gradient vector must also
lie outside the shaded red region.  [Damrong Guoy, personal
communication]}
\label{fig:vectordiagram}
\end{figure}

Erickson \etal{} introduced so-called \emph{progress constraints} on
the front at each stage.  Causality limits the magnitude of the
gradient of every simplex on each front.  The progress constraint is a
gradient constraint on certain lower dimensional faces.

The progress constraint imposed on $\tau$ is a gradient constraint on
certain edges of the front $\tau$.  For every local minimum vertex $p$
of the front $\tau$, the progress constraint limits the gradient of
the edges of $\link(p)$.  Figure~\ref{fig:vectordiagram} is the
\emph{vector diagram} that indicates the progress constraint imposed
on a single triangle of the front $\tau$.

\parasc*{Interpreting the vector diagram}

Consider a single triangle $\fp\fq\fr$ of the front $\tau$.  Let $pqr$
be the spatial projection of $\triangle{\fp\fq\fr}$.  Triangle
$\fp\fq\fr$ defines a plane in spacetime which is the graph of the
linear function $\rest{\tau}{pqr} : \mathbb{E}^2 \to \Real$.
The gradient of $\rest{\tau}{pqr}$, denoted by
$\grad\rest{\tau}{pqr}$ is a vector in $\Real^2$.  The direction
of $\grad\rest{\tau}{pqr}$ is the direction of steepest ascent
along the plane of $\triangle{\fp\fq\fr}$ and its magnitude is the
slope of this plane.

Figure~\ref{fig:vectordiagram} shows the spatial projection
$\triangle{pqr}$ on the left and the vector space $\Real^2$ on the
right.  A vector in $\Real^2$ can be indicated by an arrow with its
tail at the origin.  In Figure~\ref{fig:vectordiagram}, we simplify
the depiction of vectors and indicate a vector by a point representing
where the head of the arrow would be.

Triangle $\fp\fq\fr$ is causal if and only if the point representing
$\rest{\tau}{pqr}$ lies strictly inside the disk centered at the
origin with radius equal to $\S(\fp\fq\fr)$.  In
Figure~\ref{fig:vectordiagram}, we indicate the global slope $\S$ by a
disk centered at the origin with radius~$\S$.

Suppose we advance $\fp$ to $\fp'$ giving the triangle $\fp'\fq\fr$ of a
new front $\tau'$.  Lifting $\fp$ keeps the gradient along $\aff{qr}$
unchanged, i.e., $\rest{\tau'}{\aff{qr}} = \rest{\tau}{\aff{qr}}$,
while increasing the magnitude of the component orthogonal to $qr$ in
the direction of~$p$.

Vectors $\vec{OP}$, $\vec{OQ}$, and $\vec{OR}$ in
Figure~\ref{fig:vectordiagram} are orthogonal to $qr$, $pr$, and $pq$
respectively and oriented in the direction of the third vertex in each
case.  These three normal vectors subdivide $\Real^2$ into three
zones.  Each zone is identified with a vertex of $\triangle{pqr}$
because whenever $\grad\rest{\tau}{pqr}$ lies in that zone the
corresponding vertex is a local minimum vertex of
$\triangle{\fp\fq\fr}$.  For instance, if the gradient vector
$\grad\rest{\tau}{pqr}$ lies in zone $p$, bounded by the vectors
$\vec{OQ}$ and $\vec{OR}$, then ${\tau(p) \le \tau(q)}$ and ${\tau(p) \le
\tau(r)}$.

Consider the situation depicted in Figure~\ref{fig:vectordiagram}(i).
Points $0$ and $1$ indicate respectively the gradient vector of
$\triangle{\fp\fq\fr}$ where $p$ is a local minimum vertex and the
gradient vector of $\triangle{\fp'\fq\fr}$ after pitching $\fp$ to
$\fp'$ so that $q$ is the new local minimum vertex.  When $\fp$ is
pitched to $\fp'$, the gradient vector advances from $0$ to $1$ in the
direction $\vec{OP}$.  Advancing from point $0$ to point $1$
represents positive progress because both points are inside the disk
(the corresponding triangles are causal) and $p$ is no longer a local
minimum.

Figure~\ref{fig:vectordiagram}(ii) indicates the possible situation in
the next iteration of the algorithm when the new local minimum $\fq$
is advanced---the gradient vector makes positive progress along
direction $\vec{OQ}$ from point $1$ to point $2$.  If the algorithm is
too greedy and advances from point $0$ to point $3$ in the current
step, as shown in Figure~\ref{fig:vectordiagram}(iii), then sufficient
progress cannot be guaranteed in the next step because the gradient
vector may leave the disk (point $4$) and violate causality.
Therefore, it is necessary to ensure the gradient vector of
$\triangle{\fp'\fq\fr}$ does not enter the shaded forbidden zone shown
in Figure~\ref{fig:vectordiagram}(iii).  The forbidden zone in
Figure~\ref{fig:vectordiagram}(iii) limits the magnitude of the
gradient of the edge $pr$ on the new front $\tau'$, i.e., it imposes
an upper bound on $\tau'(p) - \tau(r)$.

Symmetrically, when the local minimum $q$ is pitched from $\fq$ to
$\fq'$, there is a corresponding forbidden zone that limits the slope
of the edge $\fq'\fr$.  See Figure~\ref{fig:vectordiagram}(iv).

Note that the forbidden zones (and therefore the progress constraint)
depends solely on the shape and orientation of $\triangle{pqr}$ and is
invariant with respect to uniform scaling.

\subsection{Progressive fronts in 2D$\times$Time}

Advancing a vertex $p$ in time increases the gradient
of simplices in $\st(p)$ but leaves unchanged the gradient of the
lower-dimensional simplices in $\link(p)$.  For each face $f \in
\link(p)$, the gradient of the front in a directional orthogonal to
$f$ increases.  Our algorithm advances an arbitrary local minimum
vertex $p$ of each front $\tau$ to get a new front $\tau'$.  The new front
$\tau'$ after advancing $p$ is causal if and only if the gradient of
$f$ is strictly less than the slope of the causal cone constraint that
limits the gradient of the simplex $F \supset f$ such that $F \in
\st(p)$.

\begin{figure}[t]
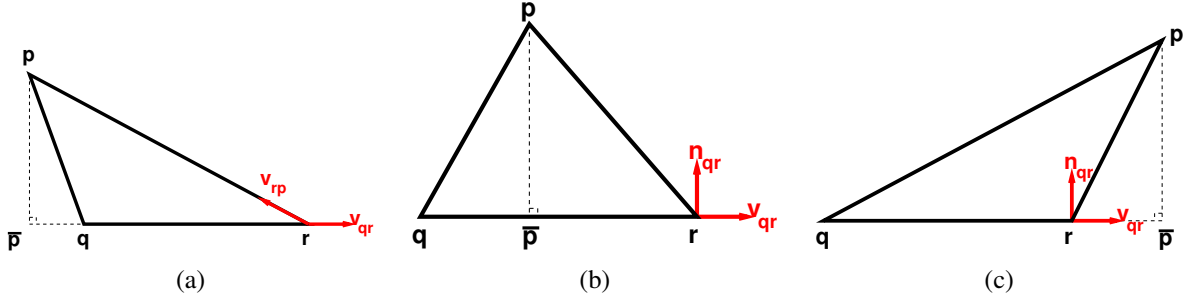
\centering
\begin{tabular}{ccc}
\includegraphics[width=0.3\linewidth]{\fig{pqr-Qobtuse}}
&
\includegraphics[width=0.3\linewidth]{\fig{pqr-QRacute}}
&
\includegraphics[width=0.3\linewidth]{\fig{pqr-Robtuse}}\\
(a) & (b) & (c)
\end{tabular}
\caption{Triangle $pqr$ of the front $\tau$ where $\tau(p) \le \tau(q) \le \tau(r)$}
\label{fig:pqr}
\label{fig:pqr-Qobtuse}
\label{fig:pqr-QRacute}
\label{fig:pqr-Robtuse}
\end{figure}

\begin{lemma}
  Let $\triangle{\fp\fq\fr}$ be a triangle of the causal front $\tau$
  with $\fp$ as its lowest vertex. Without loss of generality, assume
  $\tau(p) \le \tau(q) \le \tau(r)$.  Let $\S > 0$ be a slope such
  that $\S \le 1/\omega(\fp\fq\fr)$ and let $d_p$ denote
  $\dist(p,\aff{qr})$.  Let $\e$ be any real number in the range $0 <
  \e < 1$.  Consider two cases:
\begin{enumerate}
\item \textbf{Case $\measuredangle{pqr} \le \pi/2$:}
  If $\norm{\grad\rest{\tau}{qr}} \le (1-\e) \S$, then the front $\tau'
  = \next(\tau,p,\dt)$ satisfies $\norm{\grad\tau'} < \S$ for every $\dt \in
  [0,\e d_p \S]$.
\item \textbf{Case $\pi/2 < \measuredangle{pqr} < \pi$:}
  If $\norm{\grad\rest{\tau}{qr}} \le (1-\e) \S \sin\angle{pqr}$, then
  the front $\tau' = \next(\tau,p,\dt)$ satisfies ${\norm{\grad\tau'}
  < \S}$ for every $\dt \in [0,\e d_p \S]$.
\end{enumerate}
\label{lemma:2d:nextiscausal}
\end{lemma}

\ifproofs\begin{proof}[Proof of Lemma~\ref{lemma:2d:nextiscausal}]
Let $\bp$ be the orthogonal projection of $p$ onto the line $\aff{qr}$.

As long as $\tau'(p) \le \tau(q)$, we have $\norm{\grad\tau'} \le
\norm{\grad\tau}$ and the statement of the lemma follows.  Hence, for
the rest of this proof, assume $\tau'(p) > \tau(q)$.

Let $\vec{n}_{qr}$ denote the unit vector normal to $qr$ such
that $\vec{n}_{qr} \cdot (\vec{p} - \vec{q}) > 0$.  Let $\vec{v}_{qr}$
be the unit vector parallel to $qr$ such that $\vec{v}_{qr} \cdot
(\vec{r} - \vec{q}) > 0$.  Then, $\{\vec{n}_{qr}, \vec{v}_{qr}\}$ form
a basis for the vector space $\Real^2$.

Lifting $p$ does not change the gradient of the time function
restricted to the opposite edge, so $\grad \tau' \cdot \vec{v}_{qr} =
\grad \tau \cdot \vec{v}_{qr}$, i.e., $\grad \rest{\tau'}{qr} = \grad
\rest{\tau}{qr}$.  The scalar product $\grad \tau' \cdot \vec{n}_{qr}$
can be written as
\[
  \grad \tau' \cdot \vec{n}_{qr} = \frac{\tau'(p)-\tau(\bp)}{\abs{p \bp}}
\]
Since $q$ is the lowest vertex of $qr$, we have $\grad \tau'
\cdot \vec{v}_{qr} \ge 0$.  Also, we are given that $\grad \tau \cdot
\vec{v}_{qr} \le (1-\e) \S < \S$.  Note that $\norm{\grad\tau} \le
\grad \tau \cdot \vec{v}_{qr}$.
Therefore, $\norm{\grad \tau'} < \S$ if and only if
\begin{equation}\boxed{%
  \frac{\tau'(p) - \tau(\bp)}{\abs{p \bp}}
<
  \root{\S^2 - \norm{\grad \rest{\tau}{qr}}^2}
\label{eqn:2d:causalityconstraint}
}\end{equation}

\noindent\textbf{Case 1: $\mathbf{\angle{pqr}}$ is non-obtuse.}
See Figure~\ref{fig:pqr-QRacute}(b)--(c).  In this case, we have
$\tau(\bp) \ge \tau(q) \ge \tau(p)$.  We have
\[
\tau'(p)
=   \tau(p) + \dt
\le \tau(\bp) + \dt
\le \tau(\bp) + \e \S \abs{p \bp}
\]
Since $0 < \e < 1$ we have $\e < \root{1 - (1-\e)^2}$.
Therefore,
\[
  \tau'(p)
< 
  \tau(\bp) + \root{1 - (1-\e)^2} \, \S \abs{p \bp}
=
  \tau(\bp) + \root{\S^2 - (1-\e)^2 \S^2} \, \abs{p \bp}
\]
Since $\norm{\grad\rest{\tau}{qr}} \le (1-\e) \S$, we have
\[
  \tau'(p)
<
  \tau(\bp) + \abs{p \bp} \root{\S^2 - \norm{\grad \rest{\tau}{qr}}^2}
\]
which is precisely the causality constraint of
Equation~\ref{eqn:2d:causalityconstraint}.

\noindent\textbf{Case 2: $\mathbf{\angle{pqr}}$ is obtuse.}
See Figure~\ref{fig:pqr-Qobtuse}(a).  In this case, we have $\tau(\bp) <
\tau(q)$.

Let $\beta = \abs{\bp q}/\abs{p \bp}$.
Since $\abs{\bp q} \ne 0$, we have
\begin{equation}
  \frac{\tau'(p) - \tau(\bp)}{\abs{p \bp}}
= 
  \frac{\tau'(p) - \tau(q)}{\abs{p \bp}} 
+ 
  \frac{\tau(q) - \tau(\bp)}{\abs{\bp q}} \frac{\abs{\bp q}}{\abs{p \bp}}
= 
  \frac{\tau'(p) - \tau(q)}{\abs{p \bp}} 
+ 
  \beta \norm{\grad \rest{\tau}{qr}}
\label{eqn:2d:strengthen}
\end{equation}
Using Equation~\ref{eqn:2d:strengthen}, the causality constraint
(Equation~\ref{eqn:2d:causalityconstraint}) can be rewritten as
\begin{equation}
  \frac{\tau'(p) - \tau(q)}{\abs{p \bp}}
<
  \root{\S^2 - \norm{\grad \rest{\tau}{qr}}^2}
- \beta \norm{\grad \rest{\tau}{qr}}
\label{eqn:2d:equiv_causalityconstraint}
\end{equation}
We are given that $\norm{\grad \rest{\tau}{qr}} \le (1-\e) \S
\sin\angle{pqr}$.  Substituting this upper bound on $\norm{\grad
\rest{\tau}{qr}}$ into
Equation~\ref{eqn:2d:equiv_causalityconstraint}, we obtain the
following constraint:
\begin{equation}
  \frac{\tau'(p) - \tau(q)}{\abs{p \bp}}
<
  \S \left( \root{1 - (1-\e)^2 \sin^2\angle{pqr}} \right)
- \S (1-\e) \beta \sin\angle{pqr}
\label{eqn:2d:strong_causalityconstraint}
\end{equation}
which implies the causality constraint of
Equation~\ref{eqn:2d:equiv_causalityconstraint}.
Since $\tau(p) \le \tau(q)$ and $\tau'(p) \le \tau(p) + \e \S
\abs{p \bp}$, we have
\[
  \frac{\tau'(p) - \tau(q)}{\abs{p \bp}}
\le 
  \frac{\tau'(p) - \tau(p)}{\abs{p \bp}}
\le
  \frac{\e \S \abs{p \bp}}{\abs{p \bp}}
=
  \e \S.
\]
Equation~\ref{eqn:2d:strong_causalityconstraint} is satisfied if
\[
  \e < \root{1 - (1-\e)^2 \sin^2\angle{pqr}} - (1-\e) \beta \sin\angle{pqr}
\]
or equivalently
\[
  \left( \e + (1-\e) \beta \sin\angle{pqr} \right)^2 
+
  (1-\e)^2 \sin^2\angle{pqr}
<
  1
\]
Let $\lambda$ denote the left-hand side of the last inequality above.

We have
\begin{align*}
\lambda
&=
  \left( \e + (1-\e) \beta \sin\angle{pqr} \right)^2 
+ (1-\e)^2 \sin^2\angle{pqr}\\
&=
  \e^2 
+ (1-\e)^2 \beta^2 \sin^2\angle{pqr} 
+ 2 \e (1-\e) \beta \sin\angle{pqr}
+ (1-\e)^2 \sin^2\angle{pqr}\\
&=
  \e^2 
+ (1-\e)^2 (\beta^2 + 1) \sin^2\angle{pqr} 
+ 2 \e (1-\e) \beta \sin\angle{pqr}
\end{align*}

Note that
\[
  \beta^2 + 1
= \frac{\abs{\bp q}^2 + \abs{p \bp}^2}{\abs{p \bp}^2}
= \frac{\abs{pq}^2}{\abs{p \bp}^2}
\]
by Pythagoras' theorem applied to $\triangle{puq}$.  Also,
$\sin\angle{pqr} = \abs{p \bp}/\abs{pq}$. Hence,
\[
  (\beta^2+1) \sin^2\angle{pqr} = 1.
\]
Additionally,
\[
  \beta \, \sin\angle{pqr}
=
  \frac{\abs{\bp q}}{\abs{p \bp}} \,
  \frac{\abs{p \bp}}{\abs{pq}}
=
  - \cos\angle{pqr}
\]

Therefore, 
\begin{align*}
\lambda
&=
  \e^2 
+ (1-\e)^2
- 2 \e (1-\e) \cos\angle{pqr}\\
&=
  \e^2 
+ 1 + \e^2 - 2\e
- 2 \e (1-\e) \cos\angle{pqr}\\
&=
  1
+ 2 \e^2 
- 2\e
- 2 \e (1-\e) \cos\angle{pqr}\\
&=
  1
- 2\e (-\e + 1 + (1-\e) \cos\angle{pqr})\\
&=
  1 - 2 \e (1-\e) (1 + \cos\angle{pqr})\\
&<
  1
\end{align*}
The last inequality follows because $\triangle{pqr}$ is
non-degenerate, so $\cos\angle{pqr} \ne -1$; also, both
$\e$ and $(1-\e)$ are positive. Therefore, the causality constraint is
satisfied.
\end{proof}\fi

\begin{figure}
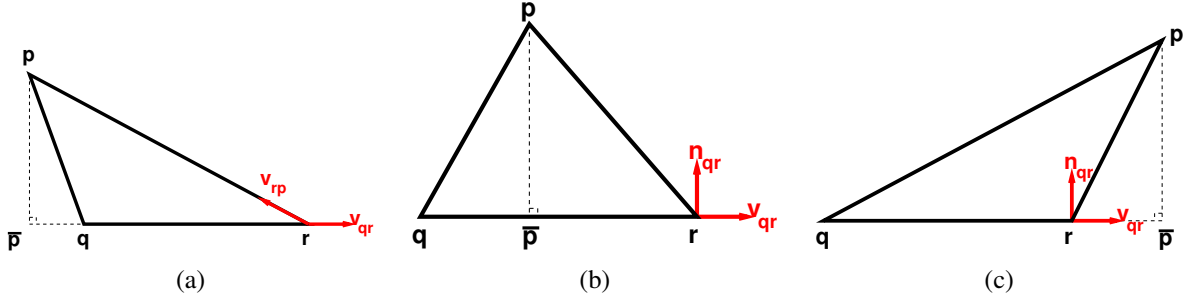
\centering
\begin{tabular}{ccc}
\includegraphics[width=0.3\linewidth]{\fig{pqr-Qobtuse}}
&
\includegraphics[width=0.3\linewidth]{\fig{pqr-QRacute}}
&
\includegraphics[width=0.3\linewidth]{\fig{pqr-Robtuse}}\\
(a) & (b) & (c)
\end{tabular}
\caption[Triangle $pqr$ with (a)~$\angle{pqr}$ obtuse;
(b)~$\angle{pqr}$ and $\angle{prq}$ acute; and (c)~$\angle{prq}$
obtuse]{Triangle $pqr$ with (a)~$\angle{pqr}$ obtuse, so $\phi_{qr} =
\sin\angle{pqr} < 1$, $\phi(pqr) = \phi_{qr}$; (b)~$\angle{pqr}$ and
$\angle{prq}$ acute, so $\phi_{qr} = 1$, $\phi(pqr) = \sin\angle{rpq}
< 1$; and (c)~$\angle{prq}$ obtuse, so $\phi_{qr} = \sin\angle{qrp} <
1$, $\phi(pqr) = \phi_{qr}$}
\label{fig:2d:phi}
\end{figure}

The progress constraint in 2D$\times$Time is motivated by the
Lemma~\ref{lemma:2d:nextiscausal} which states that if
$\triangle{\fp\fq\fr}$ is causal and satisfies progress constraint
$\S=1/\omega(\fp\fq\fr)$ then $\triangle{\fp'\fq\fr}$ obtained by
advancing a lowest vertex $\fp$ in time by $\e\,\dist(p,\aff{qr})\,\S$
is also causal.

For each simplex (an edge or a triangle) $f$ in $\st(p) \cup
\link(p)$, causality and the progress constraint limit the gradient of
$f$, separately for each simplex $f$.  Therefore, without loss of
generality, we consider each such simplex $f$ separately while
defining and computing gradient constraints on $f$.

The progress constraint is parameterized by $\e$ and a wavespeed $\S$.
We are free to choose any value for $\e$ in the allowed range.  The
bound on the gradient of $\tau$ due to the progress constraint is
proportional to the local geometry of the spatial projection of
$\tau$.

\begin{definition}
  For a triangle $pqr$, for any arbitrary edge of the triangle, say
  the edge $qr$, define $\phi_{qr} := 1$ if both $\angle{pqr}$ and
  $\angle{qrp}$ are non-obtuse; otherwise let $\phi_{qr}$ denote the
  sine of the obtuse angle of $\triangle{pqr}$, i.e., $\phi_{qr}
  := \max\,\{\,\sin\angle{pqr},\allowbreak\, \sin\angle{qrp}\,\}$
  Define $\phi(pqr)$ to be the minimum
  $\phi_e$ over every edge $e$ of the triangle.  Note that ${0 <
  \phi(pqr) \le 1}$ and $\phi(pqr) < 1$ if and only if $\triangle{pqr}$
  is obtuse (Figure~\ref{fig:2d:phi}).
\label{def:2d:phi}
\end{definition}

\begin{definition}[Progress constraint $\S$]
  Fix $\e$ in the range $0 < \e < 1$.
  Let $\fp\fq\fr$ be an arbitrary triangle of a front $\tau$. We say
  that the triangle $\fp\fq\fr$ satisfies \emph{progress constraint
  $\S$} if and only if for every lowest vertex $\fp$, equivalently for
  every highest edge $\fq\fr$, we have
  \[
    \norm{\grad \rest{\tau}{qr}}
  :=
    \frac{\tau(r) - \tau(q)}{\abs{qr}}
  \le
    (1-\e) \S \phi_{qr}.
  \]
\label{def:2d:linear:progressconstraint}
\end{definition}

We say that a triangle $\fp\fq\fr$ is \emph{progressive} if it satisfies
progress constraint $\S$.  We say that a front $\tau$ is
\emph{progressive} if every triangle of $\tau$ is progressive.
Note that every progressive triangle or front is also causal.

Suppose a lowest vertex $\fp$ is being advanced.  As long as $\fp$ is
the lowest vertex of $\triangle{\fp\fq\fr}$, the progress constraint limits
$\norm{\grad \rest{\tau'}{qr}}$ but $\norm{\grad \rest{\tau'}{qr}} =
\norm{\grad \rest{\tau}{qr}}$.  Assume without loss of generality that
$\tau(p) \le \tau(q) \le \tau(r)$.  When ${\tau'(p) > \tau(q)}$, the new
lowest vertex is $q$, so the progress constraint limits $\norm{\grad
\rest{\tau'}{rp}}$.  (We can interpret the progress constraint
inductively as a causality constraint on the $1$-dimensional facet
$pr$ opposite $q$ where the relevant slope is $(1-\e) \S \phi_{rp}$.)

In the parallel setting, we repeatedly choose an independent set of
local minima of the front $\tau_i$, equal to the number of processors,
to be advanced in time simultaneously.  The resulting patches can be
solved independently.  If a patch is accepted, the local neighborhood
of the front is advanced without any conflicts with other patches.

In subsequent sections, we will successively strengthen the progress
constraint and correspondingly strengthen our notion of progressive
fronts.  When we refer to a front $\tau$ simply as a progressive
front, we will implicitly understand this to mean that $\tau$ is
causal and $\tau$ satisfies the progress constraint as defined in the
most recent definition.

\begin{figure*}\centering
\begin{quote}
\textbf{Input:}  A bounded interval $M \subset \Real$\\
Fix $\e \in (0,1)$, a parameter to the algorithm\\[1em]

\textbf{Output:} A triangular mesh $\spt$ of $M \times [0,\infty)$\\[1em]

The initial front $\tau_0$ is $\sp \times \{0\}$, 
corresponding to time $t=0$ everywhere in space.\\[1em]

Repeat the following sequence of steps for $i = 0, 1, 2, \ldots$:

\begin{enumerate}

\item Advance in time an arbitrary local minimum vertex $\fp=(p,\tau_i(p))$
of the current front $\tau_i$ to $\fp'=(p,\tau_{i+1}(p))$ such that
$\tau_{i+1}$ is \emph{progressive} and $\tau_{i+1}(p)$ is maximized.

\item Partition the spacetime volume between $\tau$ and $\tau_{i+1}$
into a patch of triangles, each sharing the tentpole edge $\fp\fp'$.

\item Call the numerical solver to compute the solution everywhere in
the spacetime volume between $\tau_i$ and $\tau_{i+1}$ as well as the
\aposteriori{} error estimate.  The solution on $\tau_i$ is the inflow
information to the solver and the solution on $\tau_{i+1}$ is the
outflow information.

\end{enumerate}
\end{quote}
\caption{Advancing front algorithm in 2D$\times$Time.}
\label{fig:2d:alg}
\hrule
\end{figure*}

The value of $\tau_{i+1}(p)$ is constrained from above by causality
and progress constraints separately for each of the simplices incident
on $p$.  The final value chosen by the algorithm must satisfy the
constraints for each such triangle.  Therefore, it is sufficient to
consider each triangle $pqr$ incident on $p$ separately while deriving
the causality and progress constraints that apply while pitching $p$.

The progress of the front at the $(i+1)$th step is defined to be
$\tau_{i+1}(p) - \tau_i(p)$.  We have already shown in
Lemma~\ref{lemma:2d:nextiscausal} a lower bound on the amount of
progress when subject to causality constraints alone.  Next, we prove
a lower bound on the amount of progress when subject to progress
constraints alone.  Thus, we obtain a lower bound on the worst-case
amount of progress when subject to all causality and progress
constraints simultaneously.

The next lemma states that if $\triangle{\fp\fq\fr}$ is causal and
satisfies progress constraint $\S=1/\omega(\fp\fq\fr)$ then
$\triangle{\fp'\fq\fr}$ obtained by advancing a lowest vertex $\fp$ in
time by $(1-\e) \, \phi(pqr) \allowbreak\, \dist(p,\aff{qr}) \, \S$
also satisfies progress constraint $\S$.

\newpage

\begin{lemma}
  Let $\fp\fq\fr$ be a triangle of the causal front~$\tau$ with~$\fp$
  as its lowest vertex. Let $\S > 0$ denote a slope and let~$d_p$
  denote $\dist(p,\aff{qr})$.  Let~$\e$ be any real number in the
  range $0 < \e < 1$.

  If $\triangle{\fp\fq\fr}$ satisfies progress constraint~$\S$, then
  for every $\dt \in [0,(1-\e) \S d_p]$ the front $\tau' =
  \next(\tau,p_0,\dt)$ satisfies progress constraint~$\S$.
\label{lemma:2d:nextisprogressive}
\end{lemma}

\ifproofs\begin{proof}[Proof of Lemma~\ref{lemma:2d:nextisprogressive}]
By Definition~\ref{def:2d:linear:progressconstraint}, the front $\tau'$
satisfies progress constraint $\S$ if and only if
\begin{eqnarray*}
    \norm{\grad\rest{\tau'}{pq}} &\le& (1-\e) \phi_{pq} \S\\
    \norm{\grad\rest{\tau'}{qr}} &\le& (1-\e) \phi_{qr} \S\\
    \norm{\grad\rest{\tau'}{rp}} &\le& (1-\e) \phi_{rp} \S
\end{eqnarray*}

Since $\triangle{\fp\fq\fr}$ satisfies progress constraint $\S$, we
have
\begin{eqnarray*}
    \norm{\grad\rest{\tau}{pq}} &\le& (1-\e) \phi_{pq} \S\\
    \norm{\grad\rest{\tau}{qr}} &\le& (1-\e) \phi_{qr} \S\\
    \norm{\grad\rest{\tau}{rp}} &\le& (1-\e) \phi_{rp} \S
\end{eqnarray*}
Also, $p$ is a lowest vertex of $\tau$; so $\tau(p) \le
\min\{\tau(q),\tau(r)\}$.

Since advancing $p$ in time does not change the time function along
$qr$, we have $\norm{\grad\rest{\tau'}{qr}} =
\norm{\grad\rest{\tau}{qr}} \le (1-\e) \phi_{qr} \S$.

Consider the constraint $\norm{\grad\rest{\tau'}{pq}} \le (1-\e)
\phi_{pr} \S$.  As long as $\tau'(p) \le \tau(q)$, we have
$\norm{\grad\rest{\tau'}{pq}} \le \norm{\grad\rest{\tau}{pq}} \le
(1-\e) \phi_{pq} \S$.  Similarly, as long as $\tau'(p) \le \tau(r)$,
the constraint $\norm{\grad\rest{\tau'}{rp}} \le (1-\e) \phi_{rp} \S$
is automatically satisfied because $\norm{\grad\rest{\tau'}{rp}} \le
\norm{\grad\rest{\tau}{rp}} \le (1-\e) \phi_{rp} \S$.

When $\tau'(p) > \tau(q)$, the constraint
$\norm{\grad\rest{\tau'}{pq}} \le (1-\e) \phi_{pq} \S$ is equivalent to
$\tau'(p) \le \tau(q) + \abs{pq} (1-\e) \phi_{pq} \S$.  We have
\begin{align*}
  \tau'(p)
&\le
  \tau(p) + (1-\e) \S d_p\\
&\le
  \tau(q) + (1-\e) \S \min\{\abs{pq},\abs{pr}\} \phi_{qr}\\
&\le
  \tau(q) + (1-\e) \phi_{pq} \S \abs{pq}\\
&\le
  \tau(q) + (1-\e) \phi(pqr) \S \abs{pq}
\end{align*}
The last inequality follows because $\phi(pqr) \le \phi_{pq}$.

Similarly, when $\tau'(p) > \tau(r)$, the constraint
$\norm{\grad\rest{\tau'}{rp}} \le (1-\e) \phi_{rp} \S$ is equivalent to
$\tau'(p) \le \tau(r) + (1-\e) \phi_{rp} \S \abs{pr}$.  We have
\begin{align*}
  \tau'(p)
&\le
  \tau(p) + (1-\e) \S d_p\\
&\le
  \tau(r) + (1-\e) \S \min\{\abs{pq},\abs{pr}\} \phi_{rp}\\
&\le
  \tau(r) + (1-\e) \phi_{rp} \S \abs{rp}\\
&\le
  \tau(r) + (1-\e) \phi(pqr) \S \abs{rp}
\end{align*}
\end{proof}\fi

\bigskip

Note that we have in fact proved the following two statements that
imply Lemmas~\ref{lemma:2d:nextiscausal}
and~\ref{lemma:2d:nextisprogressive}.

Let $\triangle{\fp\fq\fr}$ be a triangle of the front $\tau$ with
$\fp$ as one of its lowest vertices. Without loss of generality,
assume $\tau(p) \le \tau(q) \le \tau(r)$.  Let $\e$ be any real number
in the range $0 < \e < 1$.  Let $\S > 0$ be any slope.

\begin{enumerate}
\item
  If $\norm{\grad\rest{\tau}{qr}} \le (1-\e) \S \phi_{qr}$, then
  $\triangle{\fp'\fq\fr}$ satisfies $\norm{\grad\tau'} < \S$ for every
  $\tau'(p)$ in the range $\tau(q) \le \tau'(p) \le \tau(q) + \e \S
  \dist(p,\aff{qr})$.

\item
  $\triangle{\fp'\fq\fr}$ satisfies 
\[
  \norm{\grad\rest{\tau'}{pq}} \le (1-\e) \S \phi_{pq}
\qquad\text{and}\qquad
  \norm{\grad\rest{\tau'}{pr}} \le (1-\e) \S \phi_{pr}
\]
  for every $\tau'(p)$ in the range $\tau(q) \le \tau'(p) \le \tau(q) 
  + (1-\e) \S \min\{\abs{pq},\abs{pr}\}$.
\end{enumerate}

The above claims are stronger than Lemmas~\ref{lemma:2d:nextiscausal}
and~\ref{lemma:2d:nextisprogressive} because we do not assume that
$\triangle{\fp\fq\fr}$ and the front $\tau$ is progressive.

\bigskip

As a corollary to Lemmas~\ref{lemma:2d:nextiscausal}
and~\ref{lemma:2d:nextisprogressive}, we conclude a
positive lower bound on the minimum tentpole height, similar to that
obtained by Erickson \etal{}~\cite{erickson02building} in the
nonadaptive case when the wavespeed everywhere is a fixed constant
$\omega$.  Let $\S$ denote the causal slope.

\begin{corollary}
  Let~$\tau$ be a causal front. Let~$\e$ be any real number in the
  range ${0 < \e < 1}$.  If every triangle $\triangle{\fp\fq\fr}$
  of~$\tau$ satisfies progress constraint~$\S$, then, for any local
  minimum~$p$ of~$\tau$, the front $\tau' = \next(\tau,p,\dt)$ is
  causal and satisfies the progress constraint~$\S$ for every $\dt \in
  [0, \min\{\e,(1-\e)\} w_p \S]$.
\label{cor:2d:positiveprogress}
\end{corollary}

\ifproofs\begin{proof}[Proof of Corollary~\ref{cor:2d:positiveprogress}]
Causality and progress constraints apply to each triangle $\fp\fq\fr$
of the front incident on $\fp$.  We show that each constraint
individually guarantees a positive lower bound of
$\min\{\e,(1-\e)\} w_p \S$ on the height $\tau'(p)-\tau(p)$
of the height of the tentpole at $p$.  Therefore, all the constraints
together ensure a tentpole height of at least as much.

Applying Lemma~\ref{lemma:2d:nextiscausal} to each triangle
$\fp\fq\fr$, we conclude that the front~$\tau'$ is causal for every
$\dt \in [0,\e w_p \S]$.

Consider an arbitrary triangle $\fp\fq\fr$ incident on~$\fp$.
Advancing~$p$ in time does not change the gradient of the front
along~$qr$.  Therefore, the gradient of the new front is limited only
along~$pq$ and~$pr$ by the progress constraint.  By
Lemma~\ref{lemma:2d:nextiscausal}, we conclude that each of the
gradient constraints along~$pq$ and~$pr$ guarantee a tentpole height
of at least ${(1-\e)} \, \dist(p,\aff{qr})
\,\S$. Applying Lemma~\ref{lemma:2d:nextiscausal} to each triangle
$\fp\fq\fr$, we conclude that the front~$\tau'$ satisfies the progress
constraint~$\S$ for every $\dt \in [0,(1-\e) w_p \S]$.
\end{proof}\fi

\begin{theorem}
  For any $i \ge 0$, if the front $\tau_i$ is progressive then $\tau_i$ is
  valid.
\label{thm:2d:progressiveisvalid}
\end{theorem}

\ifproofs\begin{proof}[Proof of Theorem~\ref{thm:2d:progressiveisvalid}]
The proof is identical to that of Theorem \ref{thm:1d:causalisvalid}
and follows because we choose $\e > 0$ so that $\minT > 0$ is bounded
away from zero.
\end{proof}\fi

We thus have the following theorem.

\begin{theorem}
  Given a triangulation $\sp$ of a bounded planar space domain where
  $\minW$ is the minimum width of a simplex of $\sp$ and $\S$ is the
  minimum slope anywhere in $\sp \times [0,\infty)$, for every $\e$
  such that $0 < \e < 1$ our algorithm constructs a simplicial mesh of
  $\sp \times [0,T]$ consisting of at most $\ceil{\frac{n (T +
  \diam(\sp) \S)}{\min\{\e,1-\e\} \S \minW} \Delta}$ spacetime
  elements for every real $T \ge 0$ in constant computation time per
  element, where $\Delta$ is the maximum vertex degree.
\label{thm:2d:main}
\end{theorem}
\ifproofs\begin{proof}[Proof of Theorem~\ref{thm:2d:main}]
  By Lemmas \ref{lemma:2d:nextiscausal} and
  \ref{lemma:2d:nextisprogressive}, it follows that the height of each
  tentpole constructed by the algorithm is at least $\minT =
  \min\{\e,1-\e\} \S \minW$.  By
  Theorem~\ref{thm:2d:progressiveisvalid}, after constructing at most
  $k \le \ceil{\frac{n (T + \diam(\sp) \S)}{\minT}}$ patches, the
  entire front $\tau_k$ is past the target time $T$.  Since each patch
  consists of at most $\Delta$ elements, where $\Delta$ is
  the maximum number of simplices in the star of any vertex of $\sp$,
  the theorem follows.
\end{proof}\fi


\section{Meshing in arbitrary dimensions $d$D$\times$Time}
\label{sec:linear:hidim}

For spatial dimension $d \ge 3$, we prove results analogous to
those in Section~\ref{sec:linear:2d}.  For each front $\tau_i$,
causality limits the gradient of $1$-dimensional simplices, and
causality and progress constraints together limit the gradient of
$k$-dimensional simplices for each $2 \le k \le d$.

Pitching vertex $p_0$, a local minimum of the front $\tau$, alters the
gradient of all faces of $\st(p_0)$ of all dimensions and each of
these faces limits the progress of $p_0$ in time because of causality
and progress constraints on their gradient in time.  Our objective is
to impose gradient constraints on each $k$-dimensional simplex such
that each constraint individually guarantees a minimum positive
progress in time for $p_0$.  Hence, all these gradient constraints
taken together also imply positive progress which in turn implies a
lower bound on the height and therefore the temporal aspect ratio of
each spacetime element.

Fix $\e \in (0,1)$.  The progress constraint is parameterized by $\e$
and the wavespeed $\S$.  We are free to choose any value for $\e$ in the
allowed range.

Consider an arbitrary $k$-dimensional face
$\fp_0\fp_1\fp_2{\ldots}\fp_k$ incident on a vertex $\fp_0$ of a front
$\tau$ where $2 \le k \le d$.  Let $p_0p_1p_2{\ldots}p_k$ denote the
spatial projection of this simplex.  The extent to which progress
constraint $\S$ limits the gradient of $\fp_0\fp_1\fp_2{\ldots}\fp_k$,
i.e., the gradient of the time function $\tau$ restricted to
$\fp_0\fp_1\fp_2{\ldots}\fp_k$, depends on the geometry of the simplex
$p_0p_1p_2{\ldots}p_k$ in space.

For each integer $i$ such that $0 \le i \le k$, let $\Delta_i$ denote
the facet $p_0p_1p_2{\ldots}p_{i-1}p_{i+1}{\ldots}p_k$ opposite $p_i$,
let $\bp_i$ denote the orthogonal projection of $p_i$ onto
$\aff{\Delta_i}$, let $u_i$ denote the point of $\Delta_i$ closest to
$\bp_i$, and let $\theta_i$ denote $\angle{p_iu_i\bp_i}$.  We say that
the simplex $p_0p_1p_2{\ldots}p_k$ is \emph{obtuse} if $\bp_i \not\in
\Delta_i$ for some $i$.

\begin{definition}
  For the $k$-dimensional simplex $p_0p_1p_2{\ldots}p_k$ define
\[
  \phi(p_0p_1p_2{\ldots}p_k) := \min_{0 \le i \le k}
                                  \{\,\sin{\theta_i}\,\}
\]
\label{def:hidim:phi}
\end{definition}

As in 2D$\times$Time, there are two key lemmas in higher dimensions
that imply a lower bound on the worst-case height of any tentpole.
Lemma~\ref{lemma:hidim:nextiscausal} was proved implicitly in the
paper by Erickson \etal{}~\cite{erickson02building} but was applied
only to the case of constant wavespeed.

The first lemma states that if $\fp_0\fp_1\fp_2{\ldots}\fp_k$ is causal and
satisfies progress constraint $\S$ then $\fp_0'\fp_1\fp_2{\ldots}\fp_k$
obtained by advancing a lowest vertex $\fp_0'$ in time by
$\e\S \dist(p_0,\aff{p_1p_2{\ldots}p_k})$ is also causal.

\begin{lemma}
  Let $\fp_0\fp_1\fp_2{\ldots}\fp_k$ be any $k$-dimensional simplex of
  the causal front $\tau$ for arbitrary $1 \le k \le d$ with $\fp_0$
  as its lowest vertex. Let $\bp_0$ denote the orthogonal projection
  of $p_0$ onto $\aff{p_1p_2{\ldots}p_k}$.  Let $\S > 0$ be a slope
  such that $\S \le 1/\omega(\fp_0\fp_1\fp_2{\ldots}\fp_k)$ and let
  $d_{p_0}$ denote $\dist(p_0,\aff{p_1p_2{\ldots}p_k}) =
  \abs{p_0\bp_0}$.  Let $\e$ be any real number in the range $0 < \e <
  1$.  Consider two cases:

\begin{enumerate}

\item \ensuremath{Case~\mathbf{\bp_0 \in p_1p_2{\ldots}p_k:}}
  If $\norm{\grad\rest{\tau}{p_1p_2{\ldots}p_k}} \le {(1-\e) \S}$, then,
  for every $\dt \in [0,\e d_{p_0} \S]$, the front $\tau' =
  \next(\tau,p_0,\dt)$ satisfies
  ${\norm{\grad\rest{\tau'}{p_0p_1p_2{\ldots}p_k}} < \S}$.

\item \ensuremath{Case~\mathbf{\bp_0 \not\in p_1p_2{\ldots}p_k:}}
  Let $u$ denote the point of the facet $p_1p_2{\ldots}p_k$ closest to
  $\bp_0$.  If $\norm{\grad\rest{\tau}{p_1p_2{\ldots}p_k}} \le {(1-\e)}
  \S \sin\angle{p_0u\bp_0}$, then, for every $\dt \in [0,\e d_{p_0}
  \S]$, the front $\tau' = \next(\tau,p_0,\dt)$ satisfies
  ${\norm{\grad\rest{\tau'}{p_0p_1p_2{\ldots}p_k}} < \S}$.

\end{enumerate}
\label{lemma:hidim:nextiscausal}
\end{lemma}

\ifproofs\begin{proof}[Proof of Lemma~\ref{lemma:hidim:nextiscausal}]
The constraint on $\fp_0'\fp_1\fp_2{\ldots}\fp_k$ is
equivalent to the following:
\[
  \frac{\tau'(p_0)-\tau(\bar{p_0})}{\abs{p_0\bar{p_0}}}
<
  \root{\S^2 - \norm{\grad\rest{\tau}{p_1p_2{\ldots}p_k}}^2}
\]

Without loss of generality, assume  $\tau(p_0) \le \tau(p_1) \le
\tau(p_2) \le \ldots \le \tau(p_k)$.

\noindent\textbf{Case 1: $\mathbf{\bp_0 \in p_1p_2{\ldots}p_k}$.}
In this case, we have $\tau(\bp_0) \ge \tau(p_1) \ge \tau(p_0)$.
Hence,
\[
  \frac{\tau'(p_0) - \tau(\bp_0)}{\abs{p_0\bp_0}}
\le
  \frac{\tau'(p_0) - \tau(p_0)}{\abs{p_0\bp_0}}
\]
Since $\tau'(p_0)-\tau(p_0) \le \e \S \abs{p_0\bp_0}$, it suffices to
show that
\[
  \e \S
<
  \root{\S^2 - \norm{\grad\rest{\tau}{p_1p_2{\ldots}p_k}}^2}
\]
Since $\norm{\grad\rest{\tau}{p_1p_2{\ldots}p_k}} \le (1-\e) \S$, it
suffices to show that
\[
  \e^2
<
  1 - (1-\e)^2
\]
i.e.,
\[
  2 \e^2
<
  2 \e
\]
which is true whenever $0 < \e < 1$.

\noindent\textbf{Case 2: $\mathbf{\bp_0 \not\in p_1p_2{\ldots}p_k}$.}

\begin{figure}\centering
\includegraphics{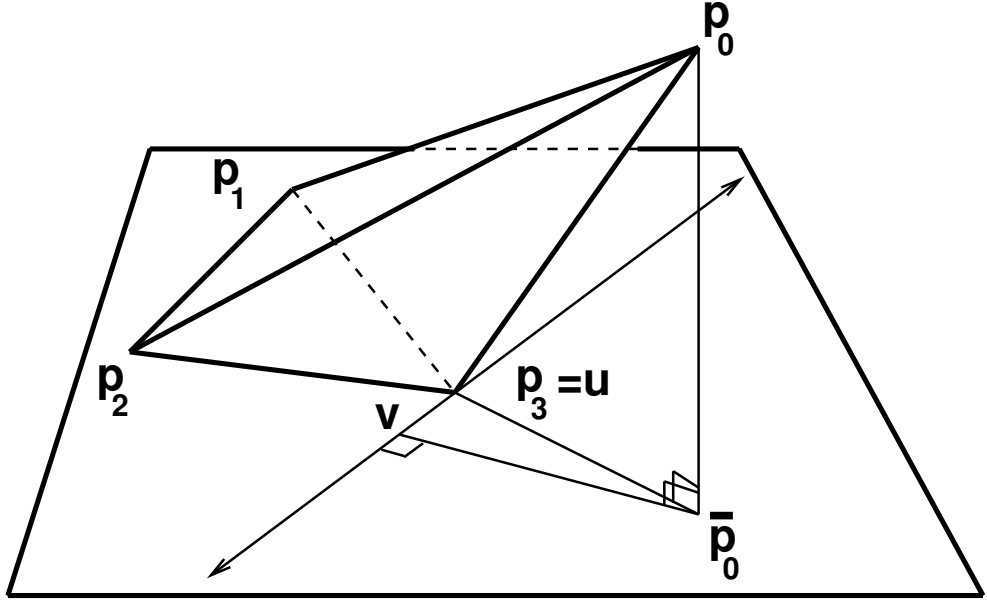}
\caption{An obtuse tetrahedron with $\bp_0 \not\in
p_1p_2{\ldots}p_k$.}
\label{fig:obtuse-tet}
\end{figure}

Note that $u \ne \bp_0$, hence, $\abs{p_0\bp_0}/\abs{p_0u} =
\sin\angle{p_0u\bp_0} < 1$.  We consider two sub-cases separately:
(a)~$\tau(\bp_0) \ge \tau(p_0)$ and (b)~$\tau(\bp_0) < \tau(p_0)$.
See Figure~\ref{fig:obtuse-tet}.

Note that we could have merged case~2(a) with case~1; however, then
the antecedent of the theorem would have depended on the current front
$\tau$.  This would have, in turn, complicated the implementation of
the algorithm and the proof of the subsequent theorems.

\noindent\textbf{Case 2(a): $\mathbf{\tau(\bp_0) \ge \tau(p_0)}$.}
In this case, since $\tau(\bp_0) \ge \tau(p_0)$, we have
\[
  \frac{\tau'(p_0) - \tau(\bp_0)}{\abs{p_0\bp_0}}
\le
  \frac{\tau'(p_0) - \tau(p_0)}{\abs{p_0\bp_0}}
\]
Since $\tau'(p_0)-\tau(p_0) \le \e \S \abs{p_0\bp_0}$, it suffices to
show that
\[
  \e \S
<
  \root{\S^2 - \norm{\grad\rest{\tau}{p_1p_2{\ldots}p_k}}^2}
\]
Since $\norm{\grad\rest{\tau}{p_1p_2{\ldots}p_k}} \le (1-\e) \S
\sin\angle{p_0u\bp_0} < (1-\e) \S$, it
suffices to show that
\[
  \e^2
<
  1 - (1-\e)^2
\]
i.e.,
\[
  2 \e^2
<
  2 \e
\]
which is true whenever $0 < \e < 1$.

\noindent\textbf{Case 2(b): $\mathbf{\tau(\bp_0) < \tau(p_0)}$.}
In fact, we will handle this case with a potentially weaker antecedent
than the one in the statement of the theorem.  Let $v$ denote the
point of $\aff{p_1p_2{\ldots}p_k}$ closest to $\bp_0$ among all points
such that $\tau(v) \ge \tau(p_0)$.  Since $u \in p_1p_2{\ldots}p_k$ we
have $\tau(\bp) \ge \tau(p_0)$; hence, $\abs{\bp_0v} \le
\abs{\bp_0u}$ and $\abs{p_0v} \le \abs{p_0u}$. Hence,
\begin{equation}
  \sin\angle{p_0v\bp_0} = \frac{\abs{p_0\bp_0}}{\abs{p_0v}}
\ge
  \frac{\abs{p_0\bp_0}}{\abs{p_0u}} = \sin\angle{p_0u\bp_0}
\label{eqn:weakerangle}
\end{equation}
We will prove the following statement which, by the inequality of
Equation~\ref{eqn:weakerangle}, implies the theorem:
\begin{quote}
  If $\norm{\grad\rest{\tau}{p_1p_2{\ldots}p_k}} \le (1-\e) \S
  \sin\angle{p_0v\bp_0}$, then, for every $\dt \in
  [0,\e\abs{p_0\bp_0}\S]$, the front $\tau' = \next(\tau,p_0,\dt)$
  satisfies $\norm{\grad\rest{\tau'}{p_0p_1p_2{\ldots}p_k}} < \S$.
\end{quote}

Let $\beta = \abs{\bp_0v}/\abs{p_0\bp_0}$. By our choice of $v$,
$\grad\rest{\tau}{p_1p_2{\ldots}p_k}$ is a vector pointing in the
direction from $\bp_0$ to $v$. Hence,
\begin{equation}
  \frac{\tau'(p_0)-\tau(\bp_0)}{\abs{p_0\bp_0}}
=
  \frac{\tau'(p_0)-\tau(v)}{\abs{p_0\bp_0}}
+ \beta \, \norm{\grad\rest{\tau}{p_1p_2{\ldots}p_k}}
\label{eqn:hidim:strengthen}
\end{equation}
Using Equation~\ref{eqn:hidim:strengthen}, the constraint on
$\fp_0'\fp_1\fp_2{\ldots}\fp_k$ can be rewritten as
\[
  \frac{\tau'(p_0) - \tau(v)}{\abs{p_0\bp_0}}
<
  \root{\S^2 - \norm{\grad \rest{\tau}{p_1p_2{\ldots}p_k}}^2}
- \beta \, \norm{\grad \rest{\tau}{p_1p_2{\ldots}p_k}}
\]

Now,
\[
  \frac{\tau'(p_0)-\tau(v)}{\abs{p_0\bp_0}}
\le
  \frac{\tau'(p_0) - \tau(p_0)}{\abs{p_0\bp_0}}
\]
Since $\tau'(p_0)-\tau(p_0) \le \e \S \abs{p_0\bp_0}$, it suffices to
show that
\[
  \e \S
<
  \root{\S^2 - \norm{\grad\rest{\tau}{p_1p_2{\ldots}p_k}}^2}
- \beta \, \norm{\grad\rest{\tau}{p_1p_2{\ldots}p_k}}
\]
By the antecedent, we have $\norm{\grad\rest{\tau}{p_1p_2{\ldots}p_k}} \le
(1-\e) \S \sin\angle{p_0v\bp_0}$. Hence, it suffices to show that
\[
  \e
<
  \root{1 - (1-\e)^2\sin^2\angle{p_0v\bp_0}}
- \beta \, (1-\e)\sin\angle{p_0v\bp_0}
\]
i.e.,
\[
  \left(\e + \beta \, (1-\e)\sin\angle{p_0v\bp_0}\right)^2
<
  1 - (1-\e)^2\sin^2\angle{p_0v\bp_0}
\]
i.e.,
\[
  \e^2
+ \left(\beta \, (1-\e)\sin\angle{p_0v\bp_0}\right)^2
+ 2 \beta \, \e \, (1-\e)\sin\angle{p_0v\bp_0}
+ (1-\e)^2\sin^2\angle{p_0v\bp_0}
<
  1
\]
i.e.,
\[
  \e^2
+ 2 \e \, (1-\e) \beta \sin\angle{p_0v\bp_0}
+ (1-\e)^2 (\beta^2+1) \sin^2\angle{p_0v\bp_0}
<
  1
\]

Note that
\begin{align*}
  \beta^2 + 1
&= \frac{\abs{\bp_0v}^2 + \abs{p_0\bp_0}^2}{\abs{p_0\bp_0}^2}\\
&= \frac{\abs{p_0v}^2}{\abs{p_0\bp_0}^2}
\end{align*}
where the last equality follows by Pythagoras' theorem applied to
$\triangle{p_0\bp_0v}$.  (See Figure~\ref{fig:obtuse-tet}.)  Also,
$\sin\angle{p_0v\bp_0} = \abs{p_0\bp_0}/\abs{p_0v}$. Hence,
$(\beta^2+1) \sin^2\angle{p_0v\bp_0} = 1$.

Additionally,
\begin{align*}
  \beta \, \sin\angle{p_0v\bp_0}
&=
  \frac{\abs{\bp_0v}}{\abs{p_0\bp_0}} \,
  \frac{\abs{p_0\bp_0}}{\abs{p_0v}}\\
&=
  \cos\angle{p_0v\bp_0}
\end{align*}
Therefore, it suffices to show that
\[
  \e^2
+ 2 \e \, (1-\e) \cos\angle{p_0v\bp_0}
+ (1-\e)^2
<
  1
\]
Observe that $\e^2 + 2 \e \, (1-\e) + (1-\e)^2 = (\e + (1-\e))^2 =
1$.  Therefore,
\[
  \e^2
+ 2 \e \, (1-\e) \cos\angle{p_0v\bp_0}
+ (1-\e)^2
=
  1
- 2 \e \, (1-\e) \, (1-\cos\angle{p_0v\bp_0})
\]
Hence, it suffices to show that
\[
  1
- 2 \e \, (1-\e) \, (1-\cos\angle{p_0v\bp_0})
<
  1
\]
Recall that $\angle{p_0v\bp_0}$ is acute, so $0 <
\cos\angle{p_0v\bp_0} < 1$, i.e., $1 - \cos\angle{p_0v\bp_0} > 0$.
Hence, the claim follows whenever $0 < \e < 1$.
\end{proof}\fi

\subsection{Progress constraint for a $k$-dimensional simplex}

Motivated by Lemma~\ref{lemma:hidim:nextiscausal}, we derive progress
constraints that are sufficient to ensure that the antecedent of
Lemma~\ref{lemma:hidim:nextiscausal} is satisfied in the \emph{next}
step by the front $\tau'$.

Consider an arbitrary $k$-dimensional face $p_0p_1p_2{\ldots}p_k$
incident on $p_0$ for any $k$ in the range $2 \le k \le d$.

\begin{definition}[Progress constraint]
  We say that a simplex $\fp_0\fp_1\fp_2{\ldots}\fp_k$ of the front
  $\tau$ satisfies the \emph{progress constraint $\S$} if and only if
  for every highest facet $\Delta_i$ where $0 \le i \le k$ we have
  $\norm{\grad{\rest{\tau}{\Delta_i}}} \le (1-\e) \,
  \phi(p_0p_1p_2{\ldots}p_k) \, \S$.
\label{def:hidim:linear:progressconstraint}
\end{definition}

Note that if $\fp_0\fp_1\fp_2{\ldots}\fp_k$ with $\fp_0$ as its lowest
vertex satisfies the progress constraint $\S$
(Definition~\ref{def:hidim:linear:progressconstraint}) then
$\norm{\grad\rest{\tau}{p_1p_2{\ldots}p_k}} \le (1-\e)
\phi(p_0p_1p_2{\ldots}p_k) \S$.

\begin{lemma}
  Let $\fp_0\fp_1\fp_2{\ldots}\fp_k$ be any $k$-dimensional simplex of
  the causal front $\tau$ for arbitrary $2 \le k \le d$ with $\fp_0$
  as its lowest vertex.  Let $\phi$ denote
  $\phi(p_0p_1p_2{\ldots}p_k)$. For each $i$ with $0 \le i \le k$, let
  $v_i$ denote the point of $\aff{(\Delta_i \cap \Delta_0)}$ closest
  to $p_0$.

  If $\fp_0\fp_1\fp_2{\ldots}\fp_k$ satisfies progress constraint
  $\S$, then for every $\dt \in [0,(1-\e) \phi \S \abs{v_ip_0}]$ the
  front $\tau' = \next(\tau,p_0,\dt)$ satisfies
  $\norm{\grad\rest{\tau'}{\Delta_i}} \le (1-\e) \phi \S$ for every
  $i$ such that $0 \le i \le k$.
\label{lemma:hidim:nextisprogressive}
\end{lemma}

\ifproofs\begin{proof}[Proof of Lemma~\ref{lemma:hidim:nextisprogressive}]
Note that $\rest{\tau'}{\Delta_0} = \rest{\tau}{\Delta_0}$; hence, the
statement is trivial for $i=0$.

Since $\fp_0\fp_1\fp_2{\ldots}\fp_k$ satisfies progress constraint
$\S$, by Definition~\ref{def:hidim:linear:progressconstraint}, we
have $\norm{\grad\rest{\tau}{\Delta_i}} \le (1-\e) \phi \S$ for
every $i$ such that $0 \le i \le k$.

Consider an arbitrary $i$ in the range $1 \le i \le k$ such that
$\Delta_i$ is a highest facet.

Note that for every simplex $f \in \link(p_0)$, the time function
$\tau$ restricted to $\aff{f}$ is unchanged by pitching $p_0$.
Therefore, pitching $p_0$ increases $\grad\rest{\tau}{\Delta_i}$ in a
direction orthogonal to $\aff{\Delta_i \cap \Delta_0}$ while keeping
the component of $\tau$ along $\Delta_i \cap \Delta_0$ unchanged.
Hence,
\[
  \norm{\grad\rest{\tau'}{\Delta_i}} - \norm{\grad\rest{\tau}{\Delta_i}}
=
  \frac{\tau'(p_0)-\tau(p_0)}{\abs{v_ip_0}}
\]

Since $\norm{\grad\rest{\tau}{\Delta_i}} \ge 0$ and $\tau'(p_0) \le
\tau(p_0) + (1-\e) \phi \S \abs{v_ip_0}$, we
have
\[
  \norm{\grad\rest{\tau'}{\Delta_i}}
\le
  \frac{\tau'(p_0)-\tau(p_0)}{\abs{v_ip_0}}
\le
  \frac{(1-\e) \phi \S \abs{v_ip_0}}{\abs{v_ip_0}}
=
  (1-\e) \phi \S.
\]
Therefore, $\fp_0'\fp_1\fp_2{\ldots}\fp_k$ satisfies the constraint in
the statement of the lemma.
\end{proof}\fi

\bigskip

Note that we have in fact proved the following two statements that
imply Lemmas~\ref{lemma:hidim:nextiscausal}
and~\ref{lemma:hidim:nextisprogressive}.

Let $\fp_0\fp_1\fp_2{\ldots}\fp_k$ be any $k$-dimensional simplex with
$\fp_0$ as its lowest vertex. Without loss of generality, assume
$\tau(p_0) \le \tau(p_1) \le \tau(p_2) \le \ldots \le \tau(p_k)$.  For
each $i$ with $0 \le i \le k$, let $v_i$ denote the point of
$\aff{(\Delta_i \cap \Delta_0)}$ closest to $p_0$.  Let $\e$ be any real
number in the range $0 < \e < 1$.  Let $\S > 0$ be any slope.

\begin{enumerate}
\item
  If $\norm{\grad\rest{\tau}{p_1p_2{\ldots}p_k}} \le (1-\e) \S
  \phi(p_1p_2{\ldots}p_k)$, then $\fp_0'\fp_1\fp_2{\ldots}\fp_k$
  satisfies $\norm{\grad\rest{\tau'}{p_0p_1p_2{\ldots}p_k}} < \S$ for
  every $\tau'(p_0)$ in the range $\tau(p_1) \le \tau'(p_0) \le
  \tau(p_1) + \e \S \dist(p_0,\aff{p_1p_2{\ldots}p_k})$.  .

\item
  $\fp_0'\fp_1\fp_2{\ldots}\fp_k$ satisfies 
\[
  \forall i, 0 \le i \le k
\quad
  \norm{\grad\rest{\tau'}{\Delta_i}} \le (1-\e) \S \phi(\Delta_i)
\]
  for every $\tau'(p_0)$ in the range $\tau(p_1) \le \tau'(p_0) \le \tau(p_1) 
  + (1-\e) \S \min_{0 \le i \le k}\{\abs{v_ip_0}\}$.
\end{enumerate}

The above claims are stronger than Lemmas~\ref{lemma:hidim:nextiscausal}
and~\ref{lemma:hidim:nextisprogressive} because we do not assume that
$\fp_0\fp_1\fp_2{\ldots}\fp_k$ and the front $\tau$ are progressive.

\bigskip

\subsection{Combining constraints for all dimensions}

Pitching vertex $p$ alters the gradients of all faces of $\st(p)$ of
all dimensions and each of these faces limits the progress of $p$ in
time.  We interpret the progress constraints mentioned in the previous
lemmas as gradient constraints on lower dimensional faces of the
front.  These gradient constraints are imposed simultaneously on all
facets of dimension from~$2$ through~$d$ of the front~$\tau_i$ at each
step.  The previous lemmas guarantee that for each~$k$, where ${1 \le
k \le d}$, the causality and progress constraint for each
$k$-dimensional face~$f$ ensures positive progress of $p$ so that the
corresponding face~$f'$ on the new front also satisfies the relevant
causality and progress constraints.

Thus, every $k$-dimensional highest face $\Delta$ has a gradient
constraint imposed on it for every $(k+1)$-dimensional highest face
$\Gamma$ such that $\Delta \subset \Gamma$.

\begin{theorem}
  For any $i \ge 0$, if the front $\tau_i$ is progressive then $\tau_i$ is
  valid.
\label{thm:hidim:progressiveisvalid}
\end{theorem}

\ifproofs\begin{proof}[Proof of Theorem~\ref{thm:hidim:progressiveisvalid}]
The proof is identical to that of Theorem \ref{thm:1d:causalisvalid}.
\end{proof}\fi

We thus have the following theorem.

\begin{theorem}
  Given a triangulation $\sp$ of a bounded planar space domain where
  $\minW$ is the minimum width of a simplex of $\sp$ and $\S$ is the
  minimum slope anywhere in $\sp \times [0,\infty)$, for every $\e$
  such that $0 < \e < 1$ our algorithm constructs a simplicial
  mesh of $\sp \times [0,T]$ consisting of at most
  $\ceil{\frac{n (T + \diam(\sp) \S)}{minT}}$ spacetime
  elements for every real $T \ge 0$ in constant computation time per
  element.
\label{thm:hidim:main}
\end{theorem}

\ifproofs\begin{proof}[Proof of Theorem~\ref{thm:hidim:main}]
  By Lemmas \ref{lemma:hidim:nextiscausal} and
  \ref{lemma:hidim:nextisprogressive}, it follows that the height of
  each tentpole constructed by the algorithm is at least $\minT = \e
  \S \minW$.  By Theorem \ref{thm:hidim:progressiveisvalid}, after
  constructing at most $k \le \ceil{\frac{\diam(\sp) \, \S}{\minT} \,
  T}$ patches, the entire front $\tau_k$ is past the target time $T$.
  Since each patch consists of at most $\Delta(\sp)$ elements, the
  theorem follows.
\end{proof}\fi


\section{Element quality}

Various definitions of quality exist for linear elements in Euclidean
space, such as ratio of circumradius to inradius, circumradius to
shortest edge length, radius of minimum enclosing ball to maximum
enclosed ball, etc..  Each definition is useful to quantify the
suitability of the element for the particular numerical technique
being used.  In $\mathbb{E}^2$, all measures of quality are
within constant factors of each other, which is not true in higher
dimensions.  We recommend the excellent survey due to Shewchuk
\cite{shewchuk02whatis} on various aspects of element quality.

The spacetime domain $\mathbb{E}^d \times \Real$ is not Euclidean
because the time axis can be scaled independently of space.  It is
therefore nontrivial to define the quality of a spacetime element.  We
consider the temporal aspect ratio of an element as a measure of its
quality.  The temporal aspect ratio of an element was defined in
Chapter \ref{sec:linear} as the ratio of its height to its duration.
We prove that our worst-case guarantees on the height of each tentpole
lead to a lower bound on the worst-case temporal aspect ratio of any
spacetime element constructed by our algorithm.

In this chapter, we have proved a lower bound on the worst-case height
of each tentpole, a progress guarantee similar to that of Erickson
\etal{}~\cite{erickson02building} and Abedi
\etal{}~\cite{abedi04spacetime}.  For the linear nonadaptive version
of our algorithm, where the slope is $\S$ everywhere, this progress
guarantee can be rephrased as follows: the height of each spacetime
tetrahedra in the tent pitched at $\fp$ is at least $\he \S w_p$ where
$\he \in (0,\half]$ is a fixed parameter and $w_p$ denotes the
distance of $p$ from the boundary of the kernel of $\link(p)$ in the
spatial projection.  Thus, the height of the tentpole at $\fp$ is
bounded from below by a positive function of $\e$, the slope $\S$, and
the shape of the triangles in $\st(p)$.

Note that the temporal aspect ratio is always in the range $(0,1]$
with a larger value corresponding to a ``better'' element.  The
duration of the tetrahedron $\fp'\fp\fq\fr$ can be at most $2\S w_p$
because both facets $\fp\fq\fr$ and $\fp'\fq\fr$ are causal. Together
with the lower bound on the height of the tetrahedron, this implies
the following theorem.

\begin{theorem}
  The temporal aspect ratio of any tetrahedron in $\spt$
  is at least $\he/2$.
\end{theorem}

When the wavespeed is not constant, the worst-case temporal aspect
ratio that we can prove is smaller by a factor $\minS/\maxS$, i.e.,
proportional to the ratio of maximum to minimum wavespeeds or the
\emph{Mach factor}.

Bisecting a triangle on the front can improve the temporal aspect
ratio of future tetrahedra because the smaller triangles may be
limited by a smaller wavespeed than their larger ancestors.  However,
newest vertex bisection does not improve the quality of the spatial
projection of front triangles, beyond a very limited amount.
Repeatedly bisecting a single triangle produces smaller triangles from
at most four different similarity classes.  To improve the temporal
aspect ratio, mesh smoothing and other generalized mesh adaptivity
operations are more useful because they improve the quality of the
spatial projection of the front.


\section{Size optimality}

The computation time of any solution strategy increases with the
number of patches---computing the solution within each patch is
expensive and, in our scenario, much more computationally expensive
than the mesh generation, especially for high polynomial orders.
Therefore, it is important to generate a mesh with close to the fewest
number of patches necessary for a given accuracy.

Our algorithm constructs groups of coupled spacetime elements inside
each tent such that the boundary of the tent is causal.  The number of
elements in the resulting patch is at most the maximum vertex degree
of the underlying space mesh.  Each element constructed by our
algorithm has both a causal inflow facet and a causal outflow facet;
additionally, the spatial projection of each element
$\fp_0'\fp_0\fp_1\fp_2{\ldots}\fp_d$ is the simplex
$p_0p_1p_2{\ldots}p_d$ in the original space mesh.  Given a
triangulation $\sp$ of the space domain and a target time $T$, we say
that a simplicial spacetime mesh of $\sp \times [0,T]$ is
\emph{solvable} if (i)~each element has both a causal inflow facet and
a causal outflow facet; and (ii)~for every point $x$ in the spatial
projection $\Delta$ of each element, the diameter of $\Delta$ does not
exceed the diameter of the simplex of $\sp$ containing $x$.

Fix an arbitrary point $x$ in space. The \emph{size} of a spacetime
mesh of $\sp \times [0,T]$ is the maximum over $x \in \sp$ of the
number of spacetime elements that intersect the temporal segment $x
\times [0,T]$.

Consider the linear nonadaptive instance of our algorithm.  The
wavespeed everywhere in spacetime is constant, equal to $\S$.

\begin{theorem}
  The size of the mesh constructed by our algorithm is
  $\tilde{O}(1/{\e^2})$ times the minimum size of any valid
  mesh of the spacetime volume $\sp \times [0,T]$.
\label{thm:sizeoptimality}
\end{theorem}

\begin{proof}
Let $D$ and $\rho$ denote the diameter and inradius respectively of
the simplex $p_0p_1p_2{\ldots}p_d$ of $\sp$ containing~$x$. Causality
limits the gradient of any edge of the $d$-simplex to less than $\S$.
Therefore, any temporal segment of duration $(d-1) \S D$ must
intersect at least two distinct elements in a solvable mesh;
therefore, the number of spacetime elements in a solvable mesh that
intersect $x \times [0,T]$ is at least $\floor{T/((d-1) \S D)}$.

Consider a minimal sequence of tent pitching steps, called a
\emph{superstep}, in which each vertex of $p_0p_1p_2{\ldots}p_d$ is lifted
at least once.  When $p_0$ is pitched, the amount of progress made by
$x$ is proportional to $\dist(x,\aff{\Delta_0})$. Since
\[
  \sum_{0 \le i \le k} \dist(x,\aff{\Delta_i})
\ge
  \rho
\]
the amount of progress made by $x$ during a superstep is at least $\e
\S \gamma \rho$, where $\gamma \in (0,1]$ denotes the minimum over all
$i$ such that $0 \le i \le k$ of $w_{p_i}/\dist(p_i,\aff{\Delta_i})$.
Hence, after at most $\ceil{T/(\e \S \gamma \rho)}$ supersteps, the
point $x$ achieves or exceeds the target time~$T$.

Causality limits the gradient of any edge of the simplex
$p_0p_1p_2{\ldots}p_d$ to less than $\S$.  Therefore, any $d-1$
vertices of $p_0p_1p_2{\ldots}p_d$ can advance less than $2(d-1)
\S D$ units of time total without also advancing the $d$th vertex.
Therefore, the number of steps in each superstep is at most
$\floor{(2(d-1) \S D)/(\e \S w)}$ where $w = \min_{0 \le i \le k}
w_{p_i}$. It follows that the number of tetrahedra in the spacetime
mesh constructed by our algorithm intersected by $x
\times [0,T]$ is at most $\ceil{T/(\e \S \gamma \rho)} \cdot 
\floor{(2(d-1) \S D)/(\e \S w)}$.

The ratio of the upper bound to the lower bound on the size is
\[
    O\left(
      d
      \frac{1}{\e^2}
      \frac{D^2}{\gamma \rho w}
    \right).
\]
\end{proof}

Note that for every fixed input space mesh $\sp$, the size of the
spacetime mesh of $\sp \times [0,T]$ constructed by our algorithm is
at most $O(1/{\e^2})$ times that of a solvable mesh of the same
spacetime volume, where the constant in the big-Oh notation depends on
$\sp$.

When the wavespeed is not constant, the size-optimality factor that we
can prove is larger by a factor $\maxS/\minS$, i.e., inversely
proportional to the \emph{Mach factor}.

When the spacetime error indicator demands different mesh resolution
in different portions of the spacetime domain, we can prove that the
mesh constructed by our algorithm is still near-optimal where the
approximation ratio is worse by a factor proportional to the ratio of
maximum to minimum element sizes allowed by the error tolerance.

Our advancing front meshing algorithms are parameterized by real
numbers like $\e$ that control the extent to which they greedily
maximize the height (and hence the temporal aspect ratio) of new
elements.  Being greedy at each step is not always a good choice for
maximizing the worst-case temporal aspect ratio and for minimizing the
total number of spacetime elements.  We have especially clear evidence
that while attempting to satisfy coplanarity constraints in order to
coarsen parts of the front it may be necessary to reduce the heights
of tentpoles.  As a result of being less greedy at each step, do we
actually increase the number of spacetime elements to an unbounded
extent?  Theorem \ref{thm:sizeoptimality} says that by being less
greedy, our advancing front algorithm is near-optimal in the number of
elements compared to a solvable mesh.


\section{Geometric primitives}

Maximizing the height of each tentpole subject to causality and
progress constraints is equivalent to shooting a ray in an arrangement
of (a small number of) planes and infinite cones.  Consider the case
where the wavespeed everywhere in spacetime is the same, i.e., the
slope of every cone of influence is equal to $\S$.

When a vertex $p$ of the causal front $\tau$ is advanced in time to
get the front $\tau'$ by pitching a tentpole from $\fp=(p,\tau(p))$ to
$\fp'=(p,\tau'(p))$, the new front $\tau'$ is causal if and only if
for every triangle $pqr$ incident on $p$ the slope of triangle
$\fp'\fq\fr$ is less than $\S$.  Let $\pi_{qr}$ denote the plane
through $\fq\fr$ making a slope of $\S$ with the space domain.  The
slope of $\triangle{\fp''\fq\fr}$ is equal to $\S$ if and only if
$\fp''=(p,\tau''(p))$ is on the plane $\pi_{qr}$ and the slope of
$\triangle{\fp'\fq\fr}$ is less than $\S$ if and only if $\tau(p) \le
\tau'(p) < \tau''(p)$.  Thus, $\tau''(p)$ can be computed by shooting
a ray in spacetime with origin at $\fp$ oriented in the positive time
direction and finding the point of intersection of this ray with the
plane $\pi_{qr}$.  The triangle $\fp'\fq\fr$ is causal only if
$\abs{\fp\fp'} < \abs{\fp\fp''}$.  Finding the supremum height of the
tentpole at $\fp$ subject to causality constraint alone is equivalent
to finding the distance along the tentpole direction before the ray
originating at $\fp$ in the tentpole direction first intersects any
plane $\pi_{qr}$ among all such planes for every edge $qr \in
\link(p)$.

The progress constraint $\norm{\grad\rest{\tau'}{pq}} \le (1-\e)
\phi_{pq} \S$ is satisfied if and only if the edge $\fp'\fq$ is below
or on the infinite right circular cone $C_q$ with apex at
$\fq=(q,\tau(q))$ and axis in the positive time direction.  Thus,
maximizing the height of the tentpole $\fp\fp'$ is equivalent to
shooting a ray with origin at $\fp$ in the tentpole direction and
determining its first intersection with $C_q$.

Thus, finding the supremum height of the tentpole at $\fp$ subject
to causality and progress constraints is equivalent to answering ray
shooting queries among an arrangement of planes, one for each edge $qr
\in \link(p)$, and infinite circular cones, one for each vertex $q$
neighboring $p$.

\parasc*{Running time} The running time is $O(\deg(p))$ for
constructing the patch at $p$, where $\deg(p)$ is the number of
simplices in $\st(p)$ of any dimension, hence $O(1)$ amortized per
spacetime element.  Additional complexity at each step comes from the
need to choose the local minimum vertex to pitch; the choice of local
minimum can be arbitrarily complicated. Advancing $p$ destroys at most
one local minimum ($p$ itself) and creates at most $\deg(p)$ new local
minima (the neighbors of $p$).  Thus, the set of local minima of each
front can be maintained efficiently.

\bigskip

If the tentpole is inclined, for instance to track a moving boundary
or shock interface, then the problem of maximizing tentpole height is
still equivalent to answering ray shooting queries.  The problem is a
little more complicated if the choice of tentpole direction is not
specified and must be chosen to optimize some geometric or numerical
quality criteria.

Note that if the wavespeed is not uniform everywhere, the problem is
more complicated by the fact that nonlocal cones can limit the height
of a tentpole and that progress constraints must anticipate future
wavespeeds.  In problems with nonuniform wavespeeds, maximizing the
height of a tentpole is not always equivalent to answering a ray
shooting query---a triangle of the new front can be tangent to a
remote cone of influence even if the corresponding tentpole is still
below the cone.

We will postpone discussions of geometric primitives for greedily
maximizing the height of each tentpole for such more general problems
to later chapters.

\subsection*{Chapter summary}

The Tent Pitcher algorithm is the first instance of an advancing front
algorithm to mesh directly in spacetime.  The algorithm is simple
because the geometric constraints that limit the height of each
tentpole are linear functions of the coordinates.  The height of each
tentpole depends on the size of the space elements incident on the
tentpole only and not on any global property of the space mesh.
Consequently, we were able to prove a lower bound on the temporal
aspect ratios of spacetime tetrahedra constructed by our algorithm.
Progress constraints are artifacts of our algorithm, necessitated by
its local advancement strategy, and not required explicitly to satisfy
causality.  Even though progress constraints limit the progress in
time at each step, we were able to prove that the number of
elements produced by our algorithm for meshing a given spacetime
volume is not much more than that in a size optimal causal mesh of the
same volume.

In this chapter, we described the constant wavespeed case where the
height of each tentpole is limited only by local constraints, as in
the case of linear PDEs.  We were able to generalize Tent Pitcher to
pitch inclined tentpoles corresponding to advancing the front in time
as well as changing the spatial projection of the front vertices.  We
were also able to incorporate additional operations such as edge
flips.  These new operations allow the algorithm to recover from a bad
quality initial front (the space mesh $\sp$).  In Chapter
\ref{sec:extensions}, we will also use these operations to track spacetime
features such as moving boundaries by aligning the tentpole direction
and/or causal facets of the front with these features.


\chapter{Meshing with nonlocal cone constraints}
\label{sec:nonlinear}

The basic advancing front algorithm of Chapter~\ref{sec:linear} relied
on a fixed lower bound $\S=\minS$ on the globally minimum slope.  The
progress constraint was a function of this minimum slope $\minS$ and
this constraint restricted the progress at every step of the
algorithm.  Due to nonlinearity, the speed at which a wave propagates
may be different in different parts of the domain.  Even at the same
point in space, the wavespeed may change with time.  In general, the
wavespeed may be a function of the solution.  In this chapter, we
consider the problem of meshing the spacetime domain when causality
limits the gradient of mesh facets differently in different parts of
the domain because the causal slope is not a constant everywhere.

In this chapter, we make our algorithm respond to changes in the
causal slope (the reciprocal of the wavespeed), even if such changes
are discontinuous and unpredictable, subject to the no-focusing
assumption.  When the wavespeed is not constant, a fast distant wave
corresponding to a shallow cone of influence can limit the amount of
local progress made by our advancing front algorithm.  Thus, when the
wavespeed is not constant, the most limiting cone constraint can be
nonlocal.  Nonlinear PDEs or even linear PDEs with discontinuous
initial conditions can exhibit such behavior.  The no-focusing
assumption, Axiom~\ref{axiom:nofocusing} and its anisotropic version
Axiom~\ref{axiom:anisotropic:nofocusing}, allows us to conservatively
estimate the slope at a point $\fp$ in spacetime where the solution
has not been computed yet, given the cone of influence of every point
on the current front.  The slope $\S(\fp)$ is bounded by the minimum
and maximum slopes of all cones of influence that contain $\fp$.  We
extend the algorithm of Chapter~\ref{sec:linear} to construct only
causal fronts, making use of the no-focusing assumption to estimate
the causality constraint in the unmeshed and unsolved portion of the
spacetime domain ahead of the current front.

\section{Problem statement}

Just like in Chapter~\ref{sec:linear}, the input to our problem is the
initial front $\tau_0$ and the initial conditions of the underlying
hyperbolic PDE in the form of the causal slope $\S(\fp)$ for every
point $\fp$ of $\tau_0$.  We want an advancing front algorithm such
that for every nonnegative real~$T$ there exists a finite integer $k
\ge 0$ such that the front $\tau_k$ after the $k$th iteration of the
algorithm satisfies $\tau_k \ge T$.  As before, we would like to
characterize \emph{valid} fronts as those that are guaranteed to make
finite progress at each step.

The main difficulty in characterizing valid fronts arises when the
causal slope at a given point in the space domain decreases over time,
i.e., the wavespeed increases.  For instance, suppose
$\triangle{\fp\fq\fr}$ is causal.  However, as soon as the local
minimum vertex, say $\fp$, is advanced in time by an arbitrarily small
positive amount to $\fp'$, the triangle $\fp'\fq\fr$ may intersect a
cone of influence with a much smaller slope, i.e., $\S(\fp'\fq\fr) \ll
\S(\fp\fq\fr)$. Consequently, $\triangle{\fp'\fq\fr}$ is not causal.
The decrease in causal slope from $\S(\fp\fq\fr)$ to $\S(\fp'\fq\fr)$
prevents the front from making nontrivial progress by advancing the
local minimum vertex $\fp$.

\parasc*{Our solution}
For one-dimensional space domains, we prove that every causal front is
valid.  We give an algorithm that given an arbitrary causal front
$\tau_i$ constructs a next front $\tau_{i+1}=\next(\tau_i,p,\dt)$ such
that $\tau_{i+1}$ is causal and $\tau_{i+1}(p)$ is maximized.

In higher dimensions, i.e., $d \ge 2$, we define \emph{progressive}
fronts and prove that if a front is progressive then it is valid.  We
give an algorithm that given any progressive front $\tau_i$ constructs
a next front $\tau_{i+1}=\next(\tau_i,p,\dt)$ such that $\tau_{i+1}$
is progressive and $\tau_{i+1}(p)$ is maximized.  Whenever $p$ is a
local minimum of $\tau_i$, the progress $\tau_{i+1}(p) - \tau_i(p)$ is
positive and bounded away from zero.

Our algorithm resolves the following conundrum.  The progress of the
front at each step $i$ is limited by the progress constraint that must
be satisfied by the next front at step $i+1$.  However, we do not know
what the next front is unless we know how much progress is possible at
step $i$.  Intuitively, the geometric constraints that apply at any
given iteration of the algorithm are predicted by simulating the $h$
next tent pitching steps for some parameter $h \ge 0$.  We initially
make conservative assumptions about the future wavespeed and
successively refine the estimate of the wavespeed encountered in the
next $h$ iterations.

Our algorithm is the first algorithm to build spacetime meshes over
arbitrary-dimensional triangulated spatial domains suitable for solving
nonlinear hyperbolic PDEs, where the wavespeed at any point in
spacetime depends on the solution and cannot be computed in advance.
Moreover, the solution can change discontinuously, for instance when a
\emph{shock} propagates through the domain.  As long as the
no-focusing condition (Axiom~\ref{axiom:nofocusing}) holds, the
resulting mesh is efficiently solvable patch-by-patch by SDG methods.

\section{Nonlocal cone constraints in 1D$\times$Time}
\label{sec:nonlinear:1d}

To complete the description of the algorithm in 1D$\times$Time, it
remains only to describe how to maximize the height
$\tau_{i+1}(p)-\tau_i(p)$ of the tentpole subject to the causality
constraint.  Even in 1D$\times$Time, this optimization problem is not
trivial in the presence of nonlinearity because of nonlocal cone
constraints.  Nonlocal cone constraints are handled in the same way in
higher dimensions but the algorithm is more complicated because of
progress constraints.

Recall that the \emph{cone of influence} of a point $\fp$ has its apex
at $\fp$ and its slope in any spatial direction is the maximum slope
of any characteristic through $\fp$ in that direction; fast waves
correspond to cones with smaller slope.

\begin{figure*}[t]
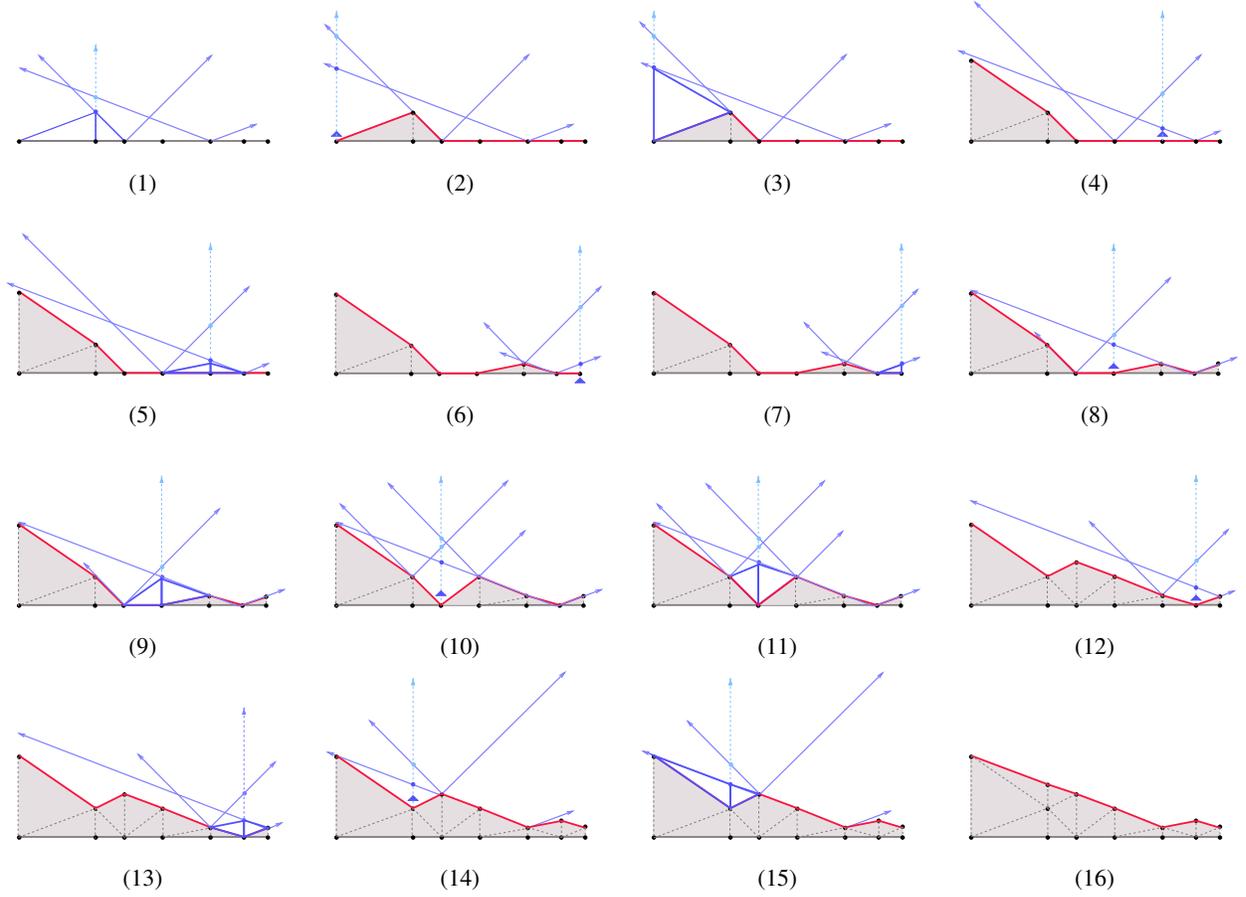
\centering\small
\begin{tabular}{cccc}
\includegraphics[width=.23\linewidth]{\fig{mesh-1}}
&
\includegraphics[width=.23\linewidth]{\fig{mesh-2}}
&
\includegraphics[width=.23\linewidth]{\fig{mesh-3}}
&
\includegraphics[width=.23\linewidth]{\fig{mesh-4}}
\\
(1) & (2) & (3) & (4)\\
\includegraphics[width=.23\linewidth]{\fig{mesh-5}}
&
\includegraphics[width=.23\linewidth]{\fig{mesh-6}}
&
\includegraphics[width=.23\linewidth]{\fig{mesh-7}}
&
\includegraphics[width=.23\linewidth]{\fig{mesh-8}}
\\
(5) & (6) & (7) & (8)\\
\includegraphics[width=.23\linewidth]{\fig{mesh-9}}
&
\includegraphics[width=.23\linewidth]{\fig{mesh-10}}
&
\includegraphics[width=.23\linewidth]{\fig{mesh-11}}
&
\includegraphics[width=.23\linewidth]{\fig{mesh-12}}
\\
(9) & (10) & (11) & (12)\\
\includegraphics[width=.23\linewidth]{\fig{mesh-13}}
&
\includegraphics[width=.23\linewidth]{\fig{mesh-14}}
&
\includegraphics[width=.23\linewidth]{\fig{mesh-15}}
&
\includegraphics[width=.23\linewidth]{\fig{mesh-16}}
\\
(13) & (14) & (15) & (16)
\end{tabular}
\caption{A sequence of tent pitching steps in
  1D$\times$Time.  Maximizing the height of each tentpole while
  staying below every cone of influence may require examining remote
  cones arbitrarily far away.}
\label{fig:1d:tents}
\end{figure*}

When the PDE is nonlinear, a distant but fast wave, i.e., a
\emph{nonlocal} cone, can overtake a slower wave and hence limit the
duration of new elements.  See Figure~\ref{fig:1d:tents}.  \emph{We
assume as a conservative estimate that the wave travels through the
space domain $\sp$ along the shortest path in the ambient Euclidean
space.}  When the medium is \emph{anisotropic}, waves travel faster in
one direction than the other, so the cones are non-circular, for
instance, with elliptical cross-sections.  We will discuss the
situation for anisotropic problems in Chapter~\ref{sec:extensions}; it
should be noted that the discussion in the current chapter is not
restricted to circular cones.

Our algorithms advance a local neighborhood $N$ of the front $\tau$ at
each step to the neighborhood $N'$ of a new causal front $\tau'$.
Hence, $\tau'$ is causal if and only if (i)~the gradient of $\tau'$ at
every point of $N'$ is less than the minimum slope anywhere in $N$,
and (ii)~the neighborhood $N'$ lies below (i.e., does not intersect)
the cone of influence $C_q$ for every point $\fq \in \tau \setminus
N$.  Each such cone of influence corresponds to a \emph{causal cone
constraint}.  In 1D$\times$Time, a cone of influence is defined by two
rays with common apex.

Thus, maximizing the progress of $\fp$, and hence the duration of new
spacetime elements, requires querying the lower hull of cones of
influence of points in $\tau \setminus N$.  After the solution is
computed on the new front $\tau'$, including the causal slope at every
point of $\tau'$, we obtain a new set of cone constraints.

For a fixed segment $pq$ incident on $p$ let $\supT(p,\fq)$ denote the
supremum upper bound on $\tau_{i+1}(p)$ such that $\fp'\fq$ on the
front $\tau_{i+1}$ is causal, i.e.,
\[
  \supT(p,\fq)
:=
  \sup \, \{ t : \fp'\fq \text{\ is causal where\ } \fp'=(p,t) \}.
\]
Let $\supT(p)$ denote the maximum $\supT(p,\fq)$ for every neighbor
$\fq$ of $\fp$.  To maximize the progress at step $i+1$, we would like
to compute $\supT(p)$.  The segment $\fp'\fq$ is causal if and only if
the slope of $\fp'\fq$ is less than the slope of the cone of influence
from every point on the front that intersects $\fp'\fq$.  In
1D$\times$Time, a cone of influence intersects $\fp'\fq$ if and only
if the cone intersects the tentpole $\fp\fp'$.

In general, a cone of influence from arbitrarily far away can
intersect the tentpole at $p$.  See Figure~\ref{fig:1d:tents}.
(Nonlocal cone constraints do not dominate local cone constraints when
the wavespeed everywhere is the same.)  Therefore, in general,
$\supT(p)$ could be determined by a cone of influence of a point
arbitrarily distant from $p$.  In this section, we give two
algorithms---one to compute $\supT(p,\fq)$ exactly using ray shooting
in an arrangement of rays (lines) and the other to approximate
$\supT(p)$ numerically using a binary search.

\parasc*{Local and nonlocal cone constraints}

Partition the front into two subsets of points: (i)~points in the star
of $\fp$ (``local'' points), and (ii)~points everywhere else on the
front (``remote'' points).  Corresponding to each subset we have two
disjoint subsets of cones of influence---$\mathcal{C}_{\text{local}}$
and $\mathcal{C}_{\text{remote}}$ respectively.  Each subset of cones
limits the new time value of $p$ and so the final time value is the
smaller of the two values for each of $\mathcal{C}_{\text{local}}$ and
$\mathcal{C}_{\text{remote}}$ taken separately.

Consider the subset $\mathcal{C}_{\text{local}}$.  Let
$\S_{\text{local}}$ denote the smallest slope among all cones of
influence in $\mathcal{C}_{\text{local}}$ The segment $\fp'\fq$ is
causal only if its slope is less than $\S_{\text{local}}$.  Let
$t_{\text{local}}$ be the supremum time value of $\fp'$ for which the
slope of $\fp'\fq$ is less than $\S_{\text{local}}$.  To compute
$t_{\text{local}}$ we substitute $\S_{\text{local}}$ in the condition
for causality of $\fp'\fq$
(Equation~\ref{eqn:1d:causalityconstraint}).

Next consider the subset $\mathcal{C}_{\text{remote}}$.  The front
$\tau_i$ is strictly below every cone in $\mathcal{C}_{\text{remote}}$
because $\tau_i$ is causal.  The segment $\fp'\fq$ is causal only if it
is also strictly below every cone in $\mathcal{C}_{\text{remote}}$.
Given a cone $C \in \mathcal{C}_{\text{remote}}$, $C$ intersects
$\fp'\fq$ if and only if $C$ intersects the tentpole $\fp\fp'$.  Let
$t_{\text{remote}}$ denote the smallest time value $T$ for which the
tentpole $\fp\fp'$ where $\fp'=(p,T)$ intersects exactly one cone in
$\mathcal{C}_{\text{remote}}$.    The segment $\fp'\fq$ is causal only
if $T < t_{\text{remote}}$.

Therefore, the progress $\tau_{i+1}(p) - \tau_i(p)$ at step $i+1$ is limited
because
\[
  \supT(p) = \min \{t_{\text{local}}, t_{\text{remote}}\}.
\]
To maximize the progress at the current step, we choose
$\tau_{i+1}(p)$ less than $\supT(p)$, e.g., equal to $\supT(p)$ minus
the machine precision.

We have the following theorem.

\begin{theorem}
  Let $\tau$ be a causal front and let $p$ be an arbitrary local
  minimum of $\tau$.  Let $w_p$ denote the spatial distance from $p$
  to its nearest neighbor.  Then, for every $\dt$ such that $0 \le \dt
  < w_p \minS$ the front $\tau' = \next(\tau,p,\dt)$ is causal.
\label{thm:1d:nonlinear:nextiscausal}
\end{theorem}

\ifproofs\begin{proof}[Proof of Theorem~\ref{thm:1d:nextiscausal}]
  Only the segments of the front incident on $\fp=(p,\tau(p))$ advance
  along with $p$.  Consider an arbitrary segment $pq$ incident on $p$.
  Since $p$ is a local minimum, we have $\tau(q) \ge \tau(p)$.  We
  have
\begin{align*}
\tau'(p)
&\le \tau(p) + \dt\\
&<   \tau(p) + w_p \minS\\
&\le \tau(q) + \abs{pq} \minS\\
&\le \tau(q) + \abs{pq} \S(\fp'\fq)
\end{align*}
  Therefore, the slope of the segment $\fp'\fq$ is less than
  $\S(\fp'\fq)$ and hence $\fp'\fq$ is causal.
\end{proof}\fi

\parasc*{Computing $\mathbf{t_{\text{remote}}}$ exactly}

Computing $t_{\text{remote}}$ is equivalent to answering a ray
shooting query in the arrangement of the cones in
$\mathcal{C}_{\text{remote}}$.  We use a bounding cone hierarchy
$\mathcal{H}$, i.e., a binary tree, obtained from a hierarchical
decomposition of the space domain to efficiently answer the ray
shooting query.  The hierarchical decomposition of the space domain
induces a corresponding hierarchical decomposition of every front.
For each element of this hierarchy, we store a cone that bounds the
cone of influence of every point of the corresponding subset of the
front.  To answer the ray shooting query, we traverse the cone
hierarchy from top to bottom starting at the root.  At every stage, we
store a subset $\mathcal{C}$ of bounding cones such that every cone in
$\mathcal{C}_{\text{remote}}$ is contained in some cone in the subset
$\mathcal{C}$.  The cones in $\mathcal{C}$ are stored in a priority
queue in non-decreasing order of the time value at which the vertical
ray at $\fp$ intersects each cone.  Initially, $\mathcal{C}$ consists
solely of the cone at the root of the hierarchy.  At every stage, if
the cone in $\mathcal{C}$ that has the earliest intersection in time
does not come from a leaf in the hierarchy then we replace it in the
priority queue with its children.  Continuing in this fashion, we
eventually determine the single facet of the front such that the cone
of influence from some point on this facet is intersected first by the
vertical ray at $\fp$.  The time coordinate of the point of
intersection is $t_{\text{remote}}$, the answer to the ray shooting
query.

Because the hierarchy is balanced its depth is $O(\log m)$, where $m$
is the number of simplices in the space mesh.  The tighter the
bounding cones, i.e., the better they approximate the actual cones of
influence, the better is the efficiency of the algorithm.  In
1D$\times$Time, we observed empirically that on average only a few
nodes in the cone hierarchy were examined by this algorithm to
determine the most constraining cone of influence.

\parasc*{Approximating $\mathbf{t_{\text{remote}}}$}

Since we know a range of values $[\tau_i(p) + \minT,
t_{\text{local}}]$ that contains $t_{\text{remote}}$, we can
approximate $t_{\text{remote}}$ up to any desired numerical accuracy
by performing a binary search in this interval.  At every iteration,
we speculatively lift~$\fp$ to the midpoint of the current search
interval.  Let~$\fp''$ be the speculative top of the tentpole
at~$\fp$.  We query the cones of influence in
$\mathcal{C}_{\text{remote}}$ to determine the minimum slope
$\S_{\text{remote}}$ among all cones that intersect~$\fp\fp''$.  If
the maximum slope of the outflow faces incident on~$\fp''$ is less
than $\S_{\text{remote}}$ then we can continue searching in the top
half of the current interval; otherwise, the binary search continues
in the bottom half of the current interval.  The search terminates
when the search interval is smaller than our desired additive error.
A bounding cone hierarchy helps in the same manner as before to
determine the minimum slope among all cones in
$\mathcal{C}_{\text{remote}}$ that intersect~$\fp\fp''$, i.e., all
cones of influence that contain~$\fp''$.

\section{Nonlocal cone constraints in 2D$\times$Time}
\label{sec:nonlinear:2d}

In 2D$\times$Time, the algorithm to maximize the progress at each step
is complicated by the presence of progress constraints in addition to
causality constraints.  In this section, we give an algorithm for
which the progress guarantee at each step is finite and bounded away
from zero.  The minimum progress guarantee is a function of the local
geometry of the spatial projection of the front $\tau_i$ and the
global minimum causal slope, analogous to the progress guarantee
proved in Chapter~\ref{sec:linear} for the constant-wavespeed case.

Our algorithm anticipates changing wavespeeds by simulating the next
few iterations at each step.  The purpose of this lookahead is to
estimate the actual causal slope encountered in the next few
iterations.  As a result, we expect that, in practice, the actual
progress is proportional to the (possibly nonlocal) slope that most
constrains causality at the current step and in the next few
iterations of the algorithm, which may be significantly larger than
the global minimum slope.  Hence, we expect our algorithm to create
spacetime elements whose sizes are proportional to the local geometry
and adapt rapidly to changing causal cone constraints.

\subsection{Looking ahead}

From Theorem~\ref{lemma:2d:nextiscausal}, we observe that the new
front $\fp'\fq\fr$ is causal if the old front $\fp\fq\fr$ satisfies
progress constraint $\S(\fp'\fq\fr)$.  We need to estimate the slope
$\S(\fp'\fq\fr)$ in the next step in order to enforce a progress
constraint in the current step.  This dependency on the future
wavespeed creates the following conundrum.  The progress of the front
at each step~$i$ is limited by the progress constraint that must be
satisfied by the next front at step~${i+1}$.  However, we do not know
what the next front is unless we know how much progress is possible at
step~$i$.  Of course, we can always use the global maximum wavespeed
as a conservative estimate of the slope $\S(\fp'\fq\fr)$; however,
this excessively conservative estimate implies a very low progress
guarantee for problems where the ratio of maximum to minimum
wavespeeds is very large.

To solve this conundrum, we develop a lookahead algorithm that
adaptively estimates future wavespeed.  For a fixed positive
integer~$h$, called the \emph{horizon}, the algorithm at the $i$th
step computes an estimated slope $\estS_h(\fp\fq\fr)$ for every
triangle $\fp\fq\fr$ of the current front.  The horizon $h$ can be
fixed or chosen adaptively at each step.

When $h=0$, we use the minimum slope $\minS$ as an estimate of the
actual causal slope on the front in the next step, so our estimate of
future slope is $\estS_0(\fp\fq\fr) = \minS$. When $h>0$, we can use
the current estimate to compute the next front and the actual slope on
this new front to refine our previous estimate.

\begin{definition}[$h$-progressive]
  Let $h$ be a nonnegative integer, called the \emph{horizon}.

  Let $\fp\fq\fr$ be a given triangle.  We inductively
  define $\triangle{\fp\fq\fr}$ as $h$-progressive as follows.

  \textbf{Base case $h=0$:} Triangle $\fp\fq\fr$ is $0$-progressive if
  and only if it is causal and satisfies progress constraint $\minS$
  (Definition~\ref{def:2d:linear:progressconstraint}).

  \textbf{Case $h > 0$:} Triangle $\fp\fq\fr$ is $h$-progressive if
  and only if all the following conditions are satisfied:

  \begin{enumerate}
  \item $\fp\fq\fr$ is causal;

  \item Let $\fp$ be an arbitrary local minimum vertex of
  $\triangle{\fp\fq\fr}$.  Let $d_p$ denote $\dist(p,\aff{qr})$ and
  let $\minT = \min\{\e,{1-\e}\} \,\allowbreak \minS d_p$.  Let $\fp'\fq\fr =
  \next(\fp\fq\fr, p, \minT)$ be the triangle obtained by advancing
  $\fp$ by $\minT$. Then, $\fp\fq\fr$ must satisfy progress constraint
  $\S(\fp'\fq\fr)$ and $\fp'\fq\fr$ must be
  $\max\{h-1,0\}$-progressive.
  \end{enumerate}

\label{def:2d:h-progressive}
\end{definition}

Note that an $h$-progressive triangle is also causal.

\begin{lemma}
  For every $h \ge 0$, if $\triangle{\fp\fq\fr}$ is $h$-progressive,
  then $\triangle{\fp\fq\fr}$ is ${(h+1)}$-progressive.
\label{lemma:2d:hprogressive:monotone}
\end{lemma}

\ifproofs\begin{proof}[Proof of Lemma~\ref{lemma:2d:hprogressive:monotone}]
If a triangle $\fp\fq\fr$ satisfies progress constraint~$\S$
(Definition~\ref{def:2d:linear:progressconstraint}), then
$\triangle{\fp\fq\fr}$ satisfies progress constraint~$\S'$ for every
${\S' \ge \S}$.  By Lemmas~\ref{lemma:2d:nextiscausal}
and~\ref{lemma:2d:nextisprogressive}, if $\triangle{\fp\fq\fr}$
satisfies progress constraint $\minS$, then the triangle $\fp'\fq\fr$
after pitching a local minimum $\fp$ by $\minT$, where $\minT$ is the
quantity in Definition~\ref{def:2d:h-progressive}, is causal and
satisfies progress constraint $\minS$; since $\minS$ is a lower bound
on the causal slope anywhere in spacetime, we conclude that triangle
$\fp'\fq\fr$ is $0$-progressive.  Therefore, by
Definition~\ref{def:2d:h-progressive}, if $\triangle{\fp\fq\fr}$ is
$0$-progressive, then it is $h$-progressive for every $h \ge 0$.
Hence, if $\triangle{\fp\fq\fr}$ is $h$-progressive, then it is
$h'$-progressive for every $h' \ge h$.
\end{proof}\fi

We are now ready to describe our advancing front algorithm in
2D$\times$Time.  Recall the causality constraint
(Equation~\ref{eqn:2d:causalityconstraint}) and the progress
constraint (Definition~\ref{def:2d:linear:progressconstraint}) that
applies at each step.

\begin{figure*}[t]\centering
\begin{quote}
\textbf{Input:}  A triangulated space domain $\sp \subset \mathbb{E}^2$\\[1em]

\textbf{Output:} A tetrahedral mesh $\spt$ of $\sp \times [0,\infty)$\\[1em]

The initial front $\tau_0$ is $\sp \times \{0\}$, 
corresponding to time $t=0$ everywhere in space.\\[1em]

Fix $h \ge 0$.\\[1em]

Repeat for $i = 0, 1, 2, \ldots$:

\begin{enumerate}

\item Advance in time an arbitrary local minimum vertex $\fp=(p,\tau_i(p))$
of the current front $\tau_i$ to $\fp'=(p,\tau_{i+1}(p))$ such that
$\tau_{i+1}$ is $h$-progressive, and $\tau_{i+1}(p)$ is
maximized.

\item Partition the spacetime volume between $\tau$ and $\tau_{i+1}$
into a patch of tetrahedra, each sharing the tentpole edge $\fp\fp'$.

\item Call the numerical solver to compute the solution in
the spacetime volume between $\tau_i$ and $\tau_{i+1}$.

\end{enumerate}
\end{quote}
\caption{Algorithm in 2D$\times$Time with nonlocal cone constraints.}
\hrule
\end{figure*}

The causal slope on the new front is computed by the solver; cone
constraints can change as a result.  In the parallel setting, nonlocal
cone constraints and updates to the cone hierarchy due to changes in
the cone constraints must be communicated across processors.

We claim that being $h$-progressive guarantees finite positive
progress at each step.  If $\triangle{\fp\fq\fr}$ is $h$-progressive
then for every $\dt$ in the range $0 \le \dt \le
\min\{\e,1-\e\} \minS \width(pqr)$ the triangle
$\fp'\fq\fr$ obtained by pitching an arbitrary local minimum vertex
$\fp$ by $\dt$, is $h$-progressive.  The amount of progress is a
function of $\triangle{pqr}$, the distribution of wavespeeds, and the
parameters $\e$ and $h$.

\begin{lemma}
  Suppose $\fp\fq\fr$ is a triangle of an $h$-progressive front $\tau$
  for some $h \ge 0$.  Let $\fp$ be an arbitrary local minimum vertex
  of $\triangle{\fp\fq\fr}$.  Let $d_p$ denote $\dist(p,\aff{qr})$ and
  let $\minT = \min\{\e,1-\e\} \minS d_p$.  Let $\fp'\fq\fr$ denote
  the corresponding triangle where $\fp'=(p,\tau(p)+\dt)$ for an
  arbitrary $\dt \in [0,\minT]$.

  Then, $\triangle{\fp'\fq\fr}$ is $h$-progressive.
\label{lemma:2d:nextishprogressive}
\end{lemma}

\ifproofs\begin{proof}[Proof of Lemma~\ref{lemma:2d:nextishprogressive}]
By Definition~\ref{def:2d:h-progressive}, triangle $\fp'\fq\fr$ is
${(h-1)}$-progressive.  By Lemma~\ref{lemma:2d:hprogressive:monotone},
triangle $\fp'\fq\fr$ is also $h$-progressive.
\end{proof}\fi

By Lemma~\ref{lemma:2d:nextishprogressive}, if the amount of progress
made by $p$ in every step is no more than the minimum $\minT =
\min\{\e,{1-\e}\} \,\allowbreak \minS \allowbreak\,\dist(p,\aff{qr})$
for every triangle $pqr$, then every front is $h$-progressive.
However, the actual progress at each step would be no more than the
progress guarantee obtained by imposing the progress constraint
$\minS$ of Definition~\ref{def:2d:linear:progressconstraint} at each
step.  It is important to minimize the number of spacetime elements by
exploiting the fact that the actual wavespeed in the next step may be
significantly smaller than the global maximum wavespeed; i.e., to take
advantage of the possibility that $\S(\fp'\fq\fr) \gg
\minS$ in Definition \ref{def:2d:h-progressive}.  In the next section,
we give an algorithm to greedily maximize the progress of local
minimum vertex~$\fp$ at each step, possibly at the expense of the
progress in subsequent steps, but still retain the minimum progress
guarantee of~$\minT$ when pitching~$\fp$.

\subsection{Greedily maximizing progress}

We want to maximize the progress at each step in a greedy
fashion, i.e., at the $i$th step given an arbitrary local minimum
vertex $p$ of the front $\tau_i$ we want to maximize
$\tau_{i+1}(p)$ where $\tau_{i+1} = \next(\tau_i,p,\dt)$ subject to
the constraint that $\tau_{i+1}$ is causal and $h$-progressive.

For a fixed triangle $pqr$ incident on $p$ let $\supT(p,\fq\fr)$ denote
\[
  \supT(p,\fq\fr)
:=
  \sup\,\{t : \text{$\fp'\fq\fr$ is causal and $h$-progressive,
              where $\fp'=(p,t)$}\}
\]
To maximize the progress at step $i$, we would like to compute
$\supT(p,\fq\fr)$.

Similar to the 1D$\times$Time case, partition the set of cones of
influence from points on the front $\tau_i$ into local and remote
subsets.  Let $\S_{\text{local}}$ denote the smallest slope among all
local cones of influence.  The simplex $\fp'\fq\fr$ is causal only if
its slope is less than~$\S_{\text{local}}$.  Let $t_{\text{local}}$ be
the supremum time value of $\fp'$ for which the slope of $\fp'\fq\fr$
is less than~$\S_{\text{local}}$.  To compute $t_{\text{local}}$ we
substitute $\S_{\text{local}}$ in the condition for causality of
$\fp'\fq\fr$ (Equation~\ref{eqn:2d:causalityconstraint}).

Unlike the 1D$\times$Time case, $\supT(p,\fq\fr)$ cannot be computed
by ray shooting queries.  In 2D$\times$Time, we need an oracle to
determine which among several cones is intersected for the smallest
$T$ by a triangle $\fp'\fq'\fr$ when the vertex $\fp$ of
$\triangle{\fp\fq\fr}$ is lifted to $\fp'=(p,T)$ while also
lifting~$\fq$ by a positive amount..  In spatial dimension $d \ge 2$,
the algorithm to compute $\supT(p,\fq\fr)$ requires queries involving
shooting ${(d-1)}$-dimensional faces of the current front.

Definition \ref{def:2d:h-progressive} immediately gives an algorithm
to answer the following question: given a triangle $\fp\fq\fr$ in
spacetime and an integer $h \ge 0$, is $\triangle{\fp\fq\fr}$
$h$-progressive?  Using the algorithm to answer this decision
question, we can approximate $\supT(p,\fq\fr)$ up to any given
numerical accuracy by performing a binary search in the interval
$(\tau_i(p), t_{\text{local}}]$ which we know contains
$\supT(p,\fq\fr)$.  The eventual height of the tentpole $\fp\fp'$ is
at least $\minT$, i.e., positive and bounded away from zero.

\def\Tstar{\ensuremath{T^*}}
\def\Tnew{\ensuremath{T}_{\text{new}}}
\def\Tnext{\ensuremath{T}_{\text{next}}}
\begin{figure}[t]\centering
\fbox{
\begin{algorithm}
\textsc{MaximizeProgress}( Front $\triangle{\fp\fq\fr}$,
                           Vertex $\fp$,
                           Integer $h > 0$ ):\\
 1.\, \textsl{Comment: $\estS_h(\fp\fq\fr)$ is the current estimate of
              the slope in the next step.}\\
 2.\, $\estS_h \leftarrow \minS$;\\
 3.\, done $\leftarrow$ false;\\
 4.\, while not done:\\
\> 4.1.\, compute the maximum time value $\Tstar$ such that\\
   \>\> $\triangle{\fp'\fq\fr}$ where $\fp' = (p,\Tstar)$\\
   \>\> is causal and satisfies progress constraint~$\estS_h$;\\
\> 4.2.\, lift $p$ to time value $\Tstar$ giving
$\triangle{\fp^*\fq\fr}$;\\
\> 4.3.\, \textsl{Comment: recursively compute $\estS_{h-1}(\fp^*\fq\fr)$.}\\
\> 4.4.\, $\S'$ $\leftarrow$ 
          \textsc{FutureSlope}($\fp^*\fq\fr$, $h-1$);\\
\> 4.5.\, done $\leftarrow$ true;\\
\> 4.6.\, let $\fp''\fq\fr$ denote the triangle after advancing $\fp$\\
\>\> so that $\fp''\fq\fr$ is causal,
     it satisfies progress constraint $\S'$,\\
\>\> and the height of $\fp\fp''$ is maximized;\\
\> 4.7.\, if $\S(\fp''\fq\fr)$ $>$
              $\estS_h$:\\
\>\> 4.7.1.\, \{\textsl{Comment: Improve the current estimate $\estS_h$.}\}\\
\>\> 4.7.2.\, $\estS_h \leftarrow \S(\fp''\fq\fr)$;\\
\>\> 4.7.3.\, done $\leftarrow$ false;\\
 5.\, return $\Tstar$;
\end{algorithm}
}\\
\vspace*{1ex}
\fbox{
\begin{algorithm}
\textsc{FutureSlope}( Front $\triangle{\fa\fb\fc}$,
                      Integer $h' \ge 0$ ):\\
 1.\, if $h' = 0$:\\
\> 1.1.\, return $\minS$;\\
 2.\, $\estS \leftarrow \infty$;\\
 3.\, for every local minimum, say $\fa$, of $\triangle{\fa\fb\fc}$:\\
\> 3.1.\, $T'$ $\leftarrow$
            \textsc{MaximizeProgress}($\triangle{\fa\fb\fc}$, \fa,
                                      $h'$);\\
\> 3.2.\, let $\triangle{\fa'\fb\fc}$ be the triangle after
          advancing $\fa$ to $\fa'=(a,T')$;\\
\> 3.3.\, $\estS \leftarrow \min\{\estS, \S(\fa'\fb\fc)\}$;\\
 4.\, return $\estS$;
\end{algorithm}
}
\caption{Algorithm to maximize height of tentpole $\fp\fp'$ subject to
$h$-progressive constraints}
\label{fig:2d:algMaximizeProgress}
\end{figure}

Algorithm \textsc{MaximizeProgress} of
Figure~\ref{fig:2d:algMaximizeProgress} computes $\supT(p,\fq\fr)$.
The correctness of the algorithm follows from the following
observation that \textsc{FutureSlope} estimates the causal slope of
the triangle in the next step after advancing sufficiently in time.
By Lemmas~\ref{lemma:2d:nextiscausal}
and~\ref{lemma:2d:nextisprogressive}, \textsc{FutureSlope} advances
the local minimum $\fa$ by at least $\minT = \min\{\e,1-\e\} \minS
\dist(a,\aff{bc})$ to $\fa'$.  Therefore, the estimated slope $\estS$
computed by \textsc{FutureSlope} is a sufficiently accurate estimate
of the future slope to ensure that the triangle $\fp'\fq\fr$ obtained
by advancing the local minimum $\fp$ to the time value $\Tstar$
computed by \textsc{MaximizeProgress} satisfies the definition of an
$h$-progressive triangle (Definition~\ref{def:2d:h-progressive}).

Maximizing the progress at the current step going from
$\triangle{\fp\fq\fr}$ to $\triangle{\fp'\fq\fr}$ is equivalent to
computing the maximum slope $\estS_h(\fp'\fq\fr)$ such that
$\triangle{\fp'\fq\fr}$ is $h$-progressive.  Algorithm
\textsc{MaximizeProgress} relies on a procedure to compute the slope
anywhere on a triangle $\triangle'$ in the future.  We need a
procedure to compute the minimum slope among all cones of influence
emanating from the current front that intersect $\triangle'$.  The
algorithm also uses geometric primitives such as ray shooting to
maximize the height of a tentpole subject to given causality and
progress constraints.  We will postpone discussion of the
implementation of these low-level subroutines to a later section.

We thus have the following theorems.

\begin{theorem}
  For every $i \ge 0$, if the front $\tau_i$ is $h$-progressive, then
  $\tau_i$ is valid.
\label{thm:2d:nonlinear:progressiveisvalid}
\end{theorem}

\ifproofs\begin{proof}[Proof of
Theorem~\ref{thm:2d:nonlinear:progressiveisvalid}] We prove the
statement by induction on $i$. The proof is almost identical to that
of Theorem~\ref{thm:1d:causalisvalid} except with the additional
complication of progress constraints in 2D$\times$Time. The initial
front $\tau_0$ is progressive by definition.  By
Lemma~\ref{lemma:2d:nextiscausal} and
Lemma~\ref{lemma:2d:nextishprogressive}, at each step $i$ the
algorithm advances a local minimum vertex $p$ of the the
$h$-progressive front $\tau_i$ to the front $\tau_{i+1}$ such that
$\tau_{i+1}$ is $h$-progressive.  Let $w_p$ denote the minimum
distance $\dist(p,\aff{qr})$ for every edge $qr$ in $\link(p)$.  By
Lemma~\ref{lemma:2d:nextishprogressive}, we know that $\tau_{i+1} \ge
\tau_i(p) + \minT$ where $\minT = \min\{\e,1-\e\} w_p \minS$.
Therefore, for every target time $T \ge 0$, the entire front achieves
or exceeds time $T$ in a finite number of steps.
\end{proof}\fi

Thus, we obtain a theorem analogous to Theorem~\ref{thm:2d:main}.

\begin{theorem}
  Given a triangulation $\sp$ of a bounded planar space domain where
  $\minW$ is the minimum width of a simplex of $\sp$ and $\S$ is the
  minimum slope anywhere in $\sp \times [0,\infty)$, for every $\e$
  such that $0 < \e < 1$ our algorithm constructs a simplicial mesh of
  $\sp \times [0,T]$ consisting of at most $\ceil{\frac{n (T +
  \diam(\sp) \maxS)}{\min\{\e,1-\e\} \minW \minS} \Delta}$ spacetime
  elements for every real $T \ge 0$, where $\Delta$ is the maximum
  vertex degree.
\label{thm:2d:nonlinear:main}
\end{theorem}
\ifproofs\begin{proof}[Proof of Theorem~\ref{thm:2d:nonlinear:main}]
  By Lemmas \ref{lemma:2d:nextiscausal} and
  \ref{lemma:2d:nextisprogressive}, it follows that the height of each
  tentpole constructed by the algorithm is at least $\minT =
  \min\{\e,1-\e\} \minW \minS$.  By
  Theorem~\ref{thm:2d:progressiveisvalid}, after constructing at most
  $k \le \ceil{\frac{n (T + \diam(\sp) \maxS)}{\minT}}$ patches, the
  entire front $\tau_k$ is past the target time $T$.  Since each patch
  consists of at most $\Delta$ elements, where $\Delta$ is
  the maximum number of simplices in the star of any vertex of $\sp$,
  the theorem follows.
\end{proof}\fi


\section{Nonlocal cone constraints in arbitrary dimensions $d$D$\times$Time}
\label{sec:nonlinear:hidim}

The algorithm and analysis of Section~\ref{sec:nonlinear:2d} extends
in a straightforward manner to higher dimensions $d \ge 3$.  The only
additional complications involve the bounding cone hierarchy and
queries to this hierarchy.  For instance, for $d=3$, whenever a vertex
$p$ of tetrahedron $pqrs$ is advanced in time, we need to query
whether the spacetime triangle corresponding to $\triangle{pqr} \in
\st(p)$ intersects any four-dimensional cone of influence.

We give in this section the definitions in arbitrary dimensions
corresponding to those in Section~\ref{sec:nonlinear:2d} and state the
equivalent theorems without repeating their proofs because they are
analogous to the corresponding theorems in~2D$\times$Time.

Consider an arbitrary $k$-dimensional face $p_0p_1p_2{\ldots}p_k$
incident on $p_0$.

\begin{definition}[$h$-progressive]
  Let $h$ be a nonnegative integer, called the \emph{horizon}.

  Let $\fp_0\fp_1\fp_2{\ldots}\fp_k$ be a given $k$-simplex.  We inductively
  define $\fp_0\fp_1\fp_2{\ldots}\fp_k$ as $h$-progressive as follows.

  \textbf{Base case $h=0$:} $\fp_0\fp_1\fp_2{\ldots}\fp_k$ is
  $0$-progressive if and only if it is causal and satisfies progress
  constraint $\minS$
  (Definition~\ref{def:hidim:linear:progressconstraint}).

  \textbf{Case $h > 0$:} $\fp_0\fp_1\fp_2{\ldots}\fp_k$ is
  $h$-progressive if and only if all the following conditions are
  satisfied:
  \begin{enumerate}
  \item $\fp_0\fp_1\fp_2{\ldots}\fp_k$ is causal;

  \item Let $\fp_0$ be an arbitrary local minimum vertex of
  $\fp_0\fp_1\fp_2{\ldots}\fp_k$.  Let $d_{p_0}$ denote
  $\dist(p_0,\aff{p_1p_2{\ldots}p_k})$ and let $\minT =
  \min\{\e,1-\e\} \minS d_{p_0}$.  Let $\fp_0'\fp_1\fp_2{\ldots}\fp_k
  = \next(\fp_0\fp_1\fp_2{\ldots}\fp_k, p_0, \minT)$ be the simplex
  obtained by advancing $\fp_0$ by $\minT$. Then,
  $\fp_0\fp_1\fp_2{\ldots}\fp_k$ must satisfy progress constraint
  $\S(\fp_0'\fp_1\fp_2{\ldots}\fp_k)$ and
  $\fp_0'\fp_1\fp_2{\ldots}\fp_k$ must be $\max\{h-1,0\}$-progressive.
  \end{enumerate}

\label{def:hidim:h-progressive}
\end{definition}

Note that an $h$-progressive simplex is also causal.

\begin{lemma}
  Suppose $\fp_0\fp_1\fp_2{\ldots}\fp_k$ is a simplex of an
  $h$-progressive front $\tau$ for some $h \ge 0$.  Let $\fp_0$ be an
  arbitrary local minimum vertex of $\fp_0\fp_1\fp_2{\ldots}\fp_k$.
  Let $d_{p_0}$ denote $\dist(p_0,\aff{p_1p_2{\ldots}p_k})$ and let
  $\minT = \min\{\e,1-\e\} \minS d_{p_0}$.  Let
  $\fp_0'\fp_1\fp_2{\ldots}\fp_k$ denote the corresponding simplex
  where $\fp_0'=(p_0,\tau(p_0)+\dt)$ for an arbitrary $\dt \in
  [0,\minT]$.

  Then, $\fp_0'\fp_1\fp_2{\ldots}\fp_k$ is $h$-progressive.
\label{lemma:hidim:nextishprogressive}
\end{lemma}

We have thus shown the finite termination of the algorithm.

\begin{theorem}
  Given a triangulation $\sp$ of a bounded $d$-dimensional space
  domain where $\minW$ is the minimum width of a simplex of $\sp$, for
  every $\e$ such that $0 < \e < 1$ our algorithm constructs a
  simplicial mesh of $\sp \times [0,T]$ consisting of at most
  $\ceil{\frac{n (T + \diam(\sp) \maxS)}{\min\{\e,1-\e\} \minW \minS}
  \Delta}$ spacetime elements for every real $T \ge 0$, where $\Delta$
  is the maximum vertex degree.
\label{thm:hidim:nonlinear:main}
\end{theorem}
\ifproofs\begin{proof}[Proof of Theorem~\ref{thm:2d:nonlinear:main}]
  By Lemmas \ref{lemma:hidim:nextiscausal} and
  \ref{lemma:hidim:nextisprogressive}, it follows that the height of each
  tentpole constructed by the algorithm is at least $\minT =
  \min\{\e,1-\e\} \minW \minS$.  By
  Theorem~\ref{thm:2d:progressiveisvalid}, after constructing at most
  $k \le \ceil{\frac{n (T + \diam(\sp) \maxS)}{\minT}}$ patches, the
  entire front $\tau_k$ is past the target time $T$.  Since each patch
  consists of at most $\Delta$ elements, where $\Delta$ is
  the maximum number of simplices in the star of any vertex of $\sp$,
  the theorem follows.
\end{proof}\fi



\section{Estimating future slope}

The algorithms of previous sections rely on efficient answers to
the following questions:
\begin{enumerate}
\item Given a triangle $\fp\fq\fr$ and a set of cones
of influence, what is the slope of the fastest cone of influence that
intersects $\triangle{\fp\fq\fr}$?
\item Given a triangle $\fp\fq\fr$ with $\fp$ as
a lowest vertex and a slope $\S$, what is the supremum height of the
tentpole $\fp\fp'$ such that $\triangle{\fp'\fq\fr}$ has slope less
than $\S$?
\end{enumerate}

Maintaining the entire arrangement of cones of influence is expensive
and unnecessary for our purpose; it suffices to obtain a cone that
bounds (tightly) the actual cone of influence at $\fp$.  In the
absence of focusing, the cone of influence of any point $\fp$ is
contained in the cone of influence of every point $\fq$ of the front
such that $\fp \in \cone^+(\fq)$.

\begin{figure*}
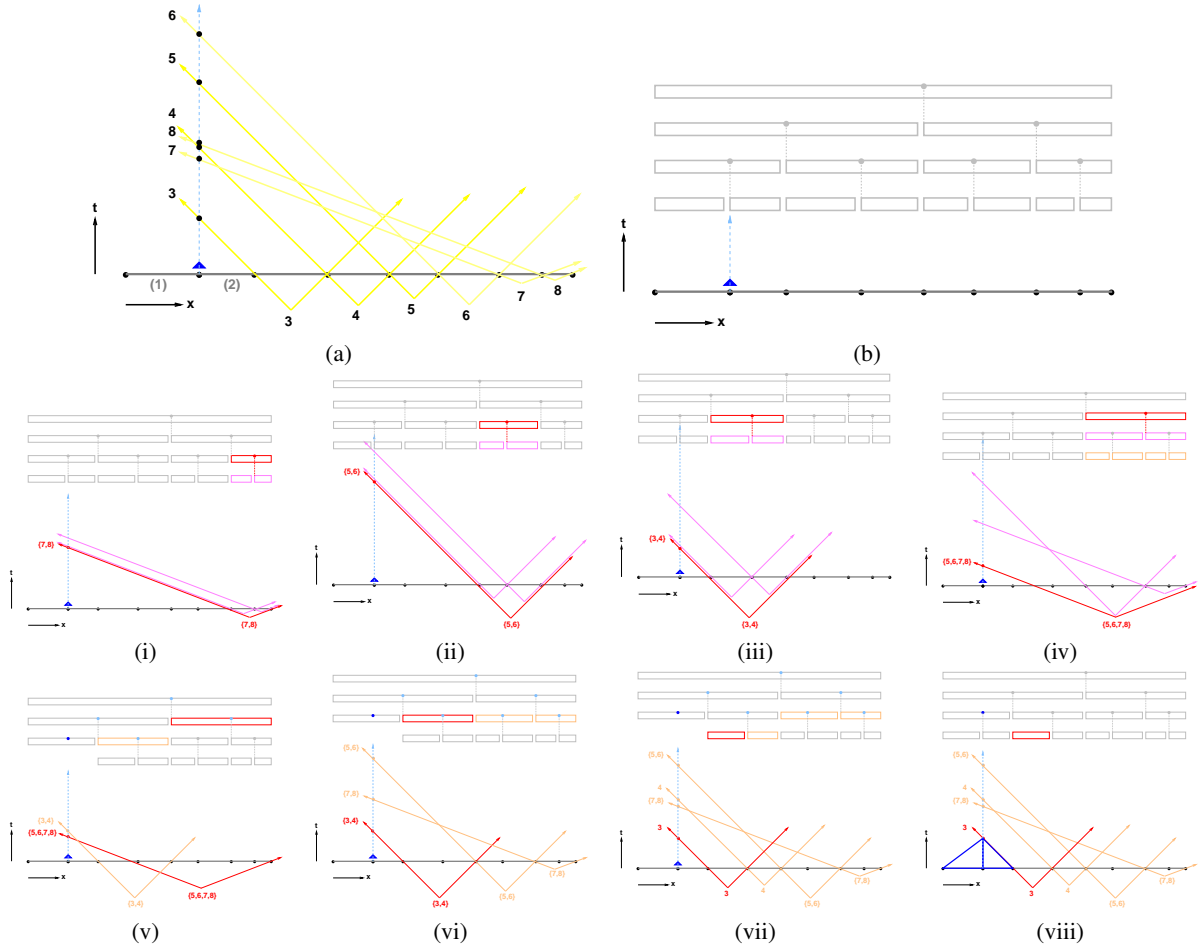
\centering\small
\begin{tabular}{cc}
\includegraphics[width=.4\textwidth]{\fig{hier-0}} &
\includegraphics[width=.4\textwidth]{\fig{hierarchy}}\\
(a) & (b)
\end{tabular}
\begin{tabular}{cccc}
\includegraphics[width=.22\textwidth]{\fig{hier-1}} &
\includegraphics[width=.22\textwidth]{\fig{hier-2}} &
\includegraphics[width=.22\textwidth]{\fig{hier-3}} &
\includegraphics[width=.22\textwidth]{\fig{hier-4}}\\
(i) & (ii) & (iii) & (iv)
\end{tabular}
\begin{tabular}{cccc}
\includegraphics[width=.22\textwidth]{\fig{hier-5}} &
\includegraphics[width=.22\textwidth]{\fig{hier-6}} &
\includegraphics[width=.22\textwidth]{\fig{hier-7}} &
\includegraphics[width=.22\textwidth]{\fig{hier-8}}\\
(v) & (vi) & (vii) & (viii)
\end{tabular}
\caption[Constructing and traversing a bounding cone
hierarchy]{Constructing and traversing a bounding cone hierarchy.
(i)~All the nonlocal cone constraints limiting the height of a
tentpole are grouped into (ii)~a cone hierarchy (a binary tree)
induced by a recursive subdivision of the front.  (i)--(iv) The cone
hierarchy is built bottom-up by merging pairs of bounding cones at
each step.  (v)--(viii) The hierarchy is expanded top-down until the
strictest cone constraint is a leaf in the hierarchy.  Only a fraction
of all cone constraints in the hierarchy are examined while traversing
the cone hierarchy.}
\label{fig:conehierarchy}
\end{figure*}

At each step, we maintain a hierarchical decomposition of the front.
We build a bounding cone hierarchy corresponding to this hierarchical
partition.  See Figure~\ref{fig:conehierarchy}.  We query the cone
hierarchy to efficiently maximize the progress in time at each tent
pitching step.  In 1D$\times$Time, this query is equivalent to
shooting a ray (the infinite extension of the tentpole into the
future) and determining its earliest intersection with any cone of
influence.  After each step, we update the cones stored in the cone
hierarchy to reflect the new wavespeeds computed on the new front.

Because of no-focusing (Axiom~\ref{axiom:nofocusing}), we can
determine (a lower bound on) the slope at a point $\fp$ in the future
by computing all cones of influence from points $\fq$ on the current
front that contain $\fp$.  The shallowest such cone determines a lower
bound on $\S(\fp)$.  It can be computationally very expensive to
determine the shallowest cone of influence that contains a given
point~$\fp$.  In particular, the shallowest cone of influence
containing~$\fp$ may correspond to a point~$\fq$ arbitrarily far
from~$\fp$.  To compute this nonlocal cone constraint efficiently, we
use a standard hierarchical decomposition, called a \emph{bounding
cone hierarchy}, of the space domain.  The elements in the hierarchy
correspond to subsets of the space domain.  For each element of the
hierarchy, we compute the minimum slope within the corresponding
subset of the space domain.  The smallest element in the hierarchy is
a single simplex.  In order to determine the strictest cone constraint
that applies locally, we traverse the hierarchy until we determine the
simplex with minimum slope whose cone of influence contains~$\fp$.  In
practice, we expect that our algorithm has to examine only a small
subset of the hierarchy; we have observed the resulting speed-up in
1D$\times$Time.  In the worst case, the algorithm has to examine every
simplex of the front, but in that case the algorithm will be at most a
constant factor slower than one that does not use a bounding cone
hierarchy.  When a patch is solved, the bounding cones are updated
with the new slopes by traversing a path from a leaf to the root of
the hierarchy.  This hierarchical approximation technique has been
applied very successfully to numerous simulation problems, such as the
Barnes-Hut divide-and-conquer method~\cite{barnes-hut86nbody} for
$N$-body simulations, as well as to collision detection in computer
graphics and robot motion planning~\cite{lin96collision} and for
indexing multi-dimensional data in geographic information
systems~\cite{guttman84rtrees}.

\subsection*{Chapter summary}

We have shown how to extend the Tent Pitcher algorithm for arbitrary
dimensions to nonlinear problems where the wavespeed is not constant.
Our expressions for the causality and progress constraints that apply
at each step make explicit the dependence on the slope of the cone of
influence most constraining the progress at that step.  This
dependence is not explicit in the formul\ae{} of Erickson \etal{}
because they assume without loss of generality that the slope is~$1$
everywhere in spacetime.  For the constant wavespeed case, the
algorithm in this paper is an alternative to the algorithm due to
Erickson \etal{} with potentially weaker progress constraints.  We can
view the algorithm of Erickson \etal{} as looking one step ahead in
the sense that the progress constraint at step $i$ guarantees that the
front constructed in step $i+1$ is causal.  Our algorithm can be
viewed as looking further---our progress constraint at step $i$
guarantees that the front constructed in step $i+h$ is causal.  It
needs to be investigated whether the extra complexity of the algorithm
for $h > 2$ or adaptively choosing $h$ at each step is justified by a
more efficient simulation overall.

The nonlocal nature of the cone constraints pose significant
challenges while implementing the algorithm in this chapter in
parallel.  Maintaining a cone hierarchy and performing queries of the
sort needed by our algorithm is also a significant challenge for
spatial dimensions $d \ge 3$.


\chapter{Adaptive refinement and coarsening}
\label{sec:adaptive}

The duration of spacetime elements constructed by our algorithm is
constrained by causality; however, all spacetime elements constructed
by pitching a simplex of the front have the same spatial diameter as
the corresponding simplex of the initial front, i.e., the space mesh
$\sp$.  In other words, every front constructed by our algorithm so
far is only as refined as the initial mesh $\sp$.  Parts of the
spacetime domain, where the solution changes rapidly, need to be
meshed with smaller elements to achieve the given error tolerance.  In
this chapter, we make our meshing algorithms adaptive to local
\textsl{a posteriori} estimates of the numerical error in response to
which we refine or coarsen the current front.  Refinement of the
current front means that spacetime elements constructed in future
steps will have smaller size, and, after a finite number of refinement
steps, reduced error.  By adapting the size of the spacetime elements
to the error estimate, we are able to compute a more efficient mesh
for a given error bound.  Without adaptivity, we would require the
initial space mesh to be as refined as the finest resolution required
at any time, which would necessitate many more elements for meshing
the same spacetime volume.

In this chapter, we concentrate on the case of constant slope (i.e.,
constant wavespeed) where $\S$ denotes the slope everywhere in
spacetime or more generally $\S = \minS$ is the reciprocal of the
globally maximum wavespeed.  In Chapter~\ref{sec:biadaptive} we will
discuss adaptive refinement and coarsening in the presence of changing
wavespeeds.

Also, in this chapter, we rectify an oversight in the proof of
correctness in our paper~\cite{abedi04spacetime} by including a case
in the proof of Theorem~\ref{thm:2d:refine:iscausal} that was missing
from the corresponding theorem in the paper.

\section{Problem statement}

Mesh adaptivity in our advancing front framework is the problem of
adapting the spatial size of front facets.  We require an algorithm
that responds to error tolerance requirements by reducing the spatial
size of front facets where necessary so that future spacetime elements
are smaller and therefore have reduced error.  When the error estimate
is sufficiently greater than the allowed error, we want the algorithm
to increase the spatial size of front facets wherever permitted so
that future spacetime elements are bigger.  Reducing the error is a
requirement for the solution strategy to advance; hence, it is
necessary for the algorithm to respond immediately to refinement
demands.  However, coarsening is a request---an excessively refined
mesh still gives a sufficiently accurate solution but it is not
efficient.  The algorithm should coarsen aggressively to be efficient
but it can decide to postpone a coarsening request until a later step.

Our main challenge is to incorporate refinement and coarsening of the
front at each step into our advancing front framework.  Each iteration
of the advancing front algorithm, i.e., each application of the
$\next$ function, is understood to be a tent pitching step followed by
zero or more refinement and coarsening operations.  The actual
operations for refining and coarsening the current front are described
in detail later in this chapter.

Mesh refinement and coarsening affects spacetime elements in the
future when some vertex of the currently refined front is pitched.
Since our algorithm coarsens only a pair of triangles that have been
previously obtained by refinement, it is advantageous to start with
the coarsest acceptable space mesh, e.g., the coarsest mesh that is
still a good enough piecewise linear approximation to the space
domain, and let the meshing algorithm adapt the mesh locally as
necessary in response to the demands for numerical accuracy.  In
Chapter \ref{sec:extensions} we discuss generalization of coarsening
beyond a simple undoing of a previous refinement.

In dimension $d \ge 2$, we saw in Chapter~\ref{sec:linear} that
nontrivial progress constraints are necessary to guarantee positive
progress at each step.  These progress constraints are functions of
the shape of the triangles on the front.  But the shape of front
facets, or more accurately the shape of the spatial projection of the
front facets, is changing as a result of refinement and coarsening!
Therefore, the challenge is to modify the progress constraint to
anticipate changes in shape due to an arbitrary amount of refinement.

\medskip

\parasc*{Our solution}

For one-dimensional space domains, we have proved in
Chapter~\ref{sec:linear} that every (causal) front is valid.
Therefore, modifying Tent Pitcher to make it adaptive in 1D$\times$Time
is very easy.  The input space mesh is a one-dimensional simplicial
complex.  To refine the front, we bisect a front segment into two
parts, say two equal halves; coarsening reverses this operation by
merging two segments that are collinear in spacetime.  Since
refinement and coarsening does not alter the gradient of the time
function restricted to the elements involved in these operations,
refinement and coarsening preserve causality.

Note that when pitching a local minimum vertex $\fp$, an interior
vertex, the smaller of the temporal aspect ratios of the resulting two
triangles is maximized if the two edges incident on $\fp$ have
approximately equal lengths of their spatial projections.  We use this
fact to guide our choice of where to subdivide a front segment.

In higher dimensions, we define \emph{progressive} fronts and prove
that if a front is progressive then it is valid.  For $d=2$, we give
an algorithm that, given any progressive front $\tau_i$, constructs a
next front $\tau_{i+1}=\next(\tau_i,p,\dt)$ such that~$\tau_{i+1}$ is
progressive and $\tau_{i+1}(p)$ is maximized.  Whenever $p$ is a local
minimum of~$\tau_i$, the progress $\tau_{i+1}(p) - \tau_i(p)$ is
guaranteed to be at least $\minT$, which is a function of the input
and bounded away from zero.  It is necessary to predict the shape of
the spatial projections of triangles on the new front $\tau_{i+1}$ and
ensure that $\tau_i$ satisfies progress constraints that anticipate
refinement and coarsening of the new front.  We choose a refinement
method, called \emph{newest vertex bisection}, that allows us to
predict all possible shapes of triangles on any front in the future
after an arbitrary number of refinement and coarsening steps.  We
incorporate constraints associated with every such shape in the
progress constraints that must be satisfied by the front at each step,
and call such a front a \emph{progressive} front.  The details and
formal definitions are in Section~\ref{sec:2d:refine:constraints}.

Our algorithm adapts to \textsl{a posteriori} estimates of numerical
error as follows. A patch is solved as soon as it is created. If the
estimated numerical error, i.e., estimated energy dissipation, for any
element in the patch is greater than some threshold $\xi_1$ then the
element is marked for refinement and the patch is rejected. If no
element is marked for refinement, then the patch is accepted.  If the
largest estimated numerical error for an element is less than some
threshold $\xi_2$, where $\xi_2 < \xi_1$, then the element is marked
as coarsenable.  If a patch is rejected, then the front is not
advanced and for every element marked for refinement, the
corresponding facet of the current front is bisected.  Since repeated
bisection decreases the spatial diameter, the size of future spacetime
elements decreases; we assume that a finite number of refinement steps
eventually reduces the numerical error.

For now, we consider only the case $d=2$, i.e., space domains $\sp$
that are 2D simplicial complexes consisting of triangles, edges, and
vertices embedded in some ambient Euclidean space.  Adaptive
refinement and coarsening in dimensions $d > 2$ remains an open
problem and we postpone a discussion of higher dimensions until
Section~\ref{sec:hidim:adaptive}.

If a patch is rejected, then for every tetrahedron $PP'Q'R'$ marked
for refinement the triangle $pqr$ in the spatial projection is
bisected.  A pair of triangles can be coarsened by merging them if the
result is a single triangle.  If the two triangles were previously
obtained by bisecting a larger triangle, then they can always be
merged whenever they are coplanar in spacetime.

\begin{figure*}[t]\centering
\begin{quote}
\textbf{Input:}  A one-dimensional space mesh $\sp \subset \mathbb{E}^1$\\[1em]

\textbf{Output:} A triangular mesh $\spt$ of $\sp \times [0,\infty)$\\[1em]

The initial front $\tau_0$ is $\sp \times \{0\}$, 
corresponding to time $t=0$ everywhere in space.\\[1em]

Repeat for $i = 0, 1, 2, \ldots$:

\begin{enumerate}

\item Advance in time an arbitrary local minimum vertex $\fp=(p,\tau_i(p))$
of the current front $\tau_i$ to $\fp'=(p,\tau_{i+1}(p))$ such that
$\tau_{i+1}$ is causal and $\tau_{i+1}(p)$ is maximized.

\begin{quote}
If all segments of $\tau_i$ incident on $\fp$ are marked as
coarsenable by the solution during the previous iteration, then choose
$\fp'$ so that the coarsenable segments become collinear in time,
unless the height $\tau_{i+1}(p)-\tau_i(p)$ of the resulting tentpole is
too small.
\end{quote}

\item Partition the spacetime volume between $\tau$ and $\tau_{i+1}$
into a patch of triangles, each sharing the tentpole edge $\fp\fp'$.

\item Compute the solution in the spacetime volume between $\tau_i$
and $\tau_{i+1}$ as well as the \aposteriori{} error estimate.

\item If the spacetime error indicator tolerates the error in the
patch, then advance the front to $\tau_{i+1}$ and merge every adjacent
pair of coarsenable collinear segments.

\item Otherwise, some segments of the front are marked for
refinement; bisect each such segment to get the new front
$\tau_{i+1}$.

\end{enumerate}
\end{quote}
\caption{Adaptive meshing algorithm in 1D$\times$Time}
\label{fig:1d:algAdaptive}
\hrule
\end{figure*}

Figure~\ref{fig:1d:algAdaptive} describes our adaptive meshing
algorithm in 1D$\times$Time; figure \ref{fig:2d:algAdaptive} describes
the adaptive algorithm in 2D$\times$Time.  Both algorithms proceed by
pitching local minima, and they refine and coarsen the front in
response to error estimates.  In the parallel setting, we repeatedly
choose an independent set of local minima of the current front, equal
to the number of processors, to be advanced in time simultaneously.
The resulting patches can be solved independently.  If a patch is
accepted, the local neighborhood of the front is advanced without any
conflicts with other patches.  Thus, the solution everywhere in
spacetime is computed patch-by-patch in an order consistent with the
partial order of dependence of patches.

\section{Meshing in 2D$\times$Time}
\label{sec:2d:adaptive}

In this section, we give new adaptive progress constraints in
2D$\times$Time that guarantee that each front, even after an arbitrary
number of refinement and coarsening steps, is able to make positive
progress by tent pitching.

First, we describe our mesh refinement procedure.

\subsection{Hierarchical front refinement}
\label{sec:2d:refinement}

Mesh refinement is needed to achieve the desired numerical accuracy.
Intuitively, the larger is the variation in the solution within a
subdomain, the smaller is the desired element size.  Where the
solution is changing slowly and smoothly, it is sufficient to use
larger elements---since the number of elements is smaller, the
solution procedure is more efficient for the given numerical accuracy.
Therefore, the mesh should be only as refined as necessary.

For tracking evolving phenomena efficiently, it is also necessary to
coarsen previously refined elements when the phenomenon recedes and the
solution is no longer changing rapidly.  Thus, coarsening is desirable
for computational efficiency.  However, over-refinement does not hurt
the quality of the solution.

Since our DG method uses a discontinuous formulation, we do not
require a smooth grading of element size.  In fact, experimental
results suggest that if the mesh successfully tracks a sharp shock,
then it is perfectly acceptable to use very coarse elements aligned
with the shock trajectory.

In the terminology of the different kinds of adaptivity that have been
studied in the literature---$p$-adaptivity (adapt the degree of the
polynomial basis functions), $h$-adaptivity (adapt the size and number
of elements in the mesh), and $r$-adaptivity (adapt the locations of
nodes in the mesh, e.g., smoothing)---we perform $h$-adaptivity to
adapt the size of spacetime elements.

\parasc*{The choice of bisection method}

There is a vast body of literature on mesh refinement for both
structured and unstructured meshes.  See the survey by Jones and
Plassman~\cite{jones97adaptive} for an introduction.  Specifically, we
are interested in hierarchical mesh refinement by recursive bisection.
Regular subdivisions of a simplex were studied by Bank
\etal~\cite{bsw-radsr-83} in 2D, and Edelsbrunner and
Grayson~\cite{eg-ess-00} in higher dimensions.  Mesh refinement in~2D
requires bisecting specified triangles to decrease their diameter as
well as propagating to neighboring triangles to maintain a
triangulation.  The longest edge refinement introduced by Rosenberg
and Stenger~\cite{rs-lbatc-75} has been later popularized by
Rivara~\cite{rivara97longestedge,rivara84algorithms,rivara96new}.
Newest vertex bisection was originally developed by
Sewell~\cite{s-agtpp-72} and later adapted by
Mitchell~\cite{m-umafe-88, m-carte-89, m-arafe-91}.  Arnold
\etal~\cite{arnold00locally}, Bey~\cite{b-tgr-95},
B\"ansch~\cite{b-lmr23-91}, Liu and Joe~\cite{lj-stb-94, lj-qlrtm-95},
Maubach~\cite{maubach95localbisection}, and Bey~\cite{b-sgrfa-00}
extended newest vertex bisection to higher dimensions.  Most
importantly, Arnold \etal{}~\cite{arnold00locally} proved that the
number of shapes generated by their recursive bisection of a simplex
is finite.

Our adaptive algorithm uses the \emph{newest vertex bisection}
refinement method, originally developed by Sewell~\cite{s-agtpp-72},
later adapted by Mitchell~\cite{m-umafe-88, m-carte-89, m-arafe-91} in
the context of multigrid methods, and still later studied and
generalized to three dimensions by B\"ansch~\cite{b-lmr23-91}.  This
method is similar to, but not identical to, \emph{longest-edge}
refinement~\cite{rs-lbatc-75,rivara84algorithms,rivara96new}.

\begin{figure}[t]\centering\small
\includegraphics[width=.9\textwidth]{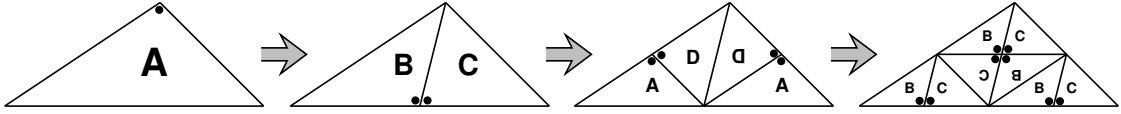}
\caption{Newest vertex refinement~\cite{abedi04spacetime}.  Newest
vertices of each triangle are marked.}
\label{fig:homothetic}
\end{figure}

We call the newest vertex of a triangle its \emph{apex} and the
opposite edge its \emph{base}.  Initially, one vertex
of each triangle in the mesh is chosen arbitrarily as its apex.
Newest vertex bisection replaces a triangle with two smaller
triangles, each with half the area of the original triangle, obtained
by bisecting along the line segment through the apex and the midpoint
of the base.  The new vertex introduced at the midpoint is the newest
vertex of both smaller triangles.  See Figure~\ref{fig:homothetic}.

The descendants of any marked triangle under newest vertex bisection
fall into only eight homothetic classes (Figure~\ref{fig:homothetic}).
There are only four directions along which the triangle or any of its
descendants can be further bisected.  These four directions are
parallel to the three edges of the triangle and the bisecting segment.
Each of the four ways to choose three of the four directions gives two
similar triangle shapes that are mirror images of each other, for a
total of eight triangle shapes, equivalent up to translation and
scaling.  Any two triangles in the same equivalence class have
corresponding newest vertices.  Refinement by three levels is
guaranteed to decrease the diameter by at least a half.

A front triangle $\fp\fq\fr$ and its grandparent
$\triangle{\fa\fb\fc}$ create homothetic tetrahedra when their
corresponding vertices are pitched.  This is because causality and
progress constraints are gradient constraints that scale with the
spatial geometry of front triangles.  Therefore, our refinement
algorithm using newest vertex bisection adapts the resolution of the
spacetime mesh but is limited in its ability to adapt the shape and
temporal aspect ratio of spacetime tetrahedra.

\parasc*{Propagation}

When we refine one triangle in a mesh, we may be forced to refine
other nearby triangles in order to maintain a conforming
triangulation.  Maintaining a triangulated front is necessary for our
algorithm---nonconforming or dangling vertices create complications
that we do not know how to solve yet.

We call an edge of the triangulation a \emph{terminal edge} if it is
the base of every triangle incident on it.

\begin{figure}
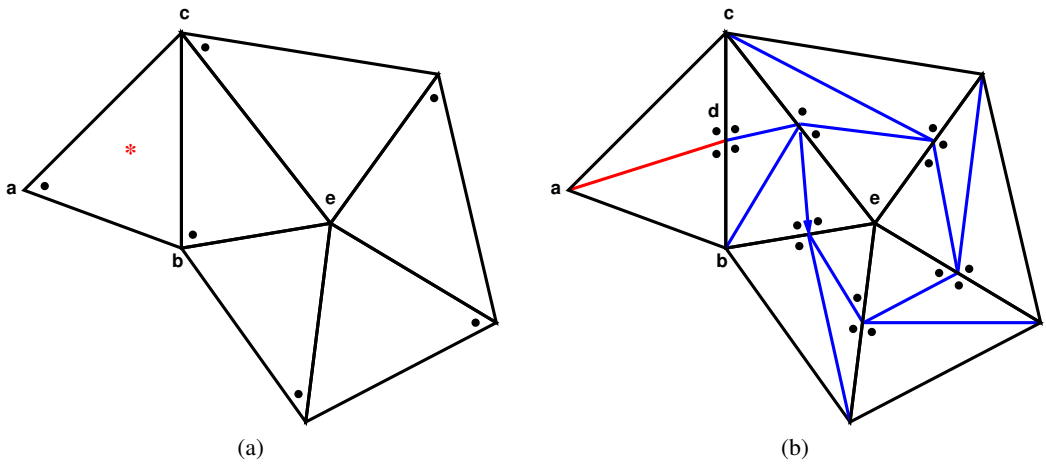
\centering\small
\begin{tabular}{c@{\qquad}c}
  \iffig\includegraphics[width=0.4\textwidth]{\fig{before-propagation}}\fi &
  \iffig\includegraphics[width=0.4\textwidth]{\fig{after-propagation}}\fi \\
 (a) & (b)
\end{tabular}\\
\caption{Refining~$\triangle{abc}$ propagates to neighboring
triangles: (a)~before and (b)~after refinement.}
\label{fig:propagation}
\end{figure}

Suppose vertex $a$ is the apex of a $\triangle{abc}$ that is bisected
(Figure~\ref{fig:propagation}).  If $bc$ is not a boundary edge then
some neighboring triangle $\triangle{cbe}$ shares the edge $bc$.  To
maintain a triangulation, $\triangle{cbe}$ must be bisected also.  If
$bc$ is not the base of $\triangle{cbe}$, then the child of
$\triangle{cbe}$ sharing edge $bc$ must be bisected as well, and the
bisection of $\triangle{cbe}$ will propagate recursively; see
Figure~\ref{fig:propagation}.

It is easy to prove that this propagation terminates, regardless of
which vertex is chosen as the apex of each triangle.  Suppose a
refinement of $\triangle{abc}$ with apex $a$ propagates to a
neighboring triangle sharing the edge $bc$.  A single newest vertex
bisection of $\triangle{abc}$ makes the edges $ab$ and $ac$ the bases
respectively of the two children of $\triangle{abc}$.  If the
propagation revisits $\triangle{abc}$, it must return by bisecting
either the edge $ab$ or the edge $ac$.  But since these edges are
now terminal edges, the propagation terminates at that step.

The propagation path touches every triangle in the worst case, but in
practice, the propagation path usually has small constant length; see
Su\'arez \etal~\cite{spc-pppil-03} for an analysis of a similar
refinement algorithm.  Because we bisect triangles first and then
propagate, instead of the other way round, our algorithm terminates,
regardless of which vertices are marked in each triangle: each
triangle in the mesh is bisected at most twice.  (Mitchell's original
head-recursive algorithm \cite{m-umafe-88, m-carte-89, m-arafe-91} can
enter an infinite loop if the initial selection of marks does not
satisfy a certain matching condition.)

\begin{figure}
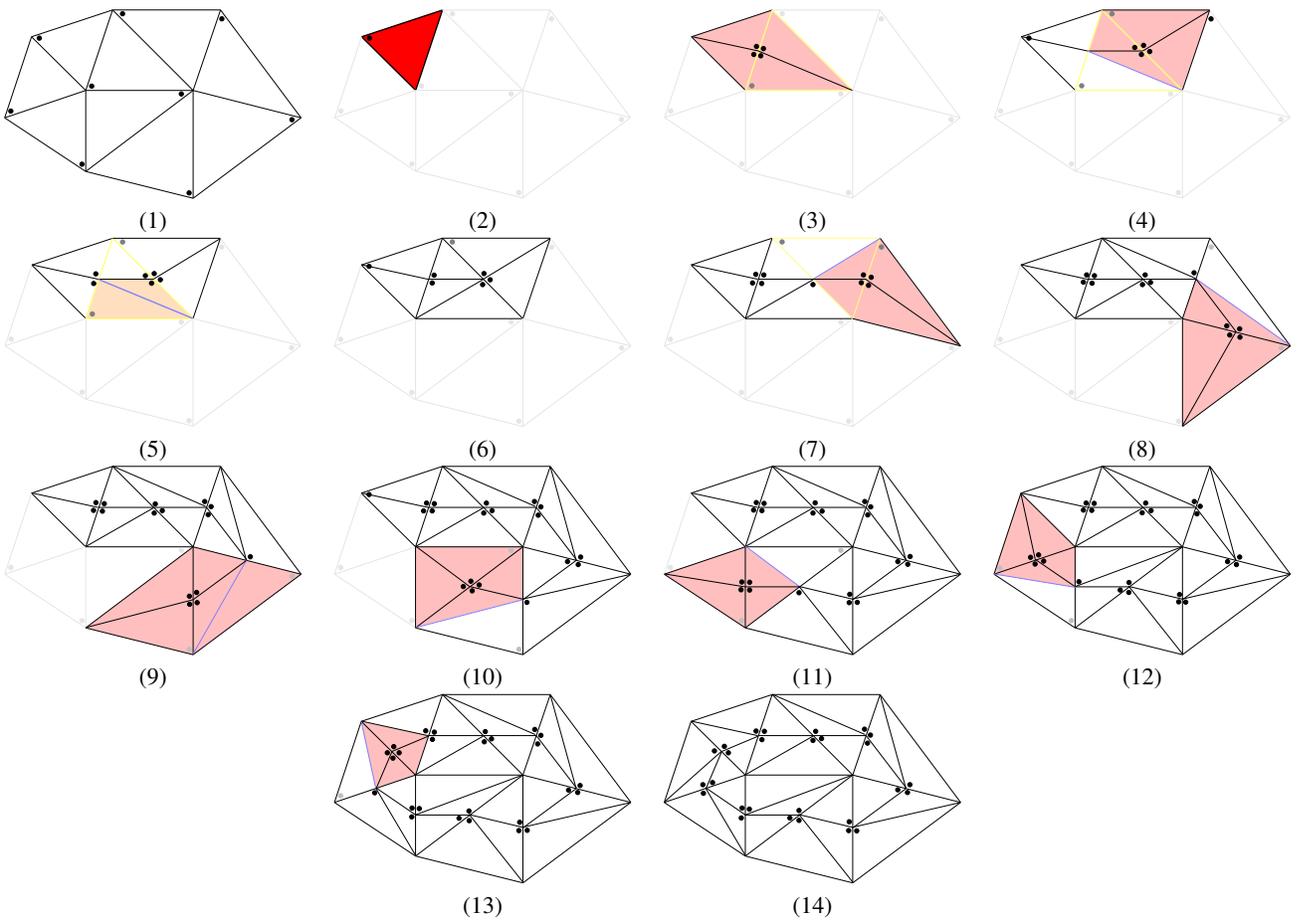
\centering\small
\begin{tabular}{cccc}
\iffig\includegraphics[width=.24\textwidth]{\fig{before-refinement}}\fi&
\iffig\includegraphics[width=.24\textwidth]{\fig{propagate-0}}\fi&
\iffig\includegraphics[width=.24\textwidth]{\fig{propagate-1}}\fi&
\iffig\includegraphics[width=.24\textwidth]{\fig{propagate-2}}\fi\\
(1) & (2) & (3) & (4)\\
\iffig\includegraphics[width=.24\textwidth]{\fig{propagate-2a}}\fi&
\iffig\includegraphics[width=.24\textwidth]{\fig{propagate-2b}}\fi&
\iffig\includegraphics[width=.24\textwidth]{\fig{propagate-3}}\fi&
\iffig\includegraphics[width=.24\textwidth]{\fig{propagate-5}}\fi\\
(5) & (6) & (7) & (8)\\
\iffig\includegraphics[width=.24\textwidth]{\fig{propagate-7}}\fi&
\iffig\includegraphics[width=.24\textwidth]{\fig{propagate-9}}\fi&
\iffig\includegraphics[width=.24\textwidth]{\fig{propagate-11}}\fi&
\iffig\includegraphics[width=.24\textwidth]{\fig{propagate-13}}\fi\\
(9) & (10) & (11) & (12)\\
&
\iffig\includegraphics[width=.24\textwidth]{\fig{propagate-15}}\fi&
\iffig\includegraphics[width=.24\textwidth]{\fig{propagate-16}}\fi\\
& (13) & (14)
\end{tabular}
\caption{Refinement propagation path terminates when it revisits a
  triangle.}
\label{fig:propagationterminates}
\end{figure}

The propagation follows a directed path in the dual graph of the
triangulation, leaving each triangle along the dual edge corresponding
to the base of the triangle.  The propagation
terminates when the base of the last triangle bisected is a terminal
edge.  The propagation either encounters a boundary edge or the
propagation revisits one of the children of a triangle that was
bisected earlier along the propagation path; see
Figure~\ref{fig:propagationterminates}.

Newest vertex bisection never subdivides any angle of a triangle more
than once.  If the apex of each triangle in the initial mesh is the
vertex with largest angle, then every vertex~$v$ is the apex of at
most~$5$ triangles incident on~$v$.  (In the degenerate case where six
equilateral triangles meet at a vertex, we can break ties using a
straightforward symbolic perturbation scheme.)  Thus, the degree of a
vertex~$v$ in the initial space mesh increases by at most~$5$ as a
result of refinement.  If~$v$ is not a vertex of the initial space
mesh then the degree of~$v$ is~$4$ when~$v$ is inserted and is at
most~$8$ at any subsequent step.  We conclude that if a triangulation
has maximum degree $\Delta$, then any refinement of that triangulation
has maximum degree at most $\max\{\Delta+5, 8\}$.

\parasc*{Basic mesh operations}

The entire refinement propagation can be expressed as a sequence of
edge bisection and edge flip operations.  Bisecting a terminal edge
achieves newest-vertex bisection of all triangles incident on it.
However, if an edge $e$ is not a terminal edge, then bisecting $e$
leaves at least one of the triangles incident on $e$ in a \emph{dirty}
state because an edge of this triangle other than its base has been
bisected.  To rectify this situation, we \emph{clean} the dirty
triangle by another edge bisection followed by an edge flip.  In this
fashion, every clean triangle is subdivided according to the rules of
newest-vertex bisection.  Dirty triangles are transient and are
cleaned before the surrounding neighborhood of the front is advanced.

\begin{figure}[t]\centering
\includegraphics[scale=0.5]{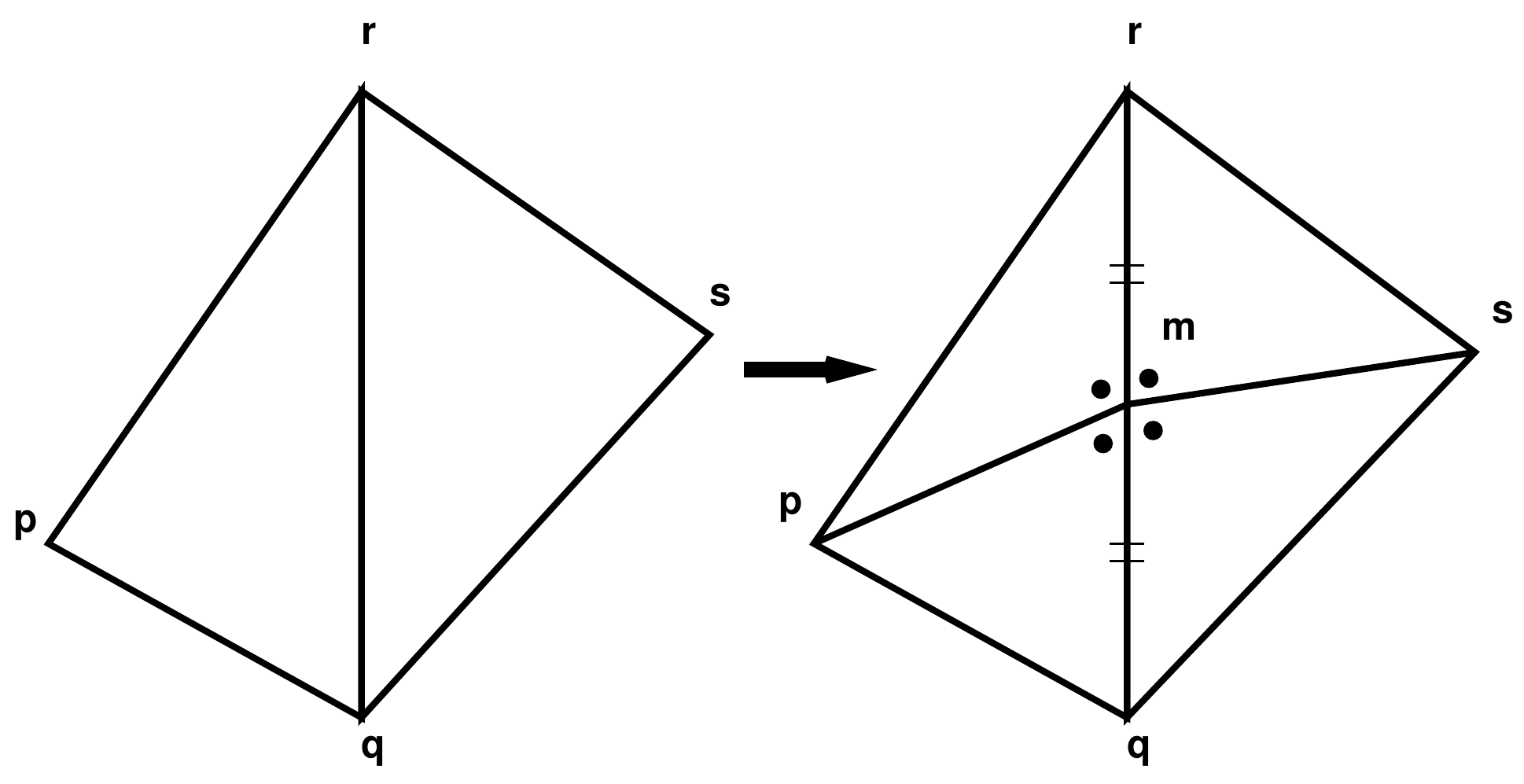}
\caption{Bisecting an edge and the two triangles incident on it}
\label{fig:bisectedge}
\end{figure}

\begin{figure}[t]\centering
\begin{algorithm}
\textsc{BisectTriangle}(Triangle $\triangle{pqr}$):\\
 1.\> Let $qr$ be the base of $\triangle{pqr}$;\\
 2.\> Let $\triangle{qrs}$ be the neighbor of $\triangle{pqr}$ sharing
the edge $qr$;\\
 3.\> Comment: If $qr$ is a boundary edge, there is no such
neighbor.\\
 4.\> Bisect the edge $qr$ by inserting a new vertex $m$ at its
midpoint;\\
 5.\> Insert the edges $pm$ and $ms$;\\
 6.\> Mark all four angles at $m$;\\
 7.\> If $qr$ is not the base of $\triangle{qrs}$:\\
 8.\>\> Mark $\triangle{qrs}$;\\
\end{algorithm}
\caption[Algorithm \textsc{BisectTriangle}]{Bisecting a triangle and its neighbor.}
\label{fig:algBisectTriangle}
\end{figure}

\noindent\textbf{Edge bisection}\quad
This operation bisects an edge shared by one or two triangles; see
Figure~\ref{fig:bisectedge}.  The new vertex is the apex of all four
new triangles.  Bisecting a triangle at \emph{level} $l$ produces two
new \emph{children} at level $l+1$; every triangle in the original
space mesh is at level zero.  The set of all triangles that ever
appear on any front form a forest of rooted binary trees with each
triangle in the initial space mesh as the root of some tree in the
forest. The \emph{height} of a node in the refinement tree is the
number of edges on a longest path from the node to any of its
descendants.

\begin{figure}[t]\centering\small
\includegraphics[scale=0.5]{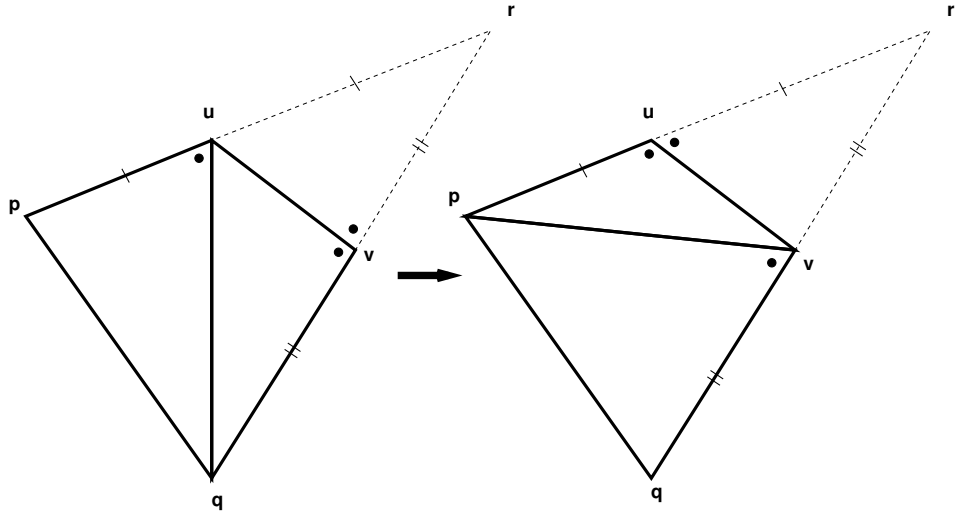}\\
\caption{Edge flip}
\label{fig:edgeflip4lazy}
\end{figure}

\noindent\textbf{Edge flip}\quad
An edge flip replaces one diagonal of a convex quadrilateral with the
other diagonal.  Note how in Figure~\ref{fig:edgeflip4lazy}, the apex of
the adjacent triangle (shown dashed) is changed.  On the left, we
see~$\triangle{pqr}$ first split into~$\triangle{pqu}$
and~$\triangle{uqr}$, then $\triangle{uqr}$ is split
into~$\triangle{uqv}$ and~$\triangle{uvr}$.  On the right, after the
flip operation, we have~$\triangle{pqr}$ first split
into~$\triangle{pqv}$ and~$\triangle{pvr}$, and then~$\triangle{pvr}$
split into~$\triangle{pvu}$ and~$\triangle{uvr}$.  Vertex~$p$ is the
apex of~$\triangle{pqr}$.  The refinement subtree rooted
at~$\triangle{pqr}$ changes as a result of the edge flip.

\parasc*{Lazy propagation}

Edge bisections caused by propagation unnecessarily create smaller
triangles in parts of the front even where the error estimate is
sufficiently below the acceptable threshold.  If the propagation were
delayed until a later step, the spacetime mesh would contain fewer
tetrahedra.  In a parallel setting, interprocessor communication is
inefficient; so, it is useful to allow the algorithm to proceed on
unrelated processors and not require all processors to wait until the
entire propagation sequence has been executed.

Therefore, we propagate refinements \emph{lazily}, by temporarily
bisecting an edge of a neighboring triangle other than its base and
marking this triangle as \emph{dirty}.  A dirty triangle can be
cleaned up later by performing one edge bisection, which in turn can
be propagated lazily, plus one edge flip.  Lazy propagation is
particularly useful for parallel implementation, where the mesh may be
distributed over multiple processors, because it minimizes both
communication costs and deadlocks.  If a processor needs to refine a
triangle in its subset of the mesh, it need not wait for the entire
propagation sequence (which may be controlled by other processors) to
be completed before proceeding with its next task.  Even in the serial
case, lazy refinement may reduce the number of changes to the mesh,
since later coarsening may stop the propagation of refinement early.

\noindent\textbf{Earnest vs.\@ lazy propagation}\quad
Algorithm \textsc{Refine} is called to refine a triangle currently on
the front.  By definition, this triangle is a leaf in the refinement
forest.  The contrast between earnest propagation
(Figure~\ref{fig:earnestAlgorithm}) and lazy propagation
(Figure~\ref{fig:lazyAlgorithm}) is shown in
Figure~\ref{fig:earnest_vs_lazy}.  Earnest propagation causes the
adjacent triangle to be split which may propagate further.  Lazy
propagation (bottom path) splits the adjacent triangle but does not
propagate further immediately---the gray arrows in
Figure~\ref{fig:earnest_vs_lazy} are not traversed until later.
Instead, the propagation is delayed until the adjacent triangle needs
to be refined or pitched.  In the interim, the triangulation may
consist of transient dirty triangles.  After cleaning up, the final
result is the same.

\begin{figure}[t]
\begin{algorithm}
\textsc{Refine}(Triangle $\triangle{pqr}$):\\
 1.\> \textsc{RefineAndPropagate}($\triangle{pqr}$);\\
\end{algorithm}

\begin{algorithm}
\textsc{RefineAndPropagate}(Triangle $\triangle{pqr}$):\\
 1.\> Let $e$ be the base of $\triangle{pqr}$;\\
 2.\> Let $\tau$ be the neighboring triangle that shares the edge $e$;\\
 3.\> \textsc{BisectTriangle}$(\triangle{pqr}$); (Figure~\ref{fig:algBisectTriangle})\\
 4.\> If $\tau$ is marked:\\
 4.1.\>\> \textsc{CleanUp}($\tau$);\\
\end{algorithm}

\begin{algorithm}
\textsc{CleanUp}(Triangle $\triangle{pqr}$):\\
 1.\> Let $p$ be the apex of $\triangle{pqr}$;\\
 2.\> Let $\triangle{pqs}$ and $\triangle{sqr}$ be the current
children of $\triangle{pqr}$;\\
 3.\> \textsl{\{Comment: The edge opposite $q$ is currently bisected.\}}\\
 4.\> \textsc{RefineAndPropagate}($\triangle{sqr}$);\\
 5.\> Flip the edge $qs$;\\
\end{algorithm}

\caption{Refinement with earnest propagation.}
\label{fig:earnestAlgorithm}
\end{figure}

\subsubsection*{Refinement with lazy propagation}

\begin{figure}[t]

\begin{algorithm}
\textsc{LazyRefine}(Triangle $\triangle{pqr}$):\\
 1.\> If $\parent(\triangle{pqr})$ exists and is marked dirty:\\
 1.1.\>\> \textsl{\{Comment: $\triangle{pqr}$ is transient and so we
cannot refine $\triangle{pqr}$.\}}\\
 1.2.\>\> \textsc{CleanUp1}($\parent(\triangle{pqr})$);\\
 2.\> Else:\\
 2.1.\>\> \textsc{LazyRefineAndPropagate}($\triangle{pqr}$);
\end{algorithm}

\begin{algorithm}
\textsc{LazyRefineAndPropagate}(Triangle $\triangle{pqr}$):\\
 1.\> Let $e$ be the base of $\triangle{pqr}$;\\
 2.\> Let $\tau$ be the neighboring triangle that shares the edge $e$;\\
 3.\> \textsc{BisectTriangle}$(\triangle{pqr}$); (Figure~\ref{fig:algBisectTriangle})\\
 4.\> If $\parent(\tau)$ exists and is marked dirty:\\
 4.1.\>\> \textsl{\{Comment: $\tau$ is transient.\}}\\
 4.2.\>\> \textsc{CleanUp2}($\parent(\tau)$);
\end{algorithm}

\begin{algorithm}
\textsc{CleanUp1}(Triangle $\triangle{pqr}$): (Figure~\ref{fig:lazy-cleanup1})\\
 1.\> \textsl{\{Comment: $\triangle{pqr}$ is marked and has height~$1$.\}}\\
 2.\> Let $p$ be the apex of $\triangle{pqr}$;\\
 3.\> Let $\triangle{pqs}$ and $\triangle{sqr}$ be the current
children of $\triangle{pqr}$;\\
 4.\> \textsl{\{Comment: The edge opposite $q$ is currently bisected.\}}\\
 5.\> \textsc{LazyRefineAndPropagate}($\triangle{sqr}$);\\
 6.\> Flip the edge $qs$;
\end{algorithm}

\begin{algorithm}
\textsc{CleanUp2}(Triangle $\triangle{pqr}$): (Figure~\ref{fig:lazy-cleanup2})\\
 1.\> \textsl{\{Comment: $\triangle{pqr}$ is marked and has height~$2$.\}}\\
 2.\> Let $p$ be the apex of $\triangle{pqr}$;\\
 3.\> Let $\triangle{pqs}$ and $\triangle{sqr}$ be the current
children of $\triangle{pqr}$;\\
 4.\> \textsl{\{Comment: The edge opposite $q$ is currently bisected.\}}\\
 5.\> If $\triangle{pqs}$ is a leaf:\\
 5.1.\>\> \textsl{\{Comment: $\triangle{sqr}$ is subdivided.\}}\\
 5.2.\>\> Flip the edge $qs$;\\
 6.\> Else if $\triangle{sqr}$ is a leaf:\\
 6.1.\>\> \textsl{\{Comment: $\triangle{pqs}$ is subdivided.\}}\\
 6.2.\>\> Let $\triangle{pst}$ and $\triangle{tsq}$ be the children of $\triangle{pqs}$;\\
 6.3.\>\> \textsc{RefineAndPropagate}($\triangle{sqr}$);\\
 6.4.\>\> Flip the edge $qs$;\\
 6.5.\>\> Flip the edge $st$;
\end{algorithm}

\caption{Refinement with lazy propagation.}
\label{fig:lazyAlgorithm}
\end{figure}

\begin{figure}[t]\centering\small
\iffig\includegraphics[scale=0.5]{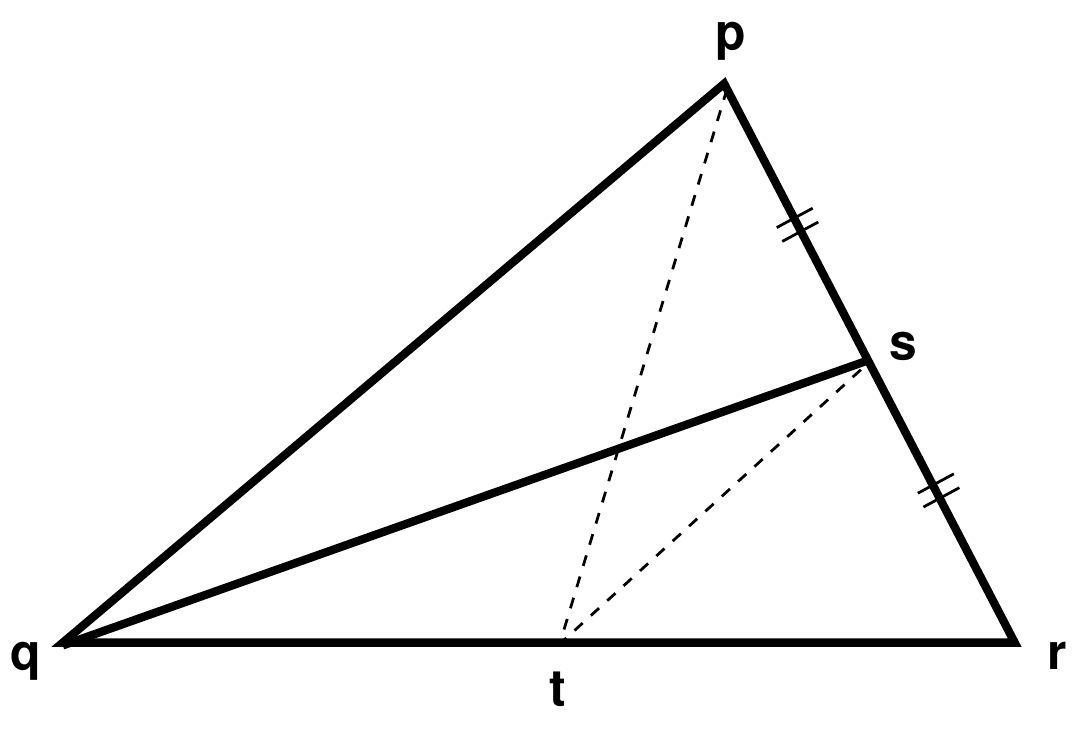}\fi
\caption[Operation of \textsc{CleanUp1}]{Before (solid) and after (dotted) one
  edge bisection and an edge flip performed by \textsc{CleanUp1}}
\label{fig:lazy-cleanup1}
\end{figure}

\begin{figure}[t]\centering\small
\begin{tabular}{cc}
\iffig\includegraphics[scale=0.5]{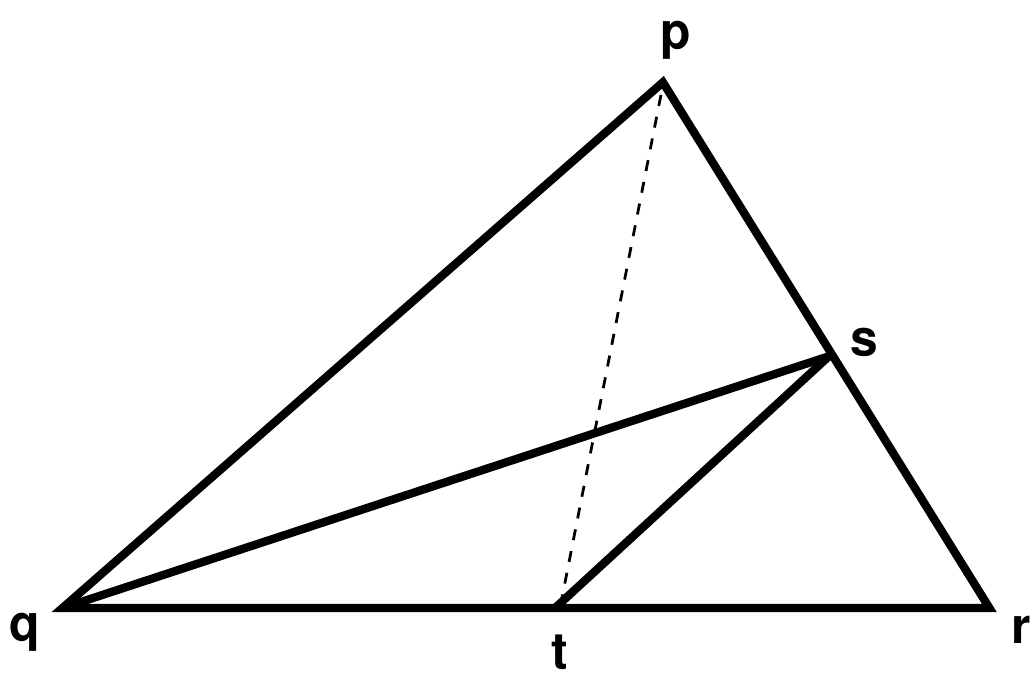}\fi
&
\iffig\includegraphics[scale=0.5]{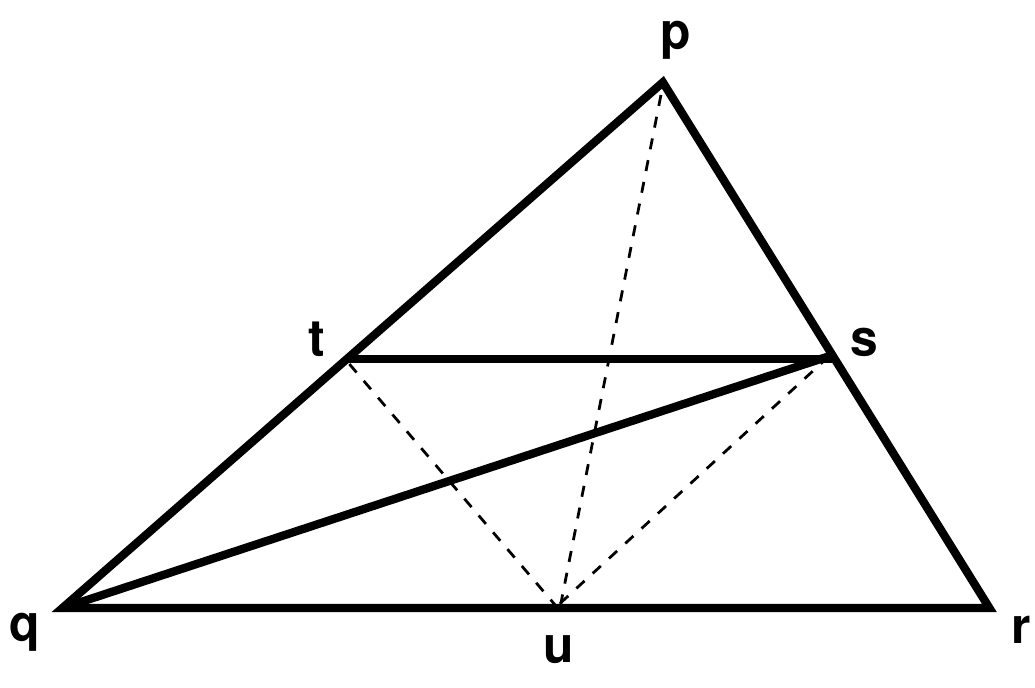}\fi\\
(a) & (b)
\end{tabular}
\caption[Operation of lazy \textsc{CleanUp2}]{Before (solid) and after
  (dotted) \textsc{CleanUp2}: (a)~only an edge flip is required;
  (b)~an edge bisection followed by two edge flips are required}
\label{fig:lazy-cleanup2}
\end{figure}

\begin{figure}[t]\centering\small
\includegraphics[scale=0.5]{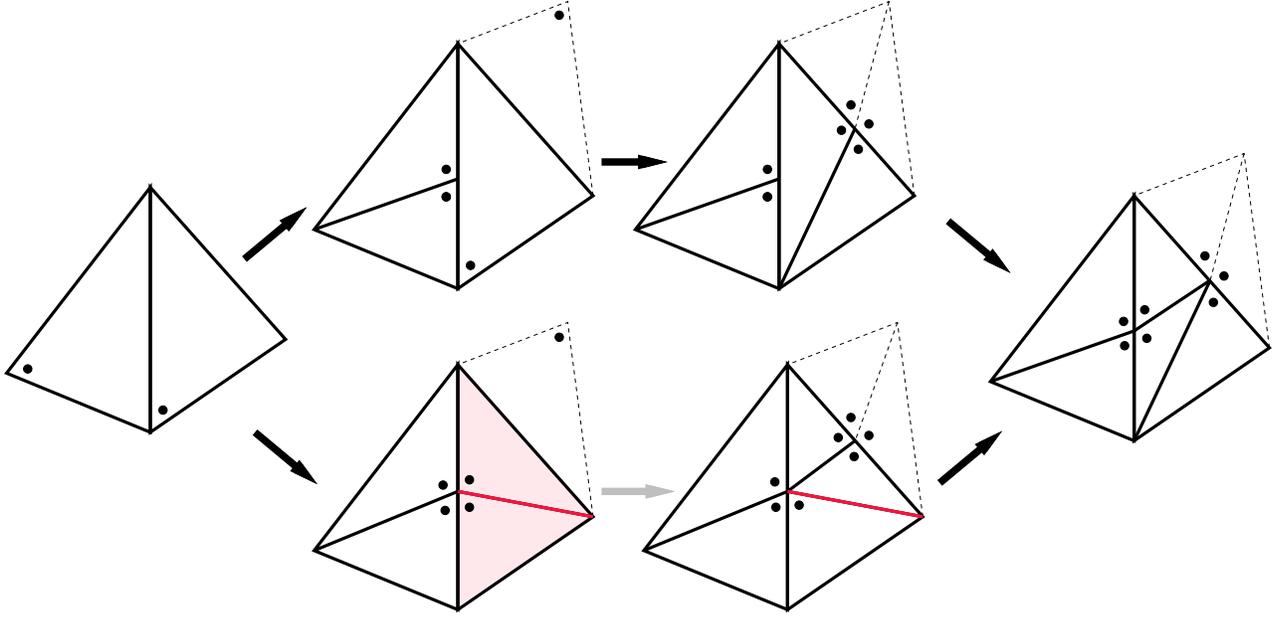}
\caption{Earnest propagation (top path) vs.\@ lazy propagation (bottom
path)}
\label{fig:earnest_vs_lazy}
\end{figure}

As long as all dirty triangles are cleaned up, refinement with lazy
propagation is identical to refinement with earnest propagation; see
Figure~\ref{fig:earnest_vs_lazy}.

\subsection{Coarsening}

Coarsening is the opposite of refinement and hence called
\emph{de-refinement}; we coarsen locally by undoing a single edge
bisection.  Unlike refinement, coarsening does not require propagation
further into the mesh to maintain a conforming triangulation, although
one coarsening step may make other coarsenings possible.  In
particular, if we refine a triangle and then immediately coarsen, we
can (but need not) coarsen along the entire refinement path.

\textsc{DeRefine} is called to coarsen a triangle currently on the
front that is a strict descendant of some triangle in the original
mesh.  By definition, the triangle being coarsened is a leaf in the
refinement forest and its parent exists.  See Figure~\ref{fig:coarsen}.

\begin{figure}
\begin{algorithm}
\textsc{DeRefine}(Triangle $\triangle{pqs}$):\\
 1.\> Let $\triangle{psr}$ be the sibling of $\triangle{pqs}$, i.e.,
edge $qr$ is currently bisected;\\
 2.\> Verify that $\triangle{pqs}$ and $\triangle{psr}$ are leaves and
are coplanar;\\
 3.\> Verify that $s$ has degree $4$;\\
 4.\> Let $u$ be the fourth neighbor of $s$ other than $\{p,q,r\}$;\\
 5.\> Verify that $\triangle{rsu}$ and $\triangle{usq}$ are leaves and
are coplanar;\\
 6.\> \textsl{\{Comment: Equivalently, $r$-$s$-$q$ must be collinear.\}}\\
 7.\> \textsl{\{Comment: All four triangles incident on $s$ need NOT be
coplanar.\}}\\
 8.\> Replace $\triangle{pqs}$ and $\triangle{psr}$ with their parent
$\triangle{pqr}$;\\
 9.\> Replace $\triangle{rsu}$ and $\triangle{usq}$ with their
parent $\triangle{rqu}$;\\
10.\> \textsl{\{Comment: The two new triangles $\triangle{pqr}$ and
$\triangle{rqu}$ are not dirty.\}}\\
11.\> If $\parent(\triangle{pqr})$ exists and is coarsenable:\\
12.\>\> \textsc{DeRefine}($\triangle{pqr}$);\\
13.\> If $\parent(\triangle{rqu})$ exists and is coarsenable:\\
14.\>\> \textsc{DeRefine}($\triangle{rqu}$);\\
\end{algorithm}
\caption{Algorithm to de-refine}
\label{fig:coarsen}
\end{figure}

\begin{figure}
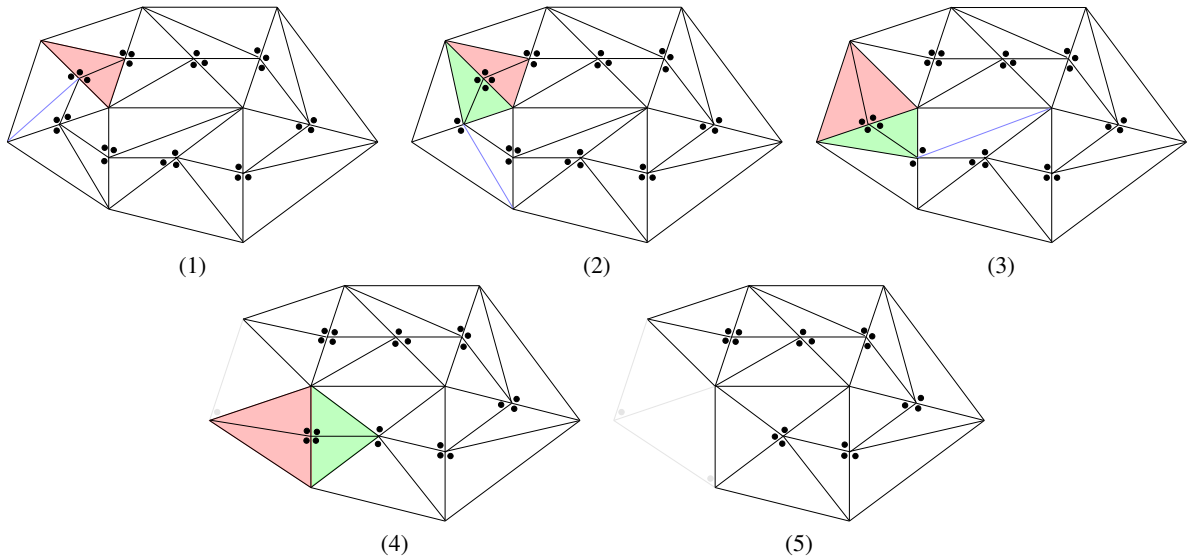
\centering\small
\begin{tabular}{ccc}
\iffig\includegraphics[width=.3\textwidth]{\fig{coarsen-0}}\fi&
\iffig\includegraphics[width=.3\textwidth]{\fig{coarsen-1}}\fi&
\iffig\includegraphics[width=.3\textwidth]{\fig{coarsen-2}}\fi\\
(1) & (2) & (3)
\end{tabular}
\begin{tabular}{cc}
\iffig\includegraphics[width=.3\textwidth]{\fig{coarsen-3}}\fi&
\iffig\includegraphics[width=.3\textwidth]{\fig{coarsen-4}}\fi\\
(4) & (5)
\end{tabular}
\caption[De-refining a loop in the refinement propagation path requires
  edge flips to create degree-$4$ vertices]{De-refining a loop in the
  refinement propagation path requires edge flips to create degree-$4$
  vertices.  To merge the first pair of triangles, an edge flip is
  necessary, (1)$\to$(2), to create a degree-$4$ vertex that can now
  be deleted.  Subsequent coarsening steps (2)$\to$(3) and (3)$\to$(4)
  each require an edge flip.}
\label{fig:coarsen_loop}
\end{figure}

When the propagation path of a refinement loops back on itself, there
will be no degree-$4$ vertex in the portion of the mesh corresponding
to this loop (Figure~\ref{fig:coarsen_loop}), and algorithm
\textsc{DeRefine} will not apply.  The solution is to perform an edge
flip to create a degree-$4$ vertex and then call \textsc{DeRefine}.
The edge flip will create a dirty triangle, and this dirty triangle
can be coarsened immediately by \textsc{DeRefine} along with its
neighbor.

\subsection{Adaptive meshing in 2D$\times$Time}

We are now ready to describe an iteration of our advancing front
algorithm.  Advance a single vertex~$p$ by a positive amount,
where~$p$ is any local minimum of the current front~$\tau_i$, to get
the new front~$\tau_{i+1}$ such that every triangle $pqr$
on~$\tau_{i+1}$ is progressive.  In the parallel setting, advance any
independent set of local minima forward in time, each subject to the
above constraint.

\begin{figure}[t]
\begin{quote}
\textbf{Input:}  A triangulation $M \subset \Real^2$\\

\textbf{Output:} A tetrahedral mesh of $M \times [0,\infty)$\\

Initial front $\tau_0$ is the space mesh at time $t=0$, i.e., $\tau_0(p) = 0$
for every $p \in V(\sp)$\\[1em]

Repeat for $i = 0, 1, 2, \ldots$:
\begin{itemize}
\item Lift a local minimum vertex $p$ of the current front $\tau_i$ to
obtain the new front $\tau_{i+1}$ such that every triangle $pqr$ on
$\tau_{i+1}$ is progressive.
\item Solve the resulting patch.
\item If the patch is rejected by solver, one or more elements in the
patch are marked for refinement.
\item For every element marked for refinement, bisect its inflow
facet; otherwise, advance the front.
\item If any pair of coarsenable siblings are coplanar, then coarsen
this pair.
\end{itemize}
\end{quote}
\caption{Adaptive Tent Pitcher algorithm with refinement and coarsening}
\label{fig:2d:algAdaptive}
\hrule
\end{figure}

To incorporate adaptivity into our meshing algorithm, we make a small
change to the main loop.  At each iteration, just as before, we choose
a local minimum vertex $\fp$ of the front, move it forward in time to
create a tent, and pass the tent and its inflow data to the spacetime
DG solver.  We assume that the solver also computes an \aposteriori{}
estimate of its own numerical error.  If the error within any element
of the tent is above some threshold, the solver \emph{rejects} the
patch, at which point the meshing algorithm throws away the tent and
refines the facets of the front whose elements had high error.
(Alternately, we could refine the facets until their diameter is
smaller than a target length scale computed by the solver.)  Note that
this refinement may propagate far outside the neighborhood of $\fp$.
We accept the numerical solution and update the front only if the
error within every element of the patch is acceptable.

On the other hand, the error estimate within an element may fall below
some second threshold, indicating that the mesh is finer than
necessary to compute the desired result.  In this case, the DG solver
marks the outflow face of that element as \emph{coarsenable}.  We can
coarsen four facets of the front into two only if they are the result
of an earlier refinement, they are all marked as coarsenable, and each
pair of triangles to be merged is coplanar.  To make coarsening
possible, our algorithm tries to make coarsenable siblings coplanar,
by lowering the top of the tent.  However, to avoid very thin
elements, we accept the lower tent only if its height is above some
threshold.  If the lower tent is accepted and its outflow faces are
still marked coarsenable, then we coarsen the front.  This means, of
course, that merging a pair of coarsenable triangles may be delayed
due to the coplanarity constraint imposing a too-small tentpole
height.  We definitely observe this in practice, leading to a more
inefficient mesh than absolutely necessary.  In fact, coarsening may
be delayed several steps.  In Chapter \ref{sec:targettime}, we give an
algorithm to ensure that every coarsenable pair of triangles will
eventually be made coplanar and merged, while still guaranteeing a
bounded minimum tentpole height.

\begin{figure}
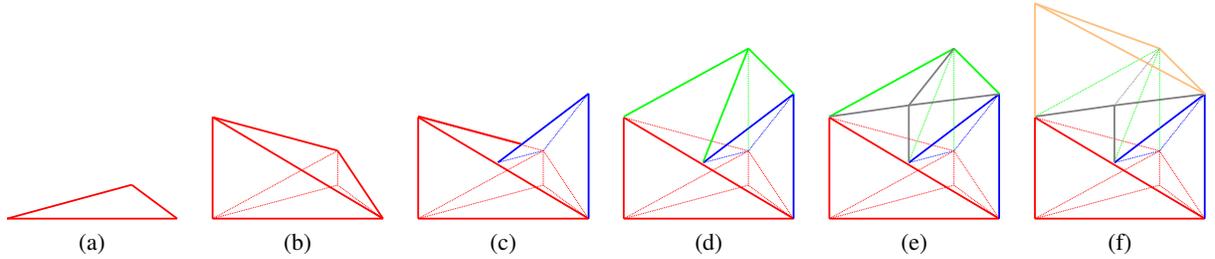
\centering\small
\begin{center}
\begin{tabular}{cccccc}
\iffig\includegraphics[width=0.14\textwidth]{\fig{pitchseq-0}}\fi
&
\iffig\includegraphics[width=0.14\textwidth]{\fig{pitchseq-1}}\fi
&
\iffig\includegraphics[width=0.14\textwidth]{\fig{pitchseq-2}}\fi
&
\iffig\includegraphics[width=0.14\textwidth]{\fig{pitchseq-3}}\fi
&
\iffig\includegraphics[width=0.14\textwidth]{\fig{pitchseq-4}}\fi
&
\iffig\includegraphics[width=0.14\textwidth]{\fig{pitchseq-5}}\fi
\\
(a) & (b) & (c) & (d) & (e) & (f)
\end{tabular}
\end{center}
\caption[A sequence of 5 tents pitched by the adaptive algorithm]{A sequence of 5 tents pitched by the adaptive algorithm:
  (a)$\to$(b)~pitch twice, (b)$\to$(c)~refine and pitch, (c)$\to$(d)~pitch,
  (d)$\to$(e)~pitch, (e)$\to$(f)~coarsen and pitch}
\label{fig:pitchseq}
\end{figure}

See Figure~\ref{fig:2d:algAdaptive} for an outline of the adaptive
Tent Pitcher algorithm.  See Figure~\ref{fig:pitchseq} for an
illustration of the effect of refine and coarsen operations
interspersed with tent pitching steps.

\parasc*{Cleaning up after a lazy propagation}

If refinement is being propagated lazily, we must ensure that the
triangles incident on $p$ are clean.

\begin{algorithm}
\textsc{CleanupBeforePitching}(Vertex $p$):\\
1.\> For every triangle $\triangle{pqr}$ incident on $p$:\\
2.\>\> If $\parent(\triangle{pqr})$ exists and is marked dirty:\\
3.\>\>\> \textsc{CleanUp1}($\parent(\triangle{pqr})$);\\
\end{algorithm}

\parasc*{Coplanarity constraints}

While pitching~$p$, we see whether coarsenable triangles
adjacent to~$p$ can be made coplanar with their siblings after
pitching.  For each triangle $\triangle{pqs}$ incident on~$p$,
consider the four triangles neighboring $\triangle{pqs}$ referred to
in \textsc{DeRefine}.  If all the four triangles are currently marked
as coarsenable, we solve for $\tau'(p)$ to compute the new value of
$\tau'(p)$ that would make~$p$ coplanar with~$q$, $r$, and~$u$.  This
is the value of $\tau'(p)$ that satisfies what we call the
\emph{coplanarity constraint}.  Satisfying the coplanarity constraint
may require a shorter tentpole at~$p$; if the shorter height is less
than some constant (such as~$\nicefrac{1}{10}$) times the height
\emph{without} the constraint, then we choose to ignore the
coplanarity constraint to avoid creating degenerate elements.
Otherwise, the final height of the tentpole at~$p$ is limited by the
coplanarity constraint and coarsenable triangles can be merged
immediately after pitching~$p$.

\subsection{Causality and Progress Constraints}
\label{sec:2d:refine:constraints}

To complete the description of our adaptive version of Tent Pitcher,
all that remains is to define \emph{progressive} triangles and fronts,
i.e., to define the progress constraints that limit the new time value
for each vertex to be pitched.  Each triangle in the space mesh
imposes constraints on the time values at each of its vertices.  When
we pitch a tent over a local minimum vertex $\fp$, the new time value
$\tau_{i+1}(p)$ is simply the largest value that satisfies the
constraints for every triangle $pqr$ incident on $p$ in the space
mesh.

One constraint is of course the causality constraint of
Equation~\ref{eqn:2d:causalityconstraint}, repeated below:
\[
  \frac{\tau_{i+1}(p) - \tau_i(\bp)}{\abs{p \bp}}
<
  \root{\S^2 - \norm{\grad \rest{\tau_i}{qr}}^2}
\]

The more subtle and important change in our algorithm is the
introduction of new progress constraints.  In the earlier non-adaptive
algorithm~\cite{erickson02building}, the progress constraint was a
function of the shape of the underlying space elements.  In our new
algorithm, the shape of the underlying element is subject to change;
each triangle in the current front may be the result of any number of
refinements since the last time any ancestor or descendant of that
triangle was pitched.  Consequently, our progress constraints must
take into account the shapes of \emph{all possible} descendants of a
triangle simultaneously.  This requirement motivates our choice of
newest vertex refinement; because the descendants of any triangle fall
into only a finite number of homothety classes, we have only a finite
number of simultaneous progress constraints.

\begin{figure*}[t]\centering\small
\includegraphics[width=0.9\textwidth]{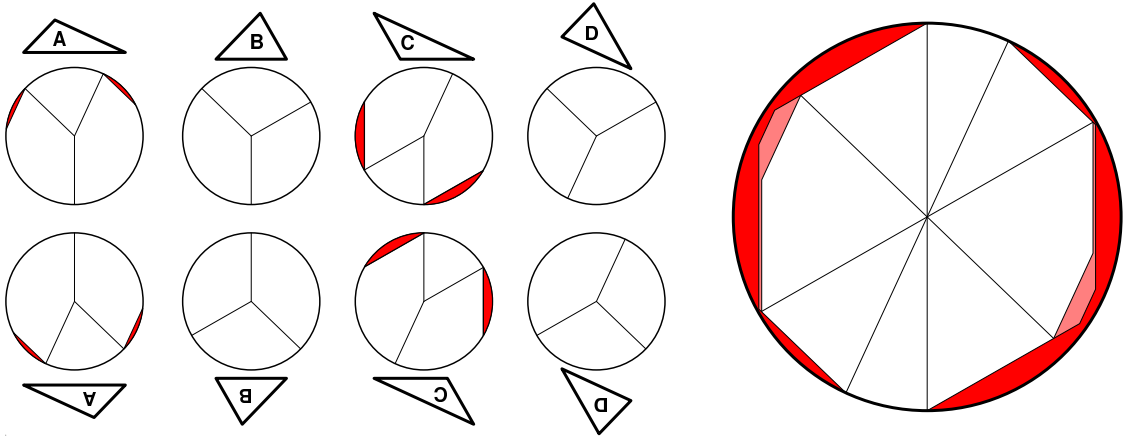}
\caption[Merging eight progress constraints into one adaptive progress
constraint]{The adaptive progress constraint is obtained from the
progress constraint for each of the eight different shapes of
descendant triangles.}
\label{F:8circles}
\end{figure*}

We can visualize these constraints by referring back to our circle
diagram; see Figure~\ref{F:8circles}.  A minimal set of progress
constraints can be obtained by simply taking the union of the
non-adaptive progress constraints of the eight shapes in the
hierarchy.  Recall that each of the eight shapes, corresponding to a
descendant of $\triangle{pqr}$, defines a forbidden zone where the
gradient vector of $\triangle{pqr}$ must not lie.  We call these the
``primary'' forbidden zones.  The progress constraint that the
corresponding triangle $\fp\fq\fr$ on the front must satisfy is the
conjunction of the progress constraint for each of these eight shapes.
In addition, we must also forbid ``secondary zones'' from which the
only alternative, due to the nature of Tent Pitcher, is to enter a
primary forbidden zone.  Staying outside the primary and secondary
forbidden zones is sufficient to guarantee that the progress at each
step is bounded away from zero.

Fix two real numbers $\e$ and $\f$ such that $0 < \e < \f < (1+\e)/2 <
1$.  For any triangle $\triangle{abc}$ with apex $a$, we define the
\emph{diminished width} of $\triangle abc$ as follows.

\begin{figure}\centering\small
\includegraphics[width=0.3\textwidth]{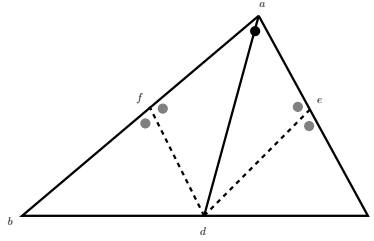}
\caption{$\triangle{abc}$ where $a$ is the apex.}
\label{fig:abc}
\end{figure}

\begin{definition}[Diminished width~\cite{abedi04spacetime}]
  Let $abc$ be an arbitrary triangle, where $a$ is the apex.  (See
  Figure~\ref{fig:abc}.)  The \emph{diminished width} of
  $\triangle{abc}$ is defined as
  \[
    \dw(abc)
  :=
   \min\,\left\{
     \begin{tabular}{l}
       $(1-\e) \dist(a,\aff bc)$,\\
       $(1-\f) \dist(b,\aff ac)$,\\
       $(1-\f) \dist(c,\aff ab)$
     \end{tabular}
   \right\}
  \]
\label{def:2d:refine:diminishedwidth}
\end{definition}

The first distance is measured from the apex to the opposite edge and
is scaled differently from the other two altitudes.  This definition
extends recursively to any descendants of $\triangle{abc}$ obtained by
newest vertex subdivision; in the interest of readability, we will
always list the vertices of any triangle with the apex first.  We
express our new progress constraints algebraically by limiting the
difference in time values along each edge in the subdivided triangle,
as follows.

\begin{definition}[Adaptive progress constraint~$\S$]
  Fix $\e$ and $\f$ such that $0 < \e < \f < (1+\e)/2 < 1$.
  Let~$\triangle{\fa\fb\fc}$ be an arbitrary triangle of a
  front~$\tau$ with apex $\fa$. Let $d$, $e$, and $f$ be midpoints of
  sides $bc$, $ac$, and $ab$ respectively (Figure~\ref{fig:abc}).  We
  say that the triangle~$\fa\fb\fc$ satisfies \emph{adaptive progress
  constraint~$\S$} if and only if
  \begin{align*}
        \abs{\tau(a) - \tau(b)} &\le 2 \dw(dca) \S \\
        \abs{\tau(a) - \tau(d)} &\le   \dw(abc) \S \\
        \abs{\tau(a) - \tau(c)} &\le 2 \dw(dab) \S \\
        \abs{\tau(b) - \tau(c)} &\le 4 \dw(ead) \S
  \end{align*}
\label{def:2d:refine:progressconstraint}
\end{definition}

The constraints in the definition of the progress constraint apply
recursively to all descendants of the triangle $\triangle abc$, but
these recursive constraints are equivalent to one of the four
constraints above.

Now suppose we are pitching a triangle $\triangle pqr$ of the front
$\tau$, where $\tau(p) \le \tau(q) \le \tau(r)$.  Let~$w$ be the
midpoint of~$qr$; let~$u$ be the midpoint of $pr$; and let~$v$ be the
midpoint of $pq$; see Figures~\ref{fig:adaptive:pqr-Qacute}
and~\ref{fig:adaptive:pqr-Qobtuse}.  Depending on which of the three
vertices is marked as the apex, the new time value $\tau'(p)$ is
bounded as a result of these progress constraints in the three
different ways enumerated below.  Notice that when~$p$ is not the
apex, lifting~$p$ also lifts either~$u$ or~$v$, so progress
constraints along edges~$qu$ or~$rv$ also indirectly limit~$\tau'(p)$.

\begin{align}
\intertext{If $p$ is the apex:}
        \tau'(p) &\le \min\left\{
                        \begin{array}{@{\,}l@{\,}}
                            \tau(q) + 2 \dw(rpw) \S,\\
                            \tau(s) + \dw(pqr) \S,\\
                            \tau(r) + 2 \dw(wpq) \S
                        \end{array}
                        \right\}
\label{eqn:2d:refine:progressconstraint:1}
\\
\intertext{If $r$ is the apex:}
        \tau'(p) &\le \min\left\{
                        \begin{array}{@{\,}l@{\,}}
                           \tau(q) + 4\dw(uvr) \S,\\
                           \tau(r) + 2\dw(vqr) \S,\\
                           2\tau(r) - \tau(q) + 2\dw(rpq) \S
                        \end{array}
                        \right\}
\label{eqn:2d:refine:progressconstraint:2}
\\
\intertext{If $q$ is the apex:}
        \tau'(p) &\le \min\left\{
                        \begin{array}{@{\,}l@{\,}}
                          \tau(q) + 2\dw(uqr) \S, \\
                          \tau(r) + 4\dw(wuq) \S, \\
                          \tau(q) - \tau(r) + 2\dw(rpq) \S
                        \end{array}
                        \right\}
\label{eqn:2d:refine:progressconstraint:3}
\end{align}

The adaptive progress constraint
(Definition~\ref{def:2d:refine:progressconstraint}) is stronger than
the non-adaptive progress constraint
(Definition~\ref{def:2d:linear:progressconstraint}); in particular,
the new progress constraint limits the gradient of all three edges of
$\triangle{\fp\fq\fr}$, not just the highest edges, and anticipates
future refinement and coarsening of triangles on the front by newest
vertex bisection by also limiting the gradient along the bisector
edge.  Of course, our algorithm works even when no refinement or
coarsening operations are actually performed.

The following theorems were proved by Abedi \etal{}.  In this section,
we rectify an oversight in the proof of correctness in the paper by
Abedi \etal{}~\cite{abedi04spacetime} by including a case in the proof
of Theorem~\ref{thm:2d:refine:iscausal} that was missing.

\begin{theorem}
  If a front $\tau_i$ is progressive, then for every local minimum
  vertex $p$, for every $\dt \in [0,\e w_p \minS]$, the front $\tau_{i+1} =
  \next(\tau_i,p,\dt)$ is causal.
\label{thm:2d:refine:iscausal}
\end{theorem}

\begin{proof}
Since only the triangles of the front incident on~$\fp$ advance
along with~$p$, we can restrict our attention to an arbitrary
triangle~$pqr$ incident on~$p$.  Let~$\tau$
and~$\tau'$ denote~$\rest{\tau_i}{pqr}$
and~$\rest{\tau_{i+1}}{pqr}$ respectively.
Let $\bp$ be the orthogonal projection of $p$ onto line $qr$.

We consider two cases separately.

\parasc*{Case 1: $\mathbf{\tau(\bp) \ge \tau(q) \ge \tau(p)}$}

\begin{figure*}[t]\centering\small
\begin{tabular}{ccc}
\includegraphics[width=.3\textwidth]{\fig{triangle-p-Qacute}} &
\includegraphics[width=.3\textwidth]{\fig{triangle-q-Qacute}} &
\includegraphics[width=.3\textwidth]{\fig{triangle-r-Qacute}}\\
(i) & (ii) & (iii)
\end{tabular}
\caption{Triangle $pqr$ with $\measuredangle{pqr} \le
\nicefrac{\pi}{2}$ with apex at $p$, $q$, and $r$ respectively.}
\label{fig:adaptive:pqr-Qacute}
\end{figure*}

See Figure~\ref{fig:adaptive:pqr-Qacute}.  In this case, we
have $\measuredangle{pqr} \le \nicefrac{\pi}{2}$.  We
will show that
\[
  \norm{\grad \rest{\tau}{qr}} \le (1-\e) \S
\]
i.e., that
\[
  \tau(r) - \tau(q) \le (1-\e) \abs{qr} \S.
\]
The proof then follows from Lemma~\ref{lemma:2d:nextiscausal}
because the antecedent of the lemma is satisfied.

We consider three cases depending on which vertex of $\triangle{pqr}$
is the apex.

If $p$ is the apex, then
\begin{align*}
\tau(r) - \tau(q)
&\le 4 \dw(upw) \S
  &\text{(by the progress constraint)}\\
&\le 4 (1-\e) \dist(u,\aff pw) \S
  &\text{(by definition of diminished width)}\\
&=   2 (1-\e) \dist(r,\aff pw) \S
  &\text{($2 \dist(u,\aff pw) = \dist(r,\aff pw)$)}\\
&\le 2 (1-\e) \abs{wr} \S
  &\text{($\dist(r,\aff pw) \le \abs{wr}$)}\\
&=   (1-\e) \abs{qr} \S
  &\text{($2 \abs{wr} = \abs{qr}$)}
\end{align*}

If $r$ is the apex, then
\begin{align*}
\tau(r) - \tau(q)
&\le 2 \dw(vrp) \S
  &\text{(by the progress constraint)}\\
&\le 2 (1-\e) \dist(v,\aff pr) \S
  &\text{(by definition of diminished width)}\\
&=   (1-\e) \dist(q,\aff pr) \S
  &\text{($2 \dist(v,\aff pr) = \dist(q,\aff pr)$)}\\
&\le (1-\e) \abs{qr} \S
  &\text{($\dist(q,\aff pr) \le \abs{qr}$)}
\end{align*}

If $q$ is the apex, then
\begin{align*}
\tau(r) - \tau(q)
&\le 2 \dw(upq) \S
  &\text{(by the progress constraint)}\\
&\le 2 (1-\e) \dist(u,\aff pq) \S
  &\text{(by definition of diminished width)}\\
&=   (1-\e) \dist(r,\aff pq) \S
  &\text{($2 \dist(u,\aff pq) = \dist(r,\aff pq)$)}\\
&\le (1-\e) \abs{qr} \S
  &\text{($\dist(r,\aff pq) \le \abs{qr}$)}
\end{align*}

\parasc*{Case 2: $\mathbf{\tau(\bp) < \tau(q)}$}

\begin{figure*}[t]\centering\small
\begin{tabular}{ccc}
\includegraphics[width=.3\textwidth]{\fig{triangle-p-Qobtuse}} &
\includegraphics[width=.3\textwidth]{\fig{triangle-q-Qobtuse}} &
\includegraphics[width=.3\textwidth]{\fig{triangle-r-Qobtuse}}\\
(i) & (ii) & (iii)
\end{tabular}
\caption{Triangle $pqr$ with $\measuredangle{pqr} > \nicefrac{\pi}{2}$
 with apex at $p$, $q$, and $r$ respectively.}
\label{fig:adaptive:pqr-Qobtuse}
\end{figure*}

See Figure~\ref{fig:adaptive:pqr-Qobtuse}.  In this case, we
have $\measuredangle{pqr} > \nicefrac{\pi}{2}$.  We
will show that
\[
  \norm{\grad \rest{\tau}{qr}} \le (1-\e) \S \sin\angle{pqr}
\]
i.e., that
\[
  \tau(r) - \tau(q) \le (1-\e) \abs{qr} \S \sin\angle{pqr}.
\]
The proof then follows from Lemma~\ref{lemma:2d:nextiscausal}
because the antecedent of the lemma is satisfied.

See Figure~\ref{fig:adaptive:pqr-Qobtuse}(a).  Let $\beta = \abs{\bp
q}/\abs{\bp p}$. Since $\abs{\bp q} \ne 0$, we have
\begin{align}
  \frac{\tau'(p) - \tau(\bp)}{\abs{\bp p}}
&= 
  \frac{\tau'(p) - \tau(q)}{\abs{\bp p}} 
+ 
  \frac{\tau(q) - \tau(\bp)}{\abs{\bp q}}
  \frac{\abs{\bp q}}{\abs{\bp p}}\nonumber\\
&= 
  \frac{\tau'(p) - \tau(q)}{\abs{\bp p}} 
+ 
  \beta \norm{\grad \rest{\tau}{qr}}
\label{eqn:2d:refine:strengthen}
\end{align}
Using Equation~\ref{eqn:2d:refine:strengthen}, the causality constraint
(Equation~\ref{eqn:2d:causalityconstraint}) can be rewritten as
\begin{align}
  \frac{\tau'(p) - \tau(q)}{\abs{\bp p}}
&\le 
  \sqrt{\S^2 - \norm{\grad \rest{\tau}{qr}}^2}\nonumber\\
&\quad {}-{} \beta \norm{\grad \rest{\tau}{qr}}
\label{eqn:2d:refine:equiv_causalityconstraint}
\end{align}

We consider three cases depending on which vertex of $\triangle{pqr}$
is the apex.

If $p$ is the apex, then
\begin{align*}
\tau(r) - \tau(q)
&\le 4 \dw(upw) \S
  &\text{(by the progress constraint)}\\
&\le 4 (1-\e) \dist(u,\aff pw) \S
  &\text{(by definition of diminished width)}\\
&=   2 (1-\e) \dist(r,\aff pw) \S
  &\text{($2 \dist(u,\aff pw) = \dist(r,\aff pw)$)}
\end{align*}
Hence,
\begin{align*}
\frac{\tau(r) - \tau(q)}{\abs{qr}}
&\le (1-\e) \frac{\dist(r,\aff pw)}{\abs{qr}/2} \S\\
&=   (1-\e) \frac{\dist(r,\aff pw)}{\abs{wr}} \S\\
&=   (1-\e) \S \sin\angle{pwr}\\
&=   (1-\e) \S \frac{\abs{\bp p}}{\abs{pw}}\\
&<   (1-\e) \S \frac{\abs{\bp p}}{\abs{pq}}
  &\text{because $\abs{pw} > \abs{pq}$}\\
&=   (1-\e) \S \sin \angle{pqr}
\end{align*}

If $r$ is the apex, then
\begin{align*}
\tau(r) - \tau(q)
&\le 2 \dw(vrp) \S
  &\text{(by the progress constraint)}\\
&\le 2 (1-\e) \dist(v,\aff pr) \S
  &\text{(by definition of diminished width)}\\
&=   (1-\e) \dist(q,\aff pr) \S
  &\text{($2 \dist(v,\aff pr) = \dist(q,\aff pr)$)}
\end{align*}
Hence,
\begin{align*}
\frac{\tau(r) - \tau(q)}{\abs{qr}}
&\le (1-\e) \S \frac{\dist(q,\aff pr)}{\abs{qr}}\\
&=   (1-\e) \S \sin \angle{qrp}\\
&=   (1-\e) \S \frac{\abs{\bp p}}{\abs{pr}}\\
&<   (1-\e) \S \frac{\abs{\bp p}}{\abs{pq}}
  &\text{because $\abs{pr} > \abs{pq}$}\\
&=   (1-\e) \S \sin \angle{pqr}
\end{align*}

If $q$ is the apex, then
\begin{align*}
\tau(r) - \tau(q)
&\le 2 \dw(upq) \S
  &\text{(by the progress constraint)}\\
&\le 2 (1-\e) \dist(u,\aff pq) \S
  &\text{(by definition of diminished width)}\\
&=   (1-\e) \dist(r,\aff pq) \S
  &\text{($2 \dist(u,\aff pq) = \dist(r,\aff pq)$)}
\end{align*}
Hence,
\begin{align*}
\frac{\tau(r) - \tau(q)}{\abs{qr}}
&\le (1-\e) \frac{\dist(r,\aff pq)}{\abs{qr}} \S\\
&=   (1-\e) \S \sin (\pi - \angle{pqr})\\
&=   (1-\e) \S \sin \angle{pqr}
\end{align*}
\end{proof}

\begin{theorem}
  If a front $\tau_i$ is progressive and if $0 < \e < \f <(1+\e)/2 <
  1$, then for every local minimum vertex $p$ there exists $\minT > 0$,
  where $\minT$ is a function of the triangle $\triangle pqr$ and the
  parameters $\e$ and $\f$, such that the front $\tau_{i+1} =
  \next(\tau_i,p,\dt)$ is progressive for every $\dt \in [0,\minT]$.
\label{thm:2d:refine:isprogressive}
\end{theorem}

\begin{proof}
Since only the triangles of the front incident on~$\fp$ advance
along with~$p$, we can restrict our attention to an arbitrary
triangle~$pqr$ incident on~$p$.  Let~$\tau$
and~$\tau'$ denote~$\rest{\tau_i}{pqr}$
and~$\rest{\tau_{i+1}}{pqr}$ respectively.
Let $\bp$ be the orthogonal projection of $p$ onto line $qr$.

Consider the progress constraints
(Equations~\ref{eqn:2d:refine:progressconstraint:1}--\ref{eqn:2d:refine:progressconstraint:3}).
Each progress constraint limits the progress $\tau'(p) - \tau(p)$ from
above.  The only progress constraint for which this limit is not
obviously positive is
Equation~\ref{eqn:2d:refine:progressconstraint:3} which applies when
$q$ is the apex of $\triangle{pqr}$, i.e.,
\[
 \tau'(p)
\le
 2 \tau(q) - \tau(r)  + 2 \dw(qrp) \S
\]
Subtracting $\tau(p)$ from both sides, because $\tau(q) \ge \tau(p)$, the above
constraint is satisfied if
\[
 \tau(r) - \tau(q)
\le
 2 \dw(qrp) \S
\]
Since $\triangle{\fp\fq\fr}$ is progressive, we have
\[
 \tau(r) - \tau(q)
\le
 2 \dw(upq) \S
\]
Therefore, it suffices to show that $\dw(qrp) - \dw(upq)$ is positive
and bounded away from zero.  Expanding the definition of diminished
width, we rewrite $X = \dw(qrp) - \dw(upq)$ as $X = \min\{A,B,C\}$ where
\begin{align*}
 A &= (1-\e) \dist(q,\aff pr) - \min\,\left\{
                                      \begin{array}{c}
                                      (1-\e) \dist(u,\aff pq),\\
                                      (1-\f) \dist(p,\aff qu),\\
                                      (1-\f) \dist(q,\aff up)
                                      \end{array}
                                      \right\}\\
   &\ge (1-\e) \dist(q,\aff pr) - (1-\f) \dist(q,\aff up)\\
   &=   (\f - \e) \dist(q,\aff pr);
\end{align*}
\begin{align*}
 B &= (1-\f) \dist(r,\aff pq) - \min\,\left\{
                                      \begin{array}{c}
                                      (1-\e) \dist(u,\aff pq),\\
                                      (1-\f) \dist(p,\aff qu),\\
                                      (1-\f) \dist(q,\aff up)
                                      \end{array}
                                      \right\}\\
   &\ge (1-\f) \dist(r,\aff pq) - (1-\e) \dist(u,\aff pq)\\
   &=   2 (1-\f) \dist(u,\aff pq) - (1-\e) \dist(u,\aff pq)\\
   &=   (1 + \e - 2 \f) \dist(u,\aff pq);
\end{align*}
\begin{align*}
 C &= (1-\f) \dist(p,\aff qr) - \min\,\left\{
                                      \begin{array}{c}
                                      (1-\e) \dist(u,\aff pq),\\
                                      (1-\f) \dist(p,\aff qu),\\
                                      (1-\f) \dist(q,\aff up)
                                      \end{array}
                                      \right\}\\
   &\ge (1-\f) (\dist(p,\aff qr) - \min\{\dist(p,\aff qu),\dist(q,\aff up)\}).
\end{align*}
Since $\e < \f < (1+\e)/2$, we have $A>0$ and $B>0$.  See
Figure~\ref{fig:adaptive:pqr-Qacute}(iii) or
Figure~\ref{fig:adaptive:pqr-Qobtuse}(iii); we also have
\begin{align*}
\dist(p,\aff qr)
&= \frac{2 \area(\triangle{pqr})}{\abs{qr}}\\
&= \frac{2 \area(\triangle{upq})}{\abs{uv}}\\
&> \frac{2 \area(\triangle{upq})}{\max\{\abs{up},\abs{uq}\}}
\end{align*}
The inequality above follows because the bisector segment~$uv$ must be
shorter than at least one of the two sides~$up$ and~$uq$.  Now,
$2\,\area(\triangle{upq})/\abs{up} = \dist(q,\aff pr) = \dist(q,\aff
up)$ and $2\,\area(\triangle{upq})/\abs{uq} = \dist(p,\aff qu)$.
Hence, it follows that~${C>0}$.
\end{proof}

\subsection{Changing the apex}
\label{sec:changemark}

Sometimes it may be advantageous to change the apex (newest vertex) of
a triangle.  For instance, if we wish to bisect a given edge without
causing the refinement to propagate to neighboring triangles, we can
make the edge a terminal edge by choosing the apexes of incident
triangles appropriately.  In Chapter~\ref{sec:extensions}, when we
discuss pitching inclined tentpoles to account for boundary motion or
to perform smoothing, the apex of a triangle may no longer correspond
to the largest angle of the triangle.  It is useful to subdivide the
largest angle, or equivalently to bisect the longest edge, to
guarantee sufficient progress in the next step.  We therefore require
some mechanism to change the apex or newest vertex of a given
triangle.

The adaptive progress constraint can be strengthened to allow for
changing the apex of any triangle.  Specifically, for each triangle
$pqr$, we impose all three types of progress constraints on
$\triangle{pqr}$ when $p$ is the apex, when $q$ is the apex, and when
$r$ is the apex simultaneously.  Thus, we say that a causal
triangle~$\fa\fb\fc$ is progressive if and it satisfies the adaptive
progress constraint of Definition~\ref{def:old:progressconstraint} for
every choice of apex of the triangle.

Hence, when pitching a vertex $p$ of $\triangle{pqr}$ we impose all
the following constraints simultaneously on the new time value
$\tau'(p)$:

\begin{align}
        \tau'(p) &\le \min\left\{
                        \begin{array}{@{\,}l@{\,}}
                            \tau(q) + 2 \dw(rpw) \S,\quad
                            \tau(s) + \dw(pqr) \S,\quad
                            \tau(r) + 2 \dw(wpq) \S,\\
                           \tau(q) + 4\dw(uvr) \S,\quad
                           \tau(r) + 2\dw(vqr) \S,\quad
                           2\tau(r) - \tau(q) + 2\dw(rpq) \S,\\
                          \tau(q) + 2\dw(uqr) \S,\quad
                          \tau(r) + 4\dw(wuq) \S,\quad
                          \tau(q) - \tau(r) + 2\dw(rpq) \S
                        \end{array}
                        \right\}
\end{align}

Since the new progress constraint is stronger,
Lemma~\ref{lemma:2d:nextiscausal} still holds.

\begin{lemma}
  If a front $\tau$ is progressive and if $0 < \e < \f <(1+\e)/2 < 1$,
  then for any local minimum vertex $p$ there exists $\Delta > 0$, where
  $\Delta$ is a function of the triangle $\triangle pqr$ and the
  parameters $\e$ and $\f$, such that the front $\tau' = \next(\tau,p,\dt)$
  is progressive for every $\dt \in [0,\Delta]$.
\label{lemma:2d:refine:isprogressive}
\end{lemma}

\begin{proof}
The proof is almost identical to that of
Lemma~\ref{lemma:2d:refine:isprogressive}, which does not rely on the
fact that the three cases, depending on which vertex of triangle $pqr$
is the apex, are mutually exclusive.
\end{proof}

\subsection{Another look at the adaptive progress constraint}

The progress constraint of Abedi \etal{} is equivalent to the
following constraint for every descendant triangle by newest vertex
bisection. Abusing notation slightly, we refer to an arbitrary
descendant as $\triangle{abc}$ with apex $a$ as in
Figure~\ref{fig:abc}.  Then, the bisector $ad$ of $\triangle{abc}$
must have gradient bounded by
\[
   \abs{\tau(a) - \tau(d)} \le \dw(abc) \S.
\]
Expanding the definition of diminished width and dividing both sides
by $\abs{ad}$, we obtain an equivalent progress constraint:
\[
  \frac{\abs{\tau(a)-\tau(d)}}{\len(ad)}
\le
   \min\,\left\{
     (1-\e) \frac{\dist(a,\aff{bc})}{\len(ad)},\,
     (1-\f) \frac{\dist(b,\aff{ac})}{\len(ad)},\,
     (1-\f) \frac{\dist(c,\aff{ab})}{\len(ad)}
   \right\}
   \S
\]
Observe the following identities: $\dist(b,\aff{ac}) = 2 \,
\dist(d,\aff{ac})$; $\dist(c,\aff{ab}) = 2 \, \dist(d,\aff{ab})$; and
$\sin\angle{adb} = \sin\angle{adc}$. Hence, we can rewrite the last
inequality as
\[
  \norm{\grad \rest{\tau}{ad}}
\le
   \min\,\left\{
     (1-\e) \sin\angle{adc},\,
     2 (1-\f) \sin\angle{dac},\,
     2 (1-\f) \sin\angle{dab}
   \right\}
   \S
\]

Therefore, the adaptive progress constraint $\S$ due to Abedi \etal{}
that limits the gradient of $\triangle{abc}$ along each of the four
edges in the subdivided triangle can be rewritten equivalently as
follows:

  Triangle~$\fa\fb\fc$ satisfies \emph{progress constraint~$\S$} if
  and only if each of the following conditions is satisfied:
\begin{enumerate}
\item
$\norm{\grad\rest{\tau}{ab}} \equiv \norm{\grad\rest{\tau}{de}}
\le \min\,\left\{
       (1-\e) \, \sin\angle{dea},\,
       2 \, (1-\f) \, \sin\angle{eda},\,
       2 \, (1-\f) \, \sin\angle{edc}
    \right\} \S$

\item
  $\norm{\grad\rest{\tau}{ad}}
\le \min\,\left\{
       (1-\e) \, \sin\angle{adc},\,
       2 \, (1-\f) \, \sin\angle{dac},\,
       2 \, (1-\f) \, \sin\angle{dab}
    \right\} \S$

\item
$\norm{\grad\rest{\tau}{ac}} \equiv \norm{\grad\rest{\tau}{df}}
\le \min\,\left\{
       (1-\e) \, \sin\angle{dfb},\,
       2 \, (1-\f) \, \sin\angle{fdb},\,
       2 \, (1-\f) \, \sin\angle{fda}
    \right\} \S$

\item
$\norm{\grad\rest{\tau}{bc}} \equiv \norm{\grad\rest{\tau}{fg}}
\le \min\,\left\{
       (1-\e) \, \sin\angle{fga},\,
       2 \, (1-\f) \, \sin\angle{gfd},\,
       2 \, (1-\f) \, \sin\angle{gfa}
    \right\} \S$
\end{enumerate}

Observing similarity and supplementarity of angles in
Figure~\ref{fig:abc}, we can rewrite the above constraints to refer
only to angles in $\triangle{abc}$ and its two children.

\begin{definition}[Restatement of adaptive progress constraint]
  Triangle~$\fa\fb\fc$ with apex $a$ satisfies the \emph{adaptive
  progress constraint} if and only if each of the following conditions
  is satisfied:
\begin{align*}
\norm{\grad\rest{\tau}{ab}}
&\le \min\,\left\{
        (1-\e) \, \sin\angle{bac},\,
        2 \, (1-\f) \, \sin\angle{bad},\,
        2 \, (1-\f) \, \sin\angle{abc}
     \right\} \S\\
\norm{\grad\rest{\tau}{ad}}
&\le \min\,\left\{
        (1-\e) \, \sin\angle{adc},\,
        2 \, (1-\f) \, \sin\angle{dac},\,
        2 \, (1-\f) \, \sin\angle{dab}
     \right\} \S\\
\norm{\grad\rest{\tau}{ac}}
&\le \min\,\left\{
        (1-\e) \, \sin\angle{bac},\,
        2 \, (1-\f) \, \sin\angle{acb},\,
        2 \, (1-\f) \, \sin\angle{dac}
     \right\} \S\\
\norm{\grad\rest{\tau}{bc}}
&\le \min\,\left\{
        (1-\e) \, \sin\angle{bda},\,
        2 \, (1-\f) \, \sin\angle{acb},\,
        2 \, (1-\f) \, \sin\angle{abc}
     \right\} \S
\end{align*}
\label{def:old:progressconstraint}
\end{definition}

\begin{lemma}
  Triangle $\triangle{\fp\fq\fr}$ satisfies the progress constraint if
  and only if its every ancestor and every descendant also satisfies
  the progress constraint.
\end{lemma}

\begin{proof}
Since $\f \le (1+\e)/2$, we have $2(1-\f) \ge (1-\e)$.  Therefore,
because of the correspondence between the angles of $\triangle{pqr}$
and those of an ancestor or a descendant, $\triangle{abc}$ satisfies
the progress constraint (Definition~\ref{def:old:progressconstraint})
if and only if every ancestor and every descendant of $\triangle{pqr}$
by newest vertex bisection also satisfies the progress constraint.
\end{proof}
\section{Mesh adaptivity in higher dimensions}
\label{sec:hidim:adaptive}

Generalizations of newest vertex bisection to higher dimensions exist
and are well-known.  For instance, the refinement method of
Maubach~\cite{maubach95localbisection,maubach96efficient} generates a
bounded number of similarity classes of simplices, independent of the
amount of refinement due to repeated bisection.  The main challenge we
confront is to extend our progress constraints in 3D$\times$Time to
account for adaptive refinement and coarsening.  We anticipate more
complications than the interplay of primary and secondary forbidden
zones which we encountered in 2D$\times$Time.

Our progress constraint still allows us to perform limited smoothing
and retriangulation of the front at each step.

Under admittedly very strong constraints on the gradient of the front,
it is possible to adapt the front to improve the mesh resolution and
temporal aspect ratio of future spacetime elements.

\begin{lemma}
  If every angle $\theta$ in the spatial projection of a front $\tau$
  satisfies $\sin \theta \ge \minSine$, and if $\norm{\grad\tau} \le
  (1-\e) \minSine \minS$, then any front $\tau'$ obtained by
  retriangulating $\tau$ is causal and satisfies progress constraint
  $\minS$ of Definition~\ref{def:hidim:linear:progressconstraint}.
\label{lemma:hidim:retriangulateisprogressive}
\end{lemma}

\ifproofs\begin{proof}
Since $\minS$ is a lower bound on the reciprocal of the fastest
wavespeed anywhere in spacetime, $\tau$ is clearly causal.

Since $\norm{\grad\tau} \le (1-\e) \minSine \minS$, for any spatial
direction vector $\vec{n}$ we have $(\grad \tau) \cdot \vec{n} \le
\norm{\grad\tau} \le (1-\e) \minSine \minS$.  Therefore, for any face
$F$ in the spatial projection of $\tau'$, we obtain
$\norm{\grad\rest{\tau}{F}} \le (1-\e) \minSine \minS$.
\end{proof}\fi

Thus, when the gradient of the front $\tau$ is bounded, essentially
any arbitrary front modification is permitted as long as the weak
simplicial complex property is maintained; in other words, a
retriangulation $\tau \to \tau'$ is permitted only if $\tau'$ conforms
to $\tau$ or if the operation can be achieved by adding new
non-degenerate tetrahedra each of which has a causal outflow facet.

\subsection*{Chapter summary}

Adaptive meshing is crucial for efficient simulations.  Unless the
size of mesh elements adapts to changing requirements, an extremely
fine mesh would be required throughout the domain.  The number of
elements required to be solved in such a fine mesh would render the
problem intractable.  In this chapter, we presented an adaptive
meshing algorithm that incorporates refinement and coarsening into our
advancing front framework.  Our algorithm adjusts, as soon as
required, the size of future spacetime elements to \aposteriori{}
estimates of the energy dissipation after solving each patch.

For the purposes of the meshing algorithm, we interpret the
dissipation-based error indicator as a black box, specifically, a
binary valued function $\mu$ that given a spacetime element $E$ either
accepts or rejects $E$.  To allow the algorithm to coarsen the mesh,
the lower and upper thresholds on either side of the target
dissipation should be sufficiently well-separated by a hysteresis
zone.

Our algorithm strives for an acceptable balance between accuracy
and efficiency.  Aggressive coarsening, for the purpose of minimizing
the number of spacetime elements, must be balanced with the minimum
acceptable element quality needed for accurate numerical analysis.

We refer the reader to the paper by Abedi
\etal{}~\cite{abedi04spacetime} for a sample application to a
scientific problem of practical complexity: the phenomenon of
crack-tip wave scattering within an elastic solid subjected to shock
loading.  As expected, our experimental results show a significant
improvement in the quality of the solution as a result of adaptivity,
especially near discontinuities in the solution or its derivative.
Also, we were able to achieve a better solution without using a fine
mesh everywhere, which would have resulted in a massive increase in
computation time.

We would like to extend all the current results to higher dimensions,
specifically to 3D$\times$Time.  For the actual mesh refinement, we
can directly use the generalization of newest vertex bisection to
three dimensions by B\"ansch \cite{b-lmr23-91} and others.  The main
theoretical hurdle in higher dimensions is deriving minimal progress
constraints that guarantee that the front can always advance.  Another
concern is to be able to describe the constraints in such a fashion
that they are easy to implement.  It seems likely that analytic
constraints should suffice in higher dimensions, so the algorithm to
maximize progress at each step can be solved exactly.


\chapter{Adaptive Meshing with Nonlocal Cone Constraints}
\label{sec:biadaptive}

In Chapter~\ref{sec:adaptive}, we gave a meshing algorithm for
2D$\times$Time that refines and coarsens the front to adapt the size
of future spacetime tetrahedra in response to numerical error
estimates.  The algorithm of Chapter~\ref{sec:adaptive} assumed a
fixed worst-case estimate of the causal slope at each step.  In
Chapter~\ref{sec:nonlinear}, we relaxed this conservative estimate.
The result was an algorithm that adapted the duration of spacetime
tetrahedra to changing wavespeeds.  In this chapter, we give an
algorithm to simultaneously adapt the mesh resolution to error
estimates and anticipate changing wavespeeds due to nonlinear
response.

We now consider the problem of supporting adaptive
refinement and coarsening in 2D$\times$Time via newest vertex
bisection in the presence of changing wavespeeds.  As noted in
Chapter~\ref{sec:adaptive}, devising necessary and sufficient progress
constraints to support refinement and coarsening via
higher-dimensional analogues of newest vertex bisection remains an
open problem.  The progress constraint at the $i$th step depends on
the geometry of the front, which may change in step $i+1$ due to
refinement and coarsening.  The amount of progress going from $\tau_i$
to $\tau_{i+1}$ is constrained by the fastest wavespeed on the
\emph{new} front $\tau_{i+1}$.  As before, we would like to
characterize the front $\tau_i$ as
\emph{valid} if and only if it is guaranteed to make finite progress
to $\tau_{i+1}$ such that the new front $\tau_{i+1}$ is also valid.
The problem is to anticipate a fast wavespeed in the future
simultaneously with changes in the front due to refinement and
coarsening.  Our contribution is to combine the progress constraints
of Chapters~\ref{sec:linear} and~\ref{sec:adaptive} with the lookahead
algorithm of Chapter~\ref{sec:nonlinear}.  We define a new notion of
\emph{progressive} fronts and prove that every progressive front is
\emph{valid}.

The front is refined in response to numerical error estimates. There
is another reason for the meshing algorithm to refine the front.  A
spacetime element created by pitching a vertex of a front $\tau$ may
encounter vastly different wavespeeds.  Note that the corresponding
triangle of $\tau$ must satisfy a progress constraint that depends on
the wavespeed encountered after pitching.  Hence, if the maximum
wavespeed on any outflow facet of the element is much larger than the
maximum wavespeed on any inflow facet, then the temporal aspect ratio
of the tetrahedron is small.  The height of the element is
proportional to the ratio of maximum to minimum wavespeed anywhere in
the element.  If the temporal aspect ratio is too small, the element
may be unsuitable for computing the numerical solution.  A spacetime
element with very small temporal aspect ratio may have a singular
stiffness matrix, which may make it impossible to solve, or may
require too many iterations for the solution to converge.  If the
solution does converge, the element will likely accumulate a large
error and will therefore be rejected; the cost of solving the element
will be incurred without any advancement of the front.  There is
empirical evidence that the accuracy of the solution within an element
is positively correlated with the quality of the element, defined as
the ratio of its inradius to its circumradius; a small decrease in the
height of the spacetime element may imply a much larger decrease in
its quality.

The computational cost of solving an element is significantly higher
than the amortized per-element cost of mesh generation.  DG solvers
for nonlinear PDEs may take several iterations for the approximate
solution to converge.  If the accuracy of the solution and its
efficiency can be improved by a better quality mesh with fewer
elements, the additional cost of mesh generation is expected to be
easily compensated by the reduced total computation time.

Therefore, for correctness as well as efficiency, it is important for
the meshing algorithm to avoid creating spacetime elements with small
temporal aspect ratio whenever possible.

In this chapter, we give a meshing algorithm that refines the front in
response to a large increase in causal wavespeed, independent of and
in addition to any error-driven refinement demanded by the solver.
While simulating the next few tent pitching steps to estimate future
wavespeed, if the algorithm predicts a bad quality element due to
increasing wavespeed, then it can preemptively refine the front in the
current step. Refining a triangle improves or maintains up to a
constant factor the temporal aspect ratio of the tetrahedron created
by pitching the triangle.  Also, more progress can be made in the
current step if the algorithm prepares for future refinement

In practice, especially for examples where the \emph{Mach
factor}---the ratio of global maximum to minimum wavespeeds---is not
too large, it may be acceptable to allow the solver alone to guide
front refinement.  We cannot yet claim an improvement in temporal
aspect ratio for arbitrary wavespeed distributions, as a result of the
algorithm in this chapter.  Regardless, we believe that this algorithm
is of theoretical interest.  The algorithm solves a two-parameter
greedy optimization problem and illustrates the tradeoff between the
two parameters---one a horizon parameter that measures future
wavespeeds and the other an adaptive lookahead parameter that
estimates the need to refine the front in the future.
Thus, the meshing algorithm presented in this chapter attempts, in
practice, to reduce the number of tetrahedra and improve their
worst-case quality.  Experiments are needed to determine whether the
complexity of a more complicated algorithm is outweighed by a better
quality and/or more efficient solution.

Recall from Chapter~\ref{sec:nonlinear} that the algorithm maintains a
bounding cone hierarchy to speed up the computation of the most
restrictive cone constraint at each step.  Whenever the front is
refined or coarsened, we update the hierarchical decomposition by
recomputing the hierarchy for a subset of the front.  For the sake of
efficiency, we rebalance the hierarchy after a significant refinement
or coarsening of any subset of the front.  Instead of updating the
slope and number of cones in the hierarchy immediately after each
step, we can update in a lazy fashion because the old cone constraints
are still valid conservative estimates of new constraints.  Lazy
updates are likely to be efficient in a parallel setting by reducing
interprocessor communication.

\section{Adaptive meshing in 1D$\times$Time with nonlocal cone constraints}
\label{sec:biadaptive:1d}

We note that it is trivial to perform adaptive refinement and
coarsening of the front in 1D$\times$Time in the presence of nonlocal
(possibly anisotropic) cone constraints.  We maximize the height
of each tentpole subject only to the constraint that each front is
causal, as described in Chapter~\ref{sec:nonlinear}.

The proof of the following theorem is almost identical to that of
Theorem~\ref{thm:1d:nextiscausal}.

\begin{theorem}
  Let $\tau$ be a causal front and let $p$ be an arbitrary local
  minimum of $\tau$.  Then, for every $\dt$ such that $0 \le \dt < w_p
  \minS$, the front $\tau' = \next(\tau,p,\dt)$ is causal.
\label{thm:1d:biadaptive:nextiscausal}
\end{theorem}

\ifproofs\begin{proof}[Proof of Theorem~\ref{thm:1d:biadaptive:nextiscausal}]
  Only the segments of the front incident on $\fp=(p,\tau(p))$ advance
  along with $p$.  Consider an arbitrary segment $pq$ incident on $p$.
  Since $p$ is a local minimum, we have $\tau(q) \ge \tau(p)$.  We
  have
\begin{align*}
\tau'(p)
&\le \tau(p) + \dt\\
&<   \tau(p) + w_p \minS\\
&\le \tau(q) + (1-\delta) \abs{pq} \S(\fp'\fq)
\end{align*}
  Hence,
\[
  \norm{\grad\rest{\tau'}{pq}}
:=
  \frac{\tau'(p)-\tau(q)}{\abs{pq}}
<
  \S(\fp'\fq).
\]
  Therefore, $\fp'\fq$ is causal.
\end{proof}\fi

Note that when pitching a local minimum vertex $\fp$, an interior
vertex, the smaller of the temporal aspect ratios of the resulting two
spacetime triangles is maximized if the two edges incident on $\fp$
have comparable spatial lengths.  Therefore, our adaptive strategy is
guided by the desire to keep the spatial lengths of adjacent front
segments roughly equal.

Because of Theorem~\ref{thm:1d:biadaptive:nextiscausal}, we obtain a
lower bound on the worst case height of any tentpole.  Thus, we have
proved Theorem~\ref{thm:1d:causalisvalid} in the situation where the
wavespeed is not constant throughout spacetime.

\section{Adapting the progress constraint to the level of
refinement in 2D$\times$Time}
\label{sec:biadaptive:2d}

In this section, we consider the adaptive meshing problem in
2D$\times$Time in the presence of changing wavespeeds.  We restrict
ourselves to mesh adaptivity by newest vertex bisection of front
triangles and the reverse operation.  See Chapter~\ref{sec:adaptive}
for a discussion of newest vertex bisection.

The key property of newest vertex bisection is that each descendant of
a triangle $\fp\fq\fr$ has its edges parallel to three of the four
directions defined by $\triangle{\fp\fq\fr}$ and its apex---the three
edges $\fp\fq$, $\fq\fr$, and $\fp\fr$ plus the bisector $\fp\fs$
through the midpoint $s$ of the base $qr$.

Recall from the statement of Lemmas~\ref{lemma:2d:nextiscausal}
and~\ref{lemma:2d:nextisprogressive} that the front at each step must
satisfy a progress constraint that depends on the wavespeed in the
next step. When the wavespeed is not constant, $\minS$ may not be a
good estimate of the future wavespeed.  We have improved the situation
somewhat by giving an algorithm to compute an estimate
$\estS_h(\fp\fq\fr)$ of the wavespeed encountered in the next step by
$\triangle{\fp\fq\fr}$ of the current front, by looking ahead $h$
steps for some horizon parameter $h$.  Thus, we can use the estimated
slope $\estS_h(\fp\fq\fr)$ as the radius of the disc in
Figure~\ref{fig:vectordiagram} when imposing adaptive progress
constraints on $\triangle{\fp\fq\fr}$.  However, this choice of
$\estS_h(\fp\fq\fr)$ may be too conservative because it uses the same
wavespeed $\estS_h(\fp\fq\fr)$ in the progress constraint on
$\triangle{\fp\fq\fr}$ as on each of its descendants after bisection.

A key property we use in deriving the new algorithm in this chapter is
that the boundary of a cone is a \emph{ruled surface}.  Recall from
Chapter~\ref{sec:nonlinear} that we partition the cone constraints
while pitching $\fa$ into local and nonlocal constraints.  Therefore,
if any triangle $\triangle{\fa\fb\fc}$ intersects a given nonlocal
cone, then this intersection is a line segment in the plane of
$\triangle{\fa\fb\fc}$.  Therefore, if the bisector edge $\fa\fd$
intersects a cone of influence, then so do at least two of the edges
of $\triangle{\fa\fb\fc}$ (Figure~\ref{fig:abc}).  Therefore, if a
fast wavespeed in the future intersects $\fa\fd$ but does not
intersect both edges $\fa\fb$ and $\fa\fc$ of $\triangle{\fa\fb\fc}$,
then it can do so only if $\triangle{\fa\fb\fc}$ has been bisected at
least once.

Thus, our contribution is to make the progress constraint imposed on a
given triangle due to the shape of its descendants adaptive to the
level of refinement of the descendant triangles.

The progress constraint that we impose on a given triangle
$\fp\fq\fr$ of a front $\tau$ depends on two parameters,~$h$ and~$l$.
The parameter $h$ is the horizon, as in
Definition~\ref{def:2d:h-progressive}.  For each descendant
$\triangle{\fa\fb\fc}$ of $\triangle{\fp\fq\fr}$ by $l$ newest vertex
bisections, we enforce a progress constraint on $\triangle{\fp\fq\fr}$
that depends on the shape of $\triangle{\fa\fb\fc}$, the level of
refinement $l$, and the horizon parameter $h$.  The greater the level of
refinement of a triangle $\triangle$, the greater is likely to be the
estimated slope $\estS_h(\triangle)$ for a fixed $h$, i.e., if
$\triangle_2$ is a descendant of $\triangle_1$ by newest vertex
bisection, then $\estS_h(\triangle_2) \ge \estS_h(\triangle_1)$.
Therefore, we anticipate up to $l$ levels of refinement where $l \ge 0$ is
another parameter.  For a triangle $\triangle_1$ at refinement
level greater than $l$, we impose a progress constraint that depends
on its ancestor $\triangle_2$ at level $l$; in other words, we draw a
figure similar to Figure~\ref{fig:vectordiagram} except with a disc of
radius $\estS_h(\triangle_2)$.  If $l=0$, then $\triangle_2 =
\triangle{\fp\fq\fr}$; the advantage of choosing $l>0$ is that in
practice we expect $\estS_h(\triangle_2) \gg
\estS_h(\fp\fq\fr)$ because $\triangle_2$ is smaller than
$\triangle{\fp\fq\fr}$.  In other words, for the same horizon
parameter~$h$, the smaller triangle $\triangle_2$ will encounter a
subset of the cone constraints in the next~$h$ steps of pitching
vertices of $\triangle_2$ as the larger
triangle~$\triangle{\fp\fq\fr}$.  For the same number of lookahead
steps~$h$, a smaller triangle is likely to advance through more tent
pitching steps than its larger ancestor because each tent pitching
step makes progress in time proportional to the size of the triangle.

To support adaptive refinement and coarsening by newest vertex
bisection, we have to enforce the adaptive progress constraint of
Definition~\ref{def:2d:refine:progressconstraint}.  We therefore
define \emph{adaptively $h$-progressive fronts}
(Definition~\ref{def:2d:h-progressive}), to refer to this stronger
adaptive progress constraint, as follows.

\begin{definition}[Adaptively $h$-progressive]
  Let $h$ be a nonnegative integer, called the \emph{horizon}.

  Let $\fp\fq\fr$ be a given triangle.  We inductively define
  $\triangle{\fp\fq\fr}$ as \emph{adaptively $h$-progressive} as
  follows.

  \textbf{Base case $h=0$:} Triangle $\fp\fq\fr$ is adaptively
  $0$-progressive if and only if it is causal and satisfies the
  adaptive progress constraint $\minS$
  (Definition~\ref{def:2d:refine:progressconstraint}).

  \textbf{Case $h > 0$:} Triangle $\fp\fq\fr$ is adaptively
  $h$-progressive if and only if all the following conditions are
  satisfied:

  \begin{enumerate}
  \item $\fp\fq\fr$ is causal;

  \item Let $\fp$ be an arbitrary local minimum vertex of
  $\triangle{\fp\fq\fr}$.  Let $d_p$ denote $\dist(p,\aff{qr})$ and
  let $\minT = \min\{\e,{1-\e}\} \,\allowbreak \minS \, d_p$.  Let
  $\fp'\fq\fr = \next(\fp\fq\fr, p, \minT)$ be the triangle obtained
  by advancing $\fp$ by $\minT$. Then, $\fp\fq\fr$ must satisfy the
  adaptive progress constraint $\S(\fp'\fq\fr)$ and $\fp'\fq\fr$ must
  be $\max\{h-1,0\}$-progressive.  \end{enumerate}

\label{def:2d:adaptive:h-progressive}
\end{definition}

We say that a front $\tau$ is \emph{adaptively $h$-progressive} if
every facet of $\tau$ is adaptively $h$-progressive
(Definition~\ref{def:2d:adaptive:h-progressive}).

The following two lemmas are analogous to
Lemmas~\ref{lemma:2d:hprogressive:monotone}
and~\ref{lemma:2d:nextishprogressive}.

\begin{lemma}
  For every $h \ge 0$, if $\triangle{\fp\fq\fr}$ is adaptively
  $h$-progressive, then $\triangle{\fp\fq\fr}$ is adaptively
  $(h+1)$-progressive.
\label{lemma:2d:adaptive:hprogressive:monotone}
\end{lemma}

\ifproofs\begin{proof}[Proof of
Lemma~\ref{lemma:2d:adaptive:hprogressive:monotone}] If a triangle
$\fp\fq\fr$ satisfies adaptive progress constraint $\S$
(Definition~\ref{def:2d:refine:progressconstraint}), then
$\triangle{\fp\fq\fr}$ satisfies progress constraint $\S'$ for every
$\S' \ge \S$.  By Theorems~\ref{thm:2d:refine:iscausal}
and~\ref{thm:2d:refine:isprogressive}, if $\triangle{\fp\fq\fr}$
satisfies the adaptive progress constraint~$\minS$, then there exists
$\minT > 0$ such that the triangle $\fp'\fq\fr$ after pitching a local
minimum~$\fp$ by~$\minT$ is causal and satisfies the adaptive progress
constraint $\minS$.  Since~$\minS$ is a lower bound on the causal
slope anywhere in spacetime, we conclude that triangle $\fp'\fq\fr$ is
adaptively $0$-progressive.  Therefore, by
Definition~\ref{def:2d:adaptive:h-progressive}, if
$\triangle{\fp\fq\fr}$ is adaptively $0$-progressive, then it is
adaptively $h$-progressive for every $h \ge 0$.  Hence, if
$\triangle{\fp\fq\fr}$ is adaptively $h$-progressive, then it is
adaptively $h'$-progressive for every $h' \ge h$.
\end{proof}\fi

\begin{lemma}
  Suppose $\fp\fq\fr$ is a triangle of an adaptively $h$-progressive
  front $\tau$ for some $h \ge 0$.  Let $\fp$ be an arbitrary local
  minimum vertex of $\triangle{\fp\fq\fr}$.  Let $d_p$ denote
  $\dist(p,\aff{qr})$.

  Then, there exists $\minT > 0$ such that $\triangle{\fp'\fq\fr}$
  where $\fp'=(p,\tau(p)+\dt)$ for an arbitrary $\dt \in [0,\minT]$ is
  adaptively $h$-progressive.
\label{lemma:2d:adaptive:nextishprogressive}
\end{lemma}

\ifproofs\begin{proof}[Proof of
Lemma~\ref{lemma:2d:adaptive:nextishprogressive}] By
Definition~\ref{def:2d:adaptive:h-progressive}, triangle $\fp'\fq\fr$
is adaptively $(h-1)$-progressive.  By
Lemma~\ref{lemma:2d:adaptive:hprogressive:monotone}, triangle
$\fp'\fq\fr$ is also adaptively $h$-progressive.
\end{proof}\fi

\begin{definition}[$(h,l)$-progressive]
  We inductively define a triangle $\triangle{\fp\fq\fr}$ as
  $(h,l)$-progressive if it is $h$-progressive 
  (Definition~\ref{def:2d:h-progressive}) and each of the two
  children obtained by newest vertex bisection of
  $\triangle{\fp\fq\fr}$ is $(h,\max\{l-1,0\})$-progressive.

  \textbf{Base case $l=0$:} $\triangle{\fp\fq\fr}$ is
  $(h,0)$-progressive if it is adaptively $h$-progressive
  (Definition~\ref{def:2d:adaptive:h-progressive}).
\label{def:2d:hl-progressive}
\end{definition}

We say that a front is \emph{$(h,l)$-progressive} if every triangle on
the front is $(h,l)$-progressive.  We say that a front is
\emph{progressive} if it is $(h,l)$-progressive for some fixed~$h$
and~$l$.

A progressive front is guaranteed to make sufficient positive progress
after at most a finite number of refinements by newest vertex
bisection of its facets.

\begin{lemma}
  If a front $\tau$ is $(h,0)$-progressive, then for every local
  minimum $p$ of $\tau$ there exists a $\minT > 0$, such that the
  front $\tau'=\next(\tau,p,\dt)$ is $(h,0)$-progressive for every
  $\dt \in [0,\minT]$.
\label{lemma:2d:biadaptive:positiveprogress}
\end{lemma}

\ifproofs\begin{proof}[Proof of
Lemma~\ref{lemma:2d:biadaptive:positiveprogress}] By definition, every
triangle $\fp\fq\fr$ of the front $\tau$ is adaptively
$h$-progressive.  By Lemma~\ref{lemma:2d:adaptive:nextishprogressive},
the triangle $\fp'\fq\fr$ obtained by advancing local minimum vertex
$\fp$ to $\fp'$ by $\minT$ is also adaptively $h$-progressive.  Hence,
there exists a $\minT > 0$ such that the front
$\tau'=\next(\tau,p,\dt)$ is $(h,0)$-progressive for arbitrary $\dt
\in [0,\minT]$.
\end{proof}\fi

By Definition~\ref{def:2d:hl-progressive}, if a triangle $\fp\fq\fr$
is $(h,l)$-progressive, then any triangle obtained by a single newest
vertex bisection of $\fp\fq\fr$ is $(h,l')$-progressive for
$l'=\max\{l-1,0\}$.

We relax the definition of a valid front
(Definition~\ref{def:validfront}) to permit refinement of the front
one or more times at each step without advancing the front.

\begin{definition}[Valid front]
  We say that a front $\tau$ is \emph{valid} if there exists a
  positive real $\dt$ bounded away from zero such that for every $T
  \in \Re^{\ge 0}$ there exists a finite sequence of fronts $\tau$,
  $\tau_1$, $\tau_2$, $\ldots$, $\tau_k$ where $\tau_k \ge T$, each
  new front~$\tau_{i+1}$ in the sequence obtained from the previous
  front~$\tau_i$ by one of two operations: either (i)~advancing some
  local neighborhood of~$\tau_i$ by~$\dt$, or (ii)~newest vertex
  bisection of some triangle of~$\tau_i$.
\label{def:2d:biadaptive:validfront}
\end{definition}

Therefore, we conclude the following theorem.

\begin{theorem}
  If a triangle $\fp\fq\fr$ is $(h,l)$-progressive, then
  $\triangle{\fp\fq\fr}$ is valid.
\label{thm:2d:biadaptive:valid}
\end{theorem}

\ifproofs\begin{proof}[Proof of Theorem~\ref{thm:2d:biadaptive:valid}]
The proof is by a very simple induction on the parameter~$l$.  If
$l=0$, then the theorem follows from
Lemma~\ref{lemma:2d:biadaptive:positiveprogress}.  Otherwise, if
$l>0$, then by the induction hypothesis, the child triangle
$\fa\fb\fc$ of $\triangle{\fp\fq\fr}$, which is $(h,l-1)$-progressive,
is valid.  Hence, by Definition~\ref{def:2d:biadaptive:validfront},
the theorem follows for the case $l>0$ also.
\end{proof}\fi
\subsection{A unified adaptive algorithm in 2D$\times$Time}

Our unified algorithm greedily maximizes the progress such that each
front is $(h,l)$-progressive for some choice of $h$ and $l$. The
algorithm can be as complicated as
desired. Definition~\ref{def:2d:hl-progressive} stresses the fact that
our algorithm can optimize the choice of $h$ and $l$, likely doing
better than the theoretical guarantee obtained by setting $h=l=0$;
however, a simple choice of $h=l=1$ may perform sufficiently well in
practice.

\begin{figure*}\centering
\begin{quote}
\textbf{Input:}  A triangulated space domain $\sp \subset \mathbb{E}^2$\\[1em]

\textbf{Output:} A tetrahedral mesh of $\sp \times [0,\infty)$\\[1em]

Initial front $\tau_0$ is the space mesh at time $t=0$\\
Fix $h$, $l$.\\[1em]

Repeat for $i = 0, 1, 2, \ldots$:

\begin{enumerate}

\item Advance in time an arbitrary local minimum vertex $\fp=(p,\tau_i(p))$
of the current front $\tau_i$ to $\fp'=(p,\tau_{i+1}(p))$ such that
$\tau_{i+1}$ is $(h,l)$-progressive, and $\tau_{i+1}(p)$ is
maximized.

\item Partition the spacetime volume between $\tau$ and $\tau_{i+1}$
into a patch of tetrahedra, each sharing the tentpole edge $\fp\fp'$.

\item Solve the resulting patch.

\item If the patch is accepted, then some outflow triangles on the new front
$\tau_{i+1}$ are marked coarsenable.  Coarsen any pair of coarsenable sibling
triangles if they are coplanar on the new front $\tau_{i+1}$, as long
as the resulting coarser front is also $(h,l)$-progressive.

\item If the patch is rejected, then one or more triangles on the
front $\tau_i$ are marked for refinement.  Perform newest vertex
bisection of every triangle marked for refinement, propagating the
bisection to neighboring triangles to maintain a triangulated front.

\end{enumerate}
\end{quote}
\caption{Unified adaptive algorithm in 2D$\times$Time.}
\label{fig:2d:algUnified}
\hrule
\end{figure*}
\subsection{A simpler algorithm in 2D$\times$Time}
\label{sec:2d:simpler}

Arguably, the algorithm presented in the previous sections is very
complicated to implement for general~$h$ and~$l$.  Is this
complication really necessary?  In the case of discontinuities in
wavespeeds due to shocks, we think the complication is unavoidable.
However, in the case that the wavespeed function is smooth, we present
a simpler algorithm that is guaranteed to terminate.  For each
triangle $\triangle$ on the front, estimate the maximum and minimum
wavespeed at any point of $\triangle$.  If the ratio of the maximum to
minimum wavespeeds is too large, say greater than $4$, then refine the
triangle; otherwise, if this ratio is too small, say less than $2$,
then mark the triangle as coarsenable.  Since the wavespeed is
continuous, this refinement will stop after a bounded number of steps.
If the wavespeed function is discontinuous, we would need to impose a
lower bound on the element size to limit refinement.

\subsection*{Chapter summary}

Refinement of a triangle on the front is easy to incorporate into the
advancing front meshing algorithm with nonlocal cone constraints
because any faster cone of influence that intersects a smaller
triangle also intersects its parent and is accounted for in the
progress constraint that must be satisfied by the parent.

Coarsening presents a challenge when the wavespeed is not constant
everywhere.  When two smaller triangles are coarsened, the maximum
wavespeed on the larger triangle is the maximum of the wavespeeds of
the original smaller triangles.  Therefore, both triangles being
coarsened must satisfy the same progress constraint before they can be
coarsened, i.e., we can merge two triangles into one if and only if
the new triangle is $h$-progressive.  When coarsening is possible only
under such constraints, we need to carefully prioritize each
coarsening step so that the front is only as refined as necessary and
not much more.  Prioritizing the various coarsening requests remains
an open problem.


\chapter{Target time conformity}
\label{sec:targettime}

For comparing the accuracy of the numerical results with those
obtained by other techniques, such as for convergence studies, we
sometimes need a spacetime mesh of the bounded volume from time $t=0$
to $t=T$ for some target time~${T > 0}$.  Also, some modifications of
the front, such as coarsening by derefinement, are permitted only when
the gradient of the front everywhere is small.  In particular, if the
entire front corresponds to a constant time function, then the
algorithm can continue exactly as if this constant-time front were the
initial input.  To make the entire front or some subset of the front
coplanar, the algorithm may have to limit the amount of progress at
each step so that the vertex being pitched is coplanar with some or
all of its neighbors.  This coplanarity constraint, in addition to
causality and progress constraints, further reduces the temporal
aspect ratio of resulting tetrahedra and may create near-degenerate
elements.  Therefore, it is important to anticipate coplanarity
constraints by limiting the amount of progress in the current step and
thus ensure that both current and future tentpole heights can still be
provably bounded from below.

In this chapter, we show how to make tent pitching less greedy about
maximizing the progress at each step so that a subset of the front can
be made coplanar in the next step.  The central idea of the algorithm
in this chapter is to trade off the amount of progress, i.e., tentpole
height, achieved in the current step and that achieved in the next
step.  This algorithm can be used in arbitrary dimensions to make the
entire front conform to a target time $T \ge 0$.  The algorithm can
also be used to coarsen the front and to retriangulate it by edge
contraction or edge dilation by first making a subset of the front
conform to a local target time. The main feature of the algorithm in
this chapter is that it retains the progress guarantee up to a
constant factor proved for the algorithms in earlier chapters.  Thus,
we are able to achieve target time conformity without significantly
sacrificing the worst-case temporal aspect ratio of spacetime
elements.  We believe that the central idea of the algorithm in this
chapter will be useful for other purposes such as when the spacetime
mesh is required to conform to a singular spacetime surface, not
necessarily a constant-time plane.

\section{Conforming to a target time}
\label{sec:targettime:reliable}

We say that a spacetime mesh $\spt$ \emph{conforms} to a target
time $T$ (or in general to any front $\Phi$) if the portion $\sp
\times T$ of the constant time plane $t=T$ is equal to the union of
causal facets of $\spt$.

In the context of pitching tents, conforming to a target time $T$
means that the height of a tentpole is further constrained so that the
top of the tentpole does not exceed time $T$.  This additional
constraint creates a problem when a vertex $\fp=(p,\tau(p))$ has
almost achieved the target time but not quite, i.e., when $\tau(p) <
T$ and $T - \tau(p)$ is very small.  In this case, advancing $p$ from
$\tau(p)$ to $T$ will create a tentpole of height $T-\tau(p)$ which is
unacceptably small and results in degenerate or poor
quality elements.  In this section, we show how to incorporate
additional constraints into our algorithm so that the advancing front
conforms to the specified target time $T$ while still proving a lower
bound on the height of each tentpole.  Thus, we retain the quality
guarantee of the algorithm within a constant factor.

Let $\gamma \in (0,\half]$ be an additional parameter to our advancing
front algorithm.

The key idea we use is that whenever the front attempts to get too
close to the target time $T$ during a single step of pitching a local
minimum $p$, we reduce the height of the current tentpole so that in
the next step the vertex $p$ will achieve the target time $T$.  We
call the current step the ``penultimate'' tent pitching step because
the next step is the last step in which vertex $p$ is pitched.  The
tradeoff we need to make is between the height of the tentpoles in the
current (penultimate) step and in the next (final) step.  We do it in
a way that in both cases the tentpole height is at least~$\gamma$
times the worst-case guarantee on the height of \emph{any} tentpole at~$p$.

Let $\tau$ denote the front at the current iteration.
Let $\fp$ be a local minimum of $\tau$ such that $\tau(p) < T$.  Such
a local minimum vertex $p$ must exist unless the entire front $\tau$
has achieved the target time $T$; in particular, the global minimum
vertex is a candidate vertex.

Denote by $w_p$ the minimum distance of $p$ from $\aff{\triangle}$ for
every simplex $\triangle$ in (the spatial projection of) the front $\tau$
incident on $p$.
\[
  w_p  :=  \min_{\triangle \in \link(p)}\,\{\,\dist(p,\aff{\triangle})\,\}
\]

Let $\S \equiv \minS$ denote the reciprocal of the global fastest
wavespeed.  For any vertex $p$, let $h_p$ denote a lower bound on the
height of the tentpole at $p$ whenever $p$ is pitched.  Note that we
require $h_p$ to be small enough to be independent of the front during
the step in which $p$ is pitched, i.e., $h_p$ can be a function of
only the spatial projection of any front.  Depending on which progress
constraint the algorithm uses, i.e., depending on the specific meshing
algorithm, $h_p$ may be a different lower bound; thus, for instance,
the algorithm of Chapter~\ref{sec:linear} guarantees a worst-case
tentpole height of
\[
  h_p := \min\{\e,1-\e\} w_p \S.
\]

The algorithm at each step enforces the following invariant on the
front $\tau$ at the beginning of step $i$: for every vertex $p$ of
$\tau$ we have:
\begin{equation}
\text{either}\qquad
  \tau(p) = T
\qquad\text{or}\qquad
  \tau(p) \le T - \gamma \, h_p
\label{eqn:invariant}
\end{equation}
We assume, of course, that the initial front $\tau_0$ satisfies this
invariant, i.e., either the space mesh $\sp$ is refined enough or the
target time $T$ is large enough to satisfy
Equation~\ref{eqn:invariant}.

Suppose the algorithm is currently pitching such a local minimum
vertex $p$ of $\tau$ such that $\tau(p) < T$.  Such a candidate
vertex must exist because a global minimum of the front is always a
candidate unless the entire front is at time $T$.

Let $\tau'(p)$ denote the time value of the top of the tentpole at
$p$, maximized to satisfy all causality and progress constraints.  The
new algorithm is allowed to lower the top by choosing a different time
value $\tau''(p)$ for the top of the tentpole such that $\tau(p) \le
\tau''(p) \le \tau'(p)$.

Let $\gamma$ be a fixed constant in the range $0 < \gamma \le \half$,
a parameter to the algorithm. For the reasons below, $\gamma=\half$
may be a good choice in practice.

Apply the following rule at each tent pitching step.

\begin{enumerate}
\item \textbf{Case 1:} If $\tau'(p) \ge T$, then choose $\tau''(p) :=
T$ as the new top of the tentpole at $p$.

\item \textbf{Case 2:} Otherwise, if $\tau'(p) \ge T - \gamma \, h_p$, then
choose $\tau''(p) := T - (1-\gamma) h_p$ as the new top of the
tentpole at $p$.

\item \textbf{Case 3:} Otherwise, we must have $\tau'(p) < T - \gamma
\, h_p$; then choose $\tau''(p) := \tau'(p)$ as the top of the tentpole at $p$.
\end{enumerate}

We prove that the height of the tentpole in the current step is
possibly reduced to $\gamma \, h_p$ instead of $h_p$; however,
the height of the tentpole the next time $p$ is pitched is guaranteed
to be at least~${(1-\gamma) h_p}$.  The choice of parameter $\gamma$
represents a tradeoff between the current step and the next step.

As a corollary, the entire front will achieve the specified target time
$T$ in a finite number of steps.

It is easy to see that the algorithm must eventually advance every
vertex that has not yet achieved the target time value $T$.  We prove
next that whenever a vertex $p$ is advanced, then the height of the
resulting tentpole at $p$ is at least~${\gamma \, h_p > 0}$.

First, we prove by induction on the number of tent pitching steps that
the invariant of Equation~\ref{eqn:invariant} is always maintained.
Let $\tau_i$ denote the front at the beginning of iteration $i$.  We
assume that the initial front $\tau_0$ automatically satisfies the
invariant. Whenever case 1 applies at step $i$, we have $\tau_{i+1}(p)
= T$; therefore, the first part of Equation~\ref{eqn:invariant} is
satisfied by the new front.  Otherwise, if case 2 applies, then
$\tau_{i+1}(p) := T - (1-\gamma) h_p \le T - \gamma \, h_p$ because
$\gamma \le 1-\gamma$.  Finally, if case 3 applies, then we have
$\tau_{i+1}(p) := \tau'(p) < T - \gamma \, h_p$, so the second part of
Equation~\ref{eqn:invariant} is satisfied by the new front.

\begin{theorem}
  Let $\tau$ be a progressive front.  Let $\gamma$ be a parameter in
  the range $(0,\half]$.  Let $T$ be a target time such that $T \ge
  \tau(p) + \gamma\, h_p$ for every vertex $p$.  Then, whenever
  our new algorithm advances a local minimum vertex $p$ of $\tau$,
  it constructs a tentpole at $p$ with height at least $\gamma \,
  h_p$.
\end{theorem}

\begin{proof}
We prove that $\tau(p) \le \tau''(p) \le \tau'(p)$ and that $\tau''(p)
- \tau(p) \ge \gamma \, h_p$.

Note that $\tau''(p) = T - (1-\gamma) h_p \le T - \gamma \,
h_p \le \tau'(p)$ because $\gamma \le \half$.

We show that lowering the top of the tentpole at $p$ from $\tau'(p)$
to $\tau''(p)$ does not create degenerate elements in the current
step.

There are three mutually exclusive possibilities that cover all
eventual outcomes.

\begin{enumerate}
\item[(a)]
Suppose case 1 applies in the current step, i.e., $\tau''(p) =
T$. Since $\tau(p) < T$, it must be the case that either case 2 or
case 3 must have applied in the \emph{previous} step.  If case 2 applied in
the previous step, then $\tau(p) = T - (1-\gamma) h_p$, so the
height of the current tentpole is $(1-\gamma) h_p \ge \gamma \,
h_p$ because $\gamma \le \half$.

This is the only (sub)case for which the claim is tight.

On the other hand, if case 3 applied in the previous step, then
$\tau(p) < T - \gamma \, h_p$, so the current tentpole height is $T -
\tau(p) > \gamma \, h_p$.

\item[(b)] Suppose case 2 applies in the current step, i.e., $\tau''(p) = T -
(1-\gamma) h_p$.  Then, the current tentpole height is
$\tau''(p)-\tau(p)$ and we have
  $\tau''(p) - \tau(p)  =  (T-(1-\gamma)h_p) - \tau(p)
                          =  (T-\tau(p)) - (1-\gamma) h_p$.

But from the tent pitcher progress guarantee, we know that
  $\tau'(p) \ge \tau(p) + h_p$.

Hence,
  $T - \tau(p)  \ge  (T-\tau'(p)) + h_p$.

We also know that $\tau'(p) < T$ because case 2 applied instead of
case 1. Hence, $T - \tau(p)  >  h_p$.

Therefore,
  $\tau''(p) - \tau(p) > h_p - (1-\gamma)h_p = \gamma \, h_p$.

\item[(c)] Suppose case 3 applies in the current step, i.e., $\tau''(p) =
\tau'(p)$.  Then, from the tent pitcher progress guarantee, we know
that $\tau'(p) = \tau''(p) \ge \tau(p) + h_p$.

\end{enumerate}
\end{proof}

As an interesting corollary, we obtain the following result for the
linear non-adaptive case.  We can construct a spacetime mesh with the
same progress guarantee that conforms to any progressive front~$\Phi$.
We can do this by running our algorithm \emph{backwards} in time,
starting from~$\Phi$ as the ``initial'' front and choosing to conform
to the constant time plane $t=0$.  (Conceptually, we can imagine this
as meshing the spacetime volume $\sp \times [-\Phi,0]$.)  It is
possible to run the algorithm backwards in time with the same progress
guarantee because the gradient constraints on the front at each step
are unchanged if all time values are replaced by their negatives.

Further, we can mesh the spacetime volume between any two progressive
fronts~$\Phi_1$ and~$\Phi_2$ over the same space domain~$\sp$, as long
as there exists a constant time plane $t=T$ between~$\Phi_1$
and~$\Phi_2$ that has a margin of at least $\gamma \, \minT$ from
either of the two fronts.  We do this by running the algorithm in two
phases: once we run the algorithm starting from the front~$\Phi_1$ and
conforming to time $t=T$; then, we run the algorithm starting from the
front~$\Phi_2$ and conforming to time $t=T$.  By adjoining the
resulting two spacetime meshes, we obtain a spacetime mesh of the
volume between~$\Phi_1$ and~$\Phi_2$.  The final spacetime mesh is a
weak simplicial complex if the spatial projections of the given two
fronts~$\Phi_1$ and~$\Phi_2$ are conforming, i.e., each triangle in
one triangulation is the union of one or more triangles of the other
triangulation.

\parasc*{Anticipating changes due to adaptive refinement and
coarsening}

For a vertex $p$, the local spatial geometry of the front can change
due to adaptive refinement and coarsening or due to some other
retriangulation of the front; thus, $w_p$ can change and therefore a
fixed~$h_p$ no longer suffices as an adequate lower bound on the
tentpole height at~$p$.  However, we can interpret the invariant of
Equation~\ref{eqn:invariant} as a constraint (an upper bound) on the
spatial size of the triangulation in the local neighborhood, $\st(p)$,
of~$p$, rather than as a lower bound on~$T$.  If the initial
front~$\tau_0$ does not satisfy the invariant, then we refine the
front until it does.  If coarsening a pair of triangles would violate
the invariant, then this coarsening is not permitted.  Any other
retriangulation that would violate the invariant is also not
permitted.  However, since refining the front is a requirement for
numerical accuracy, we anticipate all possible changes in shape of the
front due to refinement in advance and alter the definition of~$w_p$
appropriately.  This is easy because newest vertex bisection generates
only a small number of homothetic triangle shapes on the front.

\section{Conforming to a target time: a simpler heuristic}
\label{sec:targettime:heuristic}

The following is another algorithm for making a vertex $p$ conform to
its target time $T_p$ while still guaranteeing a lower bound on the
height of each tentpole height. The following logic is easier to
implement than the previous one, because it does not involve the
progress guarantee computation and does not involve the additional
parameter $\gamma$ (even though it can be modified to depend on such
a parameter).

Let $p$ be a local minimum of the front $\tau$; let $\tau(p)$ denote
its time coordinate before lifting and let $\tau'(p)$ denote the time
coordinate of the proposed tentpole at $p$, maximized subject to
causality and progress constraints only.  Thus, $h:=\tau'(p)-\tau(p)$
denotes the proposed height of the tentpole at $p$ before imposing any
target time constraint.

Let $T_p$ denote the target time for $p$.  We will compute a new time
value $\tau''(p)$ for the tentpole top, instead of $\tau'(p)$.

Apply the following rules at each tent pitching step:

\begin{enumerate}
\item \textbf{Case 1: $\mathbf{\tau'(p) \ge T_p}$.} Choose $\tau''(p) :=
T_p$.
\item \textbf{Case 2: $\mathbf{\tau'(p) < T_p}$.}
If $T - \tau'(p) < h$, then $T_p < 2\tau'(p)-\tau(p)$; in this case,
choose $\tau''(p) := (T_p+t(p))/2$.  Otherwise, we have $\tau'(p) \le
T - h$; in this case, let $\tau''(p):=\tau'(p)$ remain.
\end{enumerate}

First, we prove that lowering the tentpole top is acceptable in the
current step.  We have $t''(p) = (T-t(p))/2$; at the same time, we are
given $0 < T-t'(p) < h$ which means
\[
  T-t(p) = T-t'(p)+t'(p)-t(p) = T-t'(p)+h > h
\]
Hence, we conclude $t''(p) = (T-t(p))/2 > h/2$.

Finally, we prove that lowering the tentpole top is acceptable in
future steps.  We have $T-t''(p) = T-(T-t(p))/2$;  we also know that
$T - t'(p) < h$, so $T-t(p) = T-t'(p)+t'(p)-t(p) = T-t'(p) + h < 2h$.
Hence, we conclude that $T-t''(p) < T - 2h/2 = T-h$.

\section{Smoothing out large variations in tentpole heights}

We would like to smooth out large variations in the heights of the
tentpoles erected at any given vertex $p$ in any two consecutive
steps.

Let $p$ be a local minimum of the front $\tau$; let $\tau(p)$ denote
its time coordinate before lifting and let $\tau'(p)$ denote the time
coordinate of the proposed tentpole at $p$, maximized subject to
causality and progress constraints only.  Thus, $h:=\tau'(p)-\tau(p)$
denotes the proposed height of the tentpole at $p$ before imposing any
target time constraint.

Let $\gamma > 0$ be a parameter.  Let $h_p$ denote a lower bound on
the height of the tentpole at $p$; thus,
\[
  h_p := \min\{\e,1-\e\} w_p \S.
\]

If $h > (1+\gamma) h_p$, then choose $h' = (h+h_p)/2$ as the new
tentpole height; otherwise, let $h'=h$ remain the chosen tentpole
height.  Clearly, this choice still retains the progress guarantee of
$h_p$.

In the context of a target time constraint, we apply the above rule
only in the last case when the original tentpole height is accepted by
the logic that aims to achieve the target time $T_p$.  If the tentpole
height has already been curtailed by a target time constraint, then we
do not reduce it any further.

\section{Coarsening}

Coarsening is always stated as a request while refinement is a
necessity.  However, for the purpose of efficiency, it is important to
coarsen aggressively any parts of the front that are marked as
coarsenable so that the number of spacetime elements, and thus the
total computation time, is reduced.

Coarsening the front by merging adjacent triangles requires the
triangles to be coplanar in spacetime.  In the previous sections, we
gave algorithms to make the entire front conform to a global target
time $T$.  We use a similar approach to give a new reliable coarsening
algorithm.  Our new algorithm guarantees that any part of the front
that is marked as coarsenable will eventually be coarsened.  This
requires that coarsenable pairs of triangles be made coplanar after a
finite number of tent pitching steps.  Often, this requires that the
actual tentpole height at a vertex of the two triangles being merged
be reduced to achieve the coplanarity constraint.  Our new algorithm
guarantees that, even with the additional coplanarity constraints, the
worst-case tentpole height, and hence the minimum temporal aspect
ratio of elements, is bounded from below.  In fact, the minimum
tentpole height at a given vertex is guaranteed to be at least
$\gamma$ times the minimum height in the absence of any coplanarity
constraints, where $\gamma \in (0,\half]$ is a parameter to the
coarsening algorithm.

Suppose $S$ is a subset of triangles that are scheduled for
coarsening.  Choose a synchronization time $T$ and assign it to each
vertex incident on a triangle of $S$.  Now, use the algorithm from the
previous section to ensure that each of these vertices achieves their
target time $T$.  When all of them are coplanar with the constant time
plane $t=T$, then perform the scheduled coarsening and then remove the
scheduling constraint $T$ from all vertices.  Recompute a new target
time for all vertices of the coarser triangles if the coarser
triangles are also coarsenable.  In fact, we also recompute the target
time for every vertex of a triangle that gets refined.

What if multiple coarsenings are scheduled at the same time?  In that
case, it is possible that some vertex $p$ has achieved its scheduled
time $T_2$ and is waiting for another vertex $s$ of a sibling triangle
to also achieve time $T_2$.  Note that $s$ need not be a neighbor of
$p$, but $s$ must be a neighbor of an immediate neighbor of $p$.  In
such a situation, $p$ cannot be pitched even though it is a local
minimum because $p$ is waiting for $s$.  However, this is not a
problem because we can prove that $s$ must eventually achieve time
$T_2$.

If $s$ has no scheduling constraint, then it will eventually be
pitched because it will eventually become a local minimum vertex.
Since $s$ is lower than time $T_2$, it will be pitched to time $T_2$.
On the other hand, if $s$ has been scheduled to achieve a target time
$T_1$ because some other triangle incident on $s$ is being coarsened,
then it must be that $T_1 < T_2$.  Hence, by induction, eventually $s$
will achieve time $T_1$ at which step the algorithm is free to pitch
$s$ further until it eventually achieves time $T_2$.

Formally, the new algorithm with coarsening constraints is given next.

\subsection{New algorithm}

Let $\S \equiv \minS$ denote the reciprocal of the global fastest
wavespeed.

Given a front $\tau_i$ and a vertex $p$ of the spatial projection of
$\tau$, denote by $w_p^{i}$ the minimum distance of $p$ from line $qr$ for
every triangle $pqr$ incident on $p$.
\[
  w_p^{i}  :=  \min_{\triangle{pqr} \ni p}\,\{\,\dist(p,\aff{qr})\,\}
\]
Let $\he$ denote $\min\{\e,1-\e\}$.  Since $0 < \e < 1$, we have $0 <
\he \le \half$.
Let $h_p^{i}$ denote $\he w_p^{i} \S$.  Note that $w_p^{i}$ and
$h_p{i}$ depend on the current front $\tau_i$.

Call a vertex $v$ a \emph{leaf vertex} if $v$ is an interior vertex
incident on four triangles, or $v$ is a boundary vertex incident on
two triangles, and if every triangle incident on $v$ is a leaf in the
refinement hierarchy.
Call a leaf vertex $v$ \emph{coarsenable} if $\level(v) > 0$ and if
all triangles incident on $v$ are marked coarsenable by the numerical
analysis.

\subsubsection{Initialize}

Fix the value of parameter $\gamma$ such that $0 < \gamma \le \half$,
e.g., $\gamma = \half$.
Initialize the front $\tau_0$.  Every vertex of $\tau_0$ is assigned
level zero.  Every triangle of $\tau_0$ is assigned a level number of
zero.

Every vertex $p$ of $\tau$ (not just a local minimum) is assigned a
target time~$T_p$.  We maintain the invariant that for every front
$\tau$, for every vertex $p$ of $\tau$ we have $\tau(p) \le T_p$.

If the problem requires a finite global target time, then let $T$
denote this target time, otherwise let $T \leftarrow \infty$.
Initially, every vertex $p$ of $\tau$ (not just a local minimum) is
assigned a target time $T_p \leftarrow T$.

Refine the current front $\tau$ as many times as needed until the
following inequality is satisfied for each vertex~$p$:
\[
  T_p
\ge
  \tau(p) + \gamma \, h_p
\]
Note that $h_p$ is halved after every two levels of newest vertex
bisection; hence, for a fixed $T_p$ the refinement process will
terminate.  The refinement process does not change $\tau(p)$, hence
the above inequality gives an upper bound on $h_p$.

\subsubsection{Tent pitching}

At the current iteration of the algorithm, let $\tau_i$ denote the
current front at the beginning of iteration $i$.

\parasc*{Step 1: Assign a target time to each vertex (not just local
minima).}

The general idea is that if a set of triangles of $\tau$ is marked as
coarsenable by the numerical analysis, then we bring all the five
vertices involved to a common target time value; then, the triangles
are coplanar and the front can be coarsened immediately.

There are three steps to computing a target time~$T_p$ for each vertex~$p$.

(1)~For each vertex p, there is a lower bound $l_p$ on the target time
$T_p$. (The target time~$T_p$ is computed in step~3.)  A lower bound
is necessitated in order to maintain the invariant of
Equation~\ref{eqn:invariant}.  This lower bound $l_p$ depends on all
the triangles on the current front incident on $p$.

The lower bound $l_p$ on the target time~$T_p$ is computed as follows:
$l_p = \tau_i(p) + \gamma \, h_p^{i}$.

(2)~We can coarsen a set of four triangles at once (as long as they
are marked for coarsening by the numerical analysis).  Call this a
coarsenable cluster $C$.  Compute a target time~$T_C$ for the cluster
as the \emph{maximum} of~$l_p$ for each vertex~$p$ in the cluster.
(There will be exactly five vertices incident on the four triangles.)

If~$s$ denotes the degree-$4$ vertex of the cluster~$C$ that will be
removed as a result of coarsening cluster $C$, then the vertex $s$
participates in only one coarsenable cluster, the cluster~$C$.
Therefore, the target time~$T_C$ is also the target time~$T_s$
assigned to vertex $s$ in step 3.

(3)~For each vertex $v$, compute the target time $T_v$ as the
\emph{minimum} of the cluster times $T_C$ for each coarsenable cluster $C$ to
which the vertex~$v$ belongs.  Since $T_C \ge l_v$ for each
cluster~$C$, we conclude that $T_v \ge l_v$ and, hence, the invariant
of Equation~\ref{eqn:invariant} is satisfied.

Note that a cluster of triangles is coarsenable only if each of the
four triangles in the cluster is a leaf, i.e., has not been subdivided
any further.  If $v$ does not belong to any coarsenable cluster, then
the target time $T_v$ is the global target time $T$ (which may be
$+\infty$).  Note that step~(1) looks at \emph{all} triangles incident
on $p$ in order to compute $l_p$, but step~(3) looks at only
\emph{coarsenable} triangles incident on $p$ in order to compute
$T_p$.

After a coarsening takes place, a new lower bound $l_p$ is computed
for each vertex $p$ of the two coarser triangles that violates the
invariant, and a new target time is computed for each coarsenable
cluster to which $p$ belongs.

\begin{figure}\centering\small
\includegraphics[height=2in]{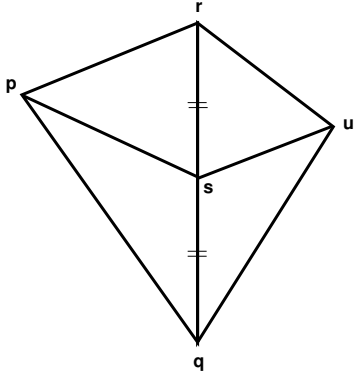}
\caption{Coarsening a set of four triangles in one step.}
\label{fig:derefine}
\end{figure}

Let~$s$ be a coarsenable vertex (see Figure~\ref{fig:derefine}) and
let $C$ denote the set of four coarsenable triangles incident on $s$.
(See figure.)  Let $V$ be the set of five vertices of the triangles
in~$S$.  (In Figure~\ref{fig:derefine}, the set $V$ is
$\{p,q,r,s,u\}$.)

For each vertex $v \in V$, compute $l_v$ such that
\[
  l_v
=
  \tau_i(v) + \gamma \, h_v^{i}
\]

Next, assign a target time $T_s$ to the coarsenable vertex $s$ (and
thus to $C$) such that $T_s := \max_{v \in V} l_v$.

Mark each vertex $p$ such that $\tau(p) < T_p$ as \emph{not ready}.

\parasc*{Step 2: Pitch a tent at a local minimum.}

Let $p$ be an arbitrary local minimum of $\tau$ such that $\tau(p) <
T_p$.  In other words, select as a candidate for tent pitching only
those local minima that have not yet achieved their respective target
times.  At least one candidate vertex must exist; in particular, a
global minimum vertex must be a candidate vertex unless the entire
front has achieved the global target time $T$.

Let $\tau'(p)$ denote the time value of the top of the tentpole at
$p$, maximized to satisfy all causality and progress constraints.  The
new algorithm chooses a possibly different time value $\tau''(p)$ for
the top of the tentpole such that $\tau(p) \le
\tau''(p) \le \tau'(p)$.

Apply the following rule at each tent pitching step.

\begin{enumerate}

\item \textbf{Case 1:} If $\tau'(p) \ge T_p$, then choose $\tau''(p) :=
T_p$ as the new top of the tentpole at $p$.  After pitching $p$, mark
$p$ as \emph{ready}.

After pitching $p$, some cluster $C$ of triangles participating in a
single coarsening step may become coplanar with the corresponding
target time $T_C$.  For each coarsenable vertex $q$ neighboring $p$,
if all neighbors of $q$ are ready, then all triangles incident on $q$
are coarsened immediately.  Note that $q$ is deleted as a result of
this coarsening step.  For each neighbor of $q$, compute a new target
time as described in Step~1.

\item \textbf{Case 2:} Otherwise, if $\tau'(p) \ge T_p - \gamma \, h_p$, then
choose $\tau''(p) := T_p - (1-\gamma) h_p$ as the new top of the
tentpole at~$p$.

\item \textbf{Case 3:} Otherwise, we must have $\tau'(p) < T_p -
\gamma \, h_p$; then choose $\tau''(p) := \tau'(p)$ as the top of the
tentpole at $p$.

\end{enumerate}

If a patch is rejected, causing refinement of the front, then each new
triangle is marked \emph{not} coarsenable.  Whenever an edge is
bisected as a result of newest vertex bisection, then assign the new
vertex (the interior point of the bisected edge) a target time equal
to the global target time $T$.

\subsubsection{A cycle in the propagation pattern}

\begin{figure}\centering\small
\includegraphics[height=3in]{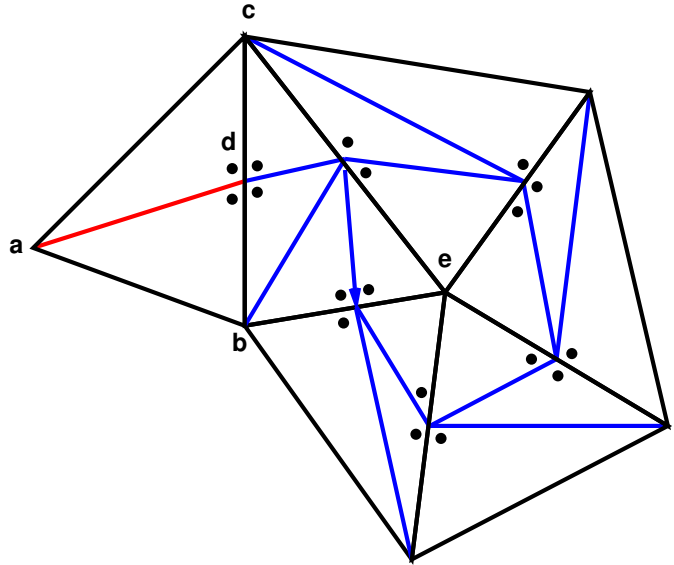}
\caption{Coarsening a set of several triangles in one step.}
\label{fig:propagationcycle}
\end{figure}

If the pattern of marked vertices is such that it causes a cycle in
the propagation pattern (see Figure~\ref{fig:propagationcycle}), then
the only reliable way to coarsen any pair of triangles in the set is
to coarsen them all simultaneously.  Fortunately, if the largest edge
of each triangle is chosen as the base, then a cyclic marking is very
unlikely to occur because it would require a cycle of isosceles
triangles.  If ties between two edges of equal length are broken
consistently, for instance using lexicographic order, then such a
cycle will \emph{never} occur.

\subsection{Proof of correctness}

It remains only to prove that the minimum height of each tentpole is
bounded.  In the previous section, we showed that for every vertex $p$
of the front $\tau_i$ the minimum height of every tentpole was at
least $\gamma \, h_p^{i}$.  Since there is a minimum element size that
is always numerically acceptable, there is a global minimum $h_p$ that
implies a corresponding positive lower bound on the minimum height of
each element.

\subsection*{Chapter summary}

Guaranteed coarsening has been an open problem ever since our adaptive
meshing result was published~\cite{abedi04spacetime}.  We need to
strike the right balance between efficiency and accuracy---the former
achieved by adaptive refinement and aggressive coarsening, the latter
achieved by theoretical guarantees on worst-case element quality.  The
coarsening algorithm in this chapter retains the lower bounds on
element temporal aspect ratios proved in the absence of coarsening,
and this represents a big step forward. However, the heuristic of
Section~\ref{sec:targettime:heuristic} seems to perform a little
better in practice.  We can imagine a few other heuristics that would
attempt to improve element quality and solution accuracy in practice,
but it seems hard to prove that they will be efficient in terms of
guaranteed coarsening of coarsenable regions.

A new coarsening operation proposed by Jeff Erickson is to delete an
interior vertex $\fu$ of degree-$4$ with neighbors $\fp$, $\fq$,
$\fr$, and $\fs$ by constructing a patch consisting of two tetrahedra:
either the two tetrahedra $\fp\fq\fr\fu$ and $\fp\fr\fs\fu$, or the
two tetrahedra $\fp\fq\fs\fu$ and $\fq\fr\fs\fu$; the first choice
creates two new triangles $\fp\fq\fr$ and $\fp\fr\fs$ on the new front
while the second choice creates the triangles $\fp\fq\fs$ and
$\fq\fr\fs$.  The coarsening operation is performed only when the
quality of each tetrahedron in the patch exceeds an acceptable
threshold.  The same idea is used to coarsen the boundary by deleting
a degree-$3$ boundary vertex and creating a single-tetrahedron patch.
The algorithm extends naturally to deleting a vertex $u$ of arbitrary
degree; the triangulation after deleting $u$ is obtained, for
instance, by contracting the edge $up$ for some neighbor $p$ of $u$.
We could prefer contracting the shortest contractible edge incident
on~$u$.  The alternative coarsening algorithm is to pitch possibly
inclined tentpoles at $u$ and/or any of its four neighbors, to make
triangles $\fp\fq\fu$ and $\fu\fq\fr$ coplanar and the triangles
$\fp\fu\fs$ and $\fs\fu\fr$ coplanar before merging them pairwise.
The former coarsening method does not guarantee non-degeneracy of the
spacetime elements but uses fewer tetrahedra (two instead of four) and
does not require that pairs of triangles be coplanar.


\chapter{Extensions}
\label{sec:extensions}

In this chapter, we describe extensions to our algorithms, in the form
of newer advancing front operations.  We show how to incorporate
operations that simultaneously advance the front in time as well as
smooth the spatial projection of the front to improve the spatial
aspect ratio of front simplices.  Front smoothing will likely increase
the amount of progress in subsequent tent pitching steps and
potentially improve the quality of future spacetime elements.  These
more general operations are potentially useful for tracking moving
boundaries.  We give a framework that employs the new advancing front
operations to make local modifications to the front in response to
changes in the geometry of the domain.  We describe our recent work
and initial progress on tracking moving domain boundaries.

Next, we discuss details of implementing our algorithms, such as the
choice of various parameters.  Our advancing front algorithms can be
tuned to perform much better in practice than the theoretical
performance guarantee by a suitable choice of parameters and
heuristics.


\section{Extensions and heuristics}

In this section, we describe extensions to our advancing front
algorithm and propose heuristics to improve the performance in
practice.

\subsection{Asymmetric cone constraints}

In the presence of anisotropy, waves travel with different speeds in
different directions.  Therefore, a cone of influence whose slope in
spatial direction $\vec{n}$ is equal to the wavespeed in that
direction must necessarily be non-circular.  The resulting cone
constraint is asymmetric because it enforces a different gradient
constraint in different spatial directions in order to satisfy
causality.

Our algorithm can be easily extended to handle asymmetric cone
constraints by enforcing the causality constraint separately for each
spatial direction.  Thus, we enforce a constraint on the gradient of
every front in every spatial direction $\vec{n}$ to satisfy the
causality constraint in the direction $\vec{n}$.  This will be our
approach in higher dimensions as well.  In the presence of asymmetric
cones of influence, we assume implicitly that each slope $\S$ is
quantified universally for every spatial direction $\vec{n}$.
Causality and progress constraints are interpreted as gradient
constraints in each spatial direction $\vec{n}$.

We can re-interpret each theorem in the anisotropic setting.  We
understand every statement to be implicitly quantified over every
spatial direction $\vec{n}$.

For instance, we can restate our theorems for 2D$\times$Time in the
anisotropic setting as follows.  Let $\triangle{\fp\fq\fr}$ be a
triangle of the front $\tau$ with $\fp$ as one of its lowest
vertices. Without loss of generality, assume $\tau(p) \le \tau(q) \le
\tau(r)$.  Let~$\e$ be any real number in the range ${0 < \e < 1}$.  Let
$\minS$ denote the global minimum causal slope in any spatial
direction.  Let $\S_{\vec{n}}(\triangle)$ denote the minimum causal
slope in spatial direction $\vec{n}$ over every point of $\triangle$,
i.e.,
\[
  \S_{\vec{n}}(\triangle) := \min_{\fp \in \triangle}
                               \{\S_{\vec{n}}(\fp)\}
\]

\begin{enumerate}
\item
  Let $\triangle{\fp'\fq\fr}$ denote the triangle of the front
  $\tau'$, obtained by pitching the local minimum $\fp$, for arbitrary
  $\tau'(p)$ in the range $\tau(q) \le \tau'(p) \le \tau(q) + \e \minS
  \dist(p,\aff{qr})$.

  For every spatial direction $\vec{n} \in \Real^2$, if
  $\norm{\grad\rest{\tau}{qr}} \le (1-\e) \phi_{qr}
  \S_{\vec{n}}(\fp'\fq\fr)$, then $\triangle{\fp'\fq\fr}$ satisfies
  $\norm{\grad\tau'} < \S_{\vec{n}}(\fp'\fq\fr)$.

\item
  Let $\triangle{\fp'\fq\fr}$ denote the triangle of the front
  $\tau'$, obtained by pitching the local minimum $\fp$, for arbitrary
  $\tau'(p)$ in the range $\tau(q) \le \tau'(p) \le \tau(q)
  + (1-\e) \minS \min\{\abs{pq},\abs{pr}\}$.

  For every spatial direction $\vec{n} \in \Real^2$,
  $\triangle{\fp'\fq\fr}$ satisfies
\[
  \norm{\grad\rest{\tau'}{pq}} \le (1-\e) \phi_{pq} \S_{\vec{n}}(\fp'\fq\fr)
\qquad\text{and}\qquad
  \norm{\grad\rest{\tau'}{pr}} \le (1-\e) \phi_{pr} \S_{\vec{n}}(\fp'\fq\fr)
\]
\end{enumerate}

Our algorithm is modified to enforce gradient constraints
simultaneously in every spatial direction.  In other words, we require
that for every triangle $\fp\fq\fr$ with local minimum vertex $\fp$,
the slope of the edge $\fq\fr$ is less than or equal to $(1-\e)
\phi_{qr} \S_{\vec{n}}(\fp\fq\fr)$.  If the wavespeed is increasing in
the future, then the lookahead process will enforce a stronger
gradient constraint parameterized by the future wavespeed
$\S'_{\vec{n}}$.  All previous theorems assure positive progress at
each step when the appropriate gradient constraint is enforced for
each spatial direction.  The correctness of our algorithm requires the
anisotropic no-focusing axiom,
Axiom~\ref{axiom:anisotropic:nofocusing}, which is also sufficient.

The asymmetric nature of cones of influence complicates the data
structures, specifically the bounding cone hierarchy.  While
constructing the hierarchy, at each step we need to compute a tight,
possibly asymmetric, cone that contains two given cones of influence.
We have several choice for representing asymmetric cones, such as
polyhedral cones (the convex hull of vectors in spacetime with the
same origin) or cones with elliptical cross-sections.  For the sake of
efficiency, we require each cone to have constant complexity.

However, a noncircular cone is still a ruled surface.  Therefore, both
our earlier algorithms, exact and approximate, to maximize the
progress at each step subject to causality continue to apply without
any changes.

Anisotropic response means that waves propagate faster in one
direction than another.  This can lead to nonlocal cone constraints.
However, our algorithm already handles nonlocal constraints.

\subsection{Non-manifold domains}

So far, we have assumed implicitly that the space domain at time $t$
is a $d$-dimensional manifold.  However, our algorithms and all
discussions in this thesis extend to non-manifold domains.  An example
of a non-manifold domain in practice would be three thin metal sheets
intersecting in a T-junction, or a mechanical assembly consisting of
parts with complicated intersections.

Let $\sp_t$ denote the space domain at time $t$.  The space domain
$\sp_t$ is a subset of Euclidean space of appropriate dimensionality
such that $\sp_t$ has a $d$-dimensional simplicial complex.  In this
case, $\sp_t$ could be a collection of triangles, edges, and vertices
forming a $2$-dimensional simplicial complex embedded in
$\mathbb{E}^3$.  However, $\sp_t$ need not be a surface; for instance,
$\sp_t$ could consist of three triangles sharing a common edge.  In
this case, we say that the space domain has dimension $d=2$.  We
measure distances between points in $\sp_t$ by taking the
corresponding distances in the ambient Euclidean space which by the
nature of the embedding is a lower bound on the geodesic distance
between the same points.

\subsection{Front smoothing}

The progress constraint at the $i$th step also depends on the geometry
of the front in step $i+1$.  As in the previous section, our aim is
also to ensure that the progress guarantee at the $i$th step is
proportional to local geometry of the spatial projection of the front
$\tau_i$.  The discussion in this section applies to arbitrary
dimensions $d \ge 2$.

The amount of progress made by a local minimum $p$ of $\tau_i$ is
proportional to
\[
  w_p := \min_{\triangle \in link(p)} \dist(p,\aff{\triangle}).
\]
If we can increase $w_p$, then we would get taller tentpoles and
therefore fewer spacetime elements overall.  In this section, we give
an algorithm that allows more general operations such as pitching
inclined tentpoles and performing edge flips as long as causality and
all progress constraints are satisfied.  Each such operation is
desirable if it improves the front locally in the sense of increasing
the minimum $w_p$ for any vertex $p$ in the local neighborhood.

\subsubsection{Inclined tentpoles}

Mesh smoothing in the neighborhood of a vertex $p$ is achieved by
pitching an inclined tentpole at $p$ from $\fp=(p,\tau(p))$ to
$\fp''=(p',\tau'(p'))$.  Advancing $p$ in time as well as moving it in
space can be thought of as a combination of two successive operations:
(i)~move $p$ in space to $p'$, and then (ii)~advance $p'$ in time from
$\fp'=(p',\tau(p'))$ to $\fp''=(p',\tau'(p'))$.  Each of the spacetime
elements in the resulting patch shares the tentpole $\fp\fp''$ and has
both a causal inflow facet on the old front $\tau$ as well as a causal
outflow facet on the new front $\tau'$.

We assume that the direction of the inclined tentpole $\fp\fp''$ is
fixed, for instance in an earlier iteration of the algorithm.  To
perform Laplacian smoothing, the direction of $\fp\fp''$ could be
chosen to move $p$ closer to the centroid of its neighbors in the
spatial projection of $\tau_i$.

Let $\tilde{\tau}$ denote the intermediate front obtained from $\tau$
after the first step. If $\tilde{\tau}$ is progressive and if $\fp'$
is a local minimum of $\tilde{\tau}$, then the final front $\tau_{i+1}
= \next(\tilde{\tau},p',\dt)$ after pitching
$\fp'=(p',\tilde{\tau}(p'))$ to $\fp''=(p',\tau_{i+1}(p'))$ is
progressive, for some $\dt > 0$.  It follows by induction on the
number of tent pitching steps that every front constructed by the
algorithm is valid.

\begin{figure}[t]\centering\small
\includegraphics[height=2in]{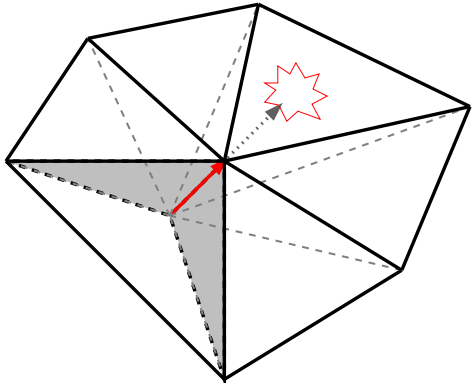}
\caption{The length of an inclined tentpole is limited so that new
front triangles are causal.}
\label{fig:inclinedtentpole}
\end{figure}

The length of the inclined tentpole $\fp\fp''$ is constrained so that
every triangle $\fp''\fq'fr$ is causal, i.e., not too steep. See
Figure~\ref{fig:inclinedtentpole}.

\subsubsection{Anticipating changing geometry}

\begin{figure}[t]\centering\small
\includegraphics[height=2in]{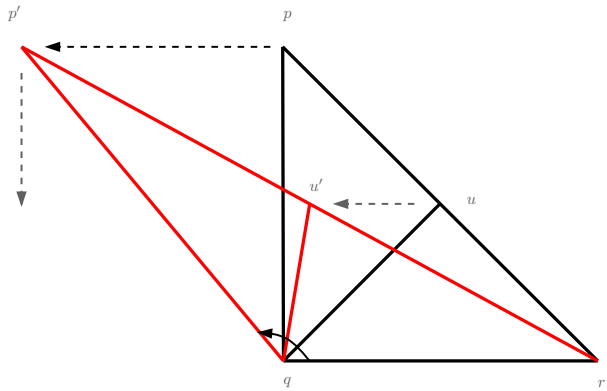}
\caption{Moving a vertex can introduce or increase an obtuse angle on
the front.}
\label{fig:motion-obtuseangle}
\end{figure}

The primary challenge we face is how to modify the progress
constraints in response to changing geometry of the spatial projection
of the front.  The need to strengthen the progress constraint is clear
from Lemma~\ref{lemma:2d:nextiscausal}---if $\phi(pqr)$ decreases
because the obtuse angle of $\triangle{pqr}$ becomes more obtuse
(Figure~\ref{fig:motion-obtuseangle}), then positive progress cannot
be guaranteed.  The factor $\phi(pqr)$ that appears in the progress
constraint of Definition~\ref{def:2d:linear:progressconstraint} must
be changed to anticipate any potential increase in the largest obtuse
angle of any triangle of the front.

We enforce a lower bound on the smallest acute angle and on the
largest obtuse angle in the spatial projection of any front. Fix
$\minSine > 0$ such that $\minSine$ is less than $\phi(pqr)$ for any
triangle $pqr$ of the front $\tau$.  (Recall
Definition~\ref{def:2d:phi}.)  We will enforce that $\sin\theta \ge
\minSine$ for every angle of every triangle on the front constructed
at each step.  We are now able to strengthen the progress constraint
to anticipate the changing geometry of the front as long as the front
does not violate the following angle bound.

\begin{definition}[Angle bound $\minSine$]
  We say that a triangle $\triangle{abc}$ satisfies the \emph{angle
  bound} $\minSine$ if the minimum sine of any angle of
  $\triangle{abc}$ is at least $\minSine$.
\label{def:2d:anglebound}
\end{definition}

In many ways, the choice of $\minSine$ is like the estimate of the
global maximum wavespeed.  We will exert a lot of effort in
Chapter~\ref{sec:nonlinear} to obtain better estimates of future
wavespeed than the global maximum.  However, for arbitrary changes to
the front due to spatial motion, we do not know yet how to adaptively
obtain a better estimate of how $\phi(pqr)$ changes.  For now, we have
to be content with a global bound on the smallest acute angle as well
as the largest obtuse angle allowed in the spatial projection of any
front.  Techniques similar to those in Chapter~\ref{sec:nonlinear} may
prove useful to adaptively estimate the changing geometry due to
motion.

Currently, we do not know how to exploit the fact that the algorithm
may freely choose the tentpole direction.

We strengthen the progress constraint by replacing, in
Definition~\ref{def:2d:linear:progressconstraint}, every occurrence of
$\phi$ by $\minSine$, where $\minSine$ is a parameter.  Since our new
progress constraint is stronger, Lemma~\ref{lemma:2d:nextiscausal}
still holds.  The price we pay for strengthening the progress
constraint is a smaller worst-case progress guarantee, smaller by a
factor of $\minSine$, due to the fact that the front in the
\emph{next} step has to satisfy a stronger gradient constraint.

\begin{definition}[Progress constraint $\S$]
  Let $\fp\fq\fr$ be an arbitrary triangle of a front $\tau$. We say
  that the triangle $\fp\fq\fr$ satisfies \emph{progress constraint
  $\S$} if and only if for every edge, say $qr$, we have
  \[
    \norm{\grad \rest{\tau}{qr}}
  :=
    \frac{\abs{\tau(r) - \tau(q)}}{\abs{qr}}
  \le
    (1-\e) \S \minSine
  \]
\label{def:2d:smoothing:progressconstraint}
\end{definition}

In arbitrary dimensions, fix $\minSine > 0$ such that $\minSine$ is
less than $\phi(p_0p_1p_2{\ldots}p_k)$ for any $k$-dimensional simplex
$p_0p_1p_2{\ldots}p_k$ of the front $\tau$.  (Recall
Definition~\ref{def:hidim:phi}.)

Consider an arbitrary $k$-dimensional face $p_0p_1p_2{\ldots}p_k$
incident on $p_0$.

\begin{definition}[Progress constraint]
  We say that a simplex $\fp_0\fp_1\fp_2{\ldots}\fp_k$ of the front
  $\tau$ satisfies the \emph{progress constraint $\S$} if and only if
  for every facet $\Delta_i$ where $0 \le i \le k$ we have
  $\norm{\grad{\rest{\tau}{\Delta_i}}} \le (1-\e) \, \S \, \minSine$.
\label{def:hidim:smoothing:progressconstraint}
\end{definition}

Note that the progress constraints of Definition
\ref{def:hidim:smoothing:progressconstraint} and Definition
\ref{def:hidim:smoothing:progressconstraint} limit the gradient of
every facet of a front simplex uniformly, unlike Definition
\ref{def:2d:linear:progressconstraint} and Definition
\ref{def:hidim:linear:progressconstraint} which limits the gradient of
only highest facets of each front simplex.

Let $p$ be a local minimum of $\tau$.  Suppose we advance $p$ in time
from $\tau(p)$ to $\tilde{\tau}(p)$ but no higher than its lowest
neighbor, i.e., $\tilde{\tau}(p) \le \tau(q)$ for every neighbor $q$
of $p$.  Additionally, we move $p$ in space to $p'$ but only so far
that $\triangle{p'qr}$ also satisfies the angle bound $\minSine$.  By
the monotonicity of the progress constraint, we know that the new
triangle $\fp'\fq\fr$ where $\fp'=(p',\tilde{\tau}(p))$ satisfies the
new progress constraint $\S$ as long as $\triangle{\fp\fq\fr}$ also
satisfies the new progress constraint $\S$.  We have thus proved the
following lemma.

\begin{lemma}
  If a front $\tau$ satisfies the progress constraint $\S$ of
  Definition~\ref{def:2d:smoothing:progressconstraint} and if $p$ is a
  local minimum of $\tau$, then the new front $\tilde{\tau}$ satisfies
  the progress constraint $\S$ of
  Definition~\ref{def:2d:linear:progressconstraint} where
  $\tilde{\tau}$ is obtained by moving $\fp=(p,\tau(p))$ to
  $\fp'=(p',\tilde{\tau}(p))$ such that $\fp'$ is also a local minimum
  of $\tilde{\tau}$, $p'$ is in the kernel of $\st(p)$, and
  $\tilde{\tau}$ satisfies angle bound $\minSine$.
\label{lemma:2d:perturbisprogressive}
\end{lemma}

We restrict the motion of $p$ so that for every triangle $pqr$
incident on $p$, the smallest angle of $\triangle{p'qr}$ is bounded
from below.  To build on earlier work which guarantees that $p'$ can
make finite positive progress if $p'$ is a local minimum, we also need
to guarantee that $\tilde{\tau}$ is progressive and that $p'$ is a
local minimum of the front $\tilde{\tau}$. It is easy to see that, as
long as $p'$ stays a local minimum, the front $\tilde{\tau}$ is
progressive if $\tau$ is progressive; this is because as long as $p'$
is a local minimum of $\tilde{\tau}$ we have
$\norm{\grad\rest{\tilde{\tau}}{\triangle}} \le
\norm{\grad\rest{\tau}{\triangle}}$ for every simplex $\triangle$.
Therefore, $p'$ must satisfy $\tau(p') \le
\tau(q)$ for every neighbor $q$ of $p$, where $p$ is a local minimum
of $\tau$.

\begin{figure}\centering
\includegraphics[width=0.8\textwidth]{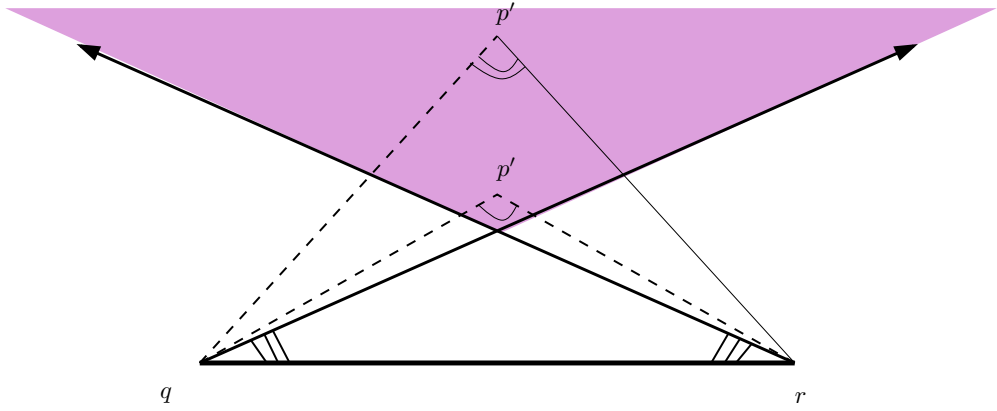}
\caption{The allowed region for $p'$ is the shaded region in the
  top: $p'$ is restricted to the allowed region to ensure that
  $\min\{\sin\angle{p'qr}, \sin\angle{p'rq}, \sin\angle{rp'q}\} \ge
  \minSine$.}
\label{fig:angleboundregion}
\end{figure}

The allowed region for $p'$ such that $\triangle{p'qr}$ satisfies the
angle bound $\minSine$ is shown in Figure~\ref{fig:angleboundregion}
when a single triangle $pqr$ is considered in isolation where $\tau(p)
\le \tau(q) \le \tau(r)$.  When all triangles incident on $p$ are
taken together and we know that the intersection of their allowed
regions is nonempty because it contains at least $p$.

\begin{lemma}
  If $\triangle{\fp\fq\fr}$ of the front $\tau$ with $\tau(p) \le
  \tau(q) \le \tau(r)$ satisfies $\norm{\grad\rest{\tau}{qr}} \le
  (1-\e) \minSine \S$, then for every $\dt \in [0,(1-\e) \minSine
  \dist(p,\aff{qr}) \S]$ the front $\tau' =
  \next(\tau,p,\dt)$ satisfies $\norm{\grad\rest{\tau'}{rp}} \le
  (1-\e) \minSine \S$.
\label{lemma:2d:smoothing:nextisprogressive}
\end{lemma}

\ifproofs\begin{proof}[Proof of
Lemma~\ref{lemma:2d:smoothing:nextisprogressive}]
By Definition~\ref{def:2d:linear:progressconstraint}, the front $\tau'$
satisfies progress constraint $\S$ if and only if
\[
  \max\{\,
    \norm{\grad\rest{\tau'}{pq}},
    \norm{\grad\rest{\tau'}{qr}},
    \norm{\grad\rest{\tau'}{rp}}\,
  \}
\le
  (1-\e) \minSine \S
\]
Since advancing $p$ in time does not change the time function along
$qr$, we have $\norm{\grad\rest{\tau'}{qr}} =
\norm{\grad\rest{\tau}{qr}} \le (1-\e) \minSine \S$.

Since $\triangle{\fp\fq\fr}$ satisfies progress constraint $\S$, we
have
\[
  \max\{\,
    \norm{\grad\rest{\tau}{pq}},
    \norm{\grad\rest{\tau}{qr}},
    \norm{\grad\rest{\tau}{rp}}\,
  \}
\le
  (1-\e) \minSine \S
\]
Also, $p$ is a lowest vertex of $\tau$; so $\tau(p) \le
\min\{\tau(q),\tau(r)\}$.

Consider the constraint $\norm{\grad\rest{\tau'}{pq}} \le (1-\e)
\minSine \S$.  As long as $\tau'(p) \le \tau(q)$, we have
$\norm{\grad\rest{\tau'}{pq}} \le \norm{\grad\rest{\tau}{pq}} \le
(1-\e) \minSine \S$.  Similarly, as long as $\tau'(p) \le \tau(r)$,
the constraint $\norm{\grad\rest{\tau'}{rp}} \le (1-\e) \minSine \S$
is automatically satisfied because $\norm{\grad\rest{\tau'}{rp}} \le
\norm{\grad\rest{\tau}{rp}} \le (1-\e) \minSine \S$.

When $\tau'(p) > \tau(q)$, the constraint
$\norm{\grad\rest{\tau'}{pq}} \le (1-\e) \minSine \S$ is equivalent to
$\tau'(p) \le \tau(q) + (1-\e) \minSine \S \abs{pq}$.  We have
\begin{align*}
  \tau'(p)
&\le
  \tau(p) + (1-\e) \minSine \S d_p\\
&\le
  \tau(q) + (1-\e) \minSine \S d_p\\
&\le
  \tau(q) + (1-\e) \minSine \S \abs{pq}
\end{align*}
where the last inequality follows because $d_p = \dist(p,\aff{qr}) \le
\abs{pq}$.

Similarly, when $\tau'(p) > \tau(r)$, the constraint
$\norm{\grad\rest{\tau'}{rp}} \le (1-\e) \minSine \S$ is equivalent to
$\tau'(p) \le \tau(r) + (1-\e) \minSine \S \abs{pr}$.  We have
\begin{align*}
  \tau'(p)
&\le
  \tau(p) + (1-\e) \minSine \S d_p\\
&\le
  \tau(r) + (1-\e) \minSine \S d_p\\
&\le
  \tau(r) + (1-\e) \minSine \S \abs{pr}
\end{align*}
where the last inequality follows because $d_p = \dist(p,\aff{qr}) \le
\abs{pr}$.
\end{proof}\fi

We have thus shown that as long as the motion of $p$ to $p'$ is
constrained to the allowed region at each step, the progress made by
pitching $p$ is positive and bounded away from zero.

\subsubsection{Edge flips}

\begin{figure}[t]\centering\small
\includegraphics[height=3.2in]{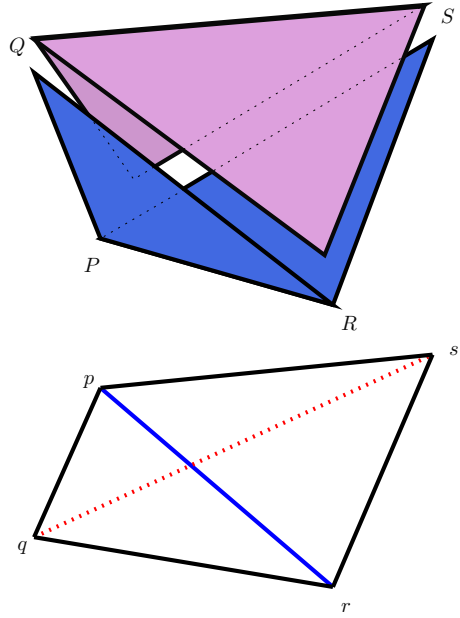}
\caption{An edge is flipped by inserting a single tetrahedron.}
\label{fig:edgeflip}
\end{figure}

Another front advancing operation is to insert a single tetrahedron
$\fp\fq\fr\fs$ with inflow facets $\triangle{\fp\fq\fr}$ and
$\triangle{\fp\fr\fs}$ and outflow facets $\triangle{\fp\fq\fs}$ and
$\triangle{\fq\fr\fs}$; see Figure~\ref{fig:edgeflip}.  If the volume
of the tetrahedron $\fp\fq\fr\fs$ is positive, this operation advances
the front.  If the solution within $\fp\fq\fr\fs$ is sufficiently
accurate, the single-element patch is accepted.

Let $\tau$ and $\tau'$ denote the fronts before and after this
operation.  We require that the front $\tau'$ be causal and progressive
to guarantee sufficient progress when a tent is pitched in the next
step.

Triangles $pqr$ and $prs$ belong to the spatial projection
of $\tau$, and triangles $pqs$ and $qrs$ belong to the spatial
projection of $\tau'$.  Therefore, the operation results in an edge
flip, replacing the diagonal $pr$ of the quadrilateral $pqrs$ with the
diagonal $qs$; of course, quadrilateral $pqrs$ must be convex.

Thus, the edge flip replacing $pr$ by $qs$ is allowed only if $pqrs$
are in convex position in the spatial projection, tetrahedron
$\fp\fq\fr\fs$ has positive volume, and all four facets of the tetrahedron
are causal and progressive.

The edge flip is desirable only if the resulting tetrahedron is
non-degenerate and has acceptable quality, and if it improves or does
not worsen the spatial aspect ratio of front triangles, i.e.,
\[
  \min\,\{\phi(pqs), \phi(qrs)\}
\ge
  \min\,\{\phi(pqr), \phi(prs)\}
\]

A similar flip operation could be defined in higher dimensions.  It is
not clear whether the resulting patch, which must sometimes consist
of more than one spacetime element, satisfies all the objectives
such as, for instance, the requirement that every element have a
causal outflow facet.

\subsection{Heuristics}

In this section, we describe more general meshing operations, in
addition to tent pitching, performed as long as each operation alters
the front only in ways such that the new front is $(h,l)$-progressive
for the chosen parameters~$h$ and~$l$.  We label these new operations
\emph{heuristics} because we cannot yet provide a guarantee that
each such operation will be possible when it becomes necessary or even
when it is desirable.  These enhancements represent a step towards an
adaptive advancing front algorithm for tracking spacetime features,
such as moving boundaries, by aligning mesh facets along these
features.  Developing such a boundary-tracking meshing algorithm is
still work in progress.  We cannot yet guarantee that any of the
operations mentioned can always be performed in response to an urgent
need to improve the quality of the front or to guarantee progress at
all.  However, we are able to give sufficient and admittedly strong
gradient constraints under which any front can be retriangulated,
subject to the angle bound, so as to achieve the generalized front
modification operations of this section.

Note that newest vertex bisection does not improve the quality of
front triangles beyond a very limited amount.  For instance, the
largest obtuse angle in a triangle also occurs in its grandchildren.
The more general operations on the other hand can improve the quality
of front triangles, and therefore the worst-case progress guarantee.
Our stronger progress constraints are sufficient gradient constraints
under which a limited amount of improvement of the front can be performed
whenever permitted by the constraints.

Our earlier work on adaptivity was restricted to adapting the mesh
resolution to an error indicator; now, we are concerned also with
adapting the mesh quality, in terms of worst-case temporal aspect
ratio of spacetime tetrahedra.

\begin{enumerate}
\item \textbf{Generalized bisection}
Our algorithm refines a triangle on the front by bisecting an edge the
triangle at a point other than its midpoint, as long as the resulting
smaller triangles satisfy the so-called \emph{angle bound}.

The bisection direction for each triangle must be chosen so that
repeated bisection retains the key property that each
descendant of a triangle $\fp\fq\fr$ has its edges parallel to three
of the four directions defined by $\triangle{\fp\fq\fr}$ and its
apex---the three edges $\fp\fq$, $\fq\fr$, and $\fp\fr$ plus the
bisector $\fp\fs$ for the appropriately chosen interior point $s$ on
the base $qr$.

Adjacent triangles must be bisected \emph{compatibly}, i.e.,
adjacent triangle must agree on how to bisect their common edge
so that refinement maintains a triangulated front.

The front is coarsened by undoing previous refinement.

The new bisection method means, for instance, that the largest angle
in any triangle on each front can be bisected.

To ensure that the front can be advanced successfully after the
bisection, each bisection is permitted only if the front after
bisection is progressive.  Note that this condition also limits the
choice of direction along which to bisect a given triangle.

\item \textbf{Mesh smoothing:}
Advancing a vertex $p$ in time takes $\fp=(p,\tau)$ to $\fp'=(p,\tau')$
where $\tau' > \tau$.  Pitching an inclined tentpole at $p$ takes
$\fp=(p,\tau)$ to $\fp'=(p',\tau')$ which simultaneously changes the spatial
projection of the front, taking $p$ to $p'$ such that $p'$ is in the
kernel of $\st(p)$.  This achieves a limited amount of smoothing of
the front, also known as $r$-refinement.

With the stronger progress constraints, we are able to allow inclined
tentpoles subject to the angle bound.  As long as the slope of each
tentpole is \emph{subsonic}, i.e., greater than the local causal
slope, we are still able to prove a positive lower bound on the
temporal aspect ratio of each spacetime tetrahedron.

\item \textbf{Flipping an edge:}
We relax the assumption that the spatial projection of each front is a
hierarchical refinement of the initial space mesh.  Flipping an
interior edge $e=(p,q)$ replaces the diagonal $e$ of a convex
quadrilateral with the other diagonal $e'=(r,s)$.  An edge flip in the
spatial projection of a front $\tau$ corresponds to adding a single
tetrahedron $\fp\fr\fq\fs$ to the spacetime mesh; this edge flip can
be performed only if the resulting tetrahedron has bounded positive
volume.

With the stronger progress constraints, we are able to allow flipping
of interior edges as long as all triangles of every front satisfy the
angle bound.  An edge flip is desirable only if the resulting
tetrahedron has bounded temporal aspect ratio.  The algorithm is free
to choose the marked vertex of each new front triangle created as a
result of an edge flip.

\end{enumerate}

By construction, since each of the above operations is performed only
when permitted by the progress constraints, the new front $\tau'$
after the operation satisfies the angle bound and is
$(h,l)$-progressive.  See Lemma \ref{lemma:hidim:retriangulateisprogressive}.

Another operation that changes both the spatial projection of a front
$\tau$ as well as the time function is the following.  Let $p$ be a
local minimum of $\tau$.  The new progress constraint limits the
gradient of each triangle $\triangle{\fp\fq\fr}$ in $\st(p)$ in four
different directions.  Suppose we advance $p$ in time from $\tau(p)$
to $\tilde{\tau}(p)$ but no higher than its lowest neighbor, i.e.,
$\tilde{\tau}(p) \le \tau(q)$ for every neighbor $q$ of $p$.
Additionally, we move $p$ in space to $p'$ but only so far that
$\triangle{p'qr}$ also satisfies the angle bound $\minSine$.  Then, by
Lemma \ref{lemma:2d:perturbisprogressive}, we know that the new front
$\tilde{\tau}$ also satisfies the new progress constraint $\S$.
Therefore, the algorithm is guaranteed to be able to pitch
$\tilde{\fp}$, a local minimum of the front $\tilde{\tau}$, by a
positive amount to make progress in time to get the new front $\tau'$
with $\fp'=(p',\tau'(p'))$.



\section{Tracking moving boundaries}

In many applications, the geometry or even the topology of the domain
changes over time.  For instance, combustion of solid rocket fuel eats
away at the boundary where the fuel burns.  The mesh must adapt by
changing the size of mesh elements, or by varying the placement of
mesh features, or both.  How to dynamically re-mesh in response to
evolving geometry is a well-studied problem~\cite{antaki00parallel}.
With other dynamic meshing algorithm, global remeshing of the domain
is performed whenever the mesh gets too distorted due to motion; at
intermediate steps, smoothing or other incremental changes are
interleaved with the solution.

In this section, we restrict our attention to meshing in response to
changes in the geometry of the domain due to boundary motion.  We
assume that the topology of the domain does not change, which means it
is sufficient to consider each connected component of the original
space mesh separately.  In our context, we always remesh locally by
anticipating and adapting to changes in the shapes of front triangles.
An external or an internal boundary (such as a material interface)
moving with time describes a surface in spacetime.  Similarly, shocks
describe singular surfaces in spacetime.  The solution jumps
discontinuously across the shock surface.
\emph{Tracking} the boundary means aligning vertices, edges, and
facets of our tetrahedral mesh with the corresponding surface
(approximately).  Boundary tracking combined with SDG methods implies
a highly accurate resolution of discontinuities in the solution field.

We incorporate spatial motion over time in our advancing front
framework by assuming either a spatial velocity vector or a direction
in spacetime for every vertex of every front.  The tentpole direction
can be inferred from either input.  Boundary vertices have the
tentpole direction assigned to them by the numerical analysis, which
is aware of the evolving nature of the domain.  For all other
vertices, the meshing algorithm is free to choose any subsonic
tentpole direction.  For instance, for the purpose of Laplacian
smoothing, we choose for every interior vertex $p$ a direction vector
in space pointing towards the centroid of the neighbors of $p$.  The
slope of the tentpole at $p$ is chosen to be at least a constant
factor greater than the reciprocal of the maximum wavespeed anywhere
in $\st(p)$.

We modify the basic framework of our interleaved mesh construction and
solution procedure to recompute vertex velocities along with or as
part of the solution.

Our motion model is sufficient to capture boundary motion that does
not alter the topology of the underlying space domain over
time---boundary vertices remain boundary vertices and interior
vertices remain interior vertices throughout.  This model therefore
seems insufficient to capture splitting apart of the domain into more
components, creation of tunnels, and other ways that increase the
topological complexity of the domain such as creation of new
non-manifold features by folding.

In previous chapters, we were meshing the spacetime domain, a prefix
of $\sp \times [0,\infty)$, for the space domain $\sp$ that did not
change over time.  In this section, our spacetime domain of interest,
$\spt$, is the volume swept through time by the space domain $\sp_t$
at time $t$.  Since different parts of a front $\tau$ may have
different time coordinates, the spatial projection of $\tau$ need not
correspond to the space domain at any time $t$.

\subsection{Tracking boundaries in 1D$\times$Time}

Meshing in 1D$\times$Time presents fewer challenges because the front
at each step is constrained only by causality.  We are free to modify
the front at each step, such as by refinement or by smoothing, as long
as the new front is causal.

\subsubsection*{Front advancing operations}

\emph{Boundary vertices} are those that have velocities prescribed by
the numerical analysis.  We assume that all prescribed velocities are
subsonic. All other vertices are \emph{interior vertices}.  The
velocity of an interior vertex is computed as the average of the
velocity vectors of its neighbors.

Our algorithm in 1D$\times$Time only performs tent pitching
operations, advancing a local minimum vertex at each step.  The major
challenge is prioritizing which vertex to pitch at each step.  Since
boundary vertices have prescribed motion, we have to adapt the
geometry of the rest of the front, especially near the boundary, to
anticipate the necessary change of geometry on the boundary.

As a first approximation, we choose in our new algorithm to prefer
pitching an interior vertex over a boundary vertex.

The goal of our algorithm is to ensure a lower bound on the temporal
aspect ratio of spacetime triangles while simultaneously satisfying a
prescribed global lower bound $\minLen$ on the length of the spatial
projection of all spacetime triangles.

Let $p$ be a local minimum vertex chosen by the algorithm to pitch
next. Let $q$ be a neighboring vertex and let $r$ be the neighbor of
$q$ other than $p$ if it exists.  The tentpole at $p$ is inclined and
its slope with respect to the space domain is equal to the reciprocal
of the speed of $p$.

The height of the tentpole at $p$ is constrained in three different
ways.  Let $h_c$ denote the supremum of the height of the tentpole
$\fp\fp'$ such that $\fp'\fq$ is causal.  Let $h_l$ denote the height
of the tentpole $\fp\fp'$ such that $\fp'\fq\fr$ are collinear in
spacetime.  Let $h_w$ denote the height of the tentpole $\fp\fp'$ such
that the length of the spatial projection of $\fp'\fq$ is equal to
$\minLen$ where $\minLen$ is the minimum allowed length of the
spatial projection of any segment of any front.

The algorithm chooses the final height of the tentpole depending on
the following cases.  The algorithm always satisfies the causality
constraint; hence, the final height $h$ of the tentpole is always less
than $h_c$.

Let $\delta$ be a positive quantity, $0 < \delta < 1$, large enough to
avoid numerical difficulties.  Hence, $(1-\delta) h_c < h_c$.

Note that $h_l \le h_c$ always because $h_l > h_c$ contradicts the
causality of $\fq\fr$.

First, consider the situation $h_w \le (1-\delta) h_c$.

\parasc*{Case 1}
Consider first the case $h_l, h_w \le (1-\delta) h_c < h_c$.

\begin{enumerate}

\item[1.1.]
\textbf{Case $h_l \le h_w \le (1-\delta) h_c$:} Choose $h := h_w$. If the
resulting patch is rejected, then bisect $pq$.

\item[1.2.]
\textbf{Case $h_w \le h_l \le (1-\delta) h_c$:} We have no choice
but to choose $h := h_w$.  If the resulting patch is rejected, then
bisect $pq$.

\end{enumerate}

\parasc*{Case 2}
Next, consider the case $h_w \le (1-\delta) h_c < h_l \le h_c$.

\begin{enumerate}

\item[2.1.]
\textbf{Case $h_w \le (1-\delta)h_c  < h_l$:} We have no choice
but to choose $h := h_w$.  If the resulting patch is rejected, then
choose $h := \half h_c$ if $\half h_c \le h_w$, otherwise bisect $pq$.

\end{enumerate}

Next, consider the situation $h_w > (1-\delta) h_c$.

\parasc*{Case 3}
Thirdly, consider the case $h_l \le (1-\delta) h_c < h_w$.  We do not
distinguish between the two subcases (i)~$h_w \le h_c$, and (ii)~$h_c
\le h_w$.

\begin{enumerate}

\item[3.1.]
\textbf{Case $h_l \le (1-\delta)h_c  < h_w$:} Choose $h := h_c$
if $h_w - h_c$ is sufficiently large, otherwise choose $h := \half
h_c$.

\end{enumerate}

\parasc*{Case 4}
Finally, consider the case $(1-\delta) h_c < h_w, h_l$.

\begin{enumerate}

\item[4.1.]
\textbf{Case $(1-\delta)h_c  < h_l \le h_w$:} 
We do not distinguish between the two subcases (i)~$h_l \le h_w \le
h_c$, and (ii)~$h_l \le h_c \le h_w$.

We have no choice but to choose $h := (1-\delta) h_c$; however, if
$h_w - h_c$ is too small, then choose $h := \half h_c$.

\item[4.2.]
\textbf{Case $(1-\delta)h_c  < h_w < h_l$:} 
We do not distinguish between the two subcases (i)~$h_w < h_l \le
h_c$, and (ii)~$h_w \le h_c \le h_l$.

We have no choice but to choose $h := (1-\delta) h_c$; however, if
$h_w - h_c$ is too small, then choose $h := \half h_c$.

\end{enumerate}

\subsubsection*{Proofs}

We would like to prove that the algorithm makes positive progress at
each step while still guaranteeing causality and a lower bound of
$\minLen$ spatial projection length.  In cases~3 and~4, the height of
the tentpole is at least a constant fraction of $h_c$.  We claim that
in any sequence of tent pitching steps, cases~1 and~2 apply only a
finite number of times before either case~3 or case~4 applies in the
next step.  Clearly, causality is always maintained because the final
height $h$ always satisfies $h \le (1-\delta) h_c < h_c$.  It is easy
to see that all four cases together ensure that the spatial projection
of every segment on each front has length at least $\minLen$.  Hence,
all requirements are satisfied.

Note that cases~1 and~2 deal with the situation $h_w \le (1-\delta)
h_c$.  If the resulting patch is rejected, then the segment $pq$ is
bisected which means that the causality constraint $h'_c$ that applies
in the next step is smaller than $h_c$ by a finite
amount. Specifically, $h'_c - h_c = \half \len(pq) \S$.  On the other
hand, the constraint $h_w$ also decreases to $h'_w$ where $h'_w - h_w
= \minLen \S$.  Therefore, if $\minLen < \len(pq)/2$, then the
constraint $h_c$ decreases faster than the constraint $h_w$ with
repeated applications of either case~1 or case~2.  Note that
$\len(pq)$ also halves with each iteration.  Therefore, that after a
bounded number of applications of cases~1 or~2, it must be the case
that either case~3 or case~4 applies.

Thus, we can show that the front advances after a finite number of
steps; however, we haven't shown yet that the front achieves a target
time $T$ in a finite number of steps, unless we also show that the
front is refined only a finite number of times before being coarsened
because $h := h_l$ is chosen at some step.

\subsubsection*{Coarsening}

Let $\fp$ be an arbitrary local minimum of a causal front $\tau$.  Let
$\fq$ and $\fr$ be neighbors of $\fp$.  We claim that in the next step
in which $\fp$ is pitched to create the new causal front $\tau'$, it
is possible to coarsen and delete at least one of the segments
$\fp\fq$ or $\fp\fr$ or both, without creating degenerate spacetime
triangles.

Without loss of generality, assume $\tau(r) \ge \tau(q)$.  First,
consider the case that $\tau(r) \ge \tau(p) + \Delta$ for some
sufficiently large $\Delta > 0$.  Then, in the next step, $\fp$ can be
pitched to $\fp'$ such that $\fq-\fp'-\fr$ are collinear in spacetime
and $\tau'(p) \ge \tau(p) + \dt$ where $\dt \ge
\frac{\len(pq)}{\len(qr)}\,\Delta$.

This suggests that we need to bound the ratio of the spatial lengths
of any two adjacent segments on the front.  For instance, if
$\frac{\len(pq)}{\len(pr)} > \frac{2}{3}$, then bisect $pq$.
Equivalently, if we start with an initial front that satisfies this
balance condition then the balance can be maintained by propagating
refinement of a segment to neighboring segments.

Next, consider the remaining case that $\tau(q), \tau(r) \approxeq
\tau(p)$.  Then, $\fp$ can be pitched to $\fp'$ such that either
$\fp'\fq$ or $\fp'\fr$ is tight against the causality constraint,
i.e., has slope equal to $(1-\e) \S$.  This reduces the number of
non-strict local minima.  Specifically, it reduces the number of local
minima for which both the incident edges are not tight against the
causality constraint.  By induction on the number of such minima,
it follows that eventually there will be a local minimum vertex such
that at least one of its incident edges has slope equal to $(1-\e)
\S$.  (Such a local minimum may be a boundary vertex and therefore
also a strict local minimum.)  In this situation, the first case above
will apply and it will result in coarsening.

\parasc*{Guaranteed coarsening}

In Chapter~\ref{sec:targettime} we give an algorithm that will
guarantee coarsening while still maintaining lower bounds on the
minimum tentpole height.  Coarsening a portion of the front takes some
time before that portion can be made coplanar and coarsenable.  As
long as the motion of the boundary permits the additional time taken
to coarsen reliably, for instance if the boundary moves very slowly,
we can guarantee that the segments near the front will also satisfy
the lower bound on the size of their spatial projections.

\subsection{Tracking boundaries in 2D$\times$Time}

In this section, we describe an algorithm in 2D$\times$Time that uses
the following advancing front operations to track moving domain
boundaries---tent pitching, smoothing, edge bisection, vertex
deletion, and edge flips.  Our algorithm does not create inverted
elements (i.e., tetrahedra with negative volume).  We cannot yet
guarantee that the algorithm will always make progress by creating
non-degenerate elements.  In the future, we could use a similar
algorithm for tracking shock fronts.  Each operation changes the
triangulation of the front.  Most operations also advance the front
and the solution.

Meshing in 2D$\times$Time to track spacetime features is challenging
because of the need to satisfy not only causality but also progress
constraints at each step.  The general framework of our algorithm in
2D$\times$Time is to prefer operations that increase the spatial
quality of front triangles without introducing refinement whenever
possible.  We devised a policy to assign priorities to various front
advancing operations with the aim of proving correctness as done for
Delaunay refinement
algorithms~\cite{edelsbrunner01geometry,cheng03quality,cheng03graded}.

First, we outline the various meshing operations, in addition to the
usual tent pitching operation which advances a single vertex in time.

\parasc*{Pitching inclined tentpoles}
We advance $\fp$ to $\fp'$ along the given tentpole direction in
spacetime, creating a tentpole $\fp\fp'$.  The tentpole direction is
the one computed by the solver; when the boundary is moving the
tentpole is inclined with respect to the vertical time axis.  The
amount of progress, i.e., the length of the inclined tentpole, is
constrained to prevent inverted elements and is limited by
causality. See Figure~\ref{fig:inclinedtentpole}.

\parasc*{Smoothing}

The meshing algorithm can freely choose the tentpole direction for
other vertices such as those created by refinement in the interior.
We choose the tentpole $\fp\fp'$ to smooth the
front and improve the spatial aspect ratio of front
triangles. Alternatively, we choose the tentpole $\fp\fp'$ to align
some implicit facets with a spacetime surface.

\parasc*{Edge bisection}

Edge bisection is the same operation as the newest vertex refinement
described in Chapter~\ref{sec:adaptive}.  Bisecting triangles on the
current front decreases the size of future spacetime elements.  As
long as the vertex incident on the largest angle is chosen as the apex
of the triangle, newest vertex refinement also decreases the largest
angle which increases the likely progress in the next step.

\parasc*{Vertex deletion}

The front is coarsened by deleting a vertex when permitted by the
numerical error estimate.  Vertex deletion is achieved by inserting a
patch of one or more tetrahedra and simultaneously advancing the
front.

\parasc*{Edge flips}

Flipping an edge of the front triangulation can increase the aspect
ratio of front triangles and improve the quality of future spacetime
tetrahedra.  An edge flip of the edge $pr$, a diagonal of the
convex quadrilateral $pqrs$, is achieved by inserting a single
tetrahedron $\fp\fq\fr\fs$ as long as the tetrahedron has positive
volume and sufficient quality.  See Figure~\ref{fig:edgeflip}.

\subsubsection*{A framework in 2D$\times$Time}

Choose the applicable operation that is first in the sequence listed
below.  If a patch created by one of the operations is rejected, then
proceed to the next operation in the sequence to try to create an
acceptable patch.  If there are several candidates for where to
perform a given operation, choose a candidate that makes it most
likely that in the next step an operation earlier in the sequence will
become possible.  If there is more than one candidate operation, then
perform these independent operations simultaneously in parallel
whenever possible.

\begin{enumerate}
\item Coarsen the front by deleting a vertex $u$

\item Flip an edge of the front if it improves the spatial aspect
ratio

\item Pitch an interior vertex to smooth local triangulation

\item Pitch a boundary vertex in prescribed direction

\end{enumerate}

In general, coarsening operations, or operations that decrease the
number of triangles on the front, are preferred over those that refine
the front because coarsening is harder to achieve in practice and to
guarantee in theory.  The front will still be at least as refined as
required by the numerical error indicator.

Additional heuristics include preferring to pitch local minima in the
interior of the domain over those on a moving boundary.  For
an interior vertex, we choose a tentpole direction that is the average
of the tentpole directions computed for its neighbors.
If an obtuse angle gets too big, we flip or bisect the opposite edge.
We bisect an edge if its endpoints have very different velocities; we choose
an average velocity for the new midpoint.

Each operation is performed only when allowed by the DG solver.  We
adapt the mesh in response to numerical error estimates as before; in
addition, we allow the algorithm to refine and coarsen the front in
response to changing local feature size.

We speculate that the above policy for choosing the various front
advancing operations and prioritizing among them is an acceptable
first approximation for very simple classes of motion.  However, a
more sophisticated policy needs to be devised.  Note that the
algorithm incurs the additional complexity of choosing which operation
to perform at each step.  Also, the more complicated the policy, the
harder it is to parallelize the meshing algorithm.

We described a framework using a wider vocabulary of advancing front
operations to mesh in spacetime.  Even without motion, we could use
our algorithm to mesh static curved features adaptively.  Theorems are
lacking, except in very restricted cases, such as 1D$\times$Time, and
uniform translation of the entire domain.  It remains a very
interesting and challenging open problem to devise a provably correct
algorithm even for a restricted but still useful class of motion.


\section{Practical issues}
\label{sec:practical}

There are several choices to be made in practice when implementing our
meshing algorithms.

\parasc*{Choice of parameter $\mathbf{\e}$}

The value of the parameter $\e$ in the range $0 < \e < 1$ controls how
greedy the algorithm is. Values of $\e$ close to $1$ strengthen the
progress constraint and therefore potentially limit the height of
tentpoles.  Choosing a very small values of $\e$ weakens the progress
constraint.  It is reasonable to expect that in practice smaller values
of $\e$ mean that most tentpole heights are limited by causality and
therefore the average tentpole height is larger.  However, the
\emph{theoretical guarantee} on tentpole height due to the causality
constraint is proportional to $\e$.  It is therefore important to
choose an appropriate value of $\e$, perhaps specific to each problem
instance, that achieves both adequate minimum tentpole height while
maximizing the average tentpole height.  A value of $\e$ around the
middle of its allowed range, i.e., $\e \approxeq \half$, is likely to
be a good choice in practice.

\parasc*{Choice of lookahead parameters}

The horizon parameter $h$ and the adaptive lookahead parameter $l$
(Chapter~\ref{sec:biadaptive}) can be fixed \apriori{} or can be
increased adaptively if greater progress is desired.  Arguably, the
algorithm for general $(h,l)$ is complicated to implement. In
practice, depending on the underlying physics, it may suffice to fix
the parameters at a small value, say $h=l=1$.

\parasc*{Choice of heuristic}

Tent Pitcher is not restricted to pitching only local minima, even
though the theoretical progress guarantee applies only to the minimum
progress made by advancing an arbitrary local minimum.  To increase
the degree of parallelism of the algorithm, it is beneficial to pitch
any vertex where some positive progress is guaranteed even though it
might not be the choice that guarantees the most progress at the
current step.

Preliminary experiments suggest that the quality and overall
efficiency of the solution strategy depend on the choice of which
local minimum vertex to advance at each step, whenever more than one
candidate is available or whenever other operations are also
available.  All algorithms so far leave this as a free choice to be
decided heuristically.  We strongly recommend making empirical studies
of how to exploit different heuristics to generate better quality and
more efficient meshes.

\parasc*{Data structures for the front}

A significant advantage of the advancing front approach is that we
need to store only the front and a collection of unsolved patches.
The front is $d$-dimensional, the same as the space domain.  The
unsolved patches are transient and can be discarded as soon as they
are solved and accepted.  We recommend separate representations of
solved elements and unsolved patches because this would allow the
solver to be oblivious of how the inflow information for a unsolved
patch was computed---whether by a previous iteration of the same
algorithm, by initial and boundary conditions, or by some entirely
different solution procedure.  Standard data
structures~\cite{weiler85halfedge} are available to represent the front in
2D$\times$Time.  The adjacency information between facets of the front
and facets of the spacetime elements stores either a many-one or a
one-many relation, because of the weak simplicial complex property,
which can be represented by pointers or arrays.  Thus, our algorithms
truly require only very basic data structures capable of representing
a weakly conforming unstructured simplicial mesh.

\parasc*{Maintaining a triangulation}

When we refine one triangle on a 2D front, we may be forced to refine
other nearby triangles in order to maintain a conforming triangulation
of the front.  See Figure~\ref{fig:propagation}.  The propagation path
touches every triangle in the worst case, but in practice, the
propagation path usually has small constant length and does not
involve loops.  This is especially the case if the largest angle in
each triangle is used to choose the marked vertex.

The algorithm need not wait for the entire propagation path to be
executed before proceeding with the next advancing step.  In fact,
only the initial bisection was required by the numerical error
indicator; all further bisections are merely artifacts of the
algorithm because the algorithm requires the front to be conforming.
Therefore, we employ a lazy propagation scheme that delays propagating
refinement to adjacent triangles unless absolutely necessary.  This
also reduces the need to synchronize and communicate across processors
in the parallel setting.

Lazy propagation splits the adjacent triangle but does not propagate
further immediately.  Instead, the propagation is delayed until the
adjacent triangle needs to be refined or pitched.  In the interim, the
triangulation may consist of transient triangles.  Eventually, the
final result is exactly the same as if the entire propagation path was
executed in one step.

\textbf{Coarsening:}
A coarsening step that nullifies the last refinement in a refinement
propagation path usually does so by deleting a degree-$4$ vertex.  An
exception to this rule is shown in Figure~\ref{fig:propagation}, where
the last refinement cannot be nullified by deleting a degree-$4$
vertex because all vertices have degree greater than $4$.  However, a
simple edge flip of one of the edges involved in the penultimate
bisection produces a degree-$4$ vertex that can be deleted to remove
the last refinement.  Thus, five triangles are involved in the
coarsening step instead of just four.  Note that this case occurs only
when a refinement propagation contains a loop.  This is not the case
when the largest angle of each triangle is marked and ties are broken
suitably.

\parasc*{Bounding cone hierarchy}

Non-constant wavespeed due to nonlinear response means that the
strictest cone constraint that limits the height of a tentpole can
belong to a point of the front arbitrarily far away.  It is expensive
to examine all cone constraints, one for each triangle of the front,
to determine the most limiting cone.  We adopt standard techniques
used in computational geometry and collision detection in robot motion
planning~\cite{latombe91robot}.  At each step, we wish to solve the
following optimization problem: maximize the height of the tentpole
subject to all cone constraints.  We group all cone constraints into a
hierarchy.  Specifically, we use a bounding cone hierarchy to
efficiently compute a hierarchical approximation of the actual
wavespeed at a point in the future.  In practice, we expect that only
a few constraints in the hierarchy need to be examined on average.

We compute a hierarchical decomposition of the front, say using
METIS~\cite{METIS,karypis98multilevel} as a mesh partitioning tool or
using a $k$d-tree~\cite{deberg00computationalgeometry}.  We build the
cone hierarchy bottom-up, combining pairs of bounding cones by
computing a tight bounding cone for them.  Each node in the hierarchy
(a binary tree) stores a cone tightly containing the cones of its
children.

The cone hierarchy is traversed top-down. Starting at the root, we
maintain a priority queue of cone constraints.  At each step, we query
the strictest constraint from the queue. If the current strictest cone
is a leaf in the hierarchy or if it guarantees sufficient progress
(say a constant fraction of the amount of progress when limited by the
causality constraint alone), then we accept the resulting tentpole
height.  Otherwise, we replace the strictest cone by its children and
repeat.

Refinement and coarsening of the front induce a corresponding
refinement and coarsening of the cone hierarchy.  Since the front
adaptivity in response to the error indicator can be be non-uniform,
the cone hierarchy can get heavily unbalanced.  To maintain
efficiency, the cone hierarchy needs to be periodically rebalanced or
recomputed.

\parasc*{Noncircular bounding cones}

Anisotropic cone constraints, i.e., cones with noncircular cross
sections, complicate the data structures used for the cone hierarchy
but do not introduce substantial algorithmic difficulties.  For
instance, if cones of influence are elliptical, then at each step
while constructing the hierarchy we need to compute an elliptical cone
that contains as tightly as possible the two elliptical cones of its
children.  Likewise, queries of the cone hierarchy are a little
different---each requires checking whether a triangle or $d$-simplex
of the front intersects an elliptical cone.  We are able to exploit
the facet that a cone of any cross-section is a ruled surface to
reduce the dimensionality of the simplex involved in checking for
intersections.

\parasc*{Parallel data structures}

The Tent Pitcher algorithm and all algorithms in this thesis are
inherently parallel---patches created by pitching non-adjacent local
minima of the same front can be solved simultaneously and
independently of each other because patch boundaries are causal.

The front is decomposed into subdomains by a hierarchical partition.
Within each subdomain, a local neighborhood $N$ can be advanced
independently of other subdomains unless $N$ is at the boundary of its
subdomain in which case interprocessor communication is required to
ensure mutual exclusion.  The bounding cone hierarchy within the
subdomain assigned to a single processor is a local sequential data
structure.  We need a parallel tree data structure to query bounding
cones stored on other processors.

Load balancing means repartitioning of the domain.  The bounding cone
hierarchy must also be recomputed.

\parasc*{Lazy updates to reduce interprocessor communication}

At every step, the wavespeed is computed as part of the solution and
the bounding cone hierarchy needs to be updated to reflect this
change.  An increase in the wavespeed somewhere may impact the
progress at a remote location.  Therefore, this change propagates
potentially to every other subdomain, requiring communication and
synchronization between processors.

We can reduce interprocessor communication by performing lazy updates
of the bounding cone hierarchy to reflect changes in the wavespeed.
Because of the no-focusing assumption, current estimates of the
wavespeed are always conservative and therefore acceptable
approximations of future wavespeed.  If sufficient progress is
guaranteed even with these conservative approximations, then there is
no immediate need to compute the new wavespeed.  Therefore, we can
queue the updates of the cone hierarchy and perform the actual update
later.


\subsection*{Chapter summary}

We have been able to use the new operations in our arsenal, such as
pitching inclined tentpoles and performing edge flips, to give
advancing front meshing algorithms in 1D- and 2D$\times$Time.
Extending the heuristics of this chapter to higher dimensions, say
3D$\times$Time would be very interesting.  These newer algorithms
remain heuristics in the sense that we are not able to prove yet that
we will be able to successfully mesh the spacetime domain of interest
with nondegenerate elements for a general class of motion of the
boundary or for tracking a fairly diverse class of spacetime features.
For instance, it remains an open question whether we can align causal
and implicit facets of the mesh with an arbitrary set of piecewise
linear constraints in spacetime.  The main challenge remains to prove
nontrivial worst-case lower bounds on the temporal aspect ratio of
spacetime elements under these additional constraints.


\chapter{Conclusion and open problems}
\label{sec:conclusions}

We extended Tent Pitcher, the advancing front algorithm due to
\Ungor{} and Sheffer~\cite{ungor00tentpitcher} and Erickson
\etal{}~\cite{erickson02building}, to adaptively compute efficient,
near-size-optimal spacetime meshes suitable for DG solutions that
adapt to \aposteriori{} spacetime error indicators as well as to
nonlinear and anisotropic physics.  Our primary motivation was to
prove that such an algorithm was at all possible with provable
theoretical guarantees.  In addition to our theoretical results, an
important contribution of our work is that the algorithms are
currently being fruitfully implemented and tested on problems of
practical complexity.  We were able to unify earlier results that
considered nonlocal cone constraints~\cite{thite04nonlocal} and mesh
adaptivity~\cite{abedi04spacetime} separately.  Our extensions to Tent
Pitcher retain the benefits of earlier algorithms such as ease of
implementation and inherent high degree of parallelism.  Additionally,
we aim to solve robustness issues and generate even better quality
meshes in practice than those guaranteed in theory.

Several theoretical problems remain open and are actively being
researched by this author and others.  We discuss them in the next and
final section.


\section{Open problems}

There are many exciting avenues to explore and many more problems are
likely to be discovered as our algorithm are used to solve
increasingly realistic problems.  In this section, we highlight some
promising directions for future research.


\parasc*{Adaptivity in arbitrary dimensions}

An important open problem is extending adaptive refinement and
coarsening to spatial dimensions $d \ge 3$.  Extensions of
newest-vertex bisection to higher dimensions are
known~\cite{arnold00locally,bansch91local}.  However, it is
challenging to devise necessary and sufficient progress constraints
that guarantee positive progress.  It is also
important that the constraints be easy to implement.  We consider the
problem of incorporating adaptive refinement and coarsening in
3D$\times$Time into our meshing algorithm to be the most interesting
and practically relevant open problem.


\parasc*{Element quality}

Consider another measure of element quality: the ratio
of inradius to circumradius.  The circumradius to inradius ratio of
an element is defined after scaling the time axis by the wavespeed.
A larger inradius:circumradius ratio means a ``better'' element.
Experiments suggest that the inradius to circumradius ratio is a
useful measure of element quality.  Our algorithms guarantee a lower
bound on the worst-case ratio as a result of causality and progress
constraints, but this bound is not very good.  At the same time, some
elements with small inradius to circumradius ratio also seem to have
small temporal aspect ratio.  It is an important research question to
devise an algorithm to explicitly maximize the smallest inradius to
circumradius ratio.

Scale the time axis relative to space by the local wavespeed.  Now,
the spacetime domain, at least locally within each patch is identical
to $(d+1)$-dimensional Euclidean space.  After this scaling, define
the \emph{quality} of a spacetime element as the ratio of its inradius
to its circumradius---a larger ratio means a better quality element.

It is clear that maximizing the tentpole height is not likely to
maximize the worst-case quality of spacetime elements.  The following
is an attempt to describe another choice of location for the top of
the tentpole so as to guarantee element quality.

Let $\tau$ be a front.  Let $\fp$ be a local minimum of $\tau$.  Let
$\fp\fq\fr$ be an arbitrary triangle incident on $\fp$.  Let $\rho$
denote the circumradius of $\triangle{\fp\fq\fr}$.  Any tetrahedron
with $\fp\fq\fr$ as a facet must have circumradius at least $\rho$.
Attempt to create a spacetime tetrahedron over $\fp\fq\fr$ with
circumradius not much larger than $\rho$ but with a guaranteed lower
bound on its inradius.

For each triangle $\fp\fq\fr$, consider its diametral circumball but
with radius scaled by $(1+\delta)$ for some $\delta \ge 0$. We wish to
place the top $\fp'$ of the tentpole inside the intersection of all
such circumballs so as not to increase the circumradii of resulting
tetrahedra by more than $(1+\delta)$.

In addition, we wish to ensure that the volume of each tetrahedron is
bounded from below.  We do this by ensuring that the distance of
$\fp'$ from the plane of $\triangle{\fp\fq\fr}$ is bounded from below.
Thus, we place $\fp'$ above the plane at an orthogonal distance of
$\gamma$ times the width of $\triangle{\fp\fq\fr}$.

Overall, we need to ensure that the intersection of all the
circumballs and all halfspaces is nonempty so that we have at least
one candidate placement of $\fp'$, the top of the tentpole at $\fp$.
Note that the final tentpole need not be vertical.


\parasc*{Empirical study of heuristics}

Even for the basic linear nonadaptive algorithm of
Chapter~\ref{sec:linear}, an experimental study is needed to try
several heuristics for which portion of the front to advance in each
step, i.e., which local minimum vertex to pitch next when several
candidates are available.  Different heuristics have been observed to
affect the mesh quality differently in 1D$\times$Time.  With the more
general advancing front operations of Chapter~\ref{sec:extensions}, it
is important to prioritize various operations that modify the front to
ensure that the quality and efficiency of the mesh is improved in
practice.


\parasc*{Provably correct boundary tracking}

We would like to devise a provably correct and complete boundary
tracking algorithm for interesting classes of motion.  In many
practical situations, the topology of the domain also changes with
time~\cite{bajaj99tetrahedral}.  For instance, a rocket fuel grain may
split into multiple components due to combustion.  We would be
interested in an advancing front algorithm, initially in
2D$\times$Time, that handles changes in the topology of the domain
over time.  In higher dimensions, a wider array of front operations is
possible and necessary for boundary tracking.  For instance, edge
flips in 2D$\times$Time generalize to $2--3$ flips that replace two
adjacent tetrahedra by three tetrahedra sharing an edge, and vice
versa.  The problem of devising a boundary tracking algorithm in
arbitrary dimensions remains open and of a great deal of interest even
in 3D$\times$Time.


\parasc*{Non-greedy algorithms}

It should be clear that the progress constraints can be
modified in different ways to meet different mesh generation
objectives.  Research so far has focused on worst-case guarantees for
temporal aspect ratio.  Much more research is needed to guarantee
worst-case quality in the sense of inradius to circumradius ratio.
Even more desirable is to guarantee a certain distribution of elements
by quality measure.  We believe that greedy algorithms will
not suffice to guarantee worst-case inradius to circumradius ratio
in theory and will not perform well, without additional heuristics, in
practice.

Another context in which greedy algorithms are unlikely to fare better
than current work is in the case of tracking moving boundaries.
Intuition suggests that we need an algorithm that constructs only
``robustly progressive'' fronts.  We say that a front $\tau$ is
\emph{robustly progressive} if $\tau$ is progressive and if $\tau$ has
a local minimum $p$ such that for every sufficiently small but finite
motion of $p$ with finite velocity the resulting front $\tilde{\tau}$
is also progressive.  For a front $\tau$ to be robustly progressive,
it suffices that $\tau$ has a local minimum $p$ such that (i)~for
every neighbor $q$ of $p$, we have $\tau(q) - \tau(p)$ is bounded away
from zero, and (ii)~the kernel of $\link(p)$ contains a disk of finite
radius centered at $p$. Deriving such an algorithm is an open
question.


\parasc*{Non-simplicial elements}

To perform more general front advancing operations, those that change
the combinatorics of the spatial projection as well as advance the
front in time, it seems necessary to allow more general spacetime
elements in the near future.  Satisfying the two requirements that the
spacetime mesh be a weak simplicial complex and that every element
have at least one causal outflow facet will be harder with the
incorporation of more general operations.  The numerical techniques
can be extended to other linear elements like hexahedra, prisms, and
pyramids with very little difficulty.  The data structures used to
represent a patch (admittedly, only a transient object) would be a
little more complicated because incidences between elements and their
facets, and adjacencies between elements will be a little more
complicated.  I anticipate no significant difficulties in extending
the advancing front framework to create and solve patches of mixed
types of linear elements.


\parasc*{More complicated patches}

Our current algorithms advance a local neighborhood $N$ of the front
$\tau$ to a new neighborhood $N'$ of the new front $\tau'$.  How
complicated can the neighborhoods $N$ and $N'$ be?  Can we triangulate
the spacetime volume between $N$ and $N'$ with a small number of
spacetime elements?  Given the spacetime volume, can we decompose it
into a minimum number of good quality spacetime elements?  There is a
tradeoff between the time to solve a single patch versus the number of
patches in the mesh for a given spacetime volume.  On the one hand, we
would like patches of small complexity so that the time to solve one
patch, even for high polynomial order, is not large.  This is
especially important for an incremental solution strategy where the
physical parameters that govern the solution are changing rapidly.
On the other hand, we would like to minimize the total number of
patches for a given accuracy to reduce the total computation time.

The number of elements in a patch created by pitching a vertex $p$ is
equal to the number of simplices in $\st(p)$.  We are already
considering operations, such as edge dilation and edge contraction,
where the number of elements in a patch could be twice as much.  The
larger is the neighborhood advanced in a single step, the less is the
degree of parallelism of the meshing algorithm.



\backmatter

\bibliographystyle{myalpha}
\bibliography{spacetime,meshing}

\newcommand{\etalchar}[1]{$^{#1}$}
\begin{thebibliography}{dBSvKO00}

\bibitem[ABE99]{amenta99optimal}
A.~B. Amenta, M.~W. Bern, and D.~Eppstein.
\newblock Optimal point placement for mesh smoothing.
\newblock {\em J. Algorithms}, 30(2):302--322, February 1999.

\bibitem[ABG{\etalchar{+}}00]{antaki00parallel}
J.~F. Antaki, G.~E. Blelloch, O.~Ghattas, I.~Malcevic, G.~L. Miller, and N.~J.
  Walkington.
\newblock A parallel dynamic-mesh lagrangian method for simulation of flows
  with dynamic interfaces.
\newblock In {\em Proc. Supercomputing (SC2000)}, 2000.

\bibitem[ACE{\etalchar{+}}04]{abedi04spacetime}
R.~Abedi, S.-H. Chung, J.~Erickson, Y.~Fan, M.~Garland, D.~Guoy, R.~Haber,
  J.~M. Sullivan, S.~Thite, and Y.~Zhou.
\newblock Spacetime meshing with adaptive refinement and coarsening.
\newblock In {\em Proc. 20th Symp. Computational Geometry}, pages 300--309,
  June 2004.

\bibitem[ACF{\etalchar{+}}04]{abedi04adaptiveDG}
R.~Abedi, S.-H. Chung, Y.~Fan, S.~Thite, J.~Erickson, and R.~B. Haber.
\newblock Adaptive discontinuous galerkin method for elastodynamics on
  unstructured spacetime grids.
\newblock In {\em Proc. XXI International Congress of Theoretical and Applied
  Mechanics (ICTAM)}, August 2004.
\newblock To appear.

\bibitem[AGR00]{amato00lineartime}
N.~M. Amato, M.~T. Goodrich, and E.~A. Ramos.
\newblock Linear-time triangulation of a simple polygon made easier via
  randomization.
\newblock In {\em Proc. 16th Symp. Computational Geometry}, pages 201--212,
  2000.

\bibitem[AHP05]{abedi05spacetime}
R.~Abedi, R.~B. Haber, and B.~Petracovici.
\newblock A spacetime discontinuous galerkin method for elastodynamics with
  element-level balance of linear momentum.
\newblock {\em Computer Methods in Applied Mechanics and Engineering}, 2005.
\newblock In press.

\bibitem[AHTE05]{abedi05adaptive}
R.~Abedi, R.~Haber, S.~Thite, and J.~Erickson.
\newblock An $h$-adaptive spacetime-discontinuous galerkin method for
  linearized elastodynamics.
\newblock {\em Revue Europ\'eenne des \'El\'ements Finis}, 2005.
\newblock Submitted.

\bibitem[AMP00]{arnold00locally}
D.~N. Arnold, A.~Mukherjee, and L.~Pouly.
\newblock Locally adaptive tetrahedral meshes using bisection.
\newblock {\em SIAM J. Sci. Comput.}, 22(2):431--448, 2000.

\bibitem[B\"91]{bansch91local}
E.~B\"ansch.
\newblock Local mesh refinement in $2$ and $3$ dimensions.
\newblock {\em Impact of Computing in Science and Engineering}, 3:181--191,
  1991.

\bibitem[B{\"a}n91]{b-lmr23-91}
E.~B{\"a}nsch.
\newblock Local mesh refinement in 2 and 3 dimensions.
\newblock {\em Impact of Computing in Science and Engineering}, 3:181--191,
  1991.

\bibitem[BEG94]{bern94provablygood}
M.~Bern, D.~Eppstein, and J.~Gilbert.
\newblock Provably good mesh generation.
\newblock {\em J. Comp. Sys. Sci.}, 48:384--409, 1994.
\newblock Special issue for FOCS'90.

\bibitem[Bey95]{b-tgr-95}
J.~Bey.
\newblock Tetrahedral grid refinement.
\newblock {\em Computing}, 55(4):355--378, 1995.

\bibitem[Bey00]{b-sgrfa-00}
J.~Bey.
\newblock Simplicial grid refinement: {On} {Freudenthal's} algorithm and the
  optimal number of congruence classes.
\newblock {\em Numer. Math.}, 85(1):1--29, 2000.

\bibitem[BH86]{barnes-hut86nbody}
J.~E. Barnes and P.~Hut.
\newblock A hierarchical $o(n log n)$ force calculation algorithm.
\newblock {\em Nature}, 324(4):446--449, December 1986.

\bibitem[BSW83]{bsw-radsr-83}
R.~E. Bank, A.~H. Sherman, and H.~Weiser.
\newblock Refinement algorithm and data structures for regular local mesh
  refinement.
\newblock In R.~Stepleman et~al., editors, {\em Scientific Computing}, pages
  3--17. IMACS/North-Holland, Amsterdam, 1983.

\bibitem[CD03]{cheng03quality}
S.-W. Cheng and T.~K. Dey.
\newblock Quality meshing with weighted {D}elaunay refinement.
\newblock {\em SIAM J. Computing}, 33:69--93, 2003.

\bibitem[CDE{\etalchar{+}}00]{cheng00sliver}
S.-W. Cheng, T.~K. Dey, H.~Edelsbrunner, M.~A. Facello, and S.-H. Teng.
\newblock Sliver exudation.
\newblock {\em J. ACM}, 47:883--904, 2000.

\bibitem[Cen03]{chalmers03fem}
C.~F.~E. Center.
\newblock Online at \url{http://www.md.chalmers.se/Centres/Phi/}, 2003.

\bibitem[Cha91]{chazelle91triangulating}
B.~Chazelle.
\newblock Triangulating a simple polygon in linear time.
\newblock {\em Discrete and Computational Geometry}, 6(5):485--524, 1991.

\bibitem[Che89]{chew89guaranteed}
L.~P. Chew.
\newblock Guaranteed-quality triangular meshes.
\newblock Technical Report TR-89-983, Department of Computer Science, Cornell
  University, 1989.

\bibitem[CKS00]{cockburn00discontinuous}
B.~Cockburn, G.~Karniadakis, and C.~Shu.
\newblock {\em Discontinuous {Galerkin} methods: theory, computation and
  applications}, volume~11 of {\em Lecture Notes in Computational Science and
  Engineering}.
\newblock Springer, 2000.

\bibitem[Coo98]{cooper98introduction}
J.~Cooper.
\newblock {\em Introduction to Partial Differential Equations with {MATLAB}}.
\newblock Birkhauser, 1998.

\bibitem[CP03]{cheng03graded}
S.-W. Cheng and S.-H. Poon.
\newblock Graded conforming {D}elaunay tetrahedralization with bounded
  radius-edge ratio.
\newblock In {\em SODA}, pages 295--304, 2003.
\newblock Version dated July~9, 2002.

\bibitem[CRMS03]{cignoni03externalmemory}
P.~Cignoni, C.~Rocchini, C.~Montani, and R.~Scopigno.
\newblock External memory management and simplification of huge meshes.
\newblock {\em IEEE Trans. on Visualization and Computer Graphics},
  9(4):525--537, 2003.

\bibitem[dBSvKO00]{deberg00computationalgeometry}
M.~de~Berg, O.~Schwarzkopf, M.~van Kreveld, and M.~Overmars.
\newblock {\em Computational Geometry: Algorithms and Applications}.
\newblock Springer-Verlag, 2nd ed. edition, 2000.

\bibitem[Ede01]{edelsbrunner01geometry}
H.~Edelsbrunner.
\newblock {\em Geometry and Topology for Mesh Generation}.
\newblock Cambridge Monographs on Applied and Computational Mathematics.
  Cambridge University Press, 2001.

\bibitem[EG00]{eg-ess-00}
H.~Edelsbrunner and D.~R. Grayson.
\newblock Edgewise subdivision of a simplex.
\newblock {\em Discrete Comput. Geom.}, 24(4):707--719, 2000.

\bibitem[EG01]{edelsbrunner01sink}
H.~Edelsbrunner and D.~Guoy.
\newblock Sink-insertion for mesh improvement.
\newblock In {\em Proc. 17th Annual Symposium on Computational Geometry}, pages
  115--123, 2001.

\bibitem[EGS{\"U}02]{erickson02building}
J.~Erickson, D.~Guoy, J.~M. Sullivan, and A.~{\"U}ng{\"o}r.
\newblock Building space-time meshes over arbitrary spatial domains.
\newblock In {\em Proc. 11th Int'l. Meshing Roundtable}, pages 391--402, 2002.

\bibitem[EGS{\"U}05]{erickson05building}
J.~Erickson, D.~Guoy, J.~M. Sullivan, and A.~{\"U}ng{\"o}r.
\newblock Building spacetime meshes over arbitrary spatial domains.
\newblock {\em Engineering with Computers}, 20:342--353, 2005.
\newblock Online at \url{http://dx.doi.org/10.1007/s00366-005-0303-0}.

\bibitem[EJL03]{eriksson03timestepping}
K.~Eriksson, C.~Johnson, and A.~Logg.
\newblock Explicit time-stepping for stiff {ODEs}.
\newblock {\em SIAM Journal on Scientific Computing}, 25(4):1142--1157, 2003.

\bibitem[Fol95]{folland95introduction}
G.~B. Folland.
\newblock {\em Introduction to Partial Differential Equations}.
\newblock Princeton University Press, 2nd ed edition, 1995.

\bibitem[fPCS99]{bajaj99tetrahedral}
T.~M. from Planar Cross~Sections.
\newblock Chandrajit bajaj and edward j. coyle and kwun-nan lin.
\newblock {\em Computer Methods in Applied Mechanics and Engineering},
  179:31--52, 1999.

\bibitem[GJPT78]{garey78triangulating}
M.~R. Garey, D.~S. Johnson, F.~P. Preparata, and R.~E. Tarjan.
\newblock Triangulating a simple polygon.
\newblock {\em Information Processing Letters}, 7(4):175--179, June 1978.

\bibitem[Gut84]{guttman84rtrees}
A.~Guttman.
\newblock A dynamic index structure for spatial searching.
\newblock In {\em Proc. ACM SIGMOD Conf. Principles Database Systems}, pages
  47--57, 1984.

\bibitem[Hea02]{heath02scientificcomputing}
M.~T. Heath.
\newblock {\em Scientific Computing: An Introductory Survey}.
\newblock McGraw-Hill, New York, 2d edition, 2002.

\bibitem[Her04]{mapletools}
J.~Herod.
\newblock Partial differential equations.
\newblock Online at \url{http://www.math.gatech.edu/~herod/conted/Async.html},
  2004.
\newblock Accompanying notes at
  \url{http://www.mapleapps.com/powertools/pdes/pdes.shtm}.

\bibitem[JJ04]{jegdic04convergence}
K.~Jegdic and R.~L. Jerrard.
\newblock Convergence of a spacetime discontinuous {Galerkin} finite element
  method for a class of hyperbolic systems.
\newblock Preprint, August 2004.

\bibitem[JP97]{jones97adaptive}
M.~T. Jones and P.~E. Plassmann.
\newblock Adaptive refinement of unstructured finite-element meshes.
\newblock {\em Finite Elements in Analysis and Design}, 25:41--60, 1997.

\bibitem[Kar]{METIS}
G.~Karypis.
\newblock {METIS}: Family of multilevel partitioning algorithms.
\newblock Online at \url{http://www-users.cs.umn.edu/~karypis/metis/}.

\bibitem[KK98]{karypis98multilevel}
G.~Karypis and V.~Kumar.
\newblock Multilevel algorithms for multi-constraint graph partitioning.
\newblock Technical Report 98-019, Department of Computer Science, University
  of Minnesota, May 1998.

\bibitem[Lat91]{latombe91robot}
J.-C. Latombe.
\newblock {\em Robot Motion Planning}.
\newblock Kluwer Academic Publishers, 1991.

\bibitem[LJ94]{lj-stb-94}
A.~Liu and B.~Joe.
\newblock On the shape of tetrahedra from bisection.
\newblock {\em Math. Comp.}, 63(207):141--154, 1994.

\bibitem[LJ95]{lj-qlrtm-95}
A.~Liu and B.~Joe.
\newblock Quality local refinement of tetrahedral meshes based on bisection.
\newblock {\em SIAM J. Sci. Comput.}, 16:1269--1291, 1995.

\bibitem[LMCG96]{lin96collision}
M.~Lin, D.~Manocha, J.~Cohen, and S.~Gottschalk.
\newblock {\em Algorithms for Robotics Motion and Manipulation}, chapter
  Collision Detection: Algorithms and Applications, pages 129--142.
\newblock A.K. Peters, 1996.

\bibitem[LS03]{labelle03anisotropic}
F.~Labelle and J.~R. Shewchuk.
\newblock Anisotropic {V}oronoi diagrams and guaranteed-quality anisotropic
  mesh generation.
\newblock In {\em Proc. 19th ACM Symp. Computational Geometry}, pages 191--200,
  2003.

\bibitem[Mau95]{maubach95localbisection}
J.~M.~L. Maubach.
\newblock Local bisection refinement for $n$-simplicial grids generated by
  reflection.
\newblock {\em SIAM J. Sci. Comput.}, 16:210--227, 1995.

\bibitem[Mau96]{maubach96efficient}
J.~M.~L. Maubach.
\newblock The efficient location of neighbors of locally refined $n$-simplicial
  grids.
\newblock In {\em Proc. 5th International Meshing Roundtable}, pages 137--156,
  October 1996.

\bibitem[Mit88]{m-umafe-88}
W.~F. Mitchell.
\newblock {\em Unified multilevel adaptive finite element methods for elliptic
  problems}.
\newblock {Ph. D.} thesis, Computer Science Department, University of Illinois,
  Urbana, IL, 1988.
\newblock Tech. Rep. UIUCDCS-R-88-1436.

\bibitem[Mit89]{m-carte-89}
W.~F. Mitchell.
\newblock A comparison of adaptive refinement techniques for elliptic problems.
\newblock {\em ACM Trans. Math. Soft}, 15:326--347, 1989.

\bibitem[Mit91]{m-arafe-91}
W.~F. Mitchell.
\newblock Adaptive refinement for arbitrary finite-element spaces with
  hierarchical bases.
\newblock {\em J. Comp. Appl. Math.}, 36:65--78, 1991.

\bibitem[MV00]{mitchell00quality}
S.~Mitchell and S.~Vavasis.
\newblock Quality mesh generation in higher dimensions.
\newblock {\em SIAM J. Comput.}, 29:1334--1370, 2000.

\bibitem[Owe99]{owen99phd}
S.~J. Owen.
\newblock {\em Non-Simplicial Unstructured Mesh Generation}.
\newblock PhD thesis, Department of Civil and Environmental Engineering,
  Carnegie Mellon University, April 1999.

\bibitem[Ric94]{Richter94}
G.~R. Richter.
\newblock An explicit finite element method for the wave equation.
\newblock {\em Applied Numerical Mathematics}, 16:65--80, 1994.

\bibitem[Riv84]{rivara84algorithms}
M.-C. Rivara.
\newblock Algorithms for refining triangular grids suitable for adaptive and
  multigrid techniques.
\newblock {\em Int. J. Numer. Meth. Eng.}, 20:745--756, 1984.

\bibitem[Riv96]{rivara96new}
M.~C. Rivara.
\newblock New mathematical tools and techniques for the refinement and/or
  improvement of unstructured triangulations.
\newblock In {\em Proc. 5th Int'l. Meshing Roundtable}, pages 77--86, 1996.

\bibitem[Riv97]{rivara97longestedge}
M.-C. Rivara.
\newblock New longest-edge algorithms for the refinement and/or improvement of
  unstructured triangulations.
\newblock {\em International Journal for Numerical Methods in Engineering},
  40:3313--3324, 1997.

\bibitem[RS75]{rs-lbatc-75}
I.~G. Rosenberg and F.~Stenger.
\newblock A lower bound on the angles of triangles constructed by bisecting the
  longest side.
\newblock {\em Math. Comput.}, 29:390--395, 1975.

\bibitem[RS92]{ruppert92difficulty}
J.~Ruppert and R.~Seidel.
\newblock On the difficulty of triangulating three-dimensional nonconvex
  polyhedra.
\newblock {\em Discrete and Computational Geometry}, 7(3):227--253, 1992.

\bibitem[Rup95]{ruppert95delaunay}
J.~Ruppert.
\newblock A {D}elaunay refinement algorithm for quality 2-dimensional mesh
  generation.
\newblock {\em J. Algorithms}, 18(3):548--585, May 1995.

\bibitem[Sev97]{seveno97towards}
E.~Seveno.
\newblock Towards an adaptive advancing front method.
\newblock In {\em Proc. 6th International Meshing Roundtable}, pages 349--360,
  1997.

\bibitem[Sew72]{s-agtpp-72}
E.~G. Sewell.
\newblock {\em Automatic generation of triangulations for piecewise polynomial
  approximation}.
\newblock {Ph. D.} thesis, Department of Mathematics, Purdue University, West
  Lafayette, IN, 1972.

\bibitem[She02]{shewchuk02whatis}
J.~R. Shewchuk.
\newblock What is a good linear element? interpolation, conditioning, and
  quality measures.
\newblock In {\em Proc. 11th International Meshing Roundtable}, pages 115--126.
  Sandia National Laboratory, September 2002.

\bibitem[SPC03]{spc-pppil-03}
J.~P. Su{\'a}rez, A.~Plaza, and G.~F. Carey.
\newblock Propagation path properties in iterative longest-edge refinemen.
\newblock In {\em Proc. 12th Internat. Meshing Roundtable}, pages 79--90, 2003.

\bibitem[Str86]{strang86appliedmath}
G.~Strang.
\newblock {\em Introduction to Applied Mathematics}.
\newblock Wellesley College, 1986.

\bibitem[ST{\"U}04]{spielman04timecomplexity}
D.~Spielman, S.-H. Teng, and A.~{\"U}ng\"or.
\newblock Time complexity of practical parallel steiner point insertion
  algorithms.
\newblock In {\em Proceedings of ACM-SPAA}, 2004.
\newblock Extended abstract.

\bibitem[Thi04]{thite04nonlocal}
S.~Thite.
\newblock Efficient spacetime meshing with nonlocal cone constraints.
\newblock In {\em Proc. 13th International Meshing Roundtable}, pages 47--58,
  September 2004.

\bibitem[Tho99]{thomas99numerical}
J.~W. Thomas.
\newblock {\em Numerical Partial Differential Equations: Conservation Laws and
  Elliptic Equations}.
\newblock Springer, 1999.

\bibitem[TW98]{tveito98introduction}
A.~Tveito and R.~Winther.
\newblock {\em Introduction to Partial Differential Equations: A Computational
  Approach}.
\newblock Springer-Verlag, 1998.

\bibitem[\"U02]{ungor02phd}
A.~\"Ung\"or.
\newblock {\em Parallel {D}elaunay Refinement and Space-Time Meshing}.
\newblock PhD thesis, University of Illinois at Urbana-Champaign, October 2002.

\bibitem[{\"U}ng04]{ungor04offcenter}
A.~{\"U}ng\"or.
\newblock Off-centers: A new type of steiner points for computing size-optimal
  guaranteed-quality {Delaunay} triangulations.
\newblock In {\em Proceedings of LATIN 2004}, pages 152--161, April 2004.

\bibitem[{\"U}S00]{ungor00tentpitcher}
A.~{\"U}ng{\"o}r and A.~Sheffer.
\newblock {T}ent-{P}itcher: A meshing algorithm for space-time discontinuous
  {Galerkin} methods.
\newblock In {\em Proc. 9th Int'l. Meshing Roundtable}, pages 111--122, 2000.

\bibitem[{\"U}S02]{us-ptstm-02}
A.~{\"U}ng{\"o}r and A.~Sheffer.
\newblock Pitching tents in space-time: {Mesh} generation for discontinuous
  {Galerkin} method.
\newblock {\em Int. J. Foundations of Computer Science}, 13(2):201--221, 2002.

\bibitem[Wei85]{weiler85halfedge}
K.~Weiler.
\newblock Edge-based data structures for solid modeling in curved-surface
  environments.
\newblock {\em IEEE Computer Graphics and Applications}, 5(1):21--40, January
  1985.

\bibitem[Wei99a]{mathworld99galerkin}
E.~W. Weisstein.
\newblock Galerkin method.
\newblock Online at \url{http://mathworld.wolfram.com/GalerkinMethod.html},
  1999.

\bibitem[Wei99b]{mathworld99pde}
E.~W. Weisstein.
\newblock Partial differential equation.
\newblock Online at
  \url{http://mathworld.wolfram.com/PartialDifferentialEquation.html}, 1999.

\bibitem[Whi74]{whitham74linear}
G.~B. Whitham.
\newblock {\em Linear and Nonlinear Waves}.
\newblock Wiley, New York, 1974.

\bibitem[YAS{\etalchar{+}}99]{YinASHT99}
L.~Yin, A.~Acharya, N.~Sobh, R.~B. Haber, and D.~A. Tortorelli.
\newblock A space-time discontinuous {Galerkin} method for elastodynamic
  analysis.
\newblock {\em Proc. of Int. Symp. on Discontinuous Galerkin Methods, Salve
  Regina University, Newport, RI}, 1999.

\bibitem[YAS{\etalchar{+}}00]{YinASHT00}
L.~Yin, A.~Acharya, N.~Sobh, R.~Haber, and D.~A. Tortorelli.
\newblock A space-time discontinuous {Galerkin} method for elastodynamic
  analysis.
\newblock In B.~Cockburn, G.~Karniadakis, and C.~Shu, editors, {\em Lecture
  Notes in Computational Science and Engineering}, volume~11, pages 459--464.
  Springer, 2000.

\bibitem[Yin02]{yin02thesis}
L.~Yin.
\newblock {\em A New Spacetime Discontinuous Galerkin Finite Element Method for
  Elastodynamic Analysis}.
\newblock PhD thesis, Department of Theoretical and Applied Mechanics,
  University of Illinois at Urbana-Champaign, 2002.

\bibitem[YLPM05]{yoon05cacheoblivious}
S.-E. Yoon, P.~Lindstrom, V.~Pascucci, and D.~Manocha.
\newblock Cache-oblivious mesh layouts.
\newblock In {\em Proc. ACM SIGGRAPH}, 2005.
\newblock Available as Lawrence Livermore National Laboratory Technical Report
  UCRL-JRNL-211774.

\bibitem[Zie95]{ziegler95lectures}
G.~M. Ziegler.
\newblock {\em Lectures on Polytopes}.
\newblock Springer, 1995.

\end{thebibliography}

\begin{cv}{\vitaname}


\noindent{\huge\textbf{\textsc{Shripad Thite}}}\\[1ex]
{\large%
Department of Computer Science\\
University of Illinois at Urbana-Champaign
\date{}}\\[1ex]

\bigskip

{%
\settowidth{\cvlabelwidth}{%
  \cvlistheadingfont Computational geometry%
}

\begin{cvlist}{Research interests}

\item[Computational geometry] Mesh generation for finite
element methods; Spacetime meshing; Optimal triangulations; Delaunay
triangulations; Combinatorial problems in geometry; Lower bounds.

\item[Robotics] Capturing and motion planning.

\item[Graph algorithms] Strong edge coloring and matching;
Optimization problems on graphs.

\item[External memory] Data structures on nonuniform
memory models of computation.

\item[Economic simulation] Computer simulations of Vickrey
auctions in deregulated electrical power markets.

\end{cvlist}
}

\begin{cvlist}{Education}

\item[2001]
    M.S., Computer Science\\
    \emph{Advisor:} Prof.\@ Michael Loui\\
    University of Illinois at Urbana-Champaign, IL, USA

\item[1997]
    B.E., Computer Engineering\\
    University of Pune, India (Government College of Engineering, Pune)

\end{cvlist}

\begin{cvlist}{Awards and honors}

\item[1997]
    Graduated with first rank in the University of Pune, Pune, India,
    among a graduating class of approximately 800. Awarded Dr.\@ D.\@
    Y.\@ Patil Gold Medal by the University of Pune.

\item[1991]
    Among 750 students selected all over India for award of the
    National Talent Search (NTS) Scholarship by the National Council
    for Educational Research and Training (NCERT), New Delhi, after a
    three-tier selection process.

\end{cvlist}

\begin{cvlist}{Theses}

\item[M.S.\@ thesis]

\textbf{Optimum Binary Search Trees on the Hierarchical Memory Model}

Modern computer architectures are modeled realistically as having
several different types of memory organized in a memory hierarchy from
fast but small caches to slow but large external storage. On the
abstract model with memory organized in a hierarchy with an arbitrary
number of levels and with arbitrary access costs at each level, we
develop algorithms for computing a minimum-cost binary search tree
(BST); these algorithms are efficient under constraints on the
parameters of the memory hierarchy. A BST is a widely used data
structure for indexing and retrieving data that is very well-studied
on the uniform memory (RAM) model of computation.

\end{cvlist}

\begin{cvlist}{Research experience}

\item[1999--]
    \textit{Research Assistant}\\
    Department of Computer Science, University of Illinois at Urbana-Champaign

    Research in computational geometry, mesh generation for finite
    element methods;
    Prof.\@ Jeff Erickson, Advisor.

\item[2001]
    M.S.\@ Thesis with Prof.\@ Michael Loui, Advisor.

\item[2001, 2002]
    \textit{Graduate Research Assistant}\\
    D-2, Los Alamos National Laboratory, Los Alamos, NM\\
    Dr.\@ Madhav Marathe, mentor

    Designed and implemented a computer program for the simulation of
    a deregulated electrical power market. The project includes being
    able to simulate power demand and consumption, market forces,
    external regulatory policies, and physical constraints of the
    power grid. The project is ongoing and program code is currently
    being written and tested.

    Research on theoretical and practical problems associated with
    wireless radio networks. This work intersects the areas of graph
    theory, computational geometry, and networking. The results of
    this research are being submitted for publication.

\item[1998]
    \textit{Graduate Research Assistant}\\
    CIC-3, Los Alamos National Laboratory, Los Alamos, NM\\
    Dr.\@ Madhav Marathe, mentor

    Extended blocking algorithms for matrices to improve
    memory-efficiency of matrix operations. Implemented and simulated
    the same on an SGI Origin multiprocessor system, and empirically
    demonstrated a significant improvement in performance. Implemented
    a blocking algorithm for graphs, as proposed by Awerbuch \textsl{et al.}

\item[1996--1997]
    \textit{Student Intern}\\
    Centre for Development of Advanced Computing (C-DAC), Pune, India

    Constructed a model in VHDL for a Fast Ethernet switch and
    performed system-level simulation and analysis using VHDL tools.

\end{cvlist}

\begin{cvlist}{Publications}

\item[2005]

    \textbf{A Unified Algorithm for Adaptive Spacetime Meshing with
    Nonlocal Cone Constraints.}
    Shripad Thite.
    In \textit{Proc.\@ 21st European Workshop on Computational
    Geometry (EWCG)}, pages 1--4, March 9--11, 2005 (Eindhoven, the
    Netherlands).
    Invited and submitted to special issue of \textit{Computational
    Geometry: Theory and Applications (CGTA)}; journal version in
    review.

\item[2004]

    \textbf{Efficient Spacetime Meshing with Nonlocal Cone Constraints.}
    Shripad Thite.
    In \textit{Proc.\@ 13th International Meshing Roundtable (IMR)},
    pages 47--58, September 19--22, 2004 (Williamsburg, Virginia).
    Invited to special issue of \textit{International Journal of
    Computational Geometry and Applications (IJCGA)}; journal special
    issue cancelled.

    \textbf{Spacetime Meshing with Adaptive Refinement and Coarsening.}
    Reza Abedi, Shuo-Heng Chung, Jeff Erickson, Yong Fan,
    Michael Garland, Damrong Guoy, Robert Haber, John M.\@ Sullivan, 
    Shripad Thite, and Yuan Zhou.
    In \textit{Proc.\@ ACM Symp.\@ on Computational Geometry (SoCG)},
    pages 300--309, June 8--11, 2004 (New York, NY);
    (33\% acceptance rate: SoCG is considered the premier
    computational geometry conference.)

    \textbf{The Distance-2 Matching Problem and its Relationship to
    the MAC-layer Capacity of Ad hoc Wireless Networks.}
    Hari Balakrishnan, Christopher L.\@ Barrett, V.\@ S.\@ Anil Kumar, 
    Madhav V.\@ Marathe, and Shripad Thite.
    IEEE Journal on Selected Areas in Communications issue on
    Fundamental Performance Limits of Wireless Sensor Networks, Volume
    22, Number 6, pages 1069--1079, August 2004.

    \textbf{Marketecture: A Simulation-Based Framework for Studying
    Experimental Deregulated Power Markets.}
    Karla Atkins, Chris Barrett, Christopher M.\@ Homan, Achla Marathe,
    Madhav Marathe, and Shripad Thite.
    \textit{6th IAEE European Energy Conference}, September 2--3, 2004
    (Zurich, Switzerland).
    (Journal version submitted to Computational Economics)

\item[2003]

    \textbf{Capturing a Convex Object with Three Discs.}
    Jeff Erickson, Shripad Thite, Fred Rothganger, and Jean Ponce.
    In \textit{Proc.\@ IEEE International Conference on Robotics and
    Automation (ICRA)}, pages 2242--2247, September 14--19, 2003
    (Taipei, Taiwan);
    (50--60\% acceptance rate: nevertheless, ICRA is
    considered the premier robotics conference.)

\item[2000]

    \textbf{Optimum Binary Search Trees on the Hierarchical Memory Model.}
    Shripad Thite.
    M.S.\@ thesis with Prof.\@ Michael Loui, Advisor.
    Department of Computer Science, University of Illinois at Urbana-Champaign,
    CSL Technical Report UILU-ENG-00-2215 ACT-142, November 2000.

\textsc{Abstracts and Manuscripts:}

\item[2005]

    \textbf{Adaptive Spacetime Meshing in 2D$\times$Time for Nonlinear
    and Anisotropic Media.}
    Shripad Thite, Jayandran Palaniappan, Jeff Erickson, and
    Robert Haber.
    At the
    \textit{8th US National Congress on Computational Mechanics},
    July 2005.

    \textbf{Meshing in 2D$\times$Time for Front-Tracking DG Methods.} 
    Shripad Thite, Jeff Erickson, Shuo-Heng Chung,
    Reza Abedi, Jayandran Palaniappan, and Robert Haber.
    At the
    \textit{8th US National Congress on Computational Mechanics},
    July 2005.

\item[2004]

    \textbf{Efficient Algorithms for Channel Assignment in Wireless
    Radio Networks.}
    Christopher L.\@ Barrett, Gabriel Istrate, V.\@ S.\@ Anil Kumar,
    Madhav V.\@ Marathe, Shripad Thite, and Sunil Thulasidasan.
    Manuscript, journal version in preparation as of October 2004.

\item[2003]

    \textbf{Efficient Spacetime Meshing with Nonlocal Cone Constraints.}
    Jeff Erickson, Robert Haber, Jayandran Palaniappan, John Sullivan,
    and Shripad Thite.
    \textit{4th Symposium on Trends in Unstructured Mesh Generation}
    at the \textit{7th US National Congress on Computational
    Mechanics}, July 27--31, 2003 (Albuquerque, NM).

    \textbf{Spacetime Meshing with Adaptive Coarsening and Refinement.}
    Reza Abedi, Shuo-Heng Chung, Jeff Erickson, Yong Fan,
    Robert Haber, John Sullivan, and Shripad Thite. 
    \textit{4th Symposium on Trends in Unstructured Mesh Generation}
    at the \textit{7th US National Congress on Computational
    Mechanics}, July 27--31, 2003 (Albuquerque, NM).

    \textbf{An Efficient Parallel Implementation of the Spacetime
    Discontinuous Galerkin Method Using Charm++.}
    L.\@ V.\@ Kale, Robert Haber, Jonathan Booth, Shripad Thite,
    and Jayandran Palaniappan.
    \textit{4th Symposium on Trends in Unstructured Mesh Generation}
    at the \textit{7th US National Congress on Computational
    Mechanics}, July 27--31, 2003 (Albuquerque, NM).

\item[2001]

    \textbf{Link Scheduling Problems in Packet Radio Networks.}
    Chris Barrett, Gabriel Istrate, Madhav Marathe, Shripad Thite, 
    and V.\@ S.\@ Anil Kumar.
    Manuscript, 2001.

    Other technical reports, manuscripts; available at
    \url{http://www.uiuc.edu/~thite/pubs/} and on request. 

\end{cvlist}

\begin{cvlist}{Conference presentations}

\item[2005]
    \textbf{Provably Good Adaptive Meshing of Spacetime.}
    50th biannual Midwest Theory Day,
    University of Illinois at Urbana-Champaign, Urbana, IL, May 2005.

    \textbf{A Unified Algorithm for Adaptive Spacetime Meshing with
    Nonlocal Cone Constraints.}
    21st European Workshop on Computational Geometry (EWCG),
    Eindhoven, the Netherlands, March 2005.

\item[2004]
    \textbf{Efficient Spacetime Meshing with Nonlocal Cone
    Constraints.}  13th International Meshing Roundtable (IMR),
    Williamsburg, VA, September 2004.

    \textbf{Spacetime Meshing with Adaptive Refinement and Coarsening.}
    ACM Symp.\@ on Computational Geometry (SoCG), Brooklyn, NY, June 2004.

\item[2003]
    \textbf{Efficient Spacetime Meshing with Nonlocal Cone
    Constraints.}  7th US National Congress on Computational
    Mechanics, Albuquerque, NM, July 2003.

    \textbf{Capturing a Convex Object with Three Discs.}
    IEEE International Conference on Robotics and Automation (ICRA),
    Taipei, Taiwan, September 2003.

\end{cvlist}

\begin{cvlist}{Professional service}

\item[Various times]
    Served as \textit{anonymous technical referee} for articles
    submitted to ACM Symposium on Computational Geometry (SoCG), ACM
    Symposium on Theory of Computing (STOC), Discrete and
    Computational Geometry, Journal of Information and Computation,
    Information Processing Letters, International Meshing Roundtable,
    IEEE Journal on Selected Areas in Communications, IEEE
    Communications Letters.

\item[2002--2003]
    Served on the \textit{Fellowships, Assistantships, and Admissions
    (FAA) Committee} of the Computer Science department. Members of
    this committee are faculty and senior graduate students, who
    evaluate entrance requirements and recommend applications for
    admissions, fellowships, and assistantships.

\item[2001]
    Organized the \textit{44th biannual Midwest Theory Day},
    University of Illinois at Urbana-Champaign, December 1,
    2001. (\url{http://www.uiuc.edu/~thite/mtd})

\end{cvlist}

\begin{cvlist}{Teaching experience}

\item[1997--1999]
    Teaching Assistant\\
    Department of Computer Science,
    University of Illinois at Urbana-Champaign

    \textit{Summer 1999}\\
    CS 273: Introduction to the Theory of Computation\\
    Duties included conducting office hours, discussion sessions, and
    helping design and grade homework assignments and exams.

    \textit{Spring 1999}\\
    CS 373: Combinatorial Algorithms

    \textit{Fall 1998:}\\
    CS 375: Automata, Formal Languages, and Computational Complexity; and\\
    CS 373: Combinatorial Algorithms

    \textit{Spring 1998:}\\
    CS 110: C++ programming laboratory.
    Duties included giving lectures, assigning and grading small- and
    medium-scale programming projects, assigning grades.

    \textit{Fall 1997:}\\
    CS 110: C programming laboratory.
    Duties included giving lectures, assigning and grading small- and
    medium-scale programming projects, assigning grades.

\end{cvlist}

\end{cv}

\end{document}